\documentclass[A4,12pt,twoside]{book}
\usepackage{amd}

\graphicspath{{Images/}{../Images/}} 
\setkeys{Gin}{width=0.85\textwidth} 
	\usepackage{xcolor}
	
	\definecolor{defscol}{HTML}{ecd8d7} 
	\definecolor{asumscol}{HTML}{ecd8d7} 
	
	\definecolor{rmkscol}{HTML}{313160} 
	\definecolor{exmscol}{HTML}{e04b52} 
	
	\definecolor{lemscol}{HTML}{2c3943} 
	\definecolor{thmscol}{HTML}{595765} 
	\definecolor{prpscol}{HTML}{9c98b1} 
	\definecolor{corscol}{HTML}{dfd9fd} 
	
	\definecolor{clmscol}{HTML}{165c58} 
	\definecolor{facscol}{HTML}{28a8a1} 

	\newtcbtheorem[number within=section]{mydefinition}{Definition}
	{
		enhanced,
		frame hidden,
		titlerule=0mm,
		toptitle=1mm,
		bottomtitle=1mm,
		fonttitle=\bfseries\large,
		coltitle=black,
		colbacktitle=defscol!40!white,
		colback=defscol!20!white,
	}{defn}

	\newtcbtheorem[use counter from=mydefinition]{myassumption}{Assumption}
	{
		enhanced,
		frame hidden,
		titlerule=0mm,
		toptitle=1mm,
		bottomtitle=1mm,
		fonttitle=\bfseries\large,
		coltitle=black,
		colbacktitle=asumscol!40!white,
		colback=asumscol!20!white,
	}{asum}

	\newtcbtheorem[use counter from=mydefinition]{mytheorem}{Theorem}
	{
		enhanced,
		frame hidden,
		titlerule=0mm,
		toptitle=1mm,
		bottomtitle=1mm,
		fonttitle=\bfseries\large,
		coltitle=black,
		colbacktitle=thmscol!40!white,
		colback=thmscol!20!white,
	}{thm}

	\newenvironment{thmpf}{
		{\noindent{\it \textbf{Proof for Theorem.}}}
		\tcolorbox[blanker,breakable,left=5mm,parbox=false,
		before upper={\parindent15pt},
		after skip=10pt,
		borderline west={1mm}{0pt}{thmscol!40!white}]
	}{
		\textcolor{thmscol!40!white}{\hbox{}\nobreak\hfill$\blacksquare$} 
		\endtcolorbox
	}

	\newtcbtheorem[use counter from=mydefinition]{mylemma}{Lemma}
	{
		enhanced,
		frame hidden,
		titlerule=0mm,
		toptitle=1mm,
		bottomtitle=1mm,
		fonttitle=\bfseries\large,
		coltitle=black,
		colbacktitle=lemscol!40!white,
		colback=lemscol!20!white,
	}{lem}

	\newenvironment{lempf}{
		{\noindent{\it \textbf{Proof for Lemma}}}
		\tcolorbox[blanker,breakable,left=5mm,parbox=false,
		before upper={\parindent15pt},
		after skip=10pt,
		borderline west={1mm}{0pt}{lemscol!40!white}]
	}{
		\textcolor{lemscol!40!white}{\hbox{}\nobreak\hfill$\blacksquare$} 
		\endtcolorbox
	}

	\newtcbtheorem[use counter from=mydefinition]{mycorollary}{Corollary}
	{
		enhanced,
		frame hidden,
		titlerule=0mm,
		toptitle=1mm,
		bottomtitle=1mm,
		fonttitle=\bfseries\large,
		coltitle=black,
		colbacktitle=corscol!40!white,
		colback=corscol!20!white,
	}{cor}

	\newenvironment{corpf}{
		{\noindent{\it \textbf{Proof for Corollary.}}}
		\tcolorbox[blanker,breakable,left=5mm,parbox=false,
		before upper={\parindent15pt},
		after skip=10pt,
		borderline west={1mm}{0pt}{corscol!40!white}]
	}{
		\textcolor{corscol!40!white}{\hbox{}\nobreak\hfill$\blacksquare$} 
		\endtcolorbox
	}

	\newtcbtheorem[use counter from=mydefinition]{myproposition}{Proposition}
	{
		enhanced,
		frame hidden,
		titlerule=0mm,
		toptitle=1mm,
		bottomtitle=1mm,
		fonttitle=\bfseries\large,
		coltitle=black,
		colbacktitle=prpscol!30!white,
		colback=prpscol!20!white,
	}{prop}

	\newenvironment{proppf}{
		{\noindent{\it \textbf{Proof for Proposition.}}}
		\tcolorbox[blanker,breakable,left=5mm,parbox=false,
		before upper={\parindent15pt},
		after skip=10pt,
		borderline west={1mm}{0pt}{prpscol!40!white}]
	}{
		\textcolor{prpscol!40!white}{\hbox{}\nobreak\hfill$\blacksquare$} 
		\endtcolorbox
	}

	\newtcbtheorem[use counter from=mydefinition]{myclaim}{Claim}
	{
		enhanced,
		frame hidden,
		titlerule=0mm,
		toptitle=1mm,
		bottomtitle=1mm,
		fonttitle=\bfseries\large,
		coltitle=black,
		colbacktitle=clmscol!40!white,
		colback=clmscol!20!white,
	}{clm}

	\newenvironment{clmpf}{
		{\noindent{\it \textbf{Proof for Claim.}}}
		\tcolorbox[blanker,breakable,left=5mm,parbox=false,
		before upper={\parindent15pt},
		after skip=10pt,
		borderline west={1mm}{0pt}{clmscol!40!white}]
	}{
		\textcolor{clmscol!40!white}{\hbox{}\nobreak\hfill$\blacksquare$} 
		\endtcolorbox
	}

	\newtcbtheorem[use counter from=mydefinition]{myfact}{Fact}
	{
		enhanced,
		frame hidden,
		titlerule=0mm,
		toptitle=1mm,
		bottomtitle=1mm,
		fonttitle=\bfseries\large,
		coltitle=black,
		colbacktitle=facscol!40!white,
		colback=facscol!20!white,
	}{fact}

	\newenvironment{myexample}{
		\tcolorbox[blanker,breakable,left=5mm,parbox=false,
		before upper={\parindent15pt},
		after skip=10pt,
		borderline west={1mm}{0pt}{clmscol!40!white}]
	}{
		\textcolor{clmscol!40!white}{\hbox{}\nobreak\hfill$\blacksquare$} 
		\endtcolorbox
	}

	\newenvironment{remark}{
		\par
		\vspace{5pt}
		\begin{minipage}{\textwidth}
			{\par\noindent{\textbf{Remark.}}}
			\tcolorbox[blanker,breakable,left=5mm,
			before skip=10pt,after skip=10pt,
			borderline west={1mm}{0pt}{rmkscol!20!white}]
		}{
			\endtcolorbox
		\end{minipage}
		\vspace{5pt}
	}

\usepackage[english]{babel} 
\usepackage{amssymb}
\usepackage{algorithm}
\usepackage{algpseudocode}
\usepackage{float}
\usepackage{afterpage}
\usepackage{etoolbox}

\usepackage{hyperref}

\setlist[itemize]{itemsep=5pt} 
\setlist[enumerate]{itemsep=5pt} 

\newcommand{\BigTitle}{A Practical Guide to Budget Pacing Algorithms in Digital Advertising}

\newcommand{\LittleTitle}{
	For Engineers
}
\title{A Pratical Guide to Budget Pacing Algorithms in Digital Advertising}

\begin{document}
	\newgeometry{top=8cm,bottom=.5in,left=2cm,right=2cm}
	
	\begin{titlepage}
		\begin{center}
			
			\textbf{\fontfamily{qhv}\selectfont\Huge \BigTitle}
			
			\par\noindent\rule{\textwidth}{4pt}\\
			
			\begin{tikzpicture}
				\shade[bottom color=lightgray,top color=white]
				(0,0) rectangle (\textwidth, 1.5)
				node[midway] {\textbf{\large \textit{\LittleTitle}}}; 
			\end{tikzpicture}
			
			
			\vspace{2cm} 
			\textbf{\Large \textit{Yuanlong Chen} }

			\vspace{\fill}
			
			\end{center}
			
	\end{titlepage}
	
	\restoregeometry

	\restoregeometry
	
	\chapter*{Preface}
	A typical real-time ad-serving funnel comprises ad targeting, conversion modeling (e.g., click-through rate prediction), budget pacing (bidding), and auction processes. While there is a wealth of research and articles on ad targeting and conversion modeling, budget pacing—a crucial component—lacks a systematic treatment specifically tailored for engineers in existing literature. This book aims to provide engineers with a practical yet relatively comprehensive introduction to budget pacing algorithms within the digital advertising domain. The book is structured as follows:
	
	In \autoref{part:some_basics}, we introduce foundational concepts in the digital advertising business, along with preliminary knowledge essential for understanding the subsequent chapters. We begin with a brief introduction to the history of digital advertising. Next, we cover some basics of programmatic ads, including the concepts of CPM, CPV, CPC, CPL and CPA ads. The entire ad-serving funnel is then briefly discussed to illustrate how ads are served in real time. The pipeline presented focuses on first-party ads (e.g., ads on Instagram or LinkedIn), though the serving pipeline for DSPs is similar. Additionally, we address basic optimization techniques, auction mechanism design, and other related preliminary topics that will be referenced throughout the book. Readers already familiar with these subjects may choose to skip this section and proceed directly to \autoref{part:pacing_algorithms}.

	 In \autoref{part:pacing_algorithms}, we discuss various pacing methods under standard second price auction. Two main bidding products, max delivery and cost cap, are used as examples to demonstrate the concepts of these pacing methods. Nevertheless, the underlying principles introduced here are applicable to other problems as well. We first provide a rigorous mathematical formulation of both the max delivery and cost cap problems. In the subsequent sections, we discuss various pacing algorithms commonly adopted in the industry, including throttling, PID controllers, MPC controllers and online adaptive optimal control. For each approach, we explain the motivation, introduce the basic background, and describe how it can be applied to bidding/pacing problems such as max delivery and cost cap. Additionally, we discuss the pros and cons of each approach, enabling readers to select the most suitable method for real-world applications based on their specific business needs. For some algorithms, pseudo-code and simple implementations are also provided to give readers a practical understanding of how to implement them in their daily work.

	 In \autoref{part:misc}, we demonstrate how the pacing frameworks introduced in \autoref{part:pacing_algorithms} can be applied to various other business scenarios. Topics include the initialization of campaign bids, bidding under different auction mechanisms (e.g., first-price auctions, where bid shading is required), bid optimization for multi-constraint problems (e.g., campaigns delivered across different placements or channels such as first-party and third-party platforms, or campaign groups where multiple campaigns share the same budget), deep funnel conversion problems (e.g., bid optimization for post-conversion events such as retention), common brand advertisements with reach and frequency requirements, and the over-delivery problem. Hopefully, these topics cover most of the tasks that a budget pacing engineer might encounter in their daily work.

	 This book is well-suited for engineers working on or interested in budget optimization and bidding algorithms in digital advertising. It is also valuable for engineers specializing in other aspects of ad serving, such as targeting and ranking, by providing insights into how downstream services in the serving funnel operate.While a basic understanding of mathematical optimization and control theory can be beneficial, it is not a prerequisite for reading this book. We also point out that the methodology discussed in this book primarily focuses on traditional control theory. For alternative approaches, such as those based on General Artificial Intelligence (GAI) methods (e.g., diffusion model-based bidding strategies), readers are encouraged to refer to the relevant research papers.

	  Budget optimization in digital advertising is a broad and complex topic. This little book primarily aims to provide engineers in the field with a comprehensive overview of the landscape of budget pacing algorithms. It does not attempt to cover every detail of budget pacing. For better readability, we omit some theoretical aspects, such as regret analysis and equilibrium analysis. Readers interested in these topics are encouraged to refer to the academic papers mentioned throughout the book for more in-depth information.
	 
	 \begin{flushright}
	 	\textbf{\textit{Y. Chen}} \\
	 	\textit{Berkeley, California}
	 \end{flushright}

	\clearpage
	\tableofcontents
	\clearpage
	
	\listofalgorithms
	\listoffigures
	\listoftables

	\chapter*{Disclaimer}
	
	This book is intended solely for \textbf{educational, academic, and informational purposes}. 
	The algorithms, methods, and technologies discussed herein, including those covered by active patents, are presented to provide a theoretical understanding and promote scholarly communication.
	
	All content in this book has been derived from publicly available sources, including research papers, articles, and patents. The inclusion of these materials is for the purpose of \textbf{summarization, comparison, and educational discussion}. No proprietary or confidential information has been used in the creation of this book.
	
	This book does not provide any authorization to reproduce, implement, or commercialize patented inventions or other proprietary technologies. Readers are advised that the use or implementation of patented algorithms or inventions may require prior permission from the respective patent holders.
	
	The author(s) do not claim any ownership of third-party intellectual property discussed in this book. Every effort has been made to attribute work appropriately and respect intellectual property rights. If there are any inaccuracies or concerns regarding attribution, please contact the author(s) for correction.

	\part{Preliminaries} 
	\label{part:some_basics}
	
	\chapter{Basics of Digital Advertising}
	
	\intro{
		In this chapter, we discuss the fundamentals of digital advertising. We begin with a brief history of digital advertising and how real-time bidding has transformed the landscape of this industry. Next, we explore the ad-serving pipeline, illustrating how ads are delivered in real time. Finally, we introduce two key campaign configurations—objective/optimization goal and charging model—and explain how different bidding products are designed based on these factors.
		
	}
	
	\section{Brief History}
	
	Digital advertising has undergone a remarkable transformation since its inception, evolving from simple banner ads to sophisticated programmatic systems powered by real-time data. At the heart of this evolution is Real-Time Bidding (RTB), an innovation that has revolutionized the way advertisers and publishers interact. RTB operates alongside key players such as Demand-Side Platforms (DSPs), Supply-Side Platforms (SSPs), and major in-house bidding systems, such as Google's and Facebook's ad-serving platforms. In this chapter, we introduce the fundamentals of digital advertising to help readers familiarize themselves with key concepts in this domain.

	\paragraph{The Early Days of Digital Advertising}  
	The first era of digital advertising began in the mid-1990s with the advent of the internet. Banner ads, such as the iconic AT\&T ad on HotWired in 1994, marked the beginning of online monetization. During this period, advertisers purchased ad space directly from publishers, with limited data available to inform decisions. 
	
	As internet adoption grew, ad networks emerged to connect advertisers with publishers more efficiently. These networks aggregated inventory but lacked sophisticated targeting capabilities, leading to inefficiencies and limited personalization.
	
	\paragraph{The Rise of Programmatic Advertising}  
	The introduction of programmatic advertising in the early 2000s addressed many of the shortcomings of traditional models. Automated systems replaced manual negotiations, enabling advertisers to target audiences based on demographic, geographic, and behavioral data. This innovation paved the way for the creation of DSPs and SSPs.
	
	\begin{itemize}
		\item \textbf{Demand-Side Platforms (DSPs):} DSPs provide advertisers with a centralized platform to manage and optimize ad campaigns across multiple channels. By leveraging advanced algorithms and real-time data, DSPs empower advertisers to bid on impressions that align with their target audience and campaign objectives.
		\item \textbf{Supply-Side Platforms (SSPs):} On the publisher side, SSPs enable the efficient management of ad inventory. SSPs connect publishers to multiple ad exchanges and DSPs, ensuring maximum revenue through competitive bidding. Together, DSPs and SSPs form the backbone of the programmatic advertising ecosystem.
	\end{itemize}
	
	\paragraph{The Emergence of Real-Time Bidding (RTB)}  
	RTB emerged in 2009 as a game-changer in programmatic advertising. Unlike earlier methods that involved bulk purchasing of ad space, RTB allows advertisers to bid on individual impressions in real time. Key milestones in RTB history include:
	
	\begin{itemize}
		\item \textbf{2009:} Google launched DoubleClick Ad Exchange, introducing the first large-scale RTB platform.
		\item \textbf{2011:} Facebook introduced Facebook Exchange (FBX), extending RTB capabilities to social media advertising.
		\item \textbf{2013:} Mobile RTB gained prominence, reflecting the rapid growth of mobile internet usage.
		\item \textbf{2015:} Header bidding strategies allowed publishers to maximize revenue by offering inventory to multiple exchanges simultaneously.
	\end{itemize}
	
	\paragraph{In-House Real-Time Bidding Systems}  
	Today, many large internet companies, such as Google, Meta(Facebook), Amazon, and LinkedIn, have developed their own in-house ad-serving systems to leverage unique data resources and monetize their vast user traffic. Advertisers can set up campaigns directly through these companies' ad management tools, and their ads are served to users across the companies' apps and websites.
	
	In this scenario, these companies effectively function as both a DSP (Demand-Side Platform) and an SSP (Supply-Side Platform) simultaneously. When a user engages with a platform, ad campaigns can be displayed in various placements, such as news feeds or search result pages. By leveraging extensive in-house user data, these platforms can optimize ad delivery more efficiently, improving targeting accuracy and maximizing engagement.
	
	\begin{figure}[H]
		\centering
		\begin{minipage}{0.48\textwidth}
			\centering
			\includegraphics[width=\linewidth]{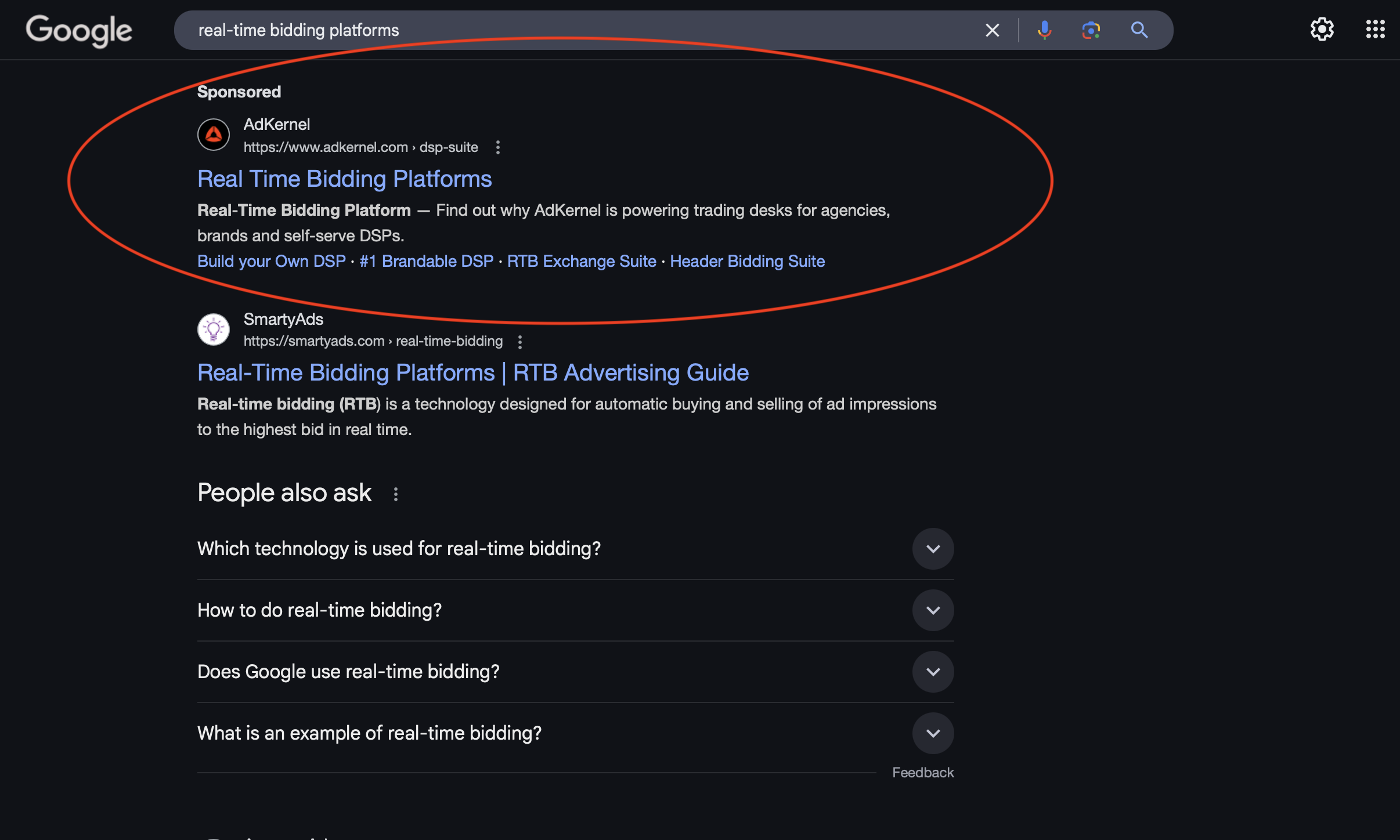}
			\caption{Google Search Ads}
		\end{minipage}
		\hfill
		\begin{minipage}{0.48\textwidth}
			\centering
			\includegraphics[width=\linewidth]{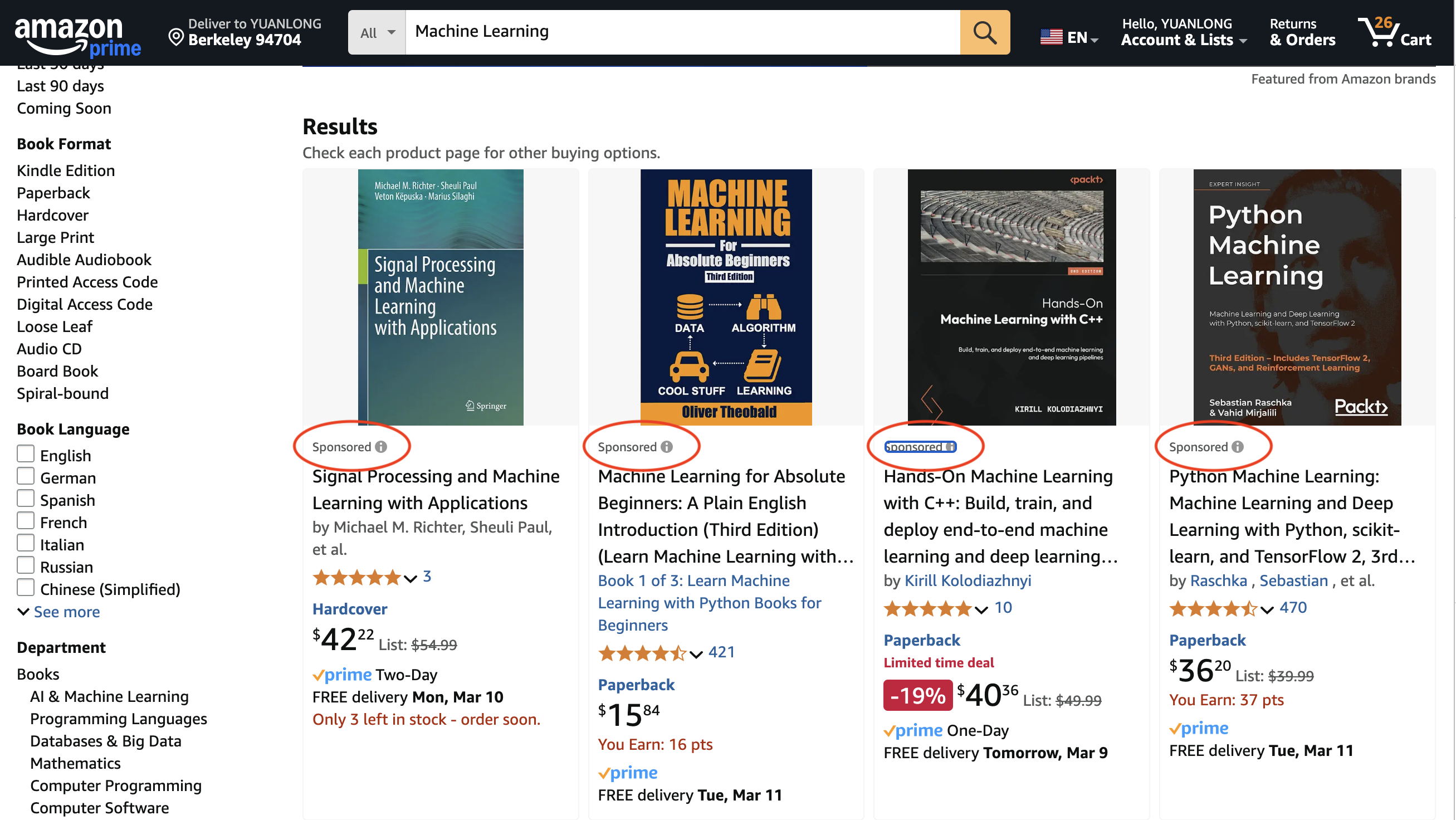}
			\caption{Amazon Ads}
		\end{minipage}
	\end{figure}
	
	The two snapshots above show typical search ads displayed when users perform searches on Google and Amazon. When a user searches for "real-time bidding platform" on Google, relevant ad campaigns related to the search keywords appear at the top of the results page. Similarly, on Amazon, when "Machine Learning" is searched, several ads for books titled "Machine Learning" are displayed among the top search results.
	
	These keywords indicate strong user intent, and aligning ads with such search queries can significantly enhance campaign efficiency. This targeting strategy increases the likelihood of users clicking on an RTB platform’s website or purchasing machine learning textbooks. Notably, in both cases, these ads are labeled as "Sponsored," signifying that they are paid advertisements.

	\begin{figure}[H]
		\centering
		\begin{minipage}{0.48\textwidth}
			\centering
			\includegraphics[width=\linewidth]{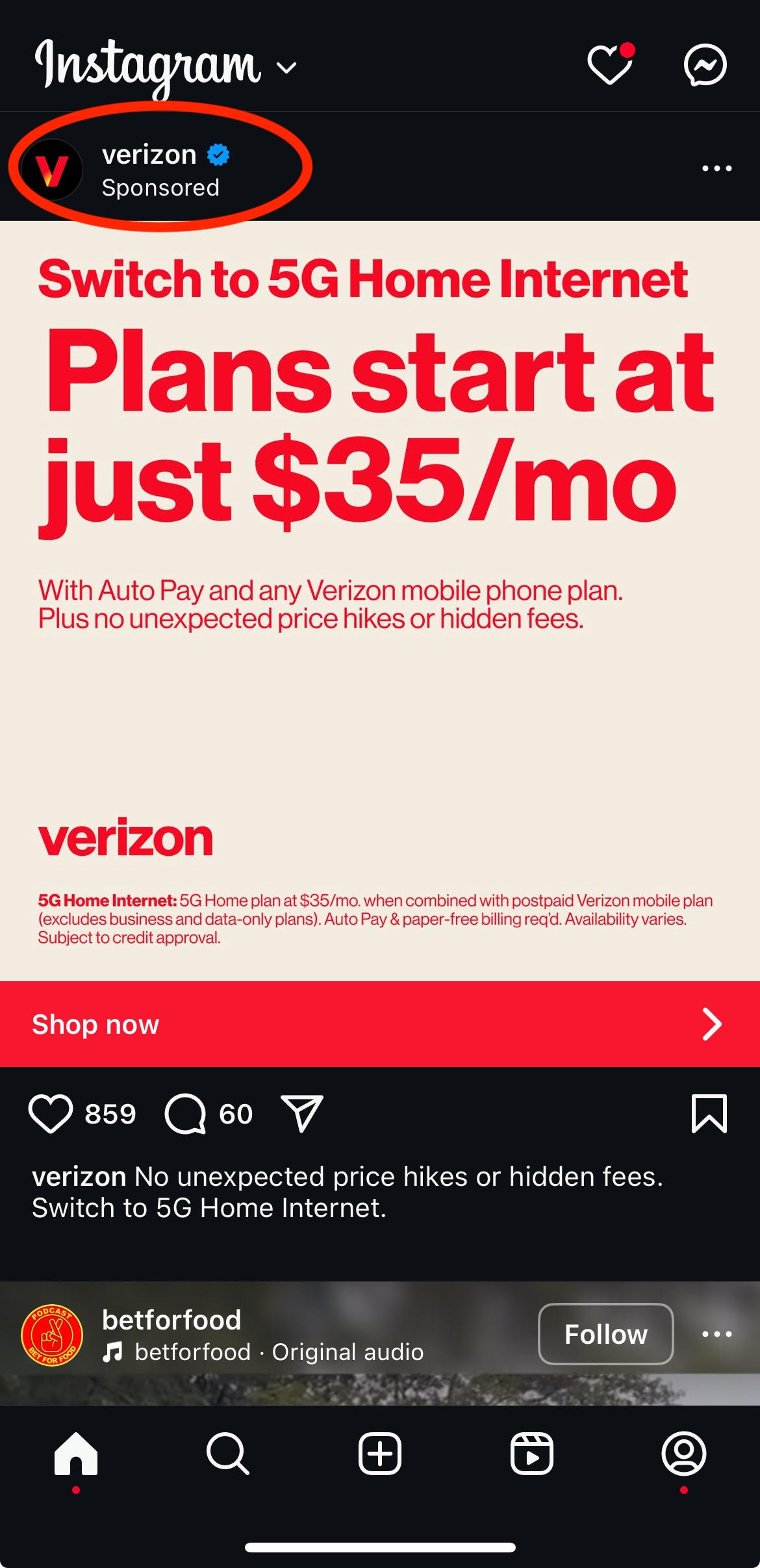}
			\caption{Instagram Ads}
		\end{minipage}
		\hfill
		\begin{minipage}{0.48\textwidth}
			\centering
			\includegraphics[width=\linewidth]{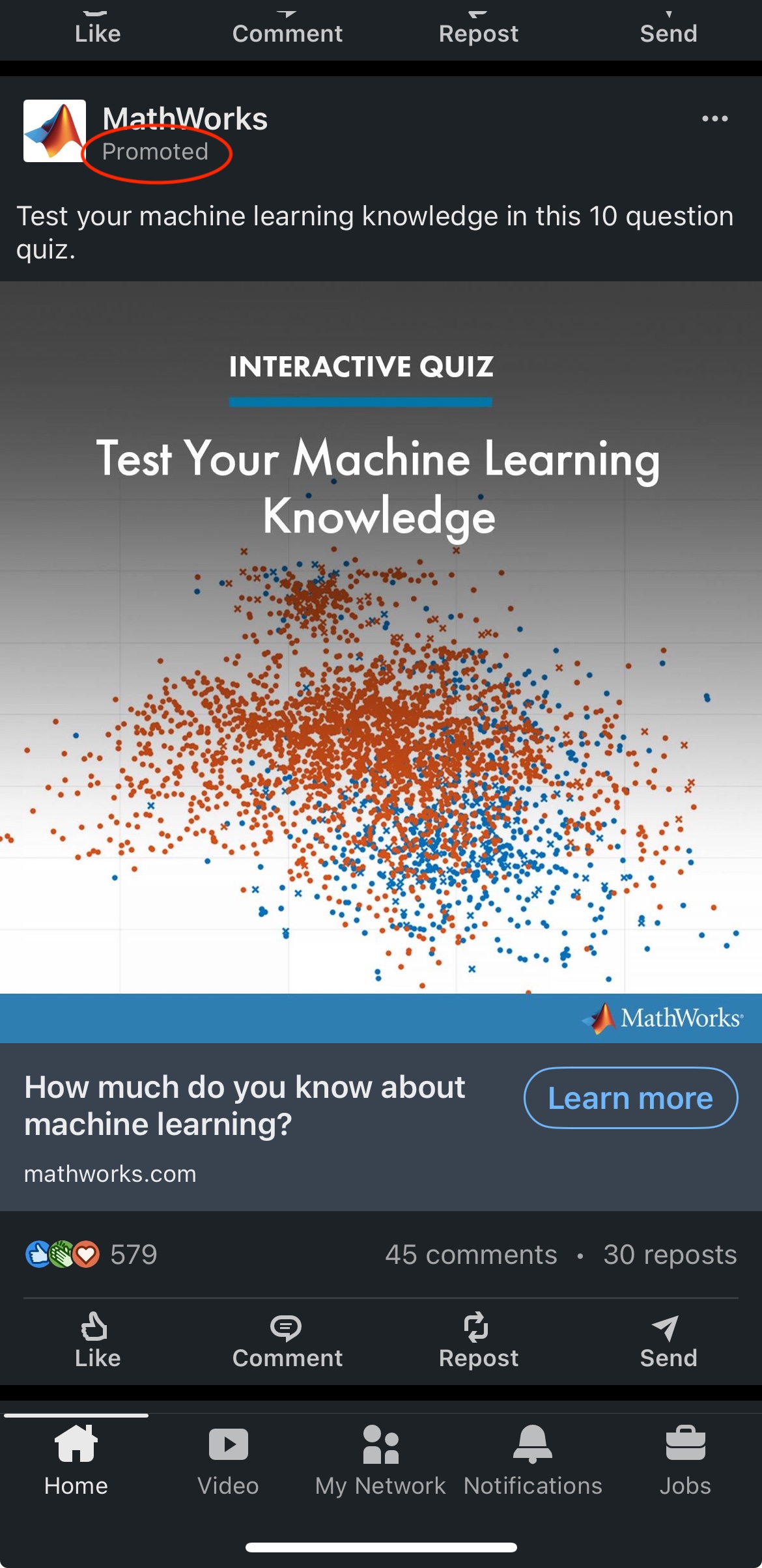}
			\caption{LinkedIn Ads}
		\end{minipage}
	\end{figure}
	
	These two snapshots illustrate a different type of ad placement. The left image shows an advertisement from Meta’s \textit{Instagram}, while the right image displays an ad from \textit{LinkedIn}. Unlike search ads, which are triggered by user search queries, these ads are generated based on a user's profile and past behavior on the platform. They are automatically inserted between organic content in the feed.
	
	For example, in LinkedIn’s ad recommendation, an advertisement from \textit{MathWorks} appears with the message: \textit{"How much do you know about machine learning?"} This ad was likely shown because the user had recently searched for topics related to machine learning, and LinkedIn’s ad delivery algorithm detected this intent.

	\section{Ads Serving Pipeline}
	
	We use LinkedIn's in-house ad delivery system as an example to illustrate how an ad campaign is served to users. Suppose an advertiser creates a campaign \( C \) through the ad management tool to maximize landing page clicks by targeting "Machine Learning Engineers." The campaign is set with a fixed budget (e.g., \$1000) and a defined start and end date. Once created, this campaign is added to LinkedIn's internal ad inventory within the ad-serving system.
	
	Whenever a user logs into LinkedIn (via mobile or desktop), an ad request \( u \) is sent to the ad-serving system to allocate available ad slots in the user's feed. The serving algorithm is then triggered to check whether the user belongs to the target audience of campaign \( C \). If the user qualifies (e.g., their profile contains the job title "Machine Learning Engineer"), the campaign is retrieved along with other eligible campaigns to participate an auction to compete for the available ad slots. The auction process for these slots depends primarily on two factors:
	
	\begin{itemize}
		\item \textbf{Ad Quality}: This is typically measured by the relevance between the campaign and the target user. For example, if a campaign is optimized for clicks, the quality score can be defined as the click-through rate (CTR)—the probability that the user will click on the ad. This score is usually predicted by a machine learning model, such as a deep neural network.
		
		\item \textbf{Bid Level}: The bid level is determined by factors such as the remaining budget, size of the target audience, and cost constraints set by the advertiser. For example, an advertiser may specify that the cost per click (CPC) must not exceed \$2. The bidding algorithm takes this constraint, along with other factors, into account to determine an optimal bid for the ad request. (This bidding process is a central topic discussed in this book.)
	\end{itemize}
	
	The auction algorithm then computes a ranking score for campaign \( C \) (along with other competing campaigns) based on both ad quality and bid level. One of the most widely used ranking scores is effective cost-per-mille (eCPM), which represents the cost per one thousand impressions. It is computed as:
		\[
		\text{eCPM} = (\text{Bid per Click}) \times \text{CTR} \times 1000.
		\]
	Since the advertiser is bidding per click, multiplying by CTR converts the expected cost into an impression-based metric.
		
	The ad slots are then assigned to the campaigns with the highest ranking scores and displayed to the user accordingly. The advertiser is charged based on the auction's pricing/charging model—either per impression (CPM), per click (CPC), or per conversion (CPA). We illustrate this procedure in the following diagram:
	
	\begin{figure}[H]
		\centering
		\includegraphics[width=0.95\textwidth]{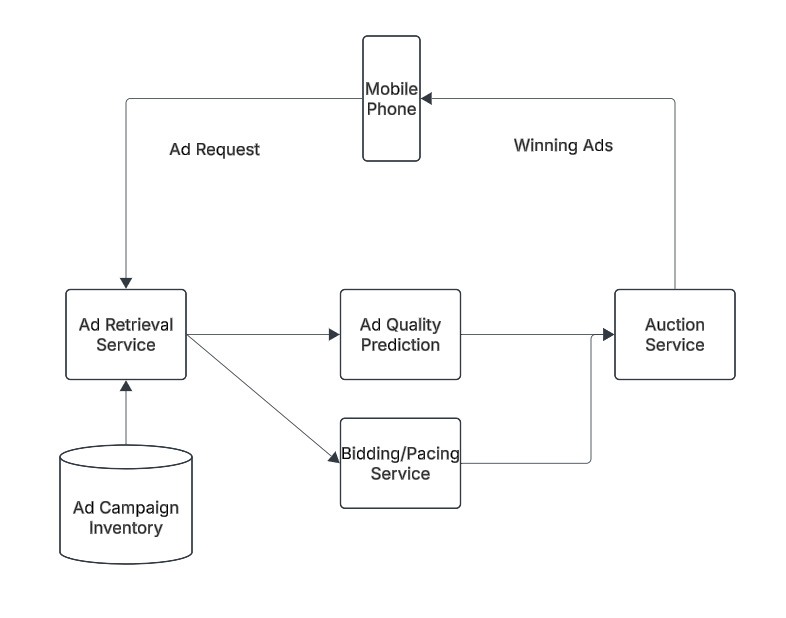}
		\caption{Illustration of the Ad Serving Pipeline}
		\label{fig:ad_serving_pipeline}
	\end{figure}
	
	The serving pipeline described here is an over-simplified version; a real-world production system is significantly more complex. For example, the enormous volume of traffic and the vast number of campaigns make it infeasible to compute ad quality scores for all campaigns simultaneously using computationally intensive machine learning models, such as deep neural networks. To address this challenge, it is common to structure the process as a multi-stage funnel, incorporating multiple stages such as candidate generation, preliminary ranking, fine-grained ranking, and pacing/auction. Each stage involves trade-offs in computational cost, latency, and accuracy to ensure efficient and scalable ad delivery. For more details on ad-serving architectures, one may refer to works such as \cite{covington2016deep}, \cite{liu2017cascade}, or \cite{wang2023empirical}.

	\section{Ad Campaign Configurations} 
	
	There are two important configurations of a campaign that help in designing the bidding and pacing strategy:
	
	\begin{itemize}
		\item \textbf{Objective and Optimization Goal}
		\item \textbf{Charging Model}
	\end{itemize}
	
	\paragraph{Objective and Optimization Goal} 
	
	The objective represents the primary outcome that advertisers seek—such as increasing brand awareness or driving conversions for a campaign. Within that objective, the optimization goal defines the specific metric that the ad platform will maximize. For example, if the objective is “Conversion”, the optimization goal might be “Website Conversion” or “Lead Generation.”
	
	When advertisers create campaigns using the ad management tool, they are required to specify both the objective and the optimization goal. The platform’s ad delivery algorithm then focuses on maximizing those specific actions to improve delivery efficiency. Different advertising platforms may define objectives and optimization goals differently.
	
\begin{figure}[H]
		\centering
		\includegraphics[width=0.95\textwidth]{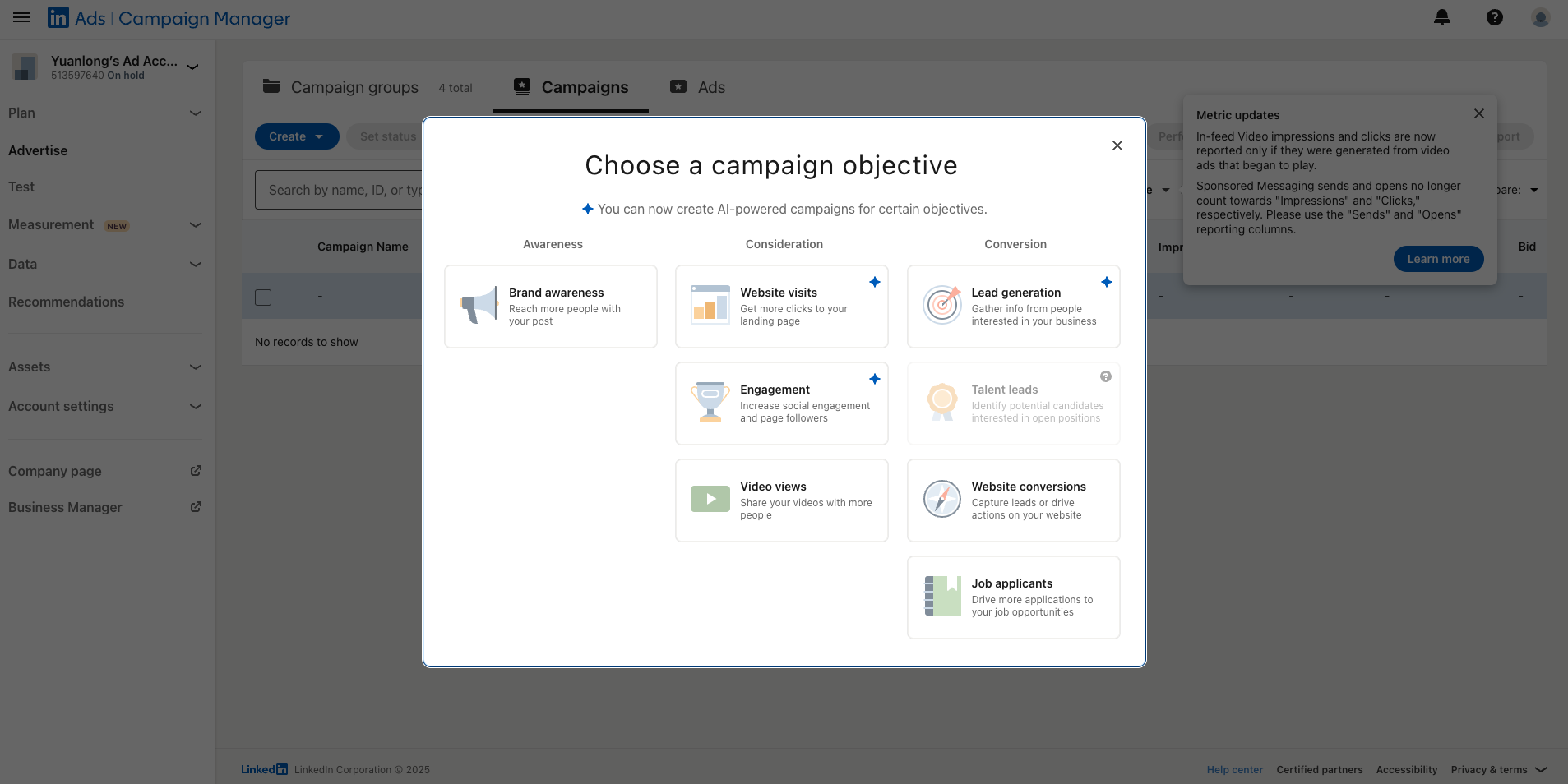}
		\caption{Objectives and Optimization Goals of LinkedIn Ads}
		\label{fig:objective_optimization_goal}
	\end{figure}

	\autoref{fig:objective_optimization_goal} illustrates the objectives and optimization goals available in LinkedIn Ads when advertisers create a new campaign using LinkedIn's Ads Campaign Manager. As shown, there are three main objectives: Awareness, Consideration, and Conversion. Within each objective, advertisers can select from multiple optimization goals.
	
	\paragraph{Charging Model}
	
	The charging model refers to how an advertiser is billed for ad placements and interactions. It determines the basis on which costs are calculated—such as per click, per thousand impressions, or per conversion. 
	
	For example, under a CPC (Cost Per Click) model, advertisers are charged each time a user clicks on the ad. Under a CPM (Cost Per Mille) model, advertisers pay for every 1,000 impressions (views), regardless of how many users engage with the ad. In an oCPM (Optimized Cost Per Mille) model, advertisers still pay for every 1,000 impressions, but the ad delivery is optimized towards achieving specific objectives, such as conversions.
	
	Different platforms may offer various charging models to advertisers. For example, Google explicitly provides pure CPC bidding (especially for Search) and a well-defined CPV (Cost Per View) model for YouTube ads. In contrast, Facebook (Meta) frequently defaults to an impression-based billing system (CPM) while leveraging advanced algorithms (oCPM) to optimize for clicks, conversions, or video views.

	\paragraph{Comparison of Different Bidding Products}
	
	A campaign's objective, optimization goal, and charging model determine how the bidding and pacing optimization strategy is designed. For example, if the optimization goal of a campaign is to maximize total web conversions and the charging model is per impression, the corresponding bidding approach is oCPM. If the optimization goal is conversion and the charging model remains the same, CPA bidding may be used. Both oCPM and CPA bidding models compute an eCPM value to participate in the auction. The eCPM calculation is as follows:
	\[
	eCPM = bid \times CVR \times CTR \times 1000.
	\]
	where \( CVR \) represents the post-click conversion rate. 
	
	If the estimates for CTR and CVR are unbiased, both oCPM and CPA bidding strategies should theoretically yield similar results. However, when prediction models contain noise, oCPM advertisers risk overpaying for conversions due to inaccurate CTR or CVR estimates, leading to higher costs per conversion. In contrast, CPA bidding mitigates this risk by ensuring that advertisers are only charged for successful conversions, making it more resilient to prediction errors. 
	
	We compare some popular bidding products in \autoref{tab:bidding_products} in terms of objective, optimization goal, and charging model.
	
	\begin{table}[h]
		\centering
		\renewcommand{\arraystretch}{2} 
		\resizebox{\textwidth}{!}{ 
			\begin{tabular}{|p{4cm}|p{4.5cm}|p{5cm}|p{4cm}|}
				\hline
				\textbf{Bidding Product} & \textbf{Objective} & \textbf{Optimization Goal} & \textbf{Charging Model} \\ 
				\hline
				\textbf{CPC (Cost Per Click)} & Drive traffic to a website or landing page. & Maximizes clicks (link clicks, website visits). & Pay per click. \\ 
				\hline
				\textbf{ECPC (Enhanced CPC)} & Increase conversions while maintaining control over CPC. & Adjusts bids dynamically to acquire clicks more likely to convert. & Pay per click (bids adjust dynamically). \\ 
				\hline
				\textbf{oCPC (Optimized CPC)} & Drive more conversions by optimizing for higher-value clicks. & Optimizes delivery for clicks likely to result in conversions. & Pay per click (optimized for conversions). \\ 
				\hline
				\textbf{CPM (Cost Per Mille)} & Maximize brand awareness and visibility. & Maximizes ad impressions (reach as many users as possible). & Pay per 1,000 impressions. \\ 
				\hline
				\textbf{oCPM (Optimized CPM)} & Drive conversions with impression-based delivery. & Optimizes impressions for users most likely to take action. & Pay per 1,000 impressions (optimized for conversions). \\ 
				\hline
				\textbf{CPA (Cost Per Action)} & Acquire conversions efficiently. & Maximizes conversions at a predefined cost per action (purchase, sign-up, etc.). & Pay per completed action. \\ 
				\hline
				\textbf{CPV (Cost Per View)} & Maximize video engagement. & Optimizes for video views (e.g., 30-second watch or full view). & Pay per completed view. \\ 
				\hline
				\textbf{CPL (Cost Per Lead \footnote{Lead(Lead Generation)refers to the process of attracting and capturing potential customers' interest in a product or service. The goal is to collect contact information, such as email addresses, phone numbers, or other details, to nurture them into paying customers.})} & Generate high-quality leads. & Maximizes lead form submissions (e.g., email sign-ups, inquiries). & Pay per lead captured. \\ 
				\hline
			\end{tabular}
		} 
		\caption{Comparison of Different Bidding Products}
		\label{tab:bidding_products}
	\end{table}

	
	\chapter{Optimization Basics}
	
	\intro{
		We introduce fundamental optimization techniques here, focusing on two main topics: the primal-dual method and the gradient descent method. Both techniques will be utilized in later chapters to derive the optimal bidding formula and design the online bid update algorithm.
	}
	
	Optimization techniques will be used throughout the remainder of this book. To ensure the book is self-contained, we introduce some fundamental concepts and useful techniques of mathematical optimization in this chapter. The topics discussed here provide the minimal necessary foundation for deriving pacing algorithms in the subsequent chapters. We do not aim to cover all aspects of optimization, as the field is vast and beyond the scope of this book. For a more comprehensive understanding, readers may refer to classical optimization textbooks such as \cite{bertsimas1997introduction}, \cite{boyd2004convex} and \cite{wolsey1999integer}.
	
	In this book, we consider the following optimization problem:
	
	\begin{equation} \label{eq:optimization}
		\begin{aligned}
			\max_{x} \quad & f(x)\\
			\text{s.t.} \quad & g_i(x) \leq 0, \quad  i = 1, \dots, n.
		\end{aligned}
	\end{equation}
	The objective of \eqref{eq:optimization} is to maximize the function \( f(x) \) subject to the constraints \( g_i(x) \leq 0 \). Many real-world problems can be formulated in this way. For example, in a pacing algorithm, one may seek to maximize the total conversions of a campaign while adhering to constraints such as budget limits, cost controls, and other operational restrictions.

	\section{Primal-Dual Method}
	A fundamental approach for solving constrained optimization problems such as \eqref{eq:optimization} is the primal-dual method, which leverages the concept of Lagrange duality. We provide a high-level overview of this method here, as it will be used frequently throughout this book. For a more comprehensive treatment of the primal-dual method and its applications, readers may refer to \cite{diewert1974applications}.
	
	A typical procedure of the primal-dual method is as follows:
	
	\begin{itemize}
		\item \textbf{Construct the Lagrangian Function}: The Lagrangian function \(\mathcal{L}(x, \lambda_i)\)\footnote{In this book, we use \(\mathcal{L}(x, \lambda_i)\) to denote \(\mathcal{L}(x, \lambda_1, \dots, \lambda_i, \dots, \lambda_n)\) when there is no risk of ambiguity.} transforms the constrained optimization problem \eqref{eq:optimization} into an unconstrained optimization problem by incorporating penalty terms via Lagrange multipliers (dual variables) \(\{\lambda_i\}_{i=1}^n\). It is defined as:
		\[
		\mathcal{L}(x, \lambda_i) = f(x) - \sum_{i=1}^{n} \lambda_i \cdot g_i(x),
		\]
		where we require \(\lambda_i \geq 0\). Notably, \(\lambda_i \cdot g_i(x) \geq 0\) when \(g_i(x)\) violates the constraints. Thus, we can interpret \(- \sum_{i=1}^{n} \lambda_i \cdot g_i(x)\) as a penalty term in the maximization of \(f(x)\). The Lagrangian function \(\mathcal{L}(x, \lambda_i)\) effectively captures the trade-off between optimality (the objective function \(f\)) and feasibility (the constraint functions \(g_i\)).
		
		\item \textbf{Derive the Dual Problem}: Based on the Lagrangian \(\mathcal{L}(x, \lambda_i)\) defined above, the dual function of \eqref{eq:optimization} is obtained by taking the supremum of \(\mathcal{L}(x, \lambda_i)\) with respect to \(x\):
		\[
		\mathcal{L}^{*}(\lambda_i) = \sup_{x} \mathcal{L}(x, \lambda_i).
		\]
		The corresponding dual problem is to minimize the dual function \(\mathcal{L}^{*}(\lambda_i)\) with respect to \(\lambda_i\):
		\begin{equation} \label{eq:optimization_dual}
			\min_{\lambda_i \geq 0} \mathcal{L}^{*}(\lambda_i).
		\end{equation}
		
		\item \textbf{Leverage the Primal-Dual Relationship}: Let \(f^*\) denote the optimal value of the primal problem \eqref{eq:optimization}, and let \(q^*\) be the optimal value of the dual problem \eqref{eq:optimization_dual}. By definition, the dual problem provides an upper bound on the primal objective, meaning that:
		\[
		f^* \leq q^*.
		\]
		Solving the dual problem thus provides an approximation of the primal solution. Under certain conditions\footnote{E.g., Slater's condition.}, strong duality holds, implying that the primal and dual solutions are equal:
		\[
		f^* = q^*.
		\]
		Furthermore, the optimal solution must satisfy the following \textbf{Karush-Kuhn-Tucker (KKT) conditions}:
		\begin{itemize}
			\item \textbf{Primal Feasibility}: 
			\[
			g_i(x^*) \leq 0.
			\]
			\item \textbf{Dual Feasibility}:
			\[
			\lambda_i^{*} \geq 0.
			\]
			\item \textbf{Complementary Slackness}:
			\[
			\lambda_i^{*} g_i(x^*) = 0.
			\]
			\item \textbf{Stationarity}:
			\[
			\nabla_{x} \mathcal{L}(x^*, \lambda_i^*) = 0.
			\]
		\end{itemize}
		Solving the KKT conditions yields an optimal solution for both the primal and dual problems when strong duality holds.
	\end{itemize}
	
	We use a concrete example to demonstrate how the primal-dual method can be applied to solve an optimization problem. Specifically, we aim to maximize a quadratic objective function subject to a linear constraint:
	
	\begin{equation*} 
		\begin{aligned}
			\max_{x} \quad &  -\frac{1}{2} x^2  + b\cdot x\\
			\text{s.t.} \quad & a\cdot x -c \leq 0, 
		\end{aligned}
	\end{equation*}
	where \(a, b, c\) are constants, and \( a \ne 0 \). 
	
	Following the procedure outlined earlier, we first construct the Lagrangian function:
	
	\[
	\mathcal{L}(x, \lambda) =  -\frac{1}{2} x^2  + b\cdot x - \lambda \cdot \left( a\cdot x -c \right). 
	\]
	
	The corresponding dual function is given by:
	
	\[
	\mathcal{L}^{*}(\lambda) = \sup_{x} \mathcal{L}(x, \lambda) = \sup_{x} \left(  -\frac{1}{2} x^2  + b\cdot x - \lambda \cdot \left( a\cdot x -c \right)\right).  
	\]
	
	Since this is a quadratic function in \(x\), it is straightforward to show that:
	
	\[
	\mathcal{L}^*({\lambda}) =  \frac{1}{2}  a^2  \lambda^2+ (c-ab) \lambda + \frac{1}{2} b^2. 
	\]
	
	This is a quadratic function in \(\lambda\). Since \(\lambda \geq 0\), it attains its minimum at:
	
	\[
	\lambda^* = \max \left(0, \frac{ab-c}{a^2} \right).
	\]
	
	\paragraph{KKT Conditions}
	
	The optimal solution must satisfy the KKT conditions:
	
	\begin{itemize}
		\item \textbf{Primal Feasibility}: 
		\[
		a\cdot x^* -c \leq 0.
		\]
		\item \textbf{Dual Feasibility}:
		\[
		\lambda^* \geq 0.
		\]
		\item \textbf{Complementary Slackness}:
		\[
		\lambda^* \cdot (a \cdot x^* -c) = 0.
		\]
		\item \textbf{Stationarity}:
		\[
		\frac{\partial}{\partial x} \mathcal{L}(x^*, \lambda^*) = 0 \quad \Rightarrow \quad x^* = b - \lambda^* \cdot a.
		\]
	\end{itemize}
	
	\paragraph{Case Analysis}
	
	There are two possible cases:
	
	\begin{itemize}
		\item \textbf{Case 1: \( \lambda^* = 0 \)}  
		
		This occurs when \( ab - c < 0 \). From the stationarity condition, we obtain:
		\[
		x^* = b - \lambda^* \cdot a = b.
		\]
		In this case, the constraint is inactive.
		
		\item \textbf{Case 2: \( \lambda^* = \frac{ab-c}{a^2} \)}  
		
		This occurs when \( ab - c  \geq 0 \). From the stationarity condition, we obtain:
		\[
		x^* = b - \lambda^* \cdot a = \frac{c}{a}. 
		\]
		In this case, the constraint is active.
	\end{itemize}

	\section{Gradient Descent Method}
	
	Gradient descent is a fundamental method for solving unconstrained mathematical optimization problems:
	\begin{equation} \label{eq:gradient_descent_formulation}
		\min_x f(x)
	\end{equation}
	 The idea is straightforward: to find the minimum value of a function \( f(x) \), we take steps in the opposite direction of the gradient at the current point, as this is the direction of steepest descent. This procedure is repeated iteratively until convergence. In this section, we provide a brief introduction to gradient descent method. For a more in-depth discussion of advanced techniques, readers may refer to \cite{ruder2016overview}.
	
	Mathematically, let \( f(x) \) be a differentiable function, and suppose we aim to minimize \( f(x) \), where \( x \in \mathbb{R}^n \) represents the parameter vector. The gradient descent method adaptively updates the parameter vector based on the following update rule until \( f(x) \) converges:
	
	\[
	x^{k+1} \gets x^k - \epsilon \cdot \nabla_x f(x^k),
	\]
	where:
	
	\begin{itemize}
		\item \( x^k \) is the parameter vector at the \( k\)-th iteration.
		\item \( \epsilon > 0 \) is the learning rate, which controls the step size.
		\item \( \nabla_x f(x^k) \) is the gradient of the function evaluated at \( x^k \).
	\end{itemize}
	
	It can be shown that, under certain regularity conditions and assuming \( f(x) \) is a convex function, gradient descent converges to a global minimum with a properly chosen learning rate.
	
\begin{figure}[H]
		\centering
		\includegraphics[width=0.95\textwidth]{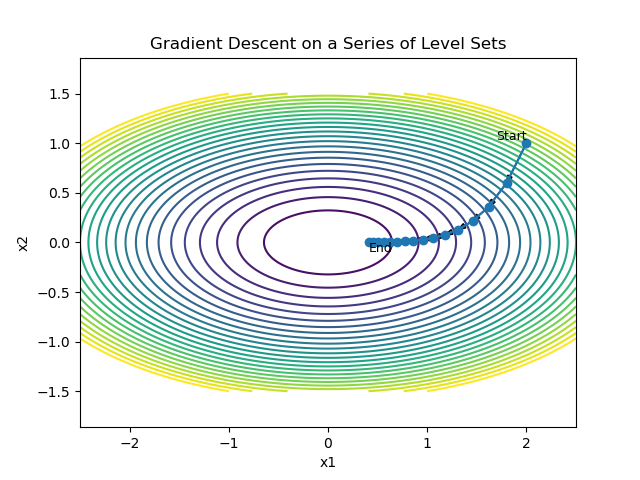}
		\caption{Illustration of the Gradient Descent Algorithm}
		\label{fig:gradient_descent}
	\end{figure}

	\paragraph{An Example}
	\autoref{fig:gradient_descent} illustrates how gradient descent works in practice to solve the following unconstrained optimization problem:
	\[
	\min_{x_1, x_2} f(x_1, x_2) = x_1^2 + 4 x_2^2. 
	\]
	Each elliptical curve in the figure represents a level set of the objective function \( f(x) \), where \( f(x) = c \) for some constant \( c \). The function \( f(x) \) is a convex quadratic function with its global minimum located at the origin \( (0,0) \). The level sets of this function are elliptical contours due to the different scaling of \( x_1 \) and \( x_2 \). Given a current iterate \( (x_1^k, x_2^k) \), the update rule for gradient descent is given by:
	\[
	x^{k+1} \gets x^k - \epsilon \cdot \nabla_{x} f(x^k),
	\]
	where \( \epsilon \) is the learning rate and \( \nabla f(x^k) \) is the gradient of the function at the current point. For the given function, the gradient is computed as:
	\[
	\nabla_{x} f(x) =
	\begin{bmatrix}
		\frac{\partial f}{\partial x_1} \\
		\frac{\partial f}{\partial x_2}
	\end{bmatrix}
	=
	\begin{bmatrix}
		2x_1 \\
		8x_2
	\end{bmatrix}.
	\]
	Thus, the gradient descent update rule becomes:
	\[
	\begin{bmatrix}
		x_1^{k+1} \\
		x_2^{k+1}
	\end{bmatrix}
	\gets
	\begin{bmatrix}
		x_1^k \\
		x_2^k
	\end{bmatrix}
	-
	\epsilon \cdot 
	\begin{bmatrix}
		2x_1^k \\
		8x_2^k
	\end{bmatrix}
	=
	\begin{bmatrix}
		x_1^k - 2 \epsilon x_1^k \\
		x_2^k - 8 \epsilon x_2^k
	\end{bmatrix}.
	\]
	This update rule shows that the values of \( x_1 \) and \( x_2 \) are reduced proportionally to their gradients, with different scaling due to the factors 2 and 8. As seen in \autoref{fig:gradient_descent}, starting from the initial point labeled "Start," gradient descent moves along the blue dotted path, progressively decreasing the function value at each step. The trajectory aligns with the steepest descent direction at each iteration. Eventually, the iterates converge to the global minimum at the point labeled "End," where \( x_1 = 0 \) and \( x_2 = 0 \), achieving the optimal solution.

	\subsection{Stochastic Gradient Descent}
	
	A useful variant of gradient descent is the so-called Stochastic Gradient Descent (SGD) method. In many optimization problems, the objective function is not a simple scalar function but rather an aggregate sum over multiple individual functions. A common form of such an objective function is:
	
	\[
	f(x) = \sum_{i=1}^{n} g(y_i; x),
	\]
	where each \( g(y_i; x) \) represents the loss function associated with a single data point \( y_i \), parameterized by the vector \( x \). The data points \( \{y_i\}_{i=1}^n \) may be drawn from an unknown i.i.d. distribution. Our goal is to find an optimal parameter \( x^* \) that minimizes the objective function \( f(x) \). In this case, the optimization problem is formulated as:
	
	\begin{equation} \label{eq:sgd_formulation}
		\min_{x}  \sum_{i=1}^{n} g(y_i; x).
	\end{equation}
	
	The standard gradient descent method computes the full gradient of \( f(x) \) at each iteration:
	
	\[
	\nabla_{x} f(x) = \sum_{i=1}^{n} \nabla_{x} g(y_i; x).
	\]
	
	The corresponding update rule is:
	
	\begin{equation} \label{eq:sgd_update}
	x^{k+1} \gets  x^k - \epsilon \cdot \sum_{i=1}^{n} \nabla_{x} g(y_i; x^k).
	\end{equation}
	
	However, in many online optimization scenarios, data arrives in a streaming manner. In production systems, waiting for all samples to be available before updating the model is impractical. Thus, batch-based gradient descent is not a suitable approach in such cases.
	
	\paragraph{Stochastic Gradient Descent (SGD)}
	
	Stochastic Gradient Descent (SGD) approximates the full gradient by using only a single function \( g(y_i; x) \) at each iteration. Instead of computing the sum over all \( n \) gradients, SGD updates the parameter using only one term at each step, allowing updates to be made in real-time:
	
	\[
	x^{k+1} \gets  x^k - \epsilon \cdot  \nabla_x g(y_k; x^k).
	\]
	
	It can be shown that, under proper conditions (e.g., i.i.d. distribution of data samples, well-chosen learning rates), the SGD algorithm exhibits guaranteed convergence properties(e.g., see \cite{bertsekas1996neuro}). We will frequently use SGD throughout this book. A typical application is in real-time bidding systems, where auction requests arrive in a continuous stream, and we need to adaptively update parameters in real-time to determine the optimal bid level.

	\section{Remarks}
	\subsection{Standard Form of an Optimization Problem}
	
	The formulation of the optimization problem in \eqref{eq:optimization} is tailored to align with our applications in bidding problems in subsequent chapters. However, it is worth noting that in many optimization textbooks (e.g., \cite{boyd2004convex}), the standard form of an optimization problem is typically presented as:
	
	\begin{equation} \label{eq:optimization_standard}
		\begin{aligned}
			\min_{x} \quad & f(x)\\
			\text{s.t.} \quad & g_i(x) \leq 0, \quad  i = 1, \dots, n, \\
			& h_j(x) = 0, \quad j = 1, \dots, q.
		\end{aligned}
	\end{equation}
	
	All these formulations are mathematically equivalent, and one can transform from one to another. For instance, to convert the maximization objective in \eqref{eq:optimization} into the minimization form in \eqref{eq:optimization_standard}, it suffices to introduce a negative sign:
	
	\[
	\max_{x} f(x) \Leftrightarrow \min_{x} \left(- f(x)\right).
	\]

	\subsection{Stochastic Approximation Algorithm}
	In this section, we discuss the Stochastic Approximation Algorithm, specifically the Robbins-Monro Algorithm, which has a close connection with the stochastic gradient descent (SGD) method introduced earlier in this chapter.
	
	The Robbins-Monro Algorithm is a fundamental stochastic approximation method introduced by Herbert Robbins and Sutton Monro in 1951. It provides an iterative procedure for finding the root of a function when only noisy observations are available.
	
	Given a function \( h(\theta) \), the goal is to determine \( \theta^* \) such that:
	
	\begin{equation*}
		h(\theta^*) = \mathbb{E}[X | \theta^*] = 0,
	\end{equation*}
	where \( X \) represents a noisy observation. Since \( X \) is not observed directly, we approximate it using a sample \( X_t \) drawn at each iteration. The Robbins-Monro algorithm updates the estimate of \( \theta \) iteratively:
	
	\begin{equation} \label{eq:robbins_monro_update}
		\theta_{t+1} = \theta_t - \epsilon_t h(\theta_t),
	\end{equation}
	where:
	\begin{itemize}
		\item \( \theta_t \) is the estimate at iteration \( t \),
		\item \( \epsilon_t \) is a sequence of step sizes (learning rates),
		\item \( h(\theta_t) \) is an unbiased estimator of the true function value.
	\end{itemize}
	
	For the algorithm to converge to \( \theta^* \) with probability 1, the step size sequence \( \{\epsilon_t\} \) must satisfy the following conditions:
	
	\begin{equation} \label{eq:step_size_condition}
		\sum_{t=1}^{\infty} \epsilon_t = \infty, \quad \sum_{t=1}^{\infty} \epsilon_t^2 < \infty.
	\end{equation}
	
	These conditions ensure that the step size is initially large enough to allow exploration but diminishes over time to facilitate convergence. In practice, a common choice is:
	\[
	\epsilon_t = \frac{1}{t}.
	\]
	For further technical details on the Robbins-Monro algorithm, refer to \cite{robbins1951stochastic}.
	
	\paragraph{Connection to Stochastic Gradient Descent}
	
	Stochastic gradient descent (SGD) can be regarded as a special case of the Robbins-Monro algorithm. Consider the optimization problem:
	
	\[
	\min_{x} \frac{1}{n} \sum_{i=1}^{n} g(y_i; x).
	\]
	When \( n \) is large, we approximate:
	\[
	\frac{1}{n} \sum_{i=1}^{n} g(y_i; x) \approx \mathbb{E}_{y \sim \mathcal{D}} \left[g(y; x) \right],
	\]
	where \( \mathcal{D} \) represents the distribution from which the random variable \( y \) is drawn.
	
	We aim to find \( x^* \) such that:
	
	\[
	\min_{x} \mathbb{E}_{y \sim \mathcal{D}} \left[g(y; x) \right] \quad  \Leftrightarrow  \quad \text{find \(x^*\) such that }  \frac{\partial}{\partial x} \mathbb{E}_{y \sim \mathcal{D}} \left[g(y; x^*) \right] = 0.
	\]
	Under mild regularity conditions, such as the dominated convergence condition, we can interchange expectation and differentiation:
	
	\[
	\frac{\partial}{\partial x} \mathbb{E}_{y \sim \mathcal{D}} \left[g(y; x^*) \right]  =  \mathbb{E}_{y \sim \mathcal{D}} \left[ \frac{\partial}{\partial x} g(y; x^*)\right].
	\]
	Setting:
	\[
	h(x^*) =  \mathbb{E}_{y \sim \mathcal{D}} \left[ \frac{\partial}{\partial x} g(y; x^*)\right],
	\]
	it follows that:
	\[
	\frac{\partial}{\partial x} g(y_t; x_t)
	\]
	is an unbiased estimator of \( h(x^*) \). Applying the Robbins-Monro update rule \eqref{eq:robbins_monro_update}, we recover the stochastic gradient descent algorithm:
	
	\begin{equation} \label{eq:sgd_update}
		x_{t+1} = x_t - \epsilon_t \frac{\partial}{\partial x} g(y_t; x_t).
	\end{equation}
	
	This connection also suggests that when selecting learning rates for SGD, we should follow the conditions in \eqref{eq:step_size_condition} to ensure convergence.

	\subsection{Online Convex Optimization}
	Another topic closely related to our earlier discussion is \textit{Online Convex Optimization} (OCO). In OCO, the goal is to design an efficient algorithm that makes decisions iteratively in scenarios where the outcome is revealed only after the decision has been made. More specifically, at each iteration, the algorithm chooses an action \(x_t\), and after this action is committed, a convex loss function \(f_t\) is revealed. The objective is to minimize the regret, defined as
	\[
	\text{Regret}_T = \sum_{t=1}^{T} f_t(x_t) - \min_{x} \sum_{t=1}^{T} f_t(x).
	\]
	
	This formulation has a deep connection to the pacing optimization problem discussed later in the book. One can think of \(f_t\) as the loss incurred after a bidding agent submits its bid to an auction (e.g., the loss from an auction opportunity). The pacing algorithm can then be viewed as an online decision-making process that aims to minimize its cumulative regret compared to the optimal fixed decision in hindsight.

	An efficient first-order algorithm for solving the OCO problem is \textit{Online Gradient Descent} (OGD), which updates the decision solely based on the gradient of the loss function \(f_t\):
	\[
	x_{t+1} = x_t - \epsilon_t \nabla f_t(x_t),
	\]
	where \(\epsilon_t\) is the step size at iteration \(t\).  
	
	It can be shown that regret bounds for OCO directly translate into convergence rates for the online stochastic gradient descent algorithm. Readers interested in a more comprehensive treatment of this topic are referred to \cite{hazan2016introduction} and \cite{zinkevich2003online}.

	\newtheorem{lemma}{Lemma}
	\newtheorem{theorem}{Theorem}
	
	\chapter{Basics of Auction Mechanisms}
	
	\intro{
		We introduce several of the most widely used auction mechanisms in the industry, including First Price Auction, Second Price Auction, VCG Auction, Myerson's Optimal Auction, and Generalized Second Price Auction. These auction mechanisms are widely adopted in real-time bidding (RTB) systems and have a profound impact on the design of bidding algorithms, which we will explore later in this book.
	}

	We provide a brief introduction to auction mechanisms in this chapter. We start by introducing single-item auctions, specifically the First Price Auction (FPA) and the Second Price Auction (SPA), with a discussion on incentive compatibility using these two auctions as examples. We then proceed to multiple-item incentive compatiple auction mechanism, the VCG auction. We also briefly discuss Myerson's classical optimal auction mechanism, which aims to maximize the total revenue of the auctioneer and is widely adopted by many companies in the industry. In the last section, we introduce one the most popular auctions in the industry: the Generalized Second Price Auction (GSP). 
	
	The discussion in this chapter serves as a primer and provides a working understanding of how auctions function in the online ad-serving funnel. For readers interested in exploring this topic further, a good introductory course on auction mechanisms is \cite{matthews1995technical}. For general auction mechanism design, see \cite{roughgarden2016twenty}, \cite{Nisan2007-gr} and \cite{hartline2013mechanism}. The classical paper on Myerson's optimal auction design is \cite{myerson1981optimal}. For the GSP auction, refer to \cite{edelman2007internet} and \cite{varian2007position}. For the VCG auction, see \cite{clarke1971multipart}, \cite{groves1973incentives}, \cite{vickrey1961counterspeculation} and \cite{varian2014vcg}.

	\section{First Price Auction}
	An auction mechanism essentially determines which bidders win the auctioned items (allocation rules) and what prices they should pay (payment rules). Suppose there is an available ad slot and $N$ ad campaigns are bidding for this slot with per click bid prices $\{b_i\}_{i=1}^N$, and their corresponding CTRs are $\{\alpha_i\}_{i=1}^N$. The per impression bids are $\{b_i \cdot \alpha_i\}_{i=1}^N$. A First Price Auction assigns the ad slot to the bidder with the highest bid per impression and charges them $b_k \cdot \alpha_k$ per impression, assuming $k$ is the index of the winning bidder (campaign). At first glance, such a mechanism appears simple and intuitive, but bidders in a First Price Auction tend to bid lower than their true values.
	
	To illustrate this, consider an example. Suppose there are two bidders (campaigns) competing for a single ad slot: bidder $B_1$ privately values each click at \$3 and bidder $B_2$ privately values each click at \$2. For simplicity, assume both bidders have a CTR of $1$. If they bid truthfully (i.e., $B_1$ bids \$3 and $B_2$ bids \$2), $B_1$ wins the ad slot and pays \$3 per click. 
	
	If we define utility $u$ as the difference between the bidder's true value $v$ and their payment $p$, i.e., 
	\[
	u = v - p,
	\]
	then in this case, the utility for $B_1$ is:
	\[
	u_1 = 3 - 3 = 0.
	\]
	Clearly, this is not an optimal strategy for $B_1$. Instead of bidding truthfully, $B_1$ could bid slightly higher than $B_2$, say \$2.01. In this case, $B_1$ still wins the auction but pays a much lower price, significantly increasing their utility. This demonstrates that in a First Price Auction, bidders have an incentive to bid below their true value, otherwise, the utility will be zero. Additionally, bidders must consider the behavior of other bidders to determine how much to lower their bid in order to maximize their utility.
	
	The example above also motivates the definition of incentive compatibility in auction mechanisms. In auction theory, incentive compatibility refers to a property of a mechanism or auction in which it is in every participant's best interest to act truthfully, i.e., to reveal their true valuation of the items or outcomes. This ensures that participants maximize their utility by being truthful, regardless of what others do. 
	
	Mathematically, let:
	\begin{itemize}
		\item \( N \): the set of participants (bidders).
		\item \( \mathcal{X} \): the set of possible outcomes of the mechanism.
		\item \( v_i: \mathcal{X} \to \mathbb{R} \): the valuation function of participant \( i \), representing their true value for each outcome.
		\item \( p_i \): the payment made by participant \( i \).
		\item \( u_i \): the utility of participant \( i \), defined as:
		\[
		u_i(x, p_i) = v_i(x) - p_i,
		\]
		where \( x \in \mathcal{X} \) is the chosen outcome.
	\end{itemize}	
	A mechanism \( (x(\cdot), p(\cdot)) \) is \textbf{incentive compatible (IC)} if, for every participant \( i \), reporting their true valuation \( v_i \) maximizes their utility:
		\[
		v_i(x(v_i, v_{-i})) - p_i(v_i, v_{-i}) \geq v_i(x(v_i', v_{-i})) - p_i(v_i', v_{-i}),
		\]
	for all \( v_i' \neq v_i \), where \( v_{-i} \) represents the valuations of all other participants.

	From the analysis above, we can conclude that the First Price Auction is not incentive compatible. We will discuss how to bid optimally under a First Price Auction later in this book.

	\begin{section} {Second Price Auction}
	Second Price Auction(or $Vickrey$ auction) is different auction format in which the highest bidder wins the auction but, different from the First Price Auction, pays the second highest bid.  Second price auction is incentive compatible. Use the example above, under second price auction, if bidder $B_1$ bids her true valuation of the ad slot, i.e.,  \$3, she only needs to pay \$2, the utility is \$3 - \$2 = \$1, there is no need for her to bid different than her private valuation. Indeed, we may prove that, under the second price auction, every bidder has a dominant strategy, that is to set her bid equal to her private valuation. More specifically, we have the following claim
	
	\highlightbox{
		A single-item second price auction is incentive compatible. If there are $N$ bidders with private evaluations of the item $\{v_i\}_{i=1}^N$, the dominant strategy of each bidder is to set the bid $b_i$ equal to $v_i$ and this truthful bidding strategy maximizes the utility $u_i$ of bidder $i$ and the utilities are non-negative if bidder bids truthfully. 
	}
	
	Fix a bidder $i$, let $B = \max_{j \neq i} b_j$ denote the highest bid among other bidders. The utility function $u_i$ of bidder $i$ is defined as 
	\[
	u_i =
	\begin{cases} 
		v_i - B & \text{if bidder \( i \) wins,} \\
		0, & \text{otherwise.}
	\end{cases}
	\]

	To prove this claim, we just need to show that the truthful bidding strategy(i.e., $b_i = v_i$) maximizes the utility of bidder $i$.  We analyze the utility for bidder \( i \) when they report truthfully (\( b_i = v_i \)) versus when they misreport (\( b_i' \neq v_i \)):
	
	\begin{itemize}
		\item \textbf{Case 1: Bidder \( i \) wins by bidding \( b_i = v_i \):}
		\begin{itemize}
			\item The bidder wins if \( v_i > = B \), and their utility is:
			\[
			u_i = v_i - B.
			\]
		\end{itemize}
		
		\item \textbf{Case 2: Bidder \( i \) wins by bidding \( b_i' > v_i \):}
		\begin{itemize}
			\item The bidder might win, but their utility remains the same:
			\[
			u_i = v_i - B.
			\]
			Overbidding does not increase their utility beyond truthfully reporting.
		\end{itemize}
		
		\item \textbf{Case 3: Bidder \( i \) loses by bidding \( b_i' < v_i \):}
		\begin{itemize}
			\item If bidder \( i \) underbids and loses (i.e., \( b_i' < B \)), their utility becomes:
			\[
			u_i = 0,
			\]
			even though they could have won the item and obtained a positive utility by bidding truthfully.
		\end{itemize}
	\end{itemize}
	
	In all cases, bidding truthfully maximizes the bidder's utility \( u_i \).  It's easy to see in all cases, the utility is non-negative. 
	
	The incentive compatibility property discussed above implies that Second Price Auctions are particularly convenient for bidders to participate in, as truthful bidding is always the dominant strategy. This ensures that no other strategy yields higher utility, regardless of the bids of others. Additionally, it avoids both the risk of losing the auction unnecessarily (as in underbidding) and the risk of overpaying (as in overbidding), making it highly appealing for real-world applications.
	
	\end{section}	
	
	\section{VCG Auction}
	The Second Price Auction works well for a single item (in the sense that it is incentive compatible). However, in real-world scenarios (e.g., in real-time bidding), it is very common to auction multiple items simultaneously. This motivates the need for a generalization that preserves desirable properties such as incentive compatibility and efficiency in such environments. The \textit{Vickrey-Clarke-Groves (VCG)} mechanism is one such generalization that achieves the following:
	
	\begin{itemize}
		\item \textbf{Efficient Allocation:} The resources are allocated to maximize total social welfare, defined as the sum of the valuations of all participants.
		\item \textbf{Incentive Compatibility:} Each participant's dominant strategy is to truthfully reveal their private valuation, even in multi-item or combinatorial settings.
	\end{itemize}
	
	The formal definition of the \textbf{\textit{VCG}} mechanism is as follows: Consider $N$ participants (bidders) and a set of possible allocations $A$. Each participant $i$ has a private valuation function $v_i(a)$ for an allocation $a \in A$. The goal of the auction is to select the allocation $a^*$ that maximizes total social welfare:
	\[
	a^* = \arg \max_{a \in A} \sum_{i=1}^N v_i(a).
	\]
	
	Once the optimal allocation $a^*$ is determined, the payment $p_i$ for each participant $i$ is computed as:
	\[
	p_i = h_i - \sum_{j \neq i} v_j(a^*),
	\]
	where $h_i$ is the hypothetical social welfare if participant $i$ were excluded:
	\[
	h_i = \max_{a \in A} \sum_{j \neq i} v_j(a).
	\]
	
	This payment structure ensures that each participant pays an amount equal to the "externality" they impose on others by participating in the auction. It can be shown that under the  \textbf{\textit{VCG}} auction, the optimal strategy for each bidder is to bid truthfully, a rigirous proof could be found in \cite{Nisan2007-gr}.

	Let's explore two concrete examples to demonstrate how the VCG mechanism works.
	
	\subsection*{Example 1: Single Item Auction}
	We apply VCG to a single-item auction. Suppose there are $N$ bidders competing for a single ad slot with private valuations $\{ v_i \}_{i=1}^N$. The allocation rule \(a^*\) is given by:
	\[
	a^* = \arg \max_{a \in A} \sum_{i=1}^N v_i(a).
	\]
	
	Since there is only one slot available, only one campaign can win the auction. The winner (e.g., bidder $k$) values the outcome at \(v_k\), while the rest of the bidders value the outcome at \(0\) as they do not win the auction. To maximize \(\sum_{i=1}^N v_i(a)\), the optimal allocation rule assigns the ad slot to the bidder with the highest valuation. Thus, bidder \(k\) with the highest valuation wins the auction.
	
	For payments, the payment for bidder \(k\) is:
	\[
	p_k(a^*) = \max_{a \in A} \sum_{j \neq k} v_j(a) - \sum_{j \neq k} v_j(a^*),
	\]
	where:
	\begin{itemize}
		\item \(\max_{a \in A} \sum_{j \neq k} v_j(a)\): The maximum value achievable by all other bidders if bidder \(k\) does not participate, which is simply the highest valuation among all other bidders, i.e., \(\max_{j \neq k} v_j\).
		\item \(\sum_{j \neq k} v_j(a^*) = 0\): In the optimal allocation, only bidder \(k\) is assigned the ad slot, and all other bidders lose the auction.
	\end{itemize}
	
	Thus:
	\[
	p_k(a^*) = \max_{j \neq k} v_j.
	\]
	
	It is clear that all other bidders pay \$0. Hence, in a single-item auction, the VCG mechanism reduces to the regular Second Price Auction.
	
	\subsection*{Example 2: Two Ad Slots with Three Bidders}
	\label{example:vcg_2_slots}
	
	Consider two ad slots with the following properties:
	\begin{itemize}
		\item Ad slot \(X\): CTR = 0.2
		\item Ad slot \(Y\): CTR = 0.1
	\end{itemize}
	
	The bidders have the following valuation per click:
	\begin{itemize}
		\item Bidder 1: \$100/click
		\item Bidder 2: \$40/click
		\item Bidder 3: \$20/click
	\end{itemize}
	
	The effective per-impression bids (CTR multiplied by the bid per click) for each slot are:
	\begin{itemize}
		\item Bidder 1: \$20 for slot \(X\), \$10 for slot \(Y\)
		\item Bidder 2: \$8 for slot \(X\), \$4 for slot \(Y\)
		\item Bidder 3: \$4 for slot \(X\), \$2 for slot \(Y\)
	\end{itemize}
	
	The total valuations for different allocations of the two slots are shown in the following table:
	
	\begin{table}[ht]
		\centering
		\begin{tabular}{ccccccc}
			\textbf{Allocation} & \(X1Y2\) & \(X1Y3\) & \(X2Y1\) & \(X2Y3\) & \(X3Y1\) & \(X3Y2\) \\
			\hline
			Bidder 1           & 20       & 20       & 10       & 0        & 10       & 0        \\
			Bidder 2           & 4        & 0        & 8        & 8        & 0        & 4        \\
			Bidder 3           & 0        & 2        & 0        & 2        & 4        & 4        \\
		\end{tabular}
		\caption{Valuations for different allocations of ad slots.}
		\label{tab:vcg_example}
	\end{table}
	\textit{Note: In this table, \text{X1Y2} means bidder $1$ wins slot $X$ and bidder $2$ wins slot $Y$, and so on.}

	\textbf{VCG Payment for Bidder 1}
	
	To compute the payment for Bidder 1, we calculate the externality they impose on others by winning their assigned slot. The payment for Bidder 1, \(p_1(a^*)\), is given by:
	\[
	p_1(a^*) = \max_{a \in A} \sum_{j \neq 1} v_j(a) - \sum_{j \neq 1} v_j(a^*),
	\]
	where:
	\begin{itemize}
		\item \(\max_{a \in A} \sum_{j \neq 1} v_j(a)\): The maximum valuation achievable by all other bidders if Bidder 1 does not participate.
		\item \(\sum_{j \neq 1} v_j(a^*)\): The total valuation of all other bidders under the current allocation \(a^*\).
	\end{itemize}
	
	For this example:
	\begin{itemize}
		\item \(	\max_{a \in A} \sum_{j \neq 1} v_j(a) = 10\): allocation \(X2Y3\): Bidder 2 wins slot \(X\) and Bidder 3 wins slot \(Y\);
		\item \(	\sum_{j \neq 1} v_j(a^*) = 4  \): current allocation \(X1Y2\): Bidder 2 wins slot \(Y\) with valuation 4, Bidder 3 gets no slot.
	\end{itemize}
	Thus:
	\[
	p_1(a^*) = 10 - 4 = 6.
	\]
	
	\textbf{VCG Payment for Bidder 2}
	
	Next, we compute the payment for Bidder 2, \(p_2(a^*)\), who wins slot \(Y\). The payment is given by:
	\[
	p_2(a^*) = \max_{a \in A} \sum_{j \neq 2} v_j(a) - \sum_{j \neq 2} v_j(a^*),
	\]
	where:
	\begin{itemize}
		\item \(\max_{a \in A} \sum_{j \neq 2} v_j(a) = 22\): Allocation \(X1Y3\): Bidder 1 wins slot \(X\) with valuation \(20\) and Bidder 3 wins slot \(Y\) with valuation \(2\).
		\item \(\sum_{j \neq 2} v_j(a^*) = 20\): Current allocation \(X1Y2\): Bidder 1 wins slot \(X\) with valuation \(20\), and Bidder 3 gets no slot.
	\end{itemize}
	Thus:
	\[
	p_2(a^*) = 22 - 20 = 2.
	\]
	
	\textbf{Conclusion:}
	\begin{itemize}
		\item Bidder 1 pays \$6 for slot \(X\).
		\item Bidder 2 pays \$2 for slot \(Y\).
	\end{itemize}

	Actually, we can prove that if there are \(k\) ad slots with CTRs \(\alpha_1 \geq \alpha_2 \geq \cdots \geq \alpha_k\), and \(N\) bidders such that the \(j\)-th bidder's value \(v_j\) satisfies \(v_1 \geq v_2 \geq \cdots \geq v_N\) (re-indexed if necessary), then:
	If we want to design a DSIC (Dominant Strategy Incentive Compatible) welfare-maximization auction mechanism, the only possible allocation is to assign the \(j\)-th highest bidder to the \(j\)-th slot for \(j = 1, 2, \cdots, k\) and the payment for the \(i\)-th highest bidder is given by:
	\begin{equation} \label{eq:vcg_pricing_formula}
	p_i = \sum_{j=1}^{k} b_{j+1}(\alpha_j - \alpha_{j+1}),
	\end{equation}
	where \(b_j = v_j\) is the truthful bid price for bidder \(j\), and we set \(\alpha_{k+1} = 0\). Readers can verify that this allocation and payment mechanism produces the same result as the one derived from the VCG auction definition in the example above. We will provide a sketch of the proof in the Remarks section, while a detailed proof can be found in \cite{roughgarden2016twenty}.

	\subsection*{A quick summary of the VCG mechanism}
	\begin{itemize}
		\item \textbf{Efficiency:} The allocation maximizes social welfare.
		\item \textbf{Incentive Compatibility:} Truthful bidding is the dominant strategy, as payments depend only on the valuations of others, not on the bidder's own valuation.
		\item \textbf{Fairness:} Participants pay only for the externality they impose, ensuring a fair payment structure.
	\end{itemize}

	\section{Myerson's Optimal Auction}

	
	The VCG mechanism is designed to maximize total welfare (i.e., the sum of participants' valuations). However, from the auctioneer's perspective, the goal may be to maximize profit instead of welfare. This raises the question:
	
	\begin{quote}
		\textbf{What does a profit-maximizing mechanism look like?}
	\end{quote}
	To answer this, we consider Myerson's Optimal Auction, which provides a framework for maximizing the auctioneer's profit while maintaining truthfulness (incentive compatibility).
	
	\subsection*{Key Concepts}
	
	If the mechanism is truthful, and fixing the bids of all other participants, the expected payment of bidder \(i\) is given by:
	\[
	p_i(b_i) = b_i x_i(b_i) - \int_{0}^{b_i} x_i(z) dz,
	\]
	where \(x_i(b_i)\) is the probability that bidder \(i\) wins the item given their bid \(b_i\).
	
	If we assume that bidder \(i\)'s valuation follows a known cumulative distribution function (CDF) \(F_i\) with probability density function (PDF) \(f_i\), we define the \textit{virtual value} for a given valuation \(v\) as:
	\[
	\phi_i(v) = v - \frac{1 - F_i(v)}{f_i(v)}.
	\]
	
	\subsection*{Myerson's Theorem}
	
	\begin{quote}
		\textbf{Theorem (Myerson, 1981):} The expected profit of any truthful mechanism is equal to its expected total virtual valuations.
	\end{quote}
	
	\textbf{Proof:}
	\[
	\mathbb{E}[p(b)] = \int_0^h p(b)f(b) db = \int_0^h b x(b) f(b) db - \int_0^h \int_0^b x(z) f(b) dz db,
	\]
	\[
	= \int_0^h b x(b) f(b) db - \int_0^h x(z) \int_z^h f(b) db dz,
	\]
	\[
	= \int_0^h b x(b) f(b) db - \int_0^h x(z) \left[1 - F(z)\right] dz,
	\]
	\[
	= \int_0^h \left[b - \frac{1 - F(b)}{f(b)}\right] x(b) f(b) db,
	\]
	\[
	= \mathbb{E}\left[\phi(b) x(b)\right].
	\]
	
	This demonstrates that maximizing expected profit is equivalent to maximizing the total virtual valuations.
	
	\subsection*{Myerson's Optimal Auction}
	
	From Myerson's Theorem, we conclude:
	\[
	\text{Profit maximization} \iff \text{Maximizing total virtual valuation}.
	\]
	This insight allows us to leverage the VCG mechanism. To design a profit-maximizing auction, we translate bids into virtual bids and run a VCG auction on these virtual bids. The steps for Myerson's Optimal Auction are as follows:
	\begin{enumerate}
		\item Given bids \(\mathbf{b}\) and value distributions \(\mathbf{F}\), compute the "virtual bids" \(\mathbf{b}'_i = \phi_i(\mathbf{b}_i)\).
		\item Run a VCG auction on the virtual bids to obtain the allocation \(\mathbf{x}'\) and payments \(\mathbf{p}'\).
		\item Output \(\mathbf{x} = \mathbf{x}'\) and \(\mathbf{p}_i = \phi^{-1}(\mathbf{p}'_i)\).
	\end{enumerate}
	
	\textbf{Note:} In VCG, it makes no sense to accept negative bid prices. Therefore, Myerson's Optimal Auction introduces a reservation price to ensure that virtual bids are non-negative.
	
	\subsection*{Example: Single Item Auction}
	
	Consider a single ad slot auction with two bidders, where the bid prices are \(b_1\) and \(b_2\). The allocation and payment rules are as follows:
	\begin{itemize}
		\item Bidder 1 wins if \(\phi_1(b_1) \geq \max\{\phi_2(b_2), 0\}\), and their payment is:
		\[
		p_1 = \inf \{b: \phi_1(b) \geq \phi_2(b_2) \text{ and } \phi_1(b) \geq 0\}.
		\]
		\item Bidder 2 wins if \(\phi_2(b_2) \geq \max\{\phi_1(b_1), 0\}\), and their payment is:
		\[
		p_2 = \inf \{b: \phi_2(b) \geq \phi_1(b_1) \text{ and } \phi_2(b) \geq 0\}.
		\]
	\end{itemize}
	
	If the value distributions are identical (\(F_1 = F_2 = F\)), then:
	\begin{itemize}
		\item Bidder 1 wins if \(b_1 \geq \max\{b_2, \phi^{-1}(0)\}\), and their payment is:
		\[
		p_1 = \max\{b_2, \phi^{-1}(0)\}.
		\]
		\item Bidder 2 wins if \(b_2 \geq \max\{b_1, \phi^{-1}(0)\}\), and their payment is:
		\[
		p_2 = \max\{b_1, \phi^{-1}(0)\}.
		\]
	\end{itemize}
	
	\subsection*{Optimal Auction for i.i.d. Distributions}
	
	For a single item auction with i.i.d. value distributions \(F\), Myerson's Optimal Auction reduces to a Vickrey Second Price Auction with a reservation price \(\phi^{-1}(0)\).
	
	\textbf{Example:} Suppose \(F\) is uniform on \([0, 1]\). Then:
	\[
	F(z) = z, \quad f(z) = 1 \implies \phi(z) = 2z - 1 \implies \phi^{-1}(0) = \frac{1}{2}.
	\]
	
	\textbf{Profit Comparison:}
	\begin{itemize}
		\item Profit without a reservation price: \(\frac{1}{3}\).
		\item Profit with a reservation price of \(\frac{1}{2}\): \(\frac{5}{12}\).
	\end{itemize}

	\subsection*{Remarks}
	One limitation of Myerson's optimal auction design is that it only applies to the so-called "single-parameter" auction environment, where each bidder can be characterized by a single number (the private value each bidder has for winning the item). 	However, in real-time bidding, there are usually multiple ad slots auctioned simultaneously. In this scenario, the "single-parameter" assumption no longer holds, and it becomes complex to extend Myerson's optimal auction mechanism design. There is active ongoing research on this topic, and readers who are interested may refer to, e.g., \cite{cai2011optimal} and \cite{manelli2007multidimensional}.

	\section{GSP Auction}
	
	Due to historical reasons, the most popular auction mechanism in the industry for multi-item auctions is the \textbf{Generalized Second Price Auction (GSP)}. It has been widely adopted by many IT companies, especially for sponsored search ads auctions and social network feed ads auctions. 
	
	The GSP auction is simple and intuitive: the highest bidder gets the top slot, the second-highest bidder gets the second slot, and so on. However, each bidder pays a price equal to the effective bid of the next-highest bidder for the slot they win.
	
	\subsection*{Example: How GSP Works}
	
	To illustrate how the GSP auction works, we revisit the example discussed in \autoref{example:vcg_2_slots}:
	
	\begin{itemize}
		\item There are two ad slots, \(X\) and \(Y\), with the following Click-Through Rates (CTR):
		\begin{itemize}
			\item Ad slot \(X\): \(\text{CTR} = 0.2\)
			\item Ad slot \(Y\): \(\text{CTR} = 0.1\)
		\end{itemize}
		
		\item There are three bidders with valuations per click:
		\begin{itemize}
			\item Bidder 1: \$100/click
			\item Bidder 2: \$40/click
			\item Bidder 3: \$20/click
		\end{itemize}
		
		\item The effective per-impression bids, computed as \(\text{CTR} \times \text{bid per click}\), are:
		\begin{itemize}
			\item Bidder 1: \$20 for slot \(X\), \$10 for slot \(Y\)
			\item Bidder 2: \$8 for slot \(X\), \$4 for slot \(Y\)
			\item Bidder 3: \$4 for slot \(X\), \$2 for slot \(Y\)
		\end{itemize}
	\end{itemize}
	
	\textbf{Allocation and Payments in GSP:}
	\begin{itemize}
		\item \textbf{Allocation:} 
		\begin{itemize}
			\item Bidder 1 wins slot \(X\).
			\item Bidder 2 wins slot \(Y\).
		\end{itemize}
		\item \textbf{Payments:}
		\begin{itemize}
			\item Bidder 1 pays \$8 (the effective bid of Bidder 2 for slot \(X\)).
			\item Bidder 2 pays \$2 (the effective bid of Bidder 3 for slot \(Y\)).
		\end{itemize}
	\end{itemize}
	
	The total payments are:
	\begin{itemize}
		\item Bidder 1: \$8
		\item Bidder 2: \$2
	\end{itemize}
	
	\section{Remarks}
	\subsection{Myerson's Lemma}
	
	We first prove Myerson's lemma and then prove \eqref{eq:vcg_pricing_formula} based on this lemma.
	 
	\begin{lemma}[Myerson's Lemma]
		\label{lem:myerson}
		Let $G(b)$ be the probability that a bidder with bid $b$ wins an auction, and let
		$H(b)$ be the expected payment that the bidder makes when bidding $b$. In any
		dominant-strategy incentive-compatible (DSIC) mechanism, the following relation
		holds:
		\[
		H(b) \;=\; b \, G(b) \;-\; \int_{0}^{b} G(z)\,\mathrm{d}z.
		\]
	\end{lemma}
	
	\begin{proof}
		\textbf{Step 1: Setup.}
		
		Consider a single bidder whose private valuation is $b$. Let $G(b)$ denote the
		probability that she \emph{wins} the auction (or is allocated the good) when she
		bids $b$, and let $H(b)$ denote her \emph{expected payment} in that situation.
		The bidder's \emph{expected utility}, when she truthfully bids her valuation $b$,
		is
		\[
		u(b) \;=\; b \, G(b) \;-\; H(b).
		\]
		
		\textbf{Step 2: Envelope Theorem Argument.}
		
		A mechanism is \emph{dominant-strategy incentive-compatible} (DSIC) if bidding
		one's true value $b$ is a best response regardless of others' bids.
		Informally, this implies that the utility $u(b)$ is the maximum possible
		utility the bidder can achieve, \emph{given} her true value $b$. Hence
		\[
		u(b) \;=\; \max_{b'} \Bigl\{ b \, G(b') \;-\; H(b') \Bigr\}.
		\]
		
		In such scenarios, the \emph{envelope theorem} says that if $u(b)$ is
		differentiable, its derivative with respect to $b$ is simply the partial
		derivative of the objective at the chosen maximizer $b' = b$. Concretely,
		\[
		\frac{\mathrm{d}}{\mathrm{d}b} \, u(b) 
		\;=\; \left. \frac{\partial}{\partial b} \Bigl( b \, G(b') 
		\;-\; H(b') \Bigr) \right|_{b'=b}.
		\]
		Since $G(b')$ and $H(b')$ depend on the \emph{bid} $b'$, rather than directly on
		the \emph{true value} $b$, they are constant with respect to $b$ when we evaluate
		the partial derivative at $b' = b$. Thus
		\[
		\frac{\mathrm{d}}{\mathrm{d}b} \, u(b)
		\;=\; G(b).
		\]
		This is the key result from the envelope theorem in the DSIC setting.
		
		\textbf{Step 3: Integrate \texorpdfstring{$u'(b) = G(b)$}{u'(b) = G(b)}.}
		
		We have established that
		\[
		u'(b) \;=\; G(b).
		\]
		Integrate both sides from $0$ to $b$:
		\[
		u(b) \;-\; u(0)
		\;=\; \int_{0}^{b} G(z) \,\mathrm{d}z.
		\]
		Typically, in auction settings, when a bidder's valuation is $0$, her utility is
		$0$. Formally, $u(0)=0$. Therefore,
		\[
		u(b) \;=\; \int_{0}^{b} G(z)\,\mathrm{d}z.
		\]
		
		\textbf{Step 4: Solve for \texorpdfstring{$H(b)$}{H(b)}.}
		
		Recall that 
		\[
		u(b) \;=\; b\,G(b) \;-\; H(b).
		\]
		Hence
		\[
		b\,G(b) \;-\; H(b)
		\;=\; \int_{0}^{b} G(z)\,\mathrm{d}z.
		\]
		Rearranging to isolate $H(b)$ gives
		\[
		H(b) 
		\;=\; b \, G(b) \;-\; \int_{0}^{b} G(z)\,\mathrm{d}z.
		\]
		This is precisely the statement of Myerson's Lemma.
		
	\end{proof}
	
	We can now proceed to prove \eqref{eq:vcg_pricing_formula}, which we summarize in the following theorem:

	\begin{theorem}[Welfare-Maximizing $k$-Slot Auction Is VCG]
		\label{thm:k-slot-dsic}
		Consider an ad auction with $k$ ad slots whose click-through rates (CTRs) satisfy
		\[
		\alpha_1 \;\ge\; \alpha_2 \;\ge\; \dots \;\ge\; \alpha_k \;\ge\; 0.
		\]
		There are $N$ bidders, each bidder $i$ having a private value $v_i$ for a single
		click.  An assignment of the $k$ slots must match exactly one slot to each of (up
		to) $k$ highest bidders.  Suppose we want a \emph{dominant-strategy
			incentive-compatible} (DSIC) mechanism that \emph{always maximizes total
			welfare} (i.e.\ sum of bidder values $\times$ CTRs). Then, up to tie-breaking,
		the \emph{only possible} DSIC welfare-maximizing allocation is:
		\[
		\text{assign the $j$-th highest bid to slot } j,
		\quad
		j=1,\ldots,k,
		\]
		and the corresponding \emph{unique} DSIC payment rule is given by
		\begin{equation}
			\label{eq:VCG-payment}
			p_i
			\;=\;
			\sum_{j=1}^{k}
			\Bigl(\alpha_j - \alpha_{j+1}\Bigr)\,b_{j+1},
			\quad
			\text{for the bidder $i$ whose bid is $b_i$ (the $j$-th highest),}
		\end{equation}
		where we set $\alpha_{k+1} = 0$ and $b_{k+1} = 0$ for notational convenience.
	\end{theorem}
	
	\begin{proof}
		
		\textbf{Step 1: Welfare Maximization Implies Sorting by CTR.}
		
		Let us index the bidders so that $b_1 \ge b_2 \ge \dots \ge b_N$ are their
		\emph{bids}.  Because the CTRs satisfy
		\(
		\alpha_1 \ge \alpha_2 \ge \dots \ge \alpha_k,
		\)
		the \emph{welfare-maximizing} assignment pairs the highest bid with
		$\alpha_1$, the second-highest bid with $\alpha_2$, and so on, up to
		the $k$-th highest bid with $\alpha_k$ (if $k \le N$).  This ensures total
		welfare
		\[
		\alpha_1 b_1 + \alpha_2 b_2 + \dots + \alpha_k b_k
		\]
		is maximized.
		
		\textbf{Step 2: DSIC Requires Monotonic Allocation.}
		
		A fundamental requirement for \emph{dominant-strategy} truthfulness is that each
		bidder's \emph{allocation} be \emph{monotone} in her own bid.  In a single-slot
		setting, that means ``if bidder $i$ increases her bid, her probability of winning
		does not decrease.''  In a multi-slot context, monotonicity translates to: if
		bidder $i$ raises her bid, she \emph{cannot} receive a \emph{lower} slot (one
		with a lower CTR).  Concretely,
		\begin{enumerate}
			\item If you bid more, your slot assignment (and hence your expected number
			of clicks) can only become \emph{better} (or stay the same), never worse.
			\item Combining this with \emph{welfare maximization} (which sorts by bids),
			one sees that the unique feasible allocation is: 
			\[
			\text{(highest bid)} \to \alpha_1,\quad
			\text{(2nd highest bid)} \to \alpha_2,\quad
			\dots,\quad
			\text{(k-th highest bid)} \to \alpha_k.
			\]
		\end{enumerate}
		
		\textbf{Step 3: Payments Are Forced by Myerson's Lemma (the Envelope Theorem).}
		
		Once the \emph{allocation} rule is fixed and is monotone, \emph{Myerson's Lemma}
		says the \emph{expected payment} function $H_i(\cdot)$ for each bidder $i$ is
		determined by the envelope condition:
		\[
		H_i(b)
		\;=\;
		b\,G_i(b)
		\;-\;
		\int_{0}^{b} G_i(z)\,\mathrm{d}z,
		\]
		where $G_i(b)$ is the \emph{(weighted) probability} that bidder $i$ receives a
		slot (or the expected CTR she obtains) when her bid is $b$.  In the $k$-slot
		environment:
		\[
		G_i(b) \;=\; \text{the CTR of the slot assigned to $i$, given $b$.}
		\]
		If bidder $i$ is the $j$-th highest bidder, then $G_i(b_i) = \alpha_j$.  As we
		\emph{raise} the bidder's bid from $0$ up to $b_i$, the bidder ``moves up the
		ladder'' of possible slots in a way determined by the sorted bids of others.
		Hence, the payment $H_i(b_i)$ is pinned down by how $G_i(\cdot)$ changes from
		$0$ to $b_i$.
		
		\textbf{Step 4: Recovering the VCG (Sum-of-Externalities) Price Formula.}
		
		Concretely, in a discrete sense, if $i$ is the $j$-th highest bidder, then to
		``get slot $j$'' instead of slot $j+1$, the bid must exceed the $(j+1)$-th
		highest bid $b_{j+1}$.  Tracking the changes in slot assignment across these
		``threshold bids'' yields exactly:
		\[
		p_i 
		\;=\; 
		\sum_{m=j}^{k-1} \Bigl(\alpha_m - \alpha_{m+1}\Bigr)\,b_{m+1},
		\quad
		\text{where $\alpha_{k+1}=0$ and $b_{k+1}=0$.}
		\]
		That is the standard generalized Vickrey--Clarke--Groves (VCG) price formula:
		each bidder $i$ pays the ``externality'' imposed on the lower‐ranked bidders,
		weighted by the slot‐quality (CTR) differences
		$\alpha_m - \alpha_{m+1}$.  One may check that this is just a discrete rewrite
		of Myerson's envelope integral.
		
		\textbf{Conclusion.}
		
		Thus, under the twin requirements of (1) maximizing total welfare and
		(2) monotonicity in each bidder's own bid (to ensure DSIC), the \emph{only} valid
		assignment is the ``highest bidder to the highest CTR'' rule, and Myerson's Lemma
		shows that there is \emph{exactly one} payment scheme that makes it DSIC:
		\[
		p_i
		\;=\;
		\sum_{j=1}^{k} \bigl(\alpha_j - \alpha_{j+1}\bigr)\,b_{j+1}.
		\]
		This completes the proof.
	\end{proof}

	\subsection{A Historical Note of GSP}
	As discussed in \autoref{example:vcg_2_slots}, the payment rule derived there is the unique one that ensures the DSIC (Dominant Strategy Incentive Compatibility) property. Therefore, we can conclude that the GSP auction is \textbf{not DSIC}. 
	
	GSP can be regarded as an incorrectly implemented version of the DSIC auction. Nevertheless, it has gained widespread popularity despite its non-DSIC property.  Google played a major role in popularizing the GSP auction. Interestingly, they considered transitioning from GSP to VCG during the summer of 2002. However, as mentioned in \cite{varian2014vcg}, three major issues prevented the change:
	\begin{enumerate}
		\item The existing GSP auction was growing rapidly and required significant engineering attention, making it challenging to develop a new auction.
		\item The VCG auction was harder to explain to advertisers.
		\item The VCG auction required advertisers to raise their bids above the levels they were accustomed to in the GSP auction.
	\end{enumerate}
	As a result, the idea of transitioning to VCG was shelved in 2002.

	\subsection{Auction Mechanism Design for Auto-Bidding}
	The mechanisms discussed in this chapter focus exclusively on single‐shot auctions. While these are widely adopted in industry, most bidding and pacing algorithms run across repeated auctions in real time. When an advertiser launches a campaign, it participates in multiple real‐time auctions under a fixed budget, which introduces new design challenges. For example, in a repeated‐auction setting, a bidder facing a tight budget may strategically shade its valuation—even under incentive‐compatible mechanisms such as second‐price auctions—in order to maximize overall performance. Designing efficient auction mechanisms for real‐world digital‐advertising scenarios under various constraints (e.g., budget caps, ROI targets) remains an active area of academic research. For more technical details, see  \cite{aggarwal2024auto}.

	\chapter{Experiment Framework}
	
	\intro{
		In this chapter, we introduce the A/B testing frameworks used in the advertising domain, which serve as powerful statistical methods for quantitatively measuring the impact of new strategies applied to ad campaigns. We discuss two frameworks: the general campaign-level A/B test and the budget-split A/B test.
	}
	
	When implementing a new algorithm, whether it is a bidding strategy or an enhancement to a prediction model, it is essential to evaluate its effectiveness and compare it against the existing baseline model. This evaluation is conducted using the A/B testing framework. 
	
	The general procedure is as follows: for the objects to which you want to apply the new strategy, \textbf{randomly} split them into two groups: the control group and the treatment group. In the control group, the baseline strategy remains unchanged, while in the treatment group, the new strategy is applied. After a certain period, relevant metrics are collected and analyzed to determine whether the new strategy outperforms the baseline.
	
	 Nearly all modifications within the ad-serving funnel must undergo this form of testing, and only those that demonstrate a positive impact on business metrics are deployed to the production system. In this chapter, we introduce two of the most commonly used A/B testing frameworks in the advertising domain: Campaign-Level A/B Testing and Budget-Split A/B Testing.

	\section{Campaign-Level A/B Test}
	
	In this section, we discuss the Campaign-Level A/B Test framework, in which the split is performed at the campaign level. Suppose there are \(N\) campaigns in the pool (\(N\) can be a large number—some major platforms may have over one million campaigns running simultaneously). Engineers develop a new pacing algorithm, \(P_1\), aimed at improving budget utilization. To evaluate its effectiveness against the existing pacing strategy, \(P_0\), the campaigns are split into two groups: the control group \(A\) and the treatment group \(B\). The baseline strategy \(P_0\) is applied to \(A\), while \(P_1\) is applied to \(B\). The experiment is then run for a certain period, during which data is collected to compare the actual budget utilization between the two groups.
	
	The key question is: how can we determine whether the results at the end of the test are statistically reliable for decision-making?
	
	To address this, we use \textit{hypothesis testing}, a statistical method that makes inferences about population parameters (such as budget utilization in our case) based on sample data. This process involves the following steps:
	
	\begin{itemize}
		\item \textbf{Define Null and Alternative Hypotheses}: The null hypothesis (\(H_0\)) assumes that the new strategy has no effect, while the alternative hypothesis (\(H_1\)) suggests otherwise. In our case, \(H_0\) states that the new pacing algorithm \(P_1\) does not alter budget utilization.
		
		\item \textbf{Choose a Significance Level \(\alpha\)\footnote{\(1-\alpha\) is referred to as the confidence level, representing how often we correctly fail to reject \(H_0\) when it is true.}}: This represents the probability of incorrectly rejecting the null hypothesis \(H_0\). For instance, selecting \(\alpha = 5\%\) implies a 5\% chance of rejecting \(H_0\) when it is actually true.
		
		\item \textbf{Select a Test Statistic}: A test statistic \(T\) is computed from the sample data collected in the experiment. In our case, we gather budget utilization data from campaigns in both the control and treatment groups and compute a test statistic to quantify how much the sample deviates from \(H_0\).
		
		\item \textbf{Compute the \(p\)-value}: Assuming \(H_0\) is true, we calculate the \(p\)-value, which represents the probability of obtaining a test result at least as extreme as the observed test statistic \(T\).
		
		\item \textbf{Make the Decision}: If \(p < \alpha\), we reject \(H_0\) because the probability of observing such an extreme result under \(H_0\) is too low (\(<\alpha\)). Otherwise, we do not reject \(H_0\).
	\end{itemize}
	
\begin{figure}[H]
		\centering
		\includegraphics[width=0.99\textwidth]{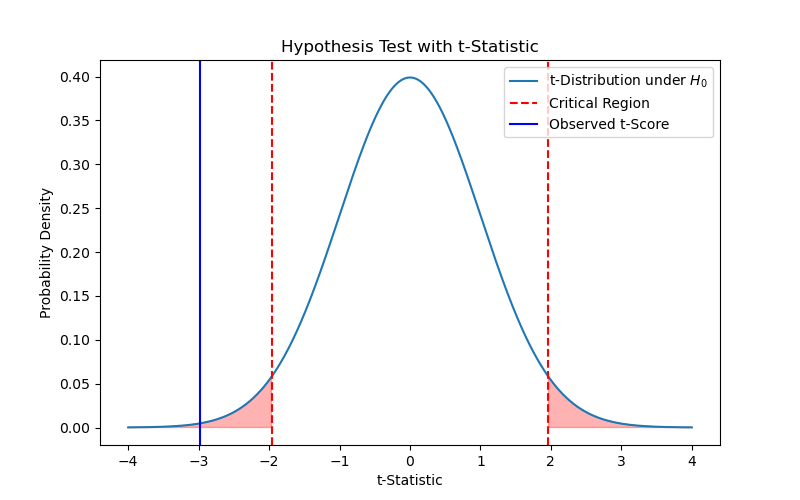}
		\caption{Hypothesis Test with \(t\)-Statistic}
		\label{fig:hypothesis_test}
	\end{figure}

	To illustrate this process, consider the example above: suppose both groups contain \(N = 100,000\) campaigns. At the end of the test, we obtain budget utilization data \(\{bu_{A,i}\}\) and \(\{bu_{B,i}\}\) from the control group \(A\) and the treatment group \(B\), respectively. We compute the average budget utilization for each group as follows:
	\[
	\text{mean}_A = \frac{1}{N} \sum_{i=1}^{N} bu_{A,i}, \quad \text{mean}_B = \frac{1}{N} \sum_{i=1}^{N} bu_{B,i}.
	\]
	Suppose we obtain \(\text{mean}_A = 95.14\%\) and \(\text{mean}_B = 97.15\%\). At first glance, budget utilization appears to have improved by approximately 2\%, but is this improvement statistically significant? To answer this, we perform a hypothesis test by defining the hypotheses:
	
	\[
	H_0 : \text{mean}_A = \text{mean}_B,  \quad H_1: \text{mean}_A \neq \text{mean}_B.
	\]
	Choosing a significance level of \(\alpha = 5\%\), we compute the Student's \(t\)-statistic as:
	
	\begin{equation} \label{eq:student_t}
	T = \frac{\text{mean}_A - \text{mean}_B}{\sqrt{ \frac{(n_A-1)\cdot std_A^2 + (n_B -1) \cdot std_B^2 }{n_A + n_B -2}} \cdot \sqrt{ \frac{1}{n_A} + \frac{1}{n_B} }}
	\end{equation}
	where \(n_A\) and \(n_B\) are the sample sizes of groups \(A\) and \(B\), respectively (in this case, \(n_A = n_B = N\)). The standard deviations \(\text{std}_A\) and \(\text{std}_B\) for each group are computed as:
	
	\[
	\text{std}_A = \sqrt{\frac{1}{n_A -1} \sum_{i=1}^{n_A} (bu_{A,i} - \text{mean}_A)^2}, \quad
	\text{std}_B = \sqrt{\frac{1}{n_B -1} \sum_{i=1}^{n_B} (bu_{B,i} - \text{mean}_B)^2}.
	\]
	We will later prove that, under the assumption that groups \(A\) and \(B\) have equal variance\footnote{For the unequal-variance case, Welch’s test can be used instead.} ,\(T\) follows a Student's \(t\)-distribution, whose probability density function is illustrated in \autoref{fig:hypothesis_test}. Suppose the computed \(t\)-statistic is \(-2.98\). In the figure, the blue line represents this observed \(t\)-statistic, which lies in the left tail of the distribution. The red dotted line marks the critical value corresponding to the significance level \(\alpha = 5\%\). Under \(H_0\), the \(t\)-statistic is expected to be centered around zero, meaning that extreme values in the red regions are unlikely to occur if \(H_0\) is true. Since our observed \(t\)-statistic falls within this critical region, we reject \(H_0\), concluding that the budget utilization increase from \(95.14\%\) to \(97.15\%\) is statistically significant and attributable to the new algorithm \(P_1\). As the impact on key metrics is positive, we can confidently deploy the new strategy to production.
	
	For further details on hypothesis testing, refer to standard statistical inference textbooks such as \cite{casella2024statistical} and \cite{rice2007mathematical}.

	\section {Budget Split A/B Test}
	The campaign-level experiment appears reasonable at first glance; however, upon closer examination of the design, several issues arise:
	
	\begin{itemize}
		\item \textbf{Skewed Budget Distribution:} The scale of budgets varies significantly across different campaigns. For small and medium-sized business (SMB) campaigns, the budget may be less than 100 dollars, whereas large enterprise branding campaigns can have budgets reaching hundreds of thousands or even exceeding one million dollars. This highly skewed distribution of campaign budgets diminishes the statistical power of the campaign-level A/B test, requiring a larger number of campaigns and a longer experiment duration to achieve statistically significant results.
		
		\item \textbf{Cannibalization Bias:} A more critical issue is known as \textit{cannibalization bias}. Consider a scenario in which the baseline pacing algorithm \(P_0\) already achieves a \(100\%\) budget utilization rate. At the same time, we aim to test a more aggressive pacing algorithm, \(P_1\), using a campaign-level A/B test. Since campaigns in both groups may participate in the same auction, the treatment group, driven by the more aggressive \(P_1\), is more likely to win the auction. Consequently, at the end of the test, we may observe a higher budget utilization rate in the treatment group than in the control group, even though \(P_0\) already achieves \(100\%\) budget utilization.
		
		The root cause of this phenomenon is that the control and treatment groups are competing against each other for auction opportunities. Some auction requests are effectively cannibalized by the more aggressive strategy \(P_1\), introducing bias into the campaign-level A/B test results.
	\end{itemize}
	
	The Budget Split testing framework was introduced to address the limitations of campaign-level A/B testing in ad marketplace experiments. The underlying idea is straightforward: instead of randomly splitting campaigns, we divide the budget within each campaign, effectively creating two identical \textit{sub-campaigns}. The control group consists of half of these sub-campaigns running the baseline strategy, while the treatment group comprises the other half, which retains identical budget settings but tests the new strategy. All auction requests are then randomly assigned and directed to one of these two groups.
	
	This approach is conceptually equivalent to creating two parallel ad marketplaces with identical budget configurations, thereby effectively eliminating \textit{cannibalization bias}. Additionally, it can be shown that the Budget Split framework is statistically more powerful than the campaign-level test, as it is capable of detecting smaller impacts of the new strategy that might be undetectable in a campaign-level experiment.
	
	For more theoretical and practical analysis of the Budget Split framework, readers may refer to \cite{liu2020trustworthy} and \cite{basse2016randomization}.

	\section{Remarks}
	
	\subsection{Proof of  Student's t-Test}
	
	Here we prove that \(T\) in \eqref{eq:student_t} follows a Stduent's t-distribution with \(n_A + n_B -2\) degrees of freedom defined as follows:

	If  \(Z \sim \mathcal{N}(0,1)\) follows a standard normal distribution and  \(W \sim \chi_k^{2}\) follows a chi-square distribution with  \(k\) degrees of freedom, then the Student's t-distribution with  degrees of freedom \(k\) is defined as:

	\begin{equation*}
		 \frac{Z}{\sqrt{W/k}} \sim t_k.
	\end{equation*}
	
	\paragraph{Test Statistic Derivation}	
	
	Let \( X_1, X_2, \dots, X_n \) be a random sample from a normal distribution \( \mathcal{N}(\mu_X, \sigma^2) \) and \( Y_1, Y_2, \dots, Y_m \) be a random sample from \( \mathcal{N}(\mu_Y, \sigma^2) \), both with unknown mean but common variance \( \sigma^2 \).
	
	The sample means and variances are given by:
	\begin{align*}
		\bar{X} &= \frac{1}{n} \sum_{i=1}^{n} X_i, & S_X^2 &= \frac{1}{n-1} \sum_{i=1}^{n} (X_i - \bar{X})^2, \\
		\bar{Y} &= \frac{1}{m} \sum_{i=1}^{m} Y_i, & S_Y^2 &= \frac{1}{m-1} \sum_{i=1}^{m} (Y_i - \bar{Y})^2.
	\end{align*}
	
	Since we assume equal variances, we use the pooled variance estimator:
	\begin{equation}
		S_p^2 = \frac{(n-1)S_X^2 + (m-1)S_Y^2}{n+m-2}.
	\end{equation}
	
	The test statistic is then defined as:
	\begin{equation}
		t = \frac{\bar{X} - \bar{Y}}{S_p \sqrt{\frac{1}{n} + \frac{1}{m}}}.
	\end{equation}
	
	\paragraph{Distribution of the Test Statistic}
	
	To prove that the test statistic follows a t-distribution, we analyze its components.
	
	\begin{itemize}
		\item \textbf{Step 1: Distribution of Sample Means}
		From properties of normal distributions,
		\begin{align*}
			\bar{X} &\sim \mathcal{N}\left( \mu_X, \frac{\sigma^2}{n} \right), \\
			\bar{Y} &\sim \mathcal{N}\left( \mu_Y, \frac{\sigma^2}{m} \right).
		\end{align*}
		Thus, their difference follows:
		\begin{equation}
			\bar{X} - \bar{Y} \sim \mathcal{N}\left( \mu_X - \mu_Y, \sigma^2 \left( \frac{1}{n} + \frac{1}{m} \right) \right).
		\end{equation}
		
		\item \textbf{Step 2: Standardization}
		Define the standardization term:
		\begin{equation}
			Z = \frac{(\bar{X} - \bar{Y}) - (\mu_X - \mu_Y)}{\sigma \sqrt{\frac{1}{n} + \frac{1}{m}}}.
		\end{equation}
		Since it is a linear transformation of a normal variable, it follows a standard normal distribution:
		\begin{equation}
			Z \sim \mathcal{N}(0,1).
		\end{equation}
		
		\item \textbf{Step 3: Distribution of the Pooled Variance}
		The pooled variance \( S_p^2 \) is the sum of two independent chi-square distributed variables:
		\begin{align*}
			(n-1) S_X^2 &\sim \sigma^2 \chi^2_{n-1}, \\
			(m-1) S_Y^2 &\sim \sigma^2 \chi^2_{m-1}.
		\end{align*}
		Thus, their sum follows a chi-square distribution with \( n+m-2 \) degrees of freedom:
		\begin{equation}
			(n+m-2) S_p^2 \sim \sigma^2 \chi^2_{n+m-2}.
		\end{equation}
		Defining
		\begin{equation}
			W = \frac{(n+m-2) S_p^2}{\sigma^2} \sim \chi^2_{n+m-2},
		\end{equation}
		we can express the test statistic as:
		\begin{equation}
			t = \frac{Z}{\sqrt{W / (n+m-2)}}.
		\end{equation}
		Since \( Z \sim \mathcal{N}(0,1) \) and \( W \sim \chi^2_{n+m-2} \), it follows that:
		\begin{equation}
			t \sim t_{n+m-2}.
		\end{equation}
	
	\end{itemize}
	This completes the proof.

%

	\part{Pacing Algorithms}
	\label{part:pacing_algorithms}
	
	\chapter{Bidding Problem Formulation}
	
	\intro{
		In this chapter, we provide a rigorous mathematical formulation of two primary bidding problems, namely max delivery and cost cap, in the context of repeated auction settings. We then employ the primal-dual method to derive the optimal bidding formulas. These results serve as the foundation for designing online control algorithms, which will be explored in the subsequent chapters.
	}
	
	In the first chapter of this part, we introduce a framework that formulates the budget pacing problem as a mathematical optimization problem through bidding. We focus specifically on the max delivery and cost cap problems, with all optimizations occurring at the campaign level unless stated otherwise. These two problems serve as examples to illustrate the core principles behind designing practical bidding algorithms.  
	
	For simplicity, unless explicitly stated otherwise, we assume throughout this book that the campaign follows a daily pacing strategy in an oCPM model, participating in a standard second-price auction where charges are incurred per impression and the objective/optimization goal is to maximize the total number of clicks.

	\begin{section}{Max Delivery} \label{sec:md_optimal}
	In the Max Delivery setting, advertisers set up a campaign with a specified budget. The objective is to optimize the clicks of the ad campaign while adhering to this budget constraint. Assuming this is a campaign operating under the standard Second Price Auction framework, a common goal is to maximize the total clicks for the campaign. Thus, the Max Delivery problem can be formulated as the following optimization problem:

	\begin{equation}  \label{eq:max_delivery}
		\begin{aligned}
			\max_{x_t \in \{0,1\}} \quad & \sum_{t=1}^T x_t \cdot r_t \\
			\text{s.t.} \quad &  \sum_{t=1}^{T} x_t \cdot c_t \leq B \\
		\end{aligned}
	\end{equation}
	where,
	\begin{itemize}
		\item \(T\) represents the total (predicted) number of auction opportunities within a day.
		\item \(r_t\) is the predicted click-through rate (CTR) for the \(t\)-th auction.
		\item \(c_t\) denotes the cost for the \(t\)-th auction. In a second-price auction, this corresponds to the highest eCPM among other bidders. 
		\item \(x_t\) is a binary decision variable indicating whether we win the \(t\)-th auction. 
	\end{itemize}
	Under the rules of a second-price auction, \(x_t = 1\) if and only if our bid per impression exceeds the highest competing bid, mathematically expressed as:
	\[
	x_t = \mathds{1}_{ \{b_t > c_t \}}
	\]
	where \(b_t\) is the bid per impression for \(t\)-th auction. We assume that both the sequences \(\{r_t\}\)  and \(\{c_t\}\) follows some unknown independent and identically distributed (\(i.i.d.\)) distribution, such as a log-normal distribution.

	\subsection {Optimal Solution to Max Delivery Problem}
		
		It is challenging to solve this problem directly. Instead of addressing it in the primal space, we apply the primal-dual method to transform it into the dual space. The Lagrangian of~\eqref{eq:max_delivery} is given by:
		\begin{equation*}
			\mathcal{L}(x_t, \lambda) = \sum_{t=1}^T x_t \cdot r_t  - \lambda \cdot \left( \sum_{t=1}^{T} x_t \cdot c_t  - B  \right)
		\end{equation*}
		The dual  is expressed as:
		\begin{equation*}
			\min_{\lambda \geq 0} \mathcal{L}^*(\lambda) = \min_{\lambda \geq 0} \max_{x_t \in \{0,1\}} \mathcal{L}(x, \lambda).
		\end{equation*}
		We can rewrite \(\mathcal{L}(x_t, \lambda)\) as:
		\begin{equation*}
			\mathcal{L}(x, \lambda) = \sum_{t=1}^T  x_t \cdot (r_t - \lambda c_t)  + \lambda B.
		\end{equation*}
		To maximize \(\mathcal{L}(x_t, \lambda)\), we can set \(x_t = 1\) whenever \(r_t - \lambda c_t > 0\), and \(x_t = 0\) otherwise. Consequently, \(\mathcal{L}^*(\lambda) = \max_{x_t \in \{0,1\}} \mathcal{L}(x_t, \lambda)\) becomes:
		\begin{equation*}
			\mathcal{L}^*(\lambda) = \sum_{t=1}^{T} (r_t - \lambda c_t)_{+}  + \lambda B,
		\end{equation*}
		where \((z)_{+} = \mathds{1}_{ \{z>0\}} \cdot z\) is the ReLU function. Therefore, the dual problem is:
		\begin{equation} \label{eq:max_delivery_dual}
			\min_{\lambda \geq 0}  \mathcal{L}^*(\lambda) = \min_{\lambda \geq 0}  \sum_{t=1}^{T} \left[ (r_t - \lambda c_t)_{+} + \lambda \cdot \frac{B}{T} \right].
		\end{equation}
	Suppose the problem is feasible and 
	\[
	\lambda^* = \argmin_{\lambda \geq 0} \mathcal{L}^*(\lambda)
	\]
	The KKT conditions indicate that if $\lambda^* > 0$ is the optimal dual variable, budget constraint must satisfy the following:
	\[
	\sum_{t=1}^{T} x_t \cdot c_t = B
	\]
	The optimal bid per impression is determined as: 
	\begin{equation*}
		b_t^* = \frac{r_t}{ \lambda^*}
	\end{equation*}
	The optimal bid per click is given by 
	\begin{equation} \label{eq:md_optimal_formula}
		b^*_{click} = \frac{1}{\lambda^*}
	\end{equation}
	As the optimal bid is constant, under the assumption that \(r_t\) and \(c_t\) are subject to some unknown i.i.d. distribution, the expected cost for each eligible auction opportunity \( \mathbb{E} \left[x_t \cdot c_t \right]\) in this campaign should also be constant. Therefore, the budget spend of the campaign within a time slot \(\Delta \tau\) should be proportional to the number of eligible auction opportunities served during that time slot.,i.e.,
	\[
	\sum_{\tau \leq t \leq \tau + \Delta \tau} x_t c_t \propto  \text{\# of auction opportunities in } (\tau, \tau + \Delta \tau).
	\]
	For more technical details, one may refer to  \cite{fernandez2017optimal}, \cite{lee2013real} and \cite{wang2017display}.

	\textbf{\\Quick summary of our main results}
	
	\highlightbox{
		The optimal bid per click for~\eqref{eq:max_delivery} in the stochastic setting is a constant bid:
		\begin{equation*}
			b_{click}^* = \frac{1}{\lambda^*}
		\end{equation*}
		Suppose supply is sufficient ($T$ big enough), the constant optimal bid  $b_{click}^*$ is the bid per click that exactly depletes the budget, it also suggests that the amount of budget depleted within a time interval is proportional to the number of auction opportunities, i.e.,
		\[
		\sum_{\tau \leq t \leq \tau + \Delta \tau} x_t c_t \propto  \text{\# of auction opportunities in } (\tau, \tau + \Delta \tau).
		\]
	}
	
	\end{section}

	\begin{section}{Cost Cap}
	Cost Cap is a product designed for price-sensitive advertisers. In addition to specifying a budget in max delivery, the advertiser also defines a cost cap, which sets an upper limit on the average cost per result. This ensures that the average cost per result does not exceed the specified cap. Using the notation from the previous section, the cost cap problem for an oCPM daily campaign with click optimization goal can be formulated as follows: 
	\begin{equation}  \label{eq:cost_cap}
		\begin{aligned}
			\max_{x_t \in \{0,1\}} \quad & \sum_{t=1}^T x_t \cdot r_t \\
			\text{s.t.} \quad &  \sum_{t=1}^{T} x_t \cdot c_t \leq B \\
			& \frac{ \sum_{t=1}^{T} x_t \cdot c_t} { \sum_{t=1}^{T} x_t \cdot r_t } \leq C \\
 		\end{aligned}
	\end{equation}
	where 
	\begin{itemize}
		\item \(T\) represents the total (predicted) number of auction opportunities within a day.
		\item \(r_t\) is the predicted click-through rate (CTR) for the \(t\)-th auction.
		\item \(c_t\) denotes the cost for the \(t\)-th auction. In a second-price auction, this corresponds to the highest eCPM 
		\item \(x_t\) is a binary decision variable indicating whether we win the \(t\)-th auction that can be expressed as \(x_t = \mathds{1}_{ \{b_t > c_t \}}\).
		\item \(C\) is the cap for average CPC specified by the advertiser.
	\end{itemize}
	We make the same assumption that the sequences  \(\{r_t\}\)  and \(\{c_t\}\) follow some unknown \(i.i.d.\) distributions.
	
	\subsection {Optimal Solution to Cost Cap Problem}
	We apply the primal-dual method to solve~\eqref{eq:cost_cap}, as was done for the maximum delivery problem. The key difference is that we now have two constraints. The Lagrangian for this problem is given by:
	\begin{equation*} 
		\mathcal{L}(x, \lambda, \mu) = \sum_{t=1}^T x_t \cdot r_t  - \lambda \cdot \left( \sum_{t=1}^{T} x_t \cdot c_t  - B  \right) - \mu \cdot \left[ \sum_{t=1}^{T} x_t \cdot c_t - C \cdot \left(\sum_{t=1}^{T} x_t \cdot r_t \right) \right] 
	\end{equation*}
	The dual  is expressed as:
	\begin{equation*}
		\min_{\lambda \geq 0, \mu \geq 0} \mathcal{L}^*(\lambda, \mu) = \min_{\lambda \geq 0, \mu \geq 0} \max_{x_t \in \{0,1\}} \mathcal{L}(x, \lambda, \mu).
	\end{equation*}
	Note that \(\mathcal{L}(x, \lambda, \mu)\) can be rewritten as:
	\begin{equation*}
		\mathcal{L}(x, \lambda, \mu) = \sum_{t=1}^T  x_t \cdot (r_t - \lambda c_t - \mu c_t + \mu C r_t)  + \lambda B.
	\end{equation*}
	Similarly, to maximize \(\mathcal{L}(x, \lambda, \mu)\), we set \(x_t = 1\) whenever \(r_t - \lambda c_t - \mu c_t + \mu C r_t  > 0\), and \(x_t = 0\) otherwise. \(\mathcal{L}^*(\lambda, \mu) = \max_{x_t \in \{0,1\}} \mathcal{L}(x, \lambda, \mu)\) then becomes:
	\begin{equation*}
		\mathcal{L}^*(\lambda, \mu) = \sum_{t=1}^{T} (r_t - \lambda c_t - \mu c_t + \mu C r_t)_{+}  + \lambda B,
	\end{equation*}
	where $(\cdot)_{+}$ again is the ReLU function.  The dual problem of ~\eqref{eq:cost_cap} is:
	\begin{equation}
		\label{eq:cost_cap_dual}
		\min_{\lambda \geq 0, \mu \geq 0}  \mathcal{L}^*(\lambda, \mu) = \min_{\lambda \geq 0, \mu \geq 0}  \sum_{t=1}^{T} \left[ (r_t - \lambda c_t - \mu c_t + \mu C r_t)_{+} + \lambda \cdot \frac{B}{T} \right].
	\end{equation}
	Suppose we have feasible solution to this problem 
	\[
	\lambda^*, \mu^* = \argmin_{\lambda \geq 0, \mu \geq 0} \mathcal{L}^*(\lambda, \mu)
	\]
	The optimal bid per impression is determined as: 
	\begin{equation*}
		b_t^* = \frac{1 + \mu^* C }{ \lambda^* + \mu^*} \cdot r_t
	\end{equation*}
	The optimal bid per click is given by 
	\begin{equation}
		\label{eq:cost_cap_bid_formula}
		b^*_{click} = \frac{1}{\lambda^* + \mu^*} + \frac{\mu^*}{\lambda^* + \mu^*} \cdot C =  \frac{\lambda^*}{\lambda^* + \mu^*} \cdot \frac{1}{\lambda^*}+ \frac{\mu^*}{\lambda^* + \mu^*} \cdot C
	\end{equation}
    Setting $\alpha = \lambda^*/(\lambda^* + \mu^*)$, we have 
    \begin{equation}
    	b^*_{click}  =  \alpha \cdot \frac{1}{\lambda^*}+ (1-\alpha) \cdot C
    \end{equation}
	Note that \(1 / \lambda^*\) is the optimal bid in max delivery without considering the cost constraint. Therefore, the optimal bid for the cost cap is simply a linear combination of the unconstrained max delivery bid and the cost cap bid. When \(\mu^* \to 0\) (i.e., \(\alpha^* \to 1\)), the cost constraint becomes invalid, and \(b^*_{\text{click}} \to 1 / \lambda^*\), reducing the problem to the max delivery problem. Conversely, when \(\lambda^* \to 0\), the budget constraint becomes invalid, and \(b^* \to  \frac{1}{\mu^{*}}C\), meaning the campaign will bid at the maximum level allowed under the cost constraint.

	\textbf{\\Quick summary of our main results}
	
	\highlightbox{
		The optimal bid per click for cost cap problem ~\eqref{eq:cost_cap} in the stochastic setting is a constant, more specifically, simply a linear combination of the unconstrained max delivery bid and the cost cap bid:
		\begin{equation*}
		b^*_{click} = \alpha \cdot \frac{1}{\lambda^*}+  (1-\alpha)\cdot C
		\end{equation*}
	where $\alpha= \frac{\lambda^*}{\lambda^* + \mu^*} $.  
	}

	\end{section}

	\begin{section}{Remarks}
		\subsection{Knapsack Problem}
		The bidding problems presented in this chapter are closely related to the well-known Knapsack problem. The name "Knapsack" originates from a scenario in which a person aims to fill a fixed-size knapsack with the most valuable items while adhering to weight constraints.
		
		Mathematically, the problem can be formulated as follows: given \(N\) items, each with a weight \(w_i\) and a value \(v_i\), the goal is to select a subset of items such that the total weight does not exceed a given knapsack capacity \(W\), while maximizing the total value of the selected items:
		
		\begin{equation} \label{eq: knapsack}
			 \begin{aligned}
			 	 \max_{x_i \in \{0,1\}} \quad & \sum_{i=1}^N x_i \cdot v_i \\ \text{s.t.} \quad & \sum_{i=1}^{N} x_i \cdot w_i \leq W \\ \end{aligned} 
		\end{equation}
		
		This is known as the \(0\text{-}1\) knapsack problem. Notably, this formulation is essentially identical to the max delivery problem \eqref{eq:max_delivery} by identifying \(v_t \equiv r_t\) (expected clicks) and \(w_t \equiv c_t\) (cost if the impression is won). In our bidding problem, the decision variable admits a concrete implementation via second-price auctions:
		\[
		x_t=\mathds{1}\{b_t>c_t\}.
		\]
		In practice, many auto-bidding/pacing algorithms use a Lagrangian relaxation of the budget constraint, which corresponds to solving the fractional (LP-relaxed) knapsack problem rather than the exact integer program. Since the primal is discrete, strong duality need not hold, and an integrality (duality) gap between the fractional optimum \(U^*\) and the \(0\text{-}1\) optimum \(P^*\) may exist.
		
		For the single-constraint knapsack, the fractional optimum has at most one fractional (marginal/split) item. As a result, the gap is bounded additively by the value of that marginal item:
		\[
		0 \le U^*-P^* \le v_{\text{split}} \le \max_t v_t,
		\]
		which in our setting implies \(U^*-P^* \le r_{\text{split}} \le \max_t r_t\). Therefore, when per-impression values are small (e.g., CTRs) or the campaign aggregates many impressions, the approximation is typically accurate in relative terms.

		For a general introduction to the Knapsack problem, readers may refer to \cite{martello1990knapsack}. A discussion on the online Knapsack problem can be found in \cite{marchetti1995stochastic} and the references cited therein.  The work of \cite{zhou2008budget} was the first to model the max delivery problem as an online Knapsack problem. Readers interested in technical details can refer to this paper for further insights.

		\subsection{Budget Allocation Based on Conversion Per Cost Distribution}
		Recall that, in the max-delivery setting, we concluded that the optimal budget allocation across time intervals is
		proportional to the number of auction opportunities $N_{\tau,\Delta\tau}$. This conclusion follows from the
		assumption that both $\{r_t\}$ (e.g., pCTR) and $\{c_t\}$ (cost) are sampled i.i.d., so that the expected spend
		within a time interval $[\tau,\tau+\Delta\tau]$ satisfies
		\[
		\mathbb{E}\!\left[\sum_{\tau \le t \le \tau+\Delta\tau} x_t c_t\right]
		=
		\mathbb{E}\!\left[\sum_{\tau \le t \le \tau+\Delta\tau} \mathds{1}_{\{r_t > c_t\cdot \lambda^*\}}\, c_t\right]
		=
		N_{\tau,\Delta\tau}\cdot \alpha,
		\]
		where
		\[
		\alpha \;=\; \mathbb{E}\!\left[\mathds{1}_{\{r_t > c_t\cdot \lambda^*\}}\, c_t\right]
		\]
		is the expected spend per auction request under the optimal threshold $\lambda^*$.
		
		In practice, however, $\{r_t\}$ and $\{c_t\}$ are rarely stationary. For example, CTR and clearing prices in the
		morning can be materially different from those later in the day. Under such distribution shifts, allocating
		budget purely based on supply is no longer optimal. Intuitively, we should move budget toward periods with
		higher marginal conversions per dollar until these marginal values equalize; once they are equal, reallocating
		budget does not increase total conversions.
		
		To formalize this, suppose the horizon is divided into $N$ episodes. Within each episode, $r_t$ and $c_t$ are
		i.i.d., but their distributions may differ across episodes. Assume further that each episode exhibits
		diminishing returns, i.e., the conversion-versus-spend curve is concave. We will show that the optimal budget
		allocation (maximizing total expected conversions) equalizes the marginal expected conversions per dollar across
		all episodes where spend is positive.
		
		Let $s_i \ge 0$ be the dollars allocated to episode $i$, and let the total budget be $B$. Denote by $C_i(s)$ the
		expected number of conversions obtained by spending $s$ dollars in episode $i$. The allocation problem can be
		written as
		\begin{equation}\label{eq: allocation_variant}
			\begin{aligned}
				\max_{s_1, s_2, \ldots, s_N} \quad & \sum_{i=1}^N C_i(s_i)\\
				\text{s.t.} \quad & \sum_{i=1}^{N} s_i  = B,\qquad s_i \ge 0.
			\end{aligned}
		\end{equation}

		We can equivalently rewrite~\eqref{eq: allocation_variant} as the following minimization problem:
		\begin{equation}\label{eq: allocation_variant_min}
			\begin{aligned}
				\min_{s_1, s_2, \ldots, s_N} \quad & -\sum_{i=1}^N C_i(s_i)\\
				\text{s.t.} \quad & \sum_{i=1}^{N} s_i = B,\qquad s_i \ge 0.
			\end{aligned}
		\end{equation}
		The Lagrangian is
		\[
		\mathcal{L}(s,\lambda,\nu)
		=
		-\sum_{i=1}^N C_i(s_i)
		+\lambda\!\left(\sum_{i=1}^N s_i - B\right)
		-\sum_{i=1}^N \nu_i s_i,
		\]
		where $\nu_i \ge 0$ are the multipliers associated with the non-negativity constraints.
		
		The KKT conditions are:
		\begin{itemize}
			\item \textbf{Primal feasibility.}
			\[
			\sum_{i=1}^N s_i^* = B,\qquad s_i^* \ge 0.
			\]
			\item \textbf{Dual feasibility.}
			\[
			\nu_i^* \ge 0,\qquad i=1,\ldots,N.
			\]
			\item \textbf{Complementary slackness.}
			\[
			\nu_i^*\, s_i^* = 0,\qquad i=1,\ldots,N.
			\]
			\item \textbf{Stationarity.}
			\[
			\frac{\partial \mathcal{L}}{\partial s_i}
			=
			-C_i'(s_i^*) + \lambda^* - \nu_i^* = 0,\qquad i=1,\ldots,N.
			\]
		\end{itemize}
		
		From complementary slackness, if $s_i^*>0$ then $\nu_i^*=0$, and stationarity gives
		\[
		C_i'(s_i^*)=\lambda^*.
		\]
		If instead $s_i^*=0$, then $\nu_i^*\ge 0$, and stationarity implies
		\[
		C_i'(0)=\lambda^*-\nu_i^* \le \lambda^*.
		\]
		This establishes the desired condition: the optimal allocation equalizes marginal expected conversions per
		dollar across all episodes with positive spend, and assigns zero spend to episodes whose initial marginal return
		is below the common level.

		Readers who are interested in this topic can check out relevant research,e .g., in \cite{kumar2022optimal} \cite{xu2015smart}.

	\end{section}

	\chapter{Throttle-based Pacing}
	
	\intro{
		In this chapter, we dicuss budget pacing algorithms via throttling. In the context of budget pacing, throttling refers to a mechanism that controls a campaign's participation in real-time auctions based on its actual budget spending relative to a target budget, which is determined by the supply pattern. This approach ensures that the ad campaign's budget is distributed in alignment with the supply pattern over its duration, thereby optimizing the campaign's performance.
	}
	
\begin{section}{Probabilistic Throttling}
	In throttle-based pacing, ad campaigns participate in online auctions with a fixed, pre-defined bid. We first consider a daily max-delivery campaign. If the fixed bid is set inappropriately and it's relatively high, it is highly likely that the campaign will win the majority of auction opportunities early on. As a result, the campaign may overspend at the start, rapidly exhausting the budget well before the end of the day. The probabilistic throttling algorithm is a mechanism used to control the participation of a campaign in real-time auctions based on its current spending relative to a target budget. The algorithm operates by dynamically adjusting a participation probability $p(t)$ that determines whether the campaign enters or skips a given auction. If the campaign is currently over-delivered (i.e., actual spending exceeds the expected spending), $p(t)$ is decreased, making it less likely to participate in the current auction, and vice versa when the campaign is under-delivered. This probabilistic control ensures that the campaign's budget is reasonably distributed over its time horizon while maximizing performance opportunities. 
	
	Modifications can be made to adapt this algorithm for a cost cap setting, where an additional performance constraint is imposed to ensure that the campaign achieves its goals within a specified cost threshold.
	
	\subsection{Throttling for Max-Delivery Campaign}
	From the discussion in \autoref{sec:md_optimal}, still assumming i.i.d. distributions of the conversion rates $\{r_t\}$ and costs $\{c_t\}$, we know that the optimal budget allocation is achieved when the budget for each duration is distributed proportionally to the total number of eligible auction opportunities (supply) available during that period. 
	
	Suppose the prediction model estimates there are \(T\) auction opportunities for this campaign within a day(in practice, \(T\) is typically derived by analyzing historical time series data, and the prediction accuracy of \(T\) at the campaign level may vary, which can degrade the performance of the pacing algorithm. Some online adjustments might be implemented to reduce the prediction noise; however, we will not discuss those techniques here. For simplicity, we assume the prediction is perfect). At the \(t\)-th auction, with a perfect pacing algorithm, the spend should be \(\alpha(t) = \frac{t}{T} \cdot B\), where \(B\) is the total budget. If the actual spend \(S(t) > \alpha(t)\), meaning pacing is ahead of schedule, we should slow down the pacing rate. An intuitive approach is to set a participation probability \(p(t)\), which determines the likelihood of the campaign participating in the \(t\)-th auction. In the case of over-delivery, we lower \(p(t)\) to reduce the chance of participating in the auction, thereby decreasing the likelihood of spending during this round. Mathematically, we can update \(p(t)\) by multiplying it by \(1 - \lambda_t\), where \(\lambda_t >0 \) is a control parameter to adjust the throttling level. Conversely, if the campaign is under-delivered, we increase \(p(t)\) by multiplying it by \(1 + \lambda_t \). Mathematically, the update rule of $p(t)$ can be expressed as follows: 
	\[
	p(t) =
	\begin{cases} 
		\min \left\{ p(t-1)\cdot (1 + \lambda_t), 1 \right\}& \text{if } S(t) \leq \alpha(t), \\
		\max \left\{   p(t-1)\cdot (1 - \lambda_t), 0 \right\} & \text{if } S(t) > \alpha(t).
	\end{cases}
	\]
	This motivates the following Algorithm~\ref{alg:throttling_pacing}:
	
		\begin{algorithm} 
			\caption{Throttling-based Budget Pacing Algorithm}
			\label{alg:throttling_pacing}
			\begin{algorithmic}[1]
				\Require $B$: Total budget of the campaign
				\Require $T$: Total number of auction opportunities
				\Require $t$: Current auction round
				\Require $S(t)$: Spend so far at  $t$-th auction
				\Require $p(t)$: Throttling probability at $t$-th auction
				\Require $\{\lambda_t\}$: Control parameters for throttling adjustment
				
				\State Initialize $p(0) \gets 1.0$ and $S(0) \gets 0.0$
				
				\For{each auction at $t$-th auction}
				\State Calculate target spend: $target\_spend \alpha(t) \gets \frac{t}{T} \times B$
				\If{$S(t) \leq \alpha(t)$}
				\State Increase throttling probability: $p(t) \gets \min \{ 1.0, p(t) \cdot (1 + \lambda_t \}$
				\Else
				\State Decrease throttling probability:  $p(t) \gets \max \{ 0.0 , p(t) \cdot (1 - \lambda_t \}$
				\EndIf
				
				\State Generate a random number $r \in [0, 1]$
				\If{$r \leq p(t)$}
				\State Participate in the auction and get the spend in current auction $c_t$
				\State Update spend: $S(t) \gets S(t-1) + c_t$
				\Else
				\State Skip the auction
				\EndIf
				\EndFor
				
			\end{algorithmic}
		\end{algorithm}
	In practice, to simplify the implementation, we may set \(\lambda_t\) as a constant, e.g. 10\%, as in \cite{LinkedInPacing}.  Also, there is no need to update $p(t)$ for every auction, we may set the update granularity to, say, 1 minute.  More technical implementation details could be found in \cite{LinkedInPacing}.  The regret analysis and the optimality of throttle-based pacing can be found in \cite{chen2024dynamic}, which also includes a comparison between throttle-based pacing and bid-based pacing, both of which we will introduce in the subsequent sections.


\end{section}

\section{Remarks}
Historically, throttle-based pacing was a natural choice for fixed-bid and manual-bidding campaigns. Under manual bidding, the advertiser specifies the bid, and in some products the advertising platform is not permitted to modify it. In this setting, the platform must control budget consumption through another decision variable. The participation probability provides a simple mechanism for doing so, because it changes the number of auctions entered without changing the bid submitted in any entered auction. Consequently, probabilistic throttling was adopted in several early advertising systems, including those described in \cite{LinkedInPacing,xu2015smart}. Nevertheless, bid modification and throttling both appeared in the early literature. It is therefore more accurate to view throttling as a common early design choice, particularly for fixed-bid campaigns, rather than as the only pacing mechanism \cite{xu2015smart,chen2024dynamic}.

As the industry shifted from manual bidding toward auto-bidding, advertising platforms obtained more flexibility to determine an auction-level bid on behalf of an advertiser. The bid can be adjusted according to the predicted value of the current auction opportunity and the advertiser's performance objective, such as maximizing conversions subject to a cost-per-action or return-on-spend constraint. In this setting, applying a pacing multiplier directly to the value-based bid is a natural design. Consequently, bid-based pacing has become increasingly prominent, and recent work has studied how budget pacing and performance-oriented auto-bidding should be coordinated or jointly optimized \cite{balseiro2024field}. This development should not be interpreted as a complete replacement of throttling by bid-based pacing. Rather, the two mechanisms expose different control interfaces and can be complementary.

Throttle-based pacing remains useful in production because it controls spending directly through auction participation. In contrast, bid-based pacing controls spending indirectly through the relationship between the submitted bid and the probability of winning. This bid-to-win relationship may be non-smooth and may change over time, so a bid adjustment does not always result in an immediate or predictable change in spend \cite{xu2015smart}. Therefore, throttling can serve as a responsive guardrail on top of an auto-bidding system. For example, the platform may reduce the participation probability when a campaign is close to exhausting its budget, when the observed supply is substantially higher than expected, when spend feedback is delayed, or when the bid optimizer is temporarily unstable. Since the throttling module is decoupled from bid calculation, it can be combined with different bidding algorithms without changing their internal logic. Recent production systems continue to use probabilistic throttling to mitigate over-delivery and improve the robustness of budget control \cite{garg2025smartfastfinish}.

Besides its role in budget control, probabilistic throttling can be useful for exploration and measurement. Because a participating campaign keeps its original bid, a campaign with a small participation probability may still occasionally compete at its full bid and explore market segments in which an aggressively reduced bid would never win \cite{chen2024dynamic}. Moreover, when the participation probabilities are logged, the random eligibility decisions provide quasi-experimental variation that can be used to estimate advertising effects \cite{gui2022auction}. In a modern advertising system, a practical architecture may therefore use bid-based pacing as the primary optimization mechanism and throttle-based pacing as a protection, fallback, or measurement layer.

	\chapter{PID Controller}
	
	\intro{
		In this chapter, we discuss how to leverage the PID controller to design pacing algorithms. We begin with a brief introduction to the fundamental principles of the PID control method. We then apply this method to design pacing algorithms for two key problems: maximum delivery and cost cap optimization.	
	}

	\section{Introduction to PID Controllers}
	
	A Proportional-Integral-Derivative (PID) controller is a widely used feedback-based control loop mechanism. It is designed to maintain a desired output by minimizing the error \( e(t) \), which is the difference between a desired setpoint \( r(t) \) and the measured process variable \( y(t) \). The PID controller achieves this by adjusting the control input \( u(t) \) based on three terms: proportional, integral, and derivative. The output of a PID controller, \( u(t) \), is given by:
	
	\begin{equation} \label{eq:pid_formula}
		u(t) = K_p e(t) + K_i \int_{0}^{t} e(\tau) \, d\tau + K_d \frac{d e(t)}{dt},
	\end{equation}
	where:
	\begin{itemize}
		\item \( e(t) = r(t) - y(t) \) is the error signal,
		\item \( K_p \) is the proportional gain, controlling the response proportional to the error,
		\item \( K_i \) is the integral gain, reducing steady-state error by integrating the error over time,
		\item \( K_d \) is the derivative gain, predicting future error by calculating the rate of change of the error.
	\end{itemize}
	
	These three terms can be interpreted as follows:
	\begin{itemize}
		\item \textbf{Proportional term} \( K_p e(t) \) provides an immediate response proportional to the current error.
		\item \textbf{Integral term} \( K_i \int_{0}^{t} e(\tau) \, d\tau \) accumulates past errors, addressing steady-state error by applying corrective action based on the error history. 
		\item \textbf{Derivative term} \( K_d \frac{d e(t)}{dt} \) predicts future errors by responding to the rate of change of the error.
	\end{itemize}
	
	In practice, the discrete form of \eqref{eq:pid_formula} is often used for implementation:
	
	\[
	u(t) = K_p e(t) + K_i \sum_{\tau=1}^{t} e(\tau) \Delta t + K_d \cdot \frac{e(t) - e(t-1)}{\Delta t}.
	\]

	\begin{section}{PID Contoller in Max Delivery}
		\subsection{Main Algorithm}
		In previous chapter, we mentioned that, optimally, the amount of budget depleted within a time interval should be proportional to the number of auction opportunities during that time interval, i.e.,

		\[
		\sum_{\tau \leq t \leq \tau + \Delta \tau} x_t c_t \propto \text{\# of auction opportunities in } (\tau, \tau + \Delta\tau).
		\]
		
		Remember that the optimal bid is the bid that just depletes the campaign's budget, assuming there is abundant supply. In practice, we can first construct the target spend per interval and use this target spend as the setpoint to apply the PID controller discussed above. Suppose we have a CPC campaign with a daily budget of \$100. We bucketize one day by choosing the bucket interval $\Delta\tau =$ 15 minutes and obtain the predicted number of eligible requests (auction opportunities) for these intervals from the prediction model, as shown in \autoref{fig:flow_ratio_15min}.
		
		\begin{figure}[H]
			\centering
			\includegraphics[width=0.99\textwidth]{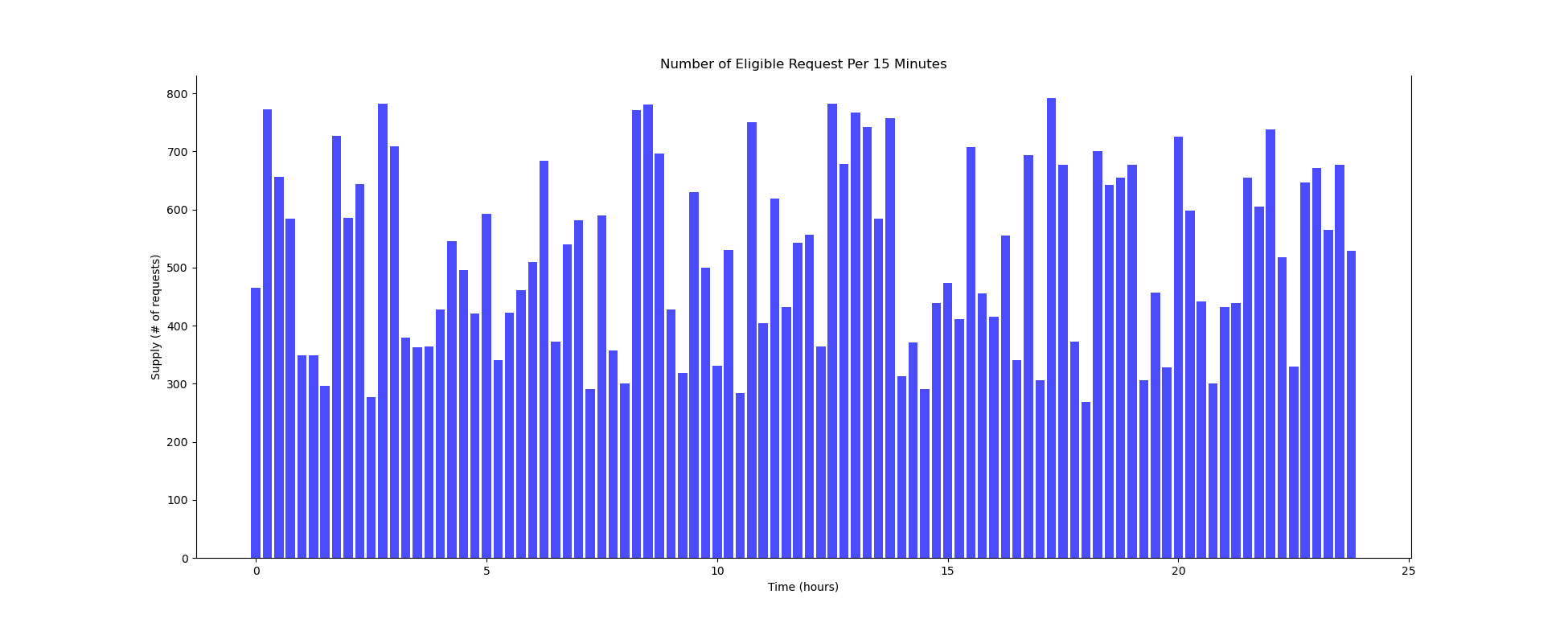}
			\caption{Supply Pattern (Number of Eligible Requests Per 15 Minutes)}
			\label{fig:flow_ratio_15min}
		\end{figure}
		
		Based on the supply pattern in \autoref{fig:flow_ratio_15min}, we can construct the target spend per bucket(15-minute interval) for this campaign. For simplicity, within each target spend interval $\Delta\tau$, we assume the budget is consumed linearly over time. Suppose \(TS(t)\) is the target spend in the \(t\)-th interval, \(NR(t)\) is the number of eligible requests in the \(t\)-th interval, \(B = 100\) is the total daily budget, and \(T = 96\) is the number of target spend intervals. The proportional budget allocation rule per interval is given by:
			\[
				TS(t) = \frac{NR(t)}{\sum_{s=1}^{T} NR(s)} \cdot B
			\]
			
		The per-interval target spend is then computed and plotted in \autoref{fig:target_spend_15min}:
		
		\begin{figure}[H]
			\centering
			\includegraphics[width=0.99\textwidth]{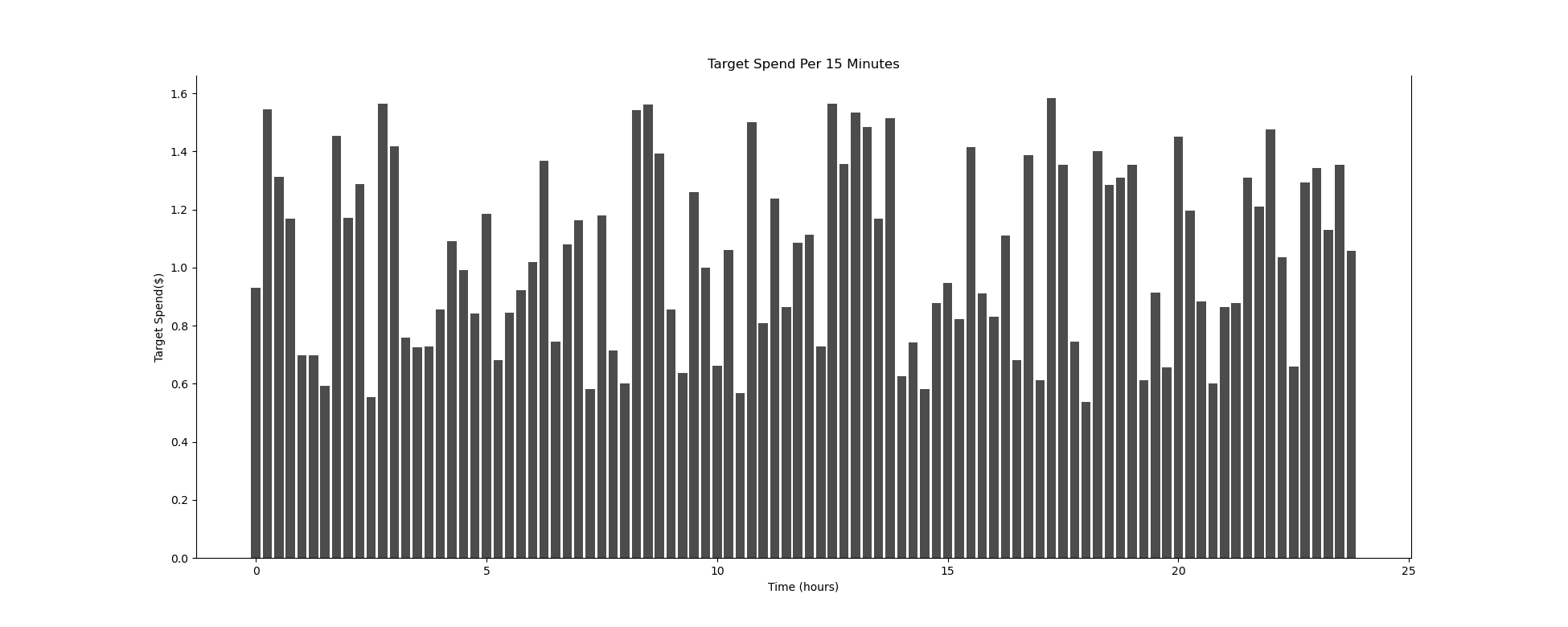}
			\caption{Target Spend Per 15 Minutes}
			\label{fig:target_spend_15min}
		\end{figure}
	
	We summarize the budget allocation algorithm as follows:
	
		\begin{algorithm}[H]
		\caption{Compute Target Budget per Bucket}
		\label{alg: budget_allocation}
		\begin{algorithmic}[1]
			\Require $B$: Total daily budget
			\Require $NR(t)$: Number of eligible requests in interval $t$
			\Require $T$: Total number of buckets in a day
			\State Initialize $TS(t) \gets 0$ for all $t = 1, \dots, T$
			\State Compute the total eligible requests:
			\[
			\text{TotalRequests} \gets \sum_{t=1}^{T} NR(t)
			\]
			\For{$t = 1$ to $T$}
			\State Compute the proportional share of the budget for bucket interval $t$:
			\[
			TS(t) \gets \frac{NR(t)}{\text{TotalRequests}} \cdot B
			\]
			\EndFor
			\State \Return $TS(t)$ for all $t = 1, \dots, T$
		\end{algorithmic}
	\end{algorithm}
	
		Suppose the bid price gets updated every \(\Delta t\) time (\(\Delta t\) is a tunable bid update interval, which we may set to, e.g., 1 minute) at \(t_0, t_1, t_2, \cdots, t_N\), where \(N = \text{MinutesInOneDay} / \Delta t\). We define the error factor and control signal as:
		
		\[
		\begin{aligned}
			e(t_k) &= r_{t_k} - s(t_k), \\
			u(t_{k}) &\gets K_p e(t_k) + K_i \sum_{j=1}^{k} e(t_j) \Delta t + K_d \frac{\Delta e(t_k)}{\Delta t},
		\end{aligned}
		\]
		
		where:
		\begin{itemize}
			\item \(r_{t_k}\): Observed spend during the \(k\)-th pacing update interval,
			\item \(s(t_k)\): Target spend derived by proportionally allocating within the target budget interval \(\Delta\tau\) that contains it, e.g., if \([t_{k-1}, t_k]\) lies within the \(t\)-th target budget interval, then 
			\[
			s(t_k) = \frac{\Delta t}{\Delta\tau} \cdot TS(t),
			\]
			where \(TS(t)\) is the target spend for the \(t\)-th interval.
			
		\end{itemize}
		
		The bid can be updated by leveraging an actuator that takes the current control signal \(u(t_k)\) to adjust the current bid price \(b(t_k)\) as:
		
		\[
		b(t_{k+1}) \gets b(t_k) \exp\{u(t_{k})\}.
		\]
		
		The PID algorithm for max delivery problem is summarized in \autoref{alg:pid_md}. For more in-depth knowledge on applying PID controllers to the max delivery pacing problem, one may refer to \cite{US20160110755A1},  \cite{wang2017display} and \cite{zhang2016feedback}.
		
		\begin{algorithm}
			\caption{PID Controller for Max Delivery}
			\label{alg:pid_md}
			\begin{algorithmic}[1]
				\Require $\Delta t$: Time interval for bid updates(e.g. 1 minute)
				\Require $T$: Total campaign duration
				\Require $B$: Total campaign budget
				\Require $\Delta \tau$: Time interval of target budget per bucket
				\Require $\{ TS(t) \}$: Target spend for each target budget interval $t$
				\Require $K_p, K_i, K_d$: PID controller gains
				\State Initialize $b(t_0) \gets \text{initial\_bid}$
				\State Initialize $u(t_0) \gets 0$
				\State Initialize cumulative error: $CE \gets 0$
				\State Initialize previous error: $PE \gets 0$
				
				\For{$k = 1$ to $N$} \Comment{$N = \text{MinutesInOneDay} / \Delta t$}
				\State Find the $t$-th target budget interval that contains $t_k$ update interval, compute:
				\[
				s(t_k) \gets \frac{\Delta t}{\Delta\tau} \cdot TS(t)
				\]
				\State Measure observed spend during the interval \(r(t_k)\)
				\State Compute the error factor:
				\[
				e(t_k) \gets  s(t_k) - r(t_k)
				\]
				\State Update the cumulative error:
				\[
					CE \gets CE + e(t_k) \cdot \Delta t
				\]
				\State Update the control signal using the PID formula:
				\[
				u(t_k) \gets K_p \cdot e(t_k) + K_i \cdot CE + K_d \cdot \frac{e(t_k) - PE}{\Delta t}
				\]
				\State Update the per click bid price:
				\[
				b(t_k) \gets b(t_{k-1}) \cdot \exp(u(t_k))
				\]
				\State Update the previous error:
				\[
					PE \gets e(t_k)
				\]
				\State Get click through rate \(r_t\) from the prediction model 
				\State Compute the per impression bid 
				\[
					b_t = b(t_k) \cdot r_t
				\]
				\EndFor
			\end{algorithmic}
		\end{algorithm}
		
		\subsection{Bidding Dynamics}
		We show here the detailed bidding dynamics for this campaign. \autoref{fig:flow_ratio_map} is a simplified plot demonstrating how the PID controller operates in a real-world dynamic environment. 
		\begin{figure}[H]
			\centering
			\includegraphics[width=0.99\textwidth]{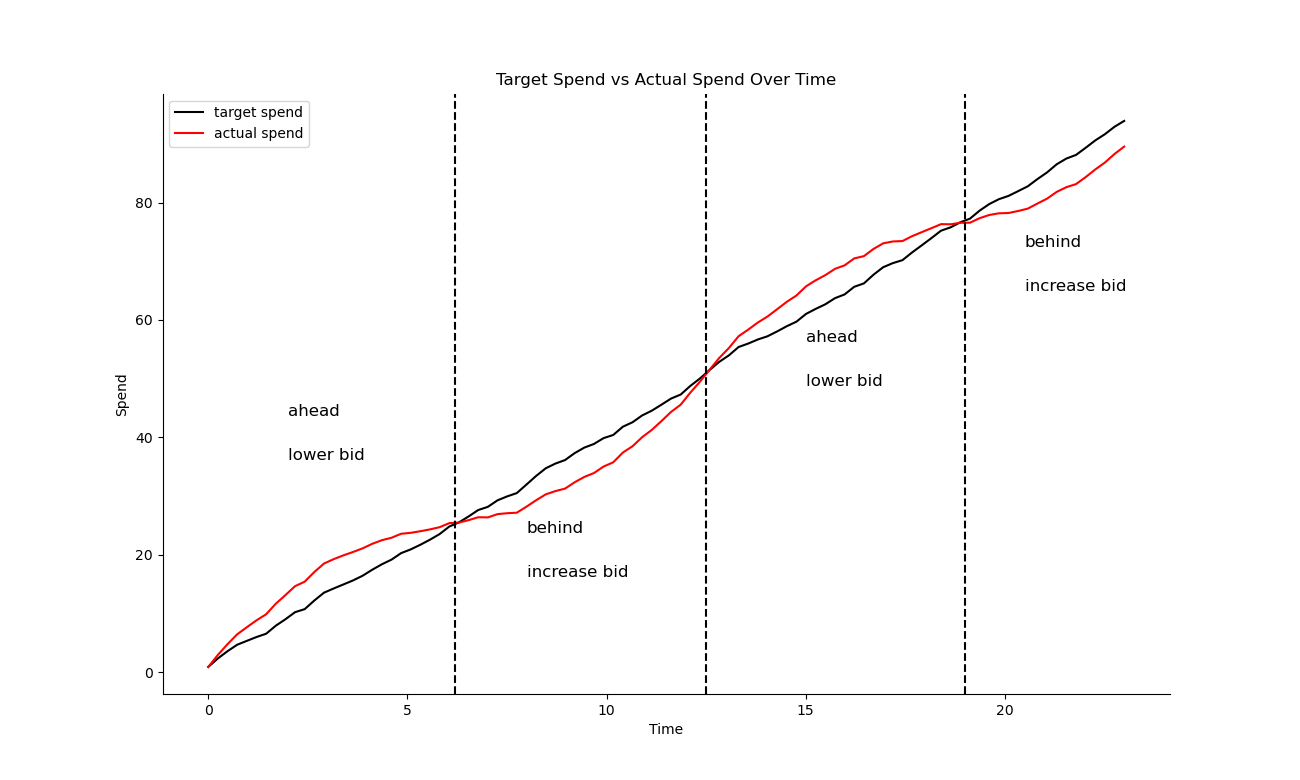}
			\caption{Target Spend vs Actual Spend Over Time}
			\label{fig:flow_ratio_map}
		\end{figure}
	The dynamics of a PID controller adjusts bids based on the relationship between the \textbf{target spend} (black line) and the \textbf{actual spend} (red line) over time. Target spend represents the ideal cumulative budget spending at any point in time to evenly distribute the budget. Actual spend reflects the real cumulative spend achieved by the campaign over time.  The graph highlights two key conditions: being \textbf{ahead of the schedule} or \textbf{behind the schedule}, and how these conditions affect bid modulation.
	
	If the delivery is behind the schedule, i.e. when the actual spend (red line) is below the target spend (black line), the campaign is under-delivering. The PID controller \textbf{increases the bid} to catch up with the target spend. Higher bids make the campaign more competitive in auctions, increasing the likelihood of winning more impressions and spending more. In the plot, the actual spend curve slopes upward more steeply after the "increase bid" label in the "behind" regions; If the delivery is ahead of the schedule, i.e. when the actual spend (red line) exceeds the target spend (black line), the campaign is over-delivering. The PID controller \textbf{lowers the bid} to slow down the spending rate and realign with the target spend. Lower bids reduce the competitiveness of the campaign in auctions, resulting in fewer impressions and slower spend. In the plot, the actual spend curve becomes less steep after the "lower bid" label in the "ahead" regions.
	
	This feedback mechanism ensures a balanced distribution of the budget over time, optimizing performance while adhering to pacing constraints.
	
	\subsection{Practical Considerations}
	We list here some practical considerations for implementing the PID controller in real-world production sytem:
	\begin{itemize}
		\item  \textbf{How to choose the error signal \(e(t_k)\) ?}  different choices of this error signals, for example, one can choose instantaneous gap of a pacing interval or choose the cumulative difference between actual spend and expected spend.  
		\item \textbf{Constructing budget allocating distribution}: aggregation across different dimensions, more details TBA.
		\item \textbf{Normalization of error signal}: The error factor \(e(t_k)\) in \autoref{alg:pid_md} is defined as \(r(t_k) - s(t_k)\), the gap between actual spend and target spend. However, the issue is that the scale of the budget varies significantly across different campaigns, requiring different controller gains to accommodate these budget fluctuations. In practice, it is challenging to maintain different controller gains for each campaign, and it is more common for all ad campaigns to share the same controller gains. In this situation, a better approach is to compute the error factor in a normalized way, i.e.,
		\[
		e(t_k) = 1 - \frac{r(t_k)}{s(t_k)},
		\]
		which normalizes the error factors for campaigns with different budget scales to a common scale.
		\item \textbf{Different Actuator}: The update rule in \autoref{alg:pid_md} is \(b(t_k) \gets b(t_{k-1}) \cdot \exp(u(t_k))\), where the \(\exp\) function is chosen as the actuator. In practice, other functions can also be used as the actuator function, such as linear functions or sigmoid functions.
		
		
		\item \textbf{Choice of \(\Delta \tau\) and \(\Delta t\)}: Some trade-offs should be considered when choosing the target spend bucket \(\Delta \tau\) and the bid update interval \(\Delta t\). 
		
		\begin{itemize}
			\item \textbf{Target Spend Bucket \(\Delta \tau\)}: 
			\begin{itemize}
				\item A small \(\Delta \tau\) provides finer-grained predictions of the supply pattern, reducing the accuracy requirements for interpolation within the bucket. However, the prediction itself may become noisier.
				\item A large \(\Delta \tau\), on the other hand, gives a more accurate estimate of the overall supply. However, the assumption of a linear distribution of supply within \(\Delta \tau\) is less likely to hold, which can negatively impact interpolation accuracy.
			\end{itemize}
			\item \textbf{Bid Update Interval \(\Delta t\)}:
			\begin{itemize}
				\item A small \(\Delta t\) results in more responsive updates to the market, but the computed error factor may contain more noise.
				\item A large \(\Delta t\) helps mitigate statistical noise in the error factor, but delayed bid updates may fail to respond to market changes in a timely manner.
			\end{itemize}
		\end{itemize}
		
		These two parameters can be tuned online. In practice, we find that setting \(\Delta \tau\) to a duration ranging from a few minutes to an hour, and setting \(\Delta t\) to a comparable range, typically works well.
		
		\item \textbf{CPC vs. oCPM}: For an oCPM campaign, spend is generated when an impression is served, so the spend signal can usually be made available to the controller shortly after the auction. For a CPC campaign, however, spend is generated only when a click is observed. Since a click occurs after the corresponding impression and the click event must be delivered through the feedback pipeline, the spend signal used by the PID controller is delayed \cite{US20160110755A1,xu2015smart}. If this delay is ignored, the controller may interpret the temporarily low observed spend as under-delivery and continue increasing the bid, which can lead to oscillation or overspending after the delayed clicks arrive. In practice, we may use a longer update interval, smaller controller gains, or an estimate of the pending spend based on recently served impressions and their predicted CTRs.
		
		\item \textbf{Uncertainty of \(T\)}: The construction of the target spend \(\{TS(t)\}\) assumes that the future supply pattern, including the total number of auction opportunities over the campaign horizon, can be predicted accurately. In practice, the realized traffic may deviate from the prediction because of day-to-day fluctuations or unexpected traffic spikes. If the remaining supply is overestimated, the controller may pace too conservatively and leave part of the budget unspent; if it is underestimated, the campaign may spend too aggressively at the beginning. Since the PID controller only tracks the given target spend, it cannot by itself correct a misspecified target. A practical system may therefore periodically update the remaining-supply prediction and reconstruct the target spend, or combine the predicted schedule with a robust online allocation rule \cite{balseiro2023online,esfandiari2015online,an2024best}.
		
		\item \textbf{Non-i.i.d. Distribution}: The proportional budget allocation discussed above is motivated by an i.i.d. assumption, under which auction opportunities have the same distribution across time. In practice, CTR, conversion rate, and market price may vary systematically by time of day, user segment, and inventory source. Therefore, two intervals with the same number of eligible requests may have very different expected values and spending potentials. A PID controller can reduce the gap between actual spend and target spend, but it cannot correct a target that is constructed from supply volume alone. In this case, the target allocation may also incorporate the predicted quality and cost of future supply, or use an online resource allocation algorithm designed for time-varying input distributions, such as the adversarial stochastic input model studied in \cite{devanur2011near}.
		
	\end{itemize}
	
	\end{section}

	\begin{section}{PID Contoller in Cost Cap}
		We present several PID control-based algorithms to solve the cost cap problem.
		\subsection{Cost-Min Alogrithm}
	The idea behind the "Cost-Min" algorithm is to adaptively compute the maximum bid that can be submitted to achieve the cost control goal and set this as the upper bound for the normal bid price, which is computed by only considering the budget constraint. Under the configuration of \autoref{eq:cost_cap}, a naive implementation of the algorithm is as follows: the total budget is $B$ and the upper bound of the average cost per click is $C$. This means we need to collect at least $N = B/C$ clicks if the budget is completely depleted. At any time $t$, suppose the actual accumulated spend is $S_t$ and the accumulated number of observed clicks is $N_t$. The remaining budget is $B - S_t$, and the remaining click goal is $N - N_t$. The new upper bound of the average cost for the remaining delivery is then:
	\[
	U_t = \frac{B - S_t}{N - N_t}.
	\]
	
	Suppose $b_t$ is the bid price derived from the max-delivery algorithm without the cost constraint. The "Cost-Min" algorithm sets the final bid as:
	\[
	\hat{b}_t = \min \{b_t, U_t\}.
	\]
	
	During each bid update interval $\Delta t$, we collect the signals for the actual spend and the number of conversions (in this case, clicks), derive the remaining budget and the remaining target number of conversions, and compute the up-to-date upper bound for the bid. At the same time, we update the delivery bid as if there were no cost constraints using the PID controller discussed in the previous section. The new bid is then set to the minimum of the delivery bid and the upper bound bid. The "Cost-Min" algorithm is summarized in \autoref{alg:cost_min}.
	
	\begin{algorithm}[H]
		\caption{Cost-Min Algorithm for Cost Cap}
		\label{alg:cost_min}
		\begin{algorithmic}[1]
			\Require $B$: Total budget, $C$: Upper bound for the average cost per click, $\Delta t$: Bid update interval
			\Require $b_t$: Bid price computed by max-delivery without cost constraint
			\Ensure $\hat{b}_t$: Final bid price considering cost cap constraint
			\State Compute the total click target: $N = \frac{B}{C}$
			\State Initialize: $U_0 \gets C$ \Comment{Initial Bid Upper Bound}
			\State Initialize: $S_0 \gets 0$, $N_0 \gets 0$ \Comment{Accumulated spend and clicks}
			\For{each bid update interval $\Delta t$}
			\State Observe the spend for the interval: $\Delta S_t$
			\State Observe the number of clicks for the interval: $\Delta N_t$
			\State Update accumulated spend: $S_t \gets S_{t-1} + \Delta S_t$
			\State Update accumulated clicks: $N_t \gets N_{t-1} + \Delta N_t$
			\State Compute remaining budget: $B_{\text{r}} = B - S_t$
			\State Compute remaining click goal: $N_{\text{r}} = N - N_t$
			\State Compute the updated upper bound for the average cost:
			\[
			U_t = \frac{B_{\text{r}}}{N_{\text{r}}}
			\]
			\State Update the delivery bid $b_t$ without considering cost constraints using PID control(e.g., \autoref{alg:pid_md})
			\State Compute the final bid price:
			\[
			\hat{b}_t = \min \{b_t, U_t\}
			\]
			\EndFor
			\State \Return $\hat{b}_t$
		\end{algorithmic}
	\end{algorithm}
	
	The upper bound \( U_t \) discussed here represents the target cost per result after time \( t \). If we bid using \( U_t \) and win the auction, then under a second-price auction mechanism, the actual payment per impression corresponds to \( eCPM_2 \), the second-highest eCPM. Effectively, the average cost per conversion is given by:
	\[
	\frac{eCPM_2}{CTR_1},
	\]
	where \( CTR_1 \) denotes the click-through rate (CTR) of this campaign. We can rewrite this expression as:
	
	\[
	\frac{eCPM_2}{CTR_1} = \frac{eCPM_2}{CTR_1 \cdot U_t} \cdot U_t = \frac{eCPM_2}{eCPM_1} \cdot U_t = \sigma \cdot U_t,
	\]
	where \( \sigma \) represents the ratio between the highest eCPM and the second-highest eCPM. Consequently, the actual cost per click (CPC) is given by:
	
	\[
	\sigma \cdot U_t,
	\]
	which is lower than the upper bound \(U_t\). If we assume that the ratio \( \sigma \) remains relatively stable over time, we can relax this constraint slightly. In the algorithm described above, we may compute \( U_t \) as:
	
	\[
	U_t  = \frac{B_r}{N_r \cdot \sigma},
	\]
	For more technical details on the "Cost-Min" algorithm, one may refer to \cite{kitts2017ad}.

		\subsection{Dynamic Cap}
	The "Cost-Min" algorithm is static and conservative in some sense. $U_t$ represents the upper bound for the average cost per conversion in the remaining delivery schedule. For a single auction opportunity, the bid could go higher as long as the average cost is controlled under the target cap. In practice, we can make the cap more dynamic and responsive to the real-time cost control quality.
	
	To understand how we should dynamically tweak the bid in real-time, we ignore the budget constraint and consider only the optimization problem with the cost control constraint. Specifically, we want to explore how the upper bound of the bid can impact conversions and cost control. Suppose $u$ is the upper bound; the number of conversions for an auction opportunity given $u$ is a random variable denoted by $M_r(u)$, and the cost per conversion for this auction is $M_c(u)$. Our goal is to solve the following optimization problem:
	\begin{equation*}
		\begin{aligned}
			\max_{u} \quad & \mathbb{E} \left[ M_r(u) \right] \\
			\text{s.t.} \quad & \mathbb{E} \left[ M_c(u) \right] \leq C
		\end{aligned}
	\end{equation*}
	where $C$ is the target cost cap.
	
	Both $\mathbb{E} \left[ M_r(u) \right]$ and $\mathbb{E} \left[ M_c(u) \right]$ are monotonically non-decreasing with respect to $u$ (i.e., the higher the bid, the more likely you are to win the auction, and the higher the price you pay). The solution to the optimization problem is given by:
	\[
	\mathbb{E} \left[ M_c(u) \right] = C.
	\]
	
	This is equivalent to solving:
	\begin{equation}
		u^* = \argmin_{u} \mathbb{E} \left[ M_c(u) - C \right]^2.
	\end{equation}
	
	To solve this, we can apply the Robbins-Monro algorithm (see \cite{robbins1951stochastic}) and iteratively update $u$ using:
	\[
	u \gets u - \epsilon \cdot \nabla_u \left[ M_c(u) - C \right]^2 = u - \epsilon \cdot 2 M_c'(u) \cdot \left[ M_c(u) - C \right],
	\]
	where $M_c'(u) \geq 0$ because $M_c(u)$ is monotonically non-decreasing. Setting $\epsilon'(u) = \epsilon \cdot 2 M_c'(u) \geq 0$, the update rule becomes:
	\[
	u \gets u - \epsilon'(u) \cdot \left[ M_c(u) - C \right].
	\]
	
	This is essentially a proportional controller (P-controller). When the actual cost per acquisition (CPA) is less than the target $C$, we increase the dynamic bid cap (upper bound); otherwise, we decrease the bid cap.
	
	Based on the analysis above, we design the algorithm for the dynamic bid cap as follows: for every update interval $\Delta t$, we collect the actual spend and actual conversions, compute the actual average cost per conversion, and based on the gap between the actual CPA and the target cost cap, we update the upper bound using a proportional controller (for simplicity, in the actual implementation, we may choose a fixed $\epsilon$ as the controller gain). The dynamic bid cap variant of the "Cost-Min" algorithm is summarized in the following \autoref{alg:dynamic_cost_min}:

	\begin{algorithm}[H]
		\caption{Dynamic Bid Cap Variant of "Cost-Min" Algorithm}
		\label{alg:dynamic_cost_min}
		\begin{algorithmic}[1]
			\Require $B$: Total budget, $C$: Target cost cap, $\Delta t$: Bid update interval
			\Require $b_t$: Delivery bid computed by max-delivery without cost constraint
			\Require $\epsilon$: Controller gain for the proportional controller
			\Ensure $\hat{b}_t$: Final bid price considering cost cap constraint
			\State Compute the total click target: $N = \frac{B}{C}$
			\State Initialize: $S_0 \gets 0$, $N_0 \gets 0$, $u \gets C$ \Comment{Accumulated spend, clicks, and initial bid cap}
			\For{each bid update interval $\Delta t$}
			\State Observe the spend for the interval: $\Delta S_t$
			\State Observe the number of conversions for the interval: $\Delta N_t$
			\State Update accumulated spend: $S_t \gets S_{t-1} + \Delta S_t$
			\State Update accumulated conversions: $N_t \gets N_{t-1} + \Delta N_t$
			\State Compute the actual CPA: $\text{CPA}_t = \frac{S_t}{N_t}$
			\State Update the dynamic bid cap using the proportional controller:
			\[
			u \gets u - \epsilon \cdot (\text{CPA}_t - C)
			\]
			\State Ensure the bid cap is non-negative: $u \gets \max(u, 0)$
			\State Compute the final bid price:
			\[
			\hat{b}_t = \min\{b_t, u\}
			\]
			\EndFor
			\State \Return $\hat{b}_t$
		\end{algorithmic}
	\end{algorithm}

		\subsection{Dual-PID}
		Recall that the optimal per-click bid for cost cap is given by
		\[
		b_{click}^{*} = \frac{1 + \mu^* C}{\lambda^* + \mu^*},
		\]
		where \(\lambda\) and \(\mu\) are the dual parameters associated with the budget and cost constraints, respectively. The KKT conditions imply that when these constraints are active, we have
		\[
		\sum_{t=1}^{T} x_t \cdot c_t = B, \quad \frac{\sum_{t=1}^{T} x_t \cdot c_t}{\sum_{t=1}^{T} x_t \cdot r_t} = C.
		\]
		The optimal dual parameters are those for which the pacing algorithm exactly depletes the total budget and the cost per click equals the cost cap. This insight motivates the design of a PID controller to update \(\lambda\) and \(\mu\).
		
		For each pacing interval \(t_k\), we collect the spend \(r(t_k)\) and the number of clicks \(n(t_k)\) for the campaign, and compute the average cost per click as
		\[
		c(t_k) = \frac{r(t_k)}{n(t_k)}.
		\]
		We then compare the actual spend \(r(t_k)\) and the actual cost per click \(c(t_k)\) to the target spend \(s(t_k)\) (computed based on the traffic pattern as described in the previous section) and the target cost per click \(C\) (the cost cap), respectively, to define the error signals for \(\lambda\) and \(\mu\):
		\begin{itemize}
			\item Budget constraint error for \(\lambda\):
			\[
			e_{\lambda}(t_k) = r(t_k) - s(t_k)
			\]
			\item Cost constraint error for \(\mu\):
			\[
			e_{\mu}(t_k) = C - \frac{r(t_k)}{n(t_k)}
			\]
		\end{itemize}
		
		The PID update equations for \(\lambda\) and \(\mu\) are then defined as
		\begin{equation*}
			\begin{aligned}
				u_{\lambda}(t_k) &= K_{p, \lambda} \, e_{\lambda}(t_k) + K_{i, \lambda} \sum_{j=0}^{k} e_{\lambda}(t_j) + K_{d, \lambda} \frac{e_{\lambda}(t_k) - e_{\lambda}(t_{k-1})}{\Delta t}, \\
				u_{\mu}(t_k) &= K_{p, \mu} \, e_{\mu}(t_k) + K_{i, \mu} \sum_{j=0}^{k} e_{\mu}(t_j) + K_{d, \mu} \frac{e_{\mu}(t_k) - e_{\mu}(t_{k-1})}{\Delta t}.
			\end{aligned}
		\end{equation*}
		The control signals \(u_{\lambda}\) and \(u_{\mu}\) are then applied to update the dual parameters:
		\[
		\lambda \gets \lambda \cdot \exp\left( u_{\lambda} \right), \quad \mu \gets \mu \cdot \exp\left( u_{\mu} \right).
		\]
		The new bid per click is computed as
		\[
		b_{click} = \frac{1 + \mu \, C}{\lambda + \mu}.
		\]
		
		We summarize this algorithm in \autoref{alg:pid_cost_cap}.

		\begin{algorithm}
		\caption{PID Controller for Cost Cap}
		\label{alg:pid_cost_cap}
		\begin{algorithmic}[1]
			\Require $\Delta t$: Time interval for bid updates(e.g., 1 minute), $\Delta \tau$: Time interval of target budget per bucket
			\Require $T$: Total campaign duration, $B$: Total campaign budget, $C$: Cost per click cap
			\Require $\{ TS(t) \}$: Target spend for each target budget interval $t$
			\Require $K_{p, \lambda}, K_{i, \lambda}, K_{d, \lambda}$: controller gains for budget constraint \(\lambda\) 
			\Require $K_{p, \mu}, K_{i, \mu}, K_{d, \mu}$: controller gains for cost constraint \(\mu\) 
			\State Initialize $\lambda(t_0) \gets \lambda_0, \mu(t_0) \gets \mu_0$
			\State Initialize cumulative errors: $CE_{\lambda} \gets 0, CE_{\mu} \gets 0$
			\State Initialize previous errors: $PE_{\lambda} \gets 0, PE_{\mu} \gets 0$
			
			\For{$k = 1$ to $N$} \Comment{$N = \text{MinutesInOneDay} / \Delta t$}
			\State Find the $t$-th target budget interval that contains $t_k$ update interval, compute:
			\[
			s(t_k) \gets \frac{\Delta t}{\Delta\tau} \cdot TS(t)
			\]
			\State Observe spend \(r(t_k)\) and click \(n(t_k)\) during the interval
			\State Compute the error factor for \(\lambda\) and \(\mu\)
			\[
			e_{\lambda}(t_k) \gets  r(t_k) - s(t_k), \quad e_{\mu}(t_k) \gets C - \frac{r(t_k)}{n(t_k)}
			\]
			\State Update the cumulative error:
			\[
			CE_{\lambda} \gets CE_{\lambda} + e_{\lambda}(t_k) \cdot \Delta t, \quad  CE_{\mu} \gets CE_{\mu} + e_{\mu}(t_k) \cdot \Delta t
			\]
			
			\State Update the control signal using the PID formula:
			\[
			u_{\lambda}(t_k) \gets K_{p, \lambda} \cdot e_{\lambda}(t_k) + K_i \cdot CE_{\lambda} + K_d \cdot \frac{e_{\lambda}(t_k) - PE_{\lambda}}{\Delta t}
			\]
			\[
			u_{\mu}(t_k) \gets K_{p, \mu} \cdot e_{\mu}(t_k) + K_i \cdot CE_{\mu} + K_d \cdot \frac{e_{\mu}(t_k) - PE_{\mu}}{\Delta t}
			\]
			\State Update the dual variables:
			\[
			\lambda(t_k) \gets \lambda(t_{k-1}) \cdot \exp(u_{\lambda}(t_k)), \quad \mu(t_k) \gets \mu(t_{k-1}) \cdot \exp(u_{\mu}(t_k))
			\]
			\State Update the previous error:
			\[
				PE_{\lambda} \gets e_{\lambda}(t_k), \quad 	PE_{\mu} \gets e_{\mu}(t_k)
			\]
			\State Get conversion rate  \(r_t\) from predction model
			
			\State Compute the new per impression bid price:
			\[
			 	b_t = \frac{1 + \mu(t_k) C}{\lambda(t_k) + \mu(t_k)} \cdot r_t
			\]
			\EndFor
		\end{algorithmic}
	\end{algorithm}


		\subsection{Remarks}
		PID-based pacing is simple, computationally efficient, and easy to implement in a large-scale advertising system. However, it does not explicitly model the nonlinear relationship between the bid, auction winning probability, spend, and conversions. For example, a small increase in the bid may have little effect when the bid remains below most competing bids, but the same increase may lead to a large change in spend after the bid crosses a dense region of the market-price distribution. Moreover, this relationship may vary over time as traffic composition, competition, and conversion rates change. Consequently, fixed controller gains that perform well under one market condition may lead to slow convergence, oscillation, or unstable bid updates under another.
		
		This issue is more pronounced in the cost cap problem, where delivery and cost control are coupled. Increasing the bid generally improves budget delivery by increasing the probability of winning auctions, but it may also increase the average cost per conversion. Conversely, decreasing the bid helps control the cost but may lead to under-delivery. In the Dual-PID algorithm, the budget controller and the cost controller update their respective dual parameters based on different error signals, but both parameters affect the same final bid. Therefore, an update intended to reduce the delivery error may worsen the cost error, and an update intended to improve cost control may move the campaign further behind its spending target. Signal delay and noise in spend and conversion measurements may further amplify these interactions. Multivariable control methods can partially address this issue by explicitly accounting for the cross-effects among multiple control objectives \cite{yang2019bid}.
		
		Another possible approach is Model Predictive Control (MPC), which we will discuss in next chapter. Instead of updating the bid solely according to the current and historical errors, MPC uses a model of the system dynamics to predict spend, conversions, and cost over a finite future horizon. At each update interval, the controller solves an optimization problem that selects a sequence of future control actions while jointly considering the remaining budget, the target spending schedule, and the cost constraint. Only the first control action is applied, and the optimization problem is solved again after new auction and performance signals become available. This receding-horizon design allows the controller to anticipate the future effect of a bid update and to explicitly balance delivery and cost control, rather than treating them as two largely independent feedback loops \cite{Camacho2004-pb,chen2026lightweight}.
		
		The additional flexibility of MPC comes at a cost. Its performance depends on the accuracy of the models used to predict the bid-to-spend and bid-to-conversion relationships, and solving a constrained optimization problem at every update interval introduces additional computational complexity. In practice, PID-based methods remain attractive when simplicity, robustness, and low serving latency are the primary concerns, while MPC or other optimization-based approaches may be preferred when the interactions among multiple constraints are sufficiently important and reliable short-term forecasts are available. A hybrid system may also use PID control as a lightweight baseline or fallback mechanism and apply MPC when more accurate models and sufficient computational resources are available.

	\end{section}

	\chapter{MPC Controller}
	
	\intro{
		In this chapter, we discuss the Model Predictive Control (MPC) framework. MPC leverages predictive modeling to plan future actions, enabling more robust control performance. We begin with a brief introduction to MPC, followed by its application to both the maximum delivery and cost cap problems. Additionally, we explore target CPA bidding, a bidding product similar to cost cap, and demonstrate how MPC can be applied to optimize this new bidding strategy.
	}
	
	\section{Introduction to Model Predictive Control}
	
	\subsection{From PID to MPC}
	In the previous chapter, we explored how to apply PID control to solve both the max delivery and cost cap problems. While PID controllers provide a simple and effective framework for designing pacing algorithms, they lack the ability for future planning, making them sometimes myopic and leading to unstable pacing dynamics. This can be detrimental to delivery performance, especially in complex problems like cost cap, where multiple constraints must be taken into account.
	
	Model Predictive Control (MPC) is a more advanced framework that addresses these challenges by incorporating predictive modeling and optimization into the control process. Unlike PID controllers, which rely solely on feedback to correct errors, MPC leverages a dynamic model of the system to predict its future states over a defined time horizon. This forward-looking capability enables MPC to make informed decisions that proactively optimize performance while respecting constraints such as daily budget limits, cost caps, and pacing targets.
	
	To illustrate the difference between PID control and MPC control, consider a simple example: imagine driving a car and aiming to maintain a speed of 60 km/h on a highway. However, the terrain is constantly changing—there are uphill slopes, downhill stretches, and areas of heavy traffic. How can we ensure that the speed stays close to the target?
	
	A PID controller works reactively. It adjusts the accelerator and brake based on the difference (error) between the current speed and the target speed (proportional term), how long this error has persisted (integral term), and how quickly the error is changing (derivative term). While effective in many situations, PID cannot predict future changes in terrain or traffic. It reacts only after the error occurs, potentially leading to delays in adjusting to a steep hill or a sudden slowdown ahead.
	
	Now consider an MPC framework. Instead of reacting only to the current speed, MPC incorporates a model of the car’s dynamics and the environment. It predicts how the car’s speed will change over the next several seconds based on factors like engine power, road gradient, and traffic conditions. With this foresight, MPC calculates the optimal sequence of accelerator and brake adjustments to keep the speed close to 60 km/h over the next few seconds while respecting constraints (e.g., not exceeding speed limits or ensuring smooth acceleration).
	
	At each step, MPC forecasts the car’s speed trajectory based on current inputs (e.g., accelerator position, road incline) and system dynamics. It then solves an optimization problem to find the best control actions (accelerator/brake adjustments) that minimize the deviation from the target speed over a finite time horizon. The first control action from the optimized sequence is applied, and the process repeats in the next time step with updated data.
	
	In the context of ad pacing, think of the "car" as the ad campaign, the "target speed" as the daily spend goal, and the "terrain" as auction dynamics and traffic fluctuations. Just as driving on a changing road requires continuous adjustments to speed, budget pacing requires continuous adjustments to bid amounts and budget allocations.	A PID controller can help maintain spend close to the target but struggles when traffic conditions change rapidly or constraints need to be enforced. MPC, with its predictive capabilities, anticipates traffic patterns and optimizes bidding strategies ahead of time, ensuring smoother and more efficient pacing.
	
	This analogy demonstrates how MPC brings a forward-looking, constraint-aware, and adaptive approach to control problems, making it a superior choice for complex scenarios like real-time ad pacing.

	\subsection{General Procedure of the MPC Method}
	
	The general procedure of MPC control is as follows: at each time step, at each time step, we solve an optimization problem  to determine a plan of action over a fixed time horizon and then apply the first input from this plan (this approach is known as Receding Horizon Control, or RHC). At the next time step, the process is repeated—solving a new optimization problem with the time horizon shifted one step forward.
	
	The key advantage of RHC is its ability to handle constraints directly while requiring significantly less parameter tuning (e.g., controller gains) compared to conventional control methods such as PID control. In essence, RHC allows constraints to be explicitly incorporated into the optimization process, whereas PID control requires extensive manual tuning of controller gains to indirectly manage constraints.
	
	For a more in-depth discussion of MPC, we encourage readers to refer to \cite{yang2019bid}, \cite{Camacho2004-pb}, or \cite{mattingley2010code}.

	\begin{section}{MPC Controller for Max Delivery}
	In PID control, supply predictions are used to compute the target trajectory (e.g., expected per bucket spend targets based on predicted supply pattern). This adjusts the desired reference signal, but the PID controller still operates reactively to minimize the error between the actual spend rate and the predicted target spend rate. The MPC approach, on the other hand, directly integrates predictions into the control process, enabling proactive adjustments that optimize campaign performance over a prediction horizon.
	
	More specifically, in the max delivery problem, we use the same ad configuration: the campaign has a daily budget $B$ and a predicted number of auction opportunities $T$. At a specific timestamp, suppose the consumed budget is $B_t$ and the observed auction count is $t$. The predicted future target spend rate for the remainder of the day can be computed as:
	\[
	TS(t) = \frac{B - B_t}{T - t}.
	\]
	
	We aim to answer the question: how should we adjust the bid so that the corresponding spend rate matches $TS(t)$? 
	
	The PID controller compares the currently observed spend rate to the reference $TS(t)$. If the observed spend rate is lower than the target, the PID controller increases the bid. However, the magnitude of this adjustment depends on the error signal and the controller gains (proportional, integral, and derivative components). 
	
	MPC, on the other hand, directly models the relationship between bid and spend rate. This model belongs to the class of \textbf{bid landscape forecasting models}, which predict the distribution of winning bid prices in online auctions. Suppose we have such a model $s = f(b)$ that maps a bid $b$ to a spend rate $s$. Given a target spend rate $TS(t)$, the bid can be computed as:
	\[
	b_t = f^{-1}\left(TS(t)\right).
	\]
	
	Next, we discuss how to model $f$ for the max delivery problem:
	
	\subsection{Online Methods}
	The online auction data (e.g., bid, spend, impressions, etc.) reflects the dynamics of both the ad campaign itself and the marketplace. Therefore, an intuitive idea is to leverage the most recent and fresh auction data from this campaign to project the future bid-spend relationship.
	
	\subsubsection{Longest Increasing Subsequence}
	Recall that, within each bid update interval $\Delta t$, the bid per click $b_t$ remains unchanged. We may collect the most recent $N$ interval bid-spend pairs $\{b_k, s_k\}$, where $s_k$ represents the spend over the fixed interval $\Delta t$ at time $k$ (and thus can be considered as the spend rate). 
	
	The bid-spend rate relationship for the next interval $\Delta t$ can then be learned from these $N$ data points, assuming that the number of auction opportunities does not change significantly over small intervals of $\Delta t$. For a specific auction, if we bid higher, the spend should increase (or at least not decrease). Therefore, $f(b)$ should be a monotonically nondecreasing function. However, $\{b_k, s_k\}$ does not necessarily form a monotonic sequence, as the data points are collected from different time intervals. To preserve the monotonicity property, the most straightforward method is to manually extract the longest increasing subsequence (LIS) from $\{b_k, s_k\}$. We may maintain a dynamic list $L$ to store the smallest ending values of increasing subsequences of different lengths, and an auxiliary array $P$ to track predecessors for LIS reconstruction. The time complexity of this approach is $O(N \log N)$. Once the LIS is extracted, we can interpolate between adjacent $b_k$ values to construct a piecewise linear monotonic function $f$. For any target spend rate $TS(t)$, the corresponding bid $b_t$ can be computed as $f^{-1}\left(TS(t)\right)$. If $TS(t)$ is outside the range of observed spend rates, a simple extrapolation can be applied to compute the target bid. The idea discussed here is summarized in the following \autoref{alg:mpc_lis}:
		
	\begin{algorithm}
		\caption{Monotonic Bid-Spend Model for MPC Using LIS}
		\label{alg:mpc_lis}
		\begin{algorithmic}[1]
			\Require $N$: Number of most recent bid-spend pairs, $\Delta t$: Bid update interval, $TS(t)$: Target spend rate
			\Ensure $b_t$: Bid value corresponding to $TS(t)$
			\State Collect the most recent $N$ bid-spend pairs $\{b_k, s_k\}$, where $s_k$ is the spend rate over interval $\Delta t$
			\State Extract the longest increasing subsequence $\{b'_k, s'_k\}$ using \textbf{Algorithm~\ref{alg:lis}}
			\State Construct the piecewise linear monotonic function $f(b)$:
			\begin{itemize}
				\item Sort $\{b'_k, s'_k\}$ in ascending order of $b'_k$
				\item For $b \in [b'_i, b'_{i+1}]$, interpolate linearly:
				\[
				f(b) = s'_i + \frac{(s'_{i+1} - s'_i)}{(b'_{i+1} - b'_i)} \cdot (b - b'_i)
				\]
			\end{itemize}
			\State Compute the bid $b_t$ by inverting $f$:
			\begin{itemize}
				\item If \(TS(t) < s'_1\): \Comment{Extrapolation below the range}
				\[
				b_t = \min \left( 0, b'_1 + \frac{TS(t) - s'_1}{s'_2 - s'_1} \cdot (b'_2 - b'_1) \right)
				\]
				\item If \(TS(t) > s'_m\) (where \(m = |\{b'_k, s'_k\}|\)): \Comment{Extrapolation above the range}
				\[
				b_t = b'_m + \frac{TS(t) - s'_m}{s'_m - s'_{m-1}} \cdot (b'_m - b'_{m-1})
				\]
				\item If \(TS(t) \in [s'_i, s'_{i+1}]\): \Comment{Interpolation within the range}
				\[
				b_t = b'_i + \frac{(b'_{i+1} - b'_i)}{(s'_{i+1} - s'_i)} \cdot (TS(t) - s'_i)
				\]
			\end{itemize}
		\end{algorithmic}
	\end{algorithm}

	\begin{algorithm}[H]
		\caption{Longest Increasing Subsequence (LIS)}
		\label{alg:lis}
		\begin{algorithmic}[1]
			\Require $\{b_k, s_k\}$: Sequence of bid-spend pairs
			\Ensure $\{b'_k, s'_k\}$: Longest increasing subsequence
			\State Initialize an empty list $L$ to store indices of the LIS and an auxiliary array $P$ of size $N$ to track predecessors
			\For{$i = 1$ to $N$}
			\State Perform binary search on $L$ to find the largest index $j$ such that $s_{L[j]} \leq s_k$
			\If{$j$ exists}
			\State Replace $L[j+1]$ with $i$
			\Else
			\State Append $i$ to $L$
			\EndIf
			\State Update $P[i]$ with the index of $L[j]$ (or set $P[i] = -1$ if $j$ does not exist)
			\EndFor
			\State Reconstruct the LIS by backtracking from the last index in $L$ using the array $P$
			\State Return the bid-spend pairs corresponding to the LIS indices in $L$: $\{b'_k, s'_k\}$
		\end{algorithmic}
	\end{algorithm}

	\subsubsection{Isotonic Regression and PAVA algorithm}
	Extracting the longest increasing subsequence (LIS) can help maintain the monotonicity properties of a bid spend-rate curve. However, this approach discards a significant number of data points, leading to data inefficiency and potentially undermining the predictive accuracy of the model. To leverage all available data while maintaining monotonicity, isotonic regression offers an effective solution.
	
	Isotonic regression is a type of regression analysis designed for situations where the target variable is expected to be non-decreasing (or non-increasing) with respect to an independent variable. It imposes an order constraint on the data, ensuring the resulting function remains monotonic.
	
	A key algorithm used for isotonic regression is the \textbf{Pool Adjacent Violators Algorithm (PAVA)}. PAVA efficiently solves isotonic regression problems by iteratively merging adjacent data points that violate the monotonicity constraint. It is computationally lightweight, operating in linear time $O(n)$, making it well-suited even for large datasets. For more technical details and implementations, one may refer to \cite{ayer1955empirical}, \cite{leeuw2009isotone}.
	
	Compared to the LIS algorithm, PAVA offers significant improvements in computational efficiency, reducing the time complexity from $O(n \log n)$ to $O(n)$. Additionally, PAVA utilizes all data points during the computation process, which reduces prediction noise and enhances model accuracy.
	
	For further technical details, refer to the method described in \cite{chen2026lightweight} and \cite{yuanlong2024method}. The steps of PAVA and how it can be applied to our problem can be summarized in the following algorithms:
	
	\begin{algorithm}[H]
		\caption{Pool Adjacent Violators Algorithm (PAVA) with Preserved Bid-Spend Pairs}
		\label{alg:pava}
		\begin{algorithmic}[1]
			\Require $\{b_k, s_k\}$: Input bid-spend pairs, where $b_k$ is the bid and $s_k$ is the spend rate, sorted such that $b_1 \leq b_2 \leq \dots \leq b_n$
			\Ensure $\{b'_k, s'_k\}$: Adjusted monotonic bid-spend pairs of the same length as the input
			\State Initialize $y'_i \gets s_i$ for all $i = 1, \dots, n$
			\State Initialize weights $w_i \gets 1$ for all $i = 1, \dots, n$
			\State Initialize $i \gets 1$
			\While{$i < n$}
			\If{$y'_i > y'_{i+1}$} \Comment{Check monotonicity}
			\State Merge $y'_i$ and $y'_{i+1}$:
			\[
			y'_i \gets \frac{w_i y'_i + w_{i+1} y'_{i+1}}{w_i + w_{i+1}}
			\]
			\State Update weights: $w_i \gets w_i + w_{i+1}$
			\State Remove $y'_{i+1}$ and $w_{i+1}$ from their respective arrays
			\State Remove $b_{i+1}$ from the bid array
			\State Adjust indices: $i \gets \max(1, i-1)$
			\Else
			\State Increment index: $i \gets i + 1$
			\EndIf
			\EndWhile
			\State Expand the adjusted spend values to match the original length:
			\For{$j = 1$ to $n$}
			\State Assign each original bid $b_j$ the adjusted spend value $y'_i$ corresponding to its current segment
			\EndFor
			\State Return $\{b_k, y'_k\}$: Adjusted monotonic bid-spend pairs
		\end{algorithmic}
	\end{algorithm}

	\begin{algorithm}[H]
		\caption{Monotonic Bid-Spend Model for MPC Using PAVA}
		\label{alg:mpc_bid_update_pava}
		\begin{algorithmic}[1]
			\Require $\{b_k, s_k\}$: Recent bid-spend pairs (bid $b_k$ and spend rate $s_k$) sorted such that $b_1 \leq b_2 \leq \dots \leq b_n$
			\Require $TS(t)$: Target spend rate, 
			\Ensure $b_t$: Bid value corresponding to $TS(t)$
			\State Apply \textbf{Algorithm~\ref{alg:pava}} to $\{b_k, s_k\}$ to compute monotonic spend rates $\{b_k, s'_k\}$
			\State Compute the bid $b_t$:
			\begin{itemize}
				\item If $TS(t) < s'_1$: \Comment{Extrapolation below range}
				\[
				b_t = b_1 + \frac{TS(t) - s'_1}{s'_2 - s'_1} \cdot (b_2 - b_1)
				\]
				\item If $TS(t) > s'_n$: \Comment{Extrapolation above range}
				\[
				b_t = b_n + \frac{TS(t) - s'_n}{s'_n - s'_{n-1}} \cdot (b_n - b_{n-1})
				\]
				\item If $TS(t) \in [s'_i, s'_{i+1}]$: \Comment{Interpolation within range}
				\[
				b_t = b_i + \frac{(b_{i+1} - b_i)}{(s'_{i+1} - s'_i)} \cdot (TS(t) - s'_i)
				\]
			\end{itemize}
			\State Return $b_t$
		\end{algorithmic}
	\end{algorithm}

	\subsection{Offline Methods}
	The online algorithms described above are lightweight and well-suited for leveraging fresh data points that represent the up-to-date bid landscape distributions. However, for ad campaigns with deep funnel conversions (e.g., CPA and CPL ads), the conversion signals are often sparse and delayed, as they need to be collected from third-party platforms.
	
	If the campaign is an oCPM campaign, there is no issue because it is charged by impressions, and the delay in spend information is negligible. However, for campaigns charged by actual results (e.g., CPA or CPL), we may encounter situations where the spend remains at zero for an extended period, followed by a sudden spike in spend when a conversion occurs. This results in highly non-smooth pacing dynamics, making it challenging to maintain consistent performance.
	
	In such cases, the online algorithms may not be the optimal strategy. Instead, it becomes necessary to leverage historical data to model the bid spend-rate curve for the max delivery problem. 
	
	\subsubsection{eCPM Distribution}
	Suppose the probability density function of the eCPM distribution of the marketplace for this campaign is $p(z)$, and the campaign-level average CTR is $r$. Under the Second Price Auction mechanism, the expected spend per impression given a bid per impression $b_i$ can be computed as:
	\begin{equation} \label{eq:ecpm_bid_cost}
		g(b_i) = \frac{\int_{0}^{b_i} z p(z) \text{d}z}{\int_{0}^{b_i} p(z) \text{d}z}.
	\end{equation}
	
	In the following \autoref{sec:mpc_remarks}, we will show that $g(\cdot)$, as defined, is monotonically non-decreasing with respect to $b$.
	
	Now, suppose the target spend for the next interval $\Delta t$ is $TS(t)$, and the number of auction opportunities within this interval is $NR(t)$. The bid spend-rate curve $f(b)$ (where $b$ is the bid per click) can be computed as:
	\[
	f(b) = g\left(b \cdot r\right) \cdot NR(t).
	\]
	To achieve the target spend $TS(t)$, set $f(b_t) = TS(t)$ and solve:
	\[
	TS(t) = g\left(b_t \cdot r\right) \cdot NR(t).
	\]
	Rearranging gives:
	\[
	b_t = \frac{g^{-1} \left( \frac{TS(t)}{NR(t)} \right)}{r}.
	\]
	
	The question now boils down to finding the eCPM distribution $p(\cdot)$. This can be achieved through statistical modeling of historical auction data, where the distribution of observed eCPM values can be estimated using techniques such as constructing empirical histograms, kernel density estimation, or parametric methods, depending on the nature of the data.
	
	To demonstrate how to derive $p(\cdot)$ for a campaign, we use the empirical histogram approach offline. First, for a given campaign, collect all auctions it participated in over the past $K$ days. For each auction, record the eCPMs from the GSP ranking stage and denote these eCPMs as $\{v_i\}$. Define the range of eCPM values:
	\[
	\overline{v} = \max_i v_i, \quad \underline{v} = \min_i v_i.
	\]
	
	Discretize the range $[\underline{v}, \overline{v}]$ into $N$ buckets, $\{[z_{j-1}, z_j]\}_{j=1}^N$, where $N$ is a large number such that each bucket is small enough to represent the distribution accurately. Count the number of eCPMs falling into each interval $[z_{j-1}, z_j]$, denoted as $n_j$. Compute the discretized probability density function (p.d.f.) for each interval:
	\[
	p_j = \frac{n_j}{\sum n_j}.
	\]
	
	Once $p_j$ is computed, $g(\cdot)$ can be calculated based on the buckets $\{[z_{j-1}, z_j]\}_{j=1}^N$. For $b \in [z_{j-1}, z_j]$, set:
	\[
	g(b) = g_j,
	\]
	where:
	\[
	g_j = \frac{\sum_{l \leq j} z_l \cdot p_l}{\sum_{l \leq j} p_l}.
	\]
	
	For any target spend $TS(t)$, to compute $b_t$, find the index $k$ such that $g_k$ is closest to $\frac{TS(t)}{NR(t)}$. Then set:
	\[
	b_t = \frac{z_k}{r}.
	\]

	We summarize the idea above in the following algorithms:
	
	\begin{algorithm}[H]
		\caption{Compute $p_j$ (Discretized p.d.f.)}
		\label{alg:compute_pj}
		\begin{algorithmic}[1]
			\Require $v$: Historical eCPM values for the campaign
			\Require $N$: Number of buckets for discretization
			\Ensure $p_j$: Discretized p.d.f., $z$: Bucket boundaries
			\State Define $\overline{v} = \max(v)$, $\underline{v} = \min(v)$
			\State Discretize $[\underline{v}, \overline{v}]$ into $N$ buckets: $\{[z_{j-1}, z_j]\}_{j=1}^N$
			\State Initialize $n_j = 0$ for all $j = 1, \dots, N$
			\For{each $v_i \in v$}
			\State Find the bucket index $j$ such that $z_{j-1} \leq v_i < z_j$
			\State Increment $n_j \gets n_j + 1$
			\EndFor
			\State Compute the total number of samples: $\text{total\_n} = \sum_{j=1}^N n_j$
			\For{each bucket $j = 1, \dots, N$}
			\State $p_j = \frac{n_j}{\text{total\_n}}$
			\EndFor
			\State \Return $p_j$, $z$
		\end{algorithmic}
	\end{algorithm}
	
	\begin{algorithm}[H]
		\caption{Compute $g(b)$ and $b_t$}
		\label{alg:compute_bt}
		\begin{algorithmic}[1]
			\Require $p_j$: Discretized p.d.f.
			\Require $z$: Bucket boundaries
			\Require $r$: Campaign-level average CTR
			\Require $TS$: Target spend rate
			\Require $NR$: Number of auction opportunities
			\Ensure $b_t$: Optimal bid per click
			\State Compute $g(b)$ for each bucket:
			\For{each bucket $j = 1, \dots, N$}
			\State $g_j = \frac{\sum_{l \leq j} z_l \cdot p_l}{\sum_{l \leq j} p_l}$
			\EndFor
			\State Find the index $k$ such that $g_k$ is closest to $\frac{TS}{NR}$
			\State Compute the optimal bid:
			\[
			b_t = \frac{z_k}{r}
			\]
			\State \Return $b_t$
		\end{algorithmic}
	\end{algorithm}

	\subsection{Practical Considerations}

	When PAVA type algorithm is implmented, the moving window is maintained in bid order, with each insertion and deletion requiring $O(\log N)$ time, thus avoiding a complete re-sort at every cycle \cite{chen2026lightweight}. Once the ordered window is updated, PAVA can be applied in $O(N)$ time; maintaining the order incrementally therefore prevents an additional $O(N \log N)$ sorting step at each pacing update and keeps the online fitting procedure lightweight for production use.
		
	\end{section}

	\begin{section} {MPC Controller for Cost Cap}
	\subsection{MPC for Cost Cap}
	
	In the previous section, we discussed the MPC control for the max delivery problem. Since the max delivery problem only has a budget constraint, solving the receding horizon problem requires only the bid-to-spend rate model \( s = f(b) \), which helps adjust the bid based on the current delivery status to maximize the objective function. 
	
	The cost cap problem, however, introduces an additional cost constraint, adding complexity because the algorithm must now balance both budget pacing and cost efficiency.
	
	\paragraph{Problem Formulation}
	
	Recall the formulation of the cost cap problem from \autoref{eq:cost_cap}:
	
	\begin{equation*} 
		\begin{aligned}
			\max_{x_t \in \{0,1\}} \quad & \sum_{t=1}^T x_t \cdot r_t \\
			\text{s.t.} \quad & \sum_{t=1}^{T} x_t \cdot c_t \leq B, \\
			& \frac{\sum_{t=1}^{T} x_t \cdot c_t}{\sum_{t=1}^{T} x_t \cdot r_t} \leq C.
		\end{aligned}
	\end{equation*}
	
	At each time step \(\tau\), with a receding horizon \(H\) (e.g., 1 hour), we solve a model predictive control (MPC) problem iteratively to optimize the objective. To do so, we must determine the new budget \(B_{\tau, H}\) and the cost per result cap \(C_{\tau, H}\) for the time interval \((\tau, \tau + H)\).
	
	\paragraph{Computing Budget and Cost Constraints}
	We compute \( B_{\tau, H} \) and \( C_{\tau, H} \) based on the delivery status at time \(\tau\):

	\begin{itemize}
	
	\item  \textbf{Computing \( B_{\tau, H} \):}  
	As discussed in \autoref{alg: budget_allocation}, the target budget per time interval is allocated proportionally to the expected number of auction requests in that interval. If at time \(\tau\), the remaining budget is \(B_{\tau, r}\), the predicted total remaining auction requests is \(TR_{\tau, r}\), and the predicted auction requests within \((\tau, \tau + H)\) is \(NR_{\tau, H}\), then the budget for this interval is:
	
	\[
	B_{\tau, H} = \frac{NR_{\tau, H}}{TR_{\tau, r}} \cdot B_{\tau, r}.
	\]
	
	\item \textbf{Computing \( C_{\tau, H} \):}  
	Suppose at time \(\tau\), we have observed \(NC_{\tau}\) conversions. To satisfy the cost cap constraint, we must collect at least \( B/C - NC_{\tau} \) conversions before the campaign ends if we aim to fully spend the total budget \(B\). The cost per result upper bound for the remainder of the campaign, \( C_{\tau, r} \), is given by:
	
	\[
	C_{\tau, r} = \frac{B_{\tau, r}}{ \frac{B}{C} - NC_{\tau} }.
	\]
	
	Assuming the conversion rate follows an i.i.d. distribution, the proportional share of the number of conversions in \((\tau, \tau + H)\) should match the proportional share of the requests in this time interval. Thus, the new cost per result cap for \((\tau, \tau + H)\) is the same as \(C_{\tau, r}\):
	
	\[
	C_{\tau, H} = \frac{B_{\tau, r}}{ \frac{B}{C} - NC_{\tau} }.
	\]
	\end{itemize}

	\paragraph{Receding Horizon Optimization Problem}
	
	In addition to the new budget and cost constraints, practical implementations may impose a lower bound \(b_l\) and an upper bound \(b_u\) on the bid to prevent extreme values. The receding horizon optimization problem for \((\tau, \tau + H)\) is formulated as:
	
	\begin{equation} \label{eq: mpc_cost_cap_receding}
		\begin{aligned}
			\max_{x_t \in \{0,1\}} \quad & \sum_{\tau \leq t \leq \tau + H} x_t \cdot r_t \\
			\text{s.t.} \quad & \sum_{\tau \leq t \leq \tau + H} x_t \cdot c_t \leq B_{\tau, H}, \\
			& \frac{\sum_{\tau \leq t \leq \tau + H} x_t \cdot c_t}{\sum_{\tau \leq t \leq \tau +H} x_t \cdot r_t} \leq C_{\tau, H}, \\
			& \qquad b_l \leq b_t \leq b_u.
		\end{aligned}
	\end{equation}
	
	From \autoref{eq:cost_cap_bid_formula}, we have shown that the optimal solution for the cost cap problem above is a constant bid. Thus, to solve the receding horizon problem, we simply need to find a constant bid between \(b_l\) and \(b_u\) such that the resulting spend and cost per result within \((\tau, \tau + H)\) satisfy the constraints in the optimization problem.
	
	For the max delivery problem, the bid-to-spend model \(s = f(b)\) is sufficient to determine the optimal bid given a target spend rate. However, for the cost cap problem, an additional bid-to-cpx model \(cpx = h(b)\) is required to capture the relationship between bid price and cost per result. This enables the algorithm to balance both budget pacing and cost efficiency for \((\tau, \tau + H)\). 
	
	To construct \(h(b)\), it suffices to model the bid-to-number-of-conversions function \(n = g(b)\) and compute:
	
	\[
	cpx = \frac{s}{n} = \frac{f(b)}{g(b)}.
	\]
	
	\paragraph{Constructing \( f(b) \) and \( g(b) \)}
	
	Recall that in the max delivery problem, \(s = f(b)\) is constructed by first collecting the most recent \(N\) interval bid-spend pairs and then applying the longest increasing subsequence algorithm (e.g., \autoref{alg:lis}) or PAVA algorithm (e.g., \autoref{alg:pava}) and interpolation methods to obtain a monotonic bid-spend sequence.
	
	The bid-to-number-of-conversions model \(n = g(b)\) can be constructed similarly:
	\begin{itemize}
		\item Collect the most recent \(N\) pacing intervals and extract \(\{b_k, n_k\}\), where \(n_k\) represents the number of conversions over the fixed interval \(k\).
		\item Apply LIS or PAVA algorithms to \(\{b_k, n_k\}\) to obtain a monotonic function \(n = g(b)\).
		\item Renormalize \(f(b)\) and \(g(b)\) to represent spend and conversions per \(H\)-time interval.
	\end{itemize}
	
	\paragraph{Finding the Optimal Bid}
	
	Since the bid-to-number-of-conversions function \(n = g(b)\) is monotonically non-decreasing, the larger the bid, the more conversions we may get. With \(h(b) = f(b) / g(b)\), the optimal bid \(b^*\) can be determined by iterating over all possible values of \(b\) and selecting the largest value satisfying the constraints:
	\[
	f(b) \leq B_{\tau, H}, \quad h(b) \leq C_{\tau, H}, \quad b_l \leq b \leq b_u.
	\]
	
	In practice, we can start with \(b_l\) and increment \(b\) by a small \(\Delta b\) at each step in the search process\footnote{Theoretically, the bid-to-spend \(f\) and bid-to-cpx \(h\) are both monotonically non-decreasing with respect to the bid price \(b\). Thus, starting from \(b_l\), we can stop the search as soon as the first \(b\) satisfies the constraints and \(b + \Delta b\) does not. This approach significantly reduces computation overhead. However, in practice, the monotonicity property does not always hold due to various factors, such as modeling errors or noisy data. As a result, enumerating all possibilities can sometimes help achieve better results, especially when the iteration process is not overly time-consuming.}. The idea discussed here is summarized in \autoref{alg:mpc_cost_cap}.
	
	\begin{algorithm}
		\caption{MPC-Based Cost Cap Bidding Algorithm}
		\label{alg:mpc_cost_cap}
		\begin{algorithmic}[1]
			\Require 
			$B_{\tau, r}$: Remaining budget; $B$: Total budget; $C$: Cost cap; \\
			$H$: Time horizon; $NC_{\tau}$: Observed conversions; $TR_{\tau, r}$: Predicted total remaining requests; \\
			$NR_{\tau, H}$: Predicted requests in $(\tau, \tau + H)$; $[b_l, b_u]$: Bid bounds; $\Delta b$: Search step size.
			\Ensure $b^*$: Optimal bid for $(\tau, \tau + H)$.
			
			\State \textbf{Step 1: Compute Budget and Cost Cap Constraints}
			\State Compute the budget allocation for the receding horizon:
			\[
			B_{\tau, H} \gets \frac{NR_{\tau, H}}{TR_{\tau, r}} \cdot B_{\tau, r}
			\]
			\State Compute the cost per result upper bound for the remaining time:
			\[
			C_{\tau, r} \gets \frac{B_{\tau, r}}{\frac{B}{C} - NC_{\tau}}
			\]
			\State Set the cost cap for the horizon:
			\[
			C_{\tau, H} \gets C_{\tau, r}
			\]
			
			\State \textbf{Step 2: Construct Models $f(b)$ and $g(b)$}
			\State Collect the most recent $N$ bid-spend pairs $\{b_k, s_k\}$ and apply LIS or PAVA to construct $f(b)$ normalized to $H$.
			\State Collect the most recent $N$ bid-conversion pairs $\{b_k, n_k\}$ and apply LIS or PAVA to construct $g(b)$ normalized to $H$.
			\State Compute $h(b)$ as:
			\[
			h(b) \gets \frac{f(b)}{g(b)}
			\]
			
			\State \textbf{Step 3: Search for Optimal Bid $b^*$}
			\State Initialize $b^* \gets b_l$.
			\For{$b$ from $b_l$ to $b_u$ with step size $\Delta b$}
			\If{$f(b) \leq B_{\tau, H}$ \textbf{and} $h(b) \leq C_{\tau, H}$}
			\State Update $b^* \gets b$.
			\EndIf
			\EndFor
			
			\State \textbf{Step 4: Return Optimal Bid}
			\State \Return $b^*$
		\end{algorithmic}
	\end{algorithm}

	\paragraph{Practical Considerations}
	We list some practical considerations here for implementing cost cap MPC controller method in real-world production system:
	
	\begin{itemize}
		\item \textbf{Data Sparsity and Signal Delay}: The MPC controller method discussed here assumes that the campaign follows an oCPM bidding strategy, where the campaign is charged per impression. It also assumes that a sufficient number of conversions can be observed within each pacing update cycle. This assumption is crucial because it allows the use of real-time data and LIS/PAVA algorithms to model the bid-to-spend function \( f \) and the bid-to-number-of-conversions function \( g \).
		
		For upper-funnel objective campaigns such as CPM and CPV, this assumption generally holds, as the time from ad exposure to conversion (e.g., impressions, video views) is only a few seconds, making any delay negligible. However, for mid-to-lower funnel objective campaigns such as CPC, CPL, or CPA, this assumption no longer holds. The delay from an impression to the final conversion (e.g., app installs) can take hours or even days. As a result, leveraging real-time conversion data to model \( g \) becomes impractical.
		
		Additionally, if the campaign is not oCPM but CPX, where charges only occur when actual conversions are realized, the cost pattern can be highly irregular. In such cases, there may be long periods with no cost, followed by sudden large charges when conversions happen. This causes the bid-to-spend function \( f \) to resemble a discontinuous "bump function," making it inappropriate to use real-time data for modeling.
		
		In these scenarios, a common approach is to leverage prediction models to construct \( f, g \), and \( h \). These predictive techniques provide a more stable and reliable estimation of bid-to-spend and bid-to-conversion relationships. We will explore these methods in greater detail in the bid landscape forecasting section in the next part.
		
		\item \textbf{Bid-to-X Model Normalization}: Recall that in the algorithm, we construct \( f \) and \( g \) based on the bid-cost and bid-conversion pairs observed during each pacing interval. As a result, \( f \) and \( g \) represent the total spend and number of conversions over the pacing interval. However, if the receding horizon \( H \) differs from the pacing interval (which is usually the case), we must normalize both functions to reflect the appropriate spend rate over \( H \).
		
		If we assume that the supply pattern remains relatively stable over short periods (i.e., when \( H \) is small), then the number of eligible requests can be considered uniformly distributed and proportional to the time duration. In such cases, we can normalize \( f \) and \( g \) by multiplying them by the ratio \( H / \Delta t \), where \( \Delta t \) is the pacing interval. This approach allows us to avoid relying on predicted supply for every pacing interval when updating the bid price.
		
		However, if \( H \) is relatively long, the assumption of a stationary supply distribution no longer holds. In this case, we must rely on the predicted number of requests and compute the normalization factor as the ratio of the number of expected requests in the upcoming \( H \) to that of the most recent pacing interval.

	\end{itemize}
	 			
		\subsection{MPC for Target CPA}
		
		Target CPA (sometimes referred to as target cost) is another popular automated bidding strategy in which advertisers set a target cost per action (conversion). The goal is to acquire as many conversions as possible at or around the specified CPA. 
		
		Compared to the cost cap strategy, which is generally considered more “strict” about not exceeding the cost threshold, target CPA is more flexible in practice. A small deviation from the specified CPA is often tolerable, allowing the algorithm to prioritize achieving a higher number of conversions while maintaining an average cost close to the target CPA. Target CPA is typically a good option for advertisers who have a clear, data-driven understanding of what the ideal CPA should be and can tolerate small deviations above the target.

		\paragraph{Problem Formulation}
			The target CPA differs from the cost cap in terms of its cost constraint. More specifically, suppose we allow the average CPA to fluctuate around the target \( C \) by a margin of \( \delta \) (e.g., \( \delta = 10\% \)). The target CPA problem can then be formulated as:
			
			\begin{equation*} 
				\begin{aligned}
					\max_{x_t \in \{0,1\}} \quad & \sum_{t=1}^T x_t \cdot r_t \\
					\text{s.t.} \quad & \sum_{t=1}^{T} x_t \cdot c_t \leq B, \\
					(1 - \delta) \cdot C \leq  & \frac{\sum_{t=1}^{T} x_t \cdot c_t}{\sum_{t=1}^{T} x_t \cdot r_t} \leq (1 + \delta) \cdot C.
				\end{aligned}
			\end{equation*}
			In practice, instead of imposing two constraints with \( (1-\delta) \cdot C \) and \( (1+\delta) \cdot C \), we can reformulate the problem to enforce that the posterior CPA is strictly equal to the target \( C \). The problem then becomes:
			
			\begin{equation} 
				\begin{aligned}
					\max_{x_t \in \{0,1\}} \quad & \sum_{t=1}^T x_t \cdot r_t \\
					\text{s.t.} \quad & \sum_{t=1}^{T} x_t \cdot c_t \leq B, \\
					& \frac{\sum_{t=1}^{T} x_t \cdot c_t}{\sum_{t=1}^{T} x_t \cdot r_t} = C.
				\end{aligned}
			\end{equation}
			The objective is now to maximize the total number of conversions while ensuring that the actual CPA remains as close as possible to the target \( C \), subject to the budget constraint \( B \). If the cost constraint is strictly satisfied—meaning that each conversion costs exactly \( C \) on average—then maximizing conversions is equivalent to maximizing overall spend delivery.
			
		\paragraph{Computing New Constraints}
		We choose the pacing lifetime of the campaign as the receding horizon. At each pacing interval starting at time \(\tau\), we need to determine the updated budget constraint \(B_{\tau}\) and cost constraint \(C_{\tau}\). 
		
		The budget constraint is straightforward: suppose at time \(\tau\), the observed spend is \(S_{\tau}\), then the remaining budget is:
		\[
		B_{\tau} = B - S_{\tau}.
		\]
		For the cost constraint, suppose we have collected \(NC_{\tau}\) conversions so far. If the average CPA is on target, the expected spend should be \(C \cdot NC_{\tau}\). We define the deviation \(D_{\tau}\) at time \(\tau\) as the difference between this expected spend and the actual spend:
		\begin{equation} \label{eq: mpc_target_cpa_debt}
			D_{\tau} = S_{\tau} - C \cdot NC_{\tau}.
		\end{equation}
		
		This represents the \textbf{spend deviation} that must be adjusted within the remaining campaign lifetime. Let \( S_{\tau, r} \) denote the projected future spend from time \( \tau \) until the end of the campaign, and let \( NC_{\tau, r} \) represent the expected number of conversions within this period. To maintain the target CPA, the remaining ad delivery must compensate for the accumulated spend deviation. This relationship can be expressed as:
		\[
		\left( S_{\tau, r} - C \cdot NC_{\tau, r} \right) = -D_{\tau}.
		\]
		Since:
		\[
		S_{\tau, r}  = \sum_{t \geq \tau} x_t \cdot c_t ,  \quad NC_{\tau, r} = \sum_{t\geq \tau} x_t \cdot r_t,
		\]
		the cost constraint can be rewritten as:
		\begin{equation} \label{eq:mpc_target_cpa_new_cost_constraint}
			\sum_{t \geq \tau} x_t \cdot c_t  - C \cdot \left(\sum_{t\geq \tau} x_t \cdot r_t \right) = - D_{\tau},
		\end{equation}
		where \(D_{\tau}\) is the accumulated spend deviation at \(\tau\) as defined in \autoref{eq: mpc_target_cpa_debt}.

		\paragraph{Receding Horizon Optimization Problem}
		Based on the discussion in the previous paragraph, we can now formulate the receding horizon optimization problem at time \(\tau\):
		
		\begin{equation*} 
			\begin{aligned}
				\max_{x_t \in \{0,1\}} \quad & \sum_{t \geq \tau} x_t \cdot r_t \\
				\text{s.t.} \quad & \sum_{t\geq \tau} x_t \cdot c_t \leq B - S_{\tau}, \\
				& \sum_{t \geq \tau} x_t \cdot c_t  - C \cdot \sum_{t\geq \tau} x_t \cdot r_t = - D_{\tau}, \\
				& b_l \leq b_t \leq b_u.
			\end{aligned}
		\end{equation*}
		
		Here, \( S_{\tau} \) represents the observed spend at time \( \tau \), and \( D_{\tau} \) is the accumulated spend deviation, as defined in \autoref{eq: mpc_target_cpa_debt},   \(b_l\) and \(b_u\) are lower and upper bounds on the bid to prevent extreme values. As mentioned earlier, as long as the cost constraint is satisfied, maximizing total conversions is equivalent to maximizing total spend. 
		
		To capture the balance between spend and conversions, we define the \textbf{repay rate function} \( \mathcal{R}(b) \) as:
		
		\[
		\mathcal{R}(b) = f(b) - C\cdot g(b),
		\]
		where:
		\begin{itemize}
			\item \( f(b) \) is the \textbf{bid-to-spend} function, representing the expected cost as a function of bid price.
			\item \( g(b) \) is the \textbf{bid-to-number-of-conversions} function, representing the expected number of conversions as a function of bid price.
		\end{itemize}
		
		Thus, our objective is to determine the optimal bid \( b_{\tau}^* \) such that \( \mathcal{R}(b_{\tau}^*) \) offsets the accumulated spend deviation \( D_{\tau} \) before the campaign ends while simultaneously maximizing total spend. 
		
		Before solving for \( b_{\tau}^* \), we first need to construct the functions \( f(b) \) and \( g(b) \).

		\paragraph{Constructing \(f(b)\) and \(g(b)\)}
		Both \(f(b)\) and \(g(b)\) can be constructed similarly to the cost cap problem. We collect the most recent \(N\) interval bid-spend and bid-conversion pairs, then apply the LIS or PAVA algorithm along with interpolation methods to obtain a monotonic bid-to-X sequence.

		\paragraph{Finding the Optimal Bid} 
		Now, we have everything ready to solve for \( b_{\tau}^* \). We define a \textbf{projected spend deviation offset function} as:
		
		\[
		\mathcal{P} (b, \Omega) = \mathcal{R}(b) \cdot \frac{\Omega}{\Delta t}  + D_{\tau},
		\]
		where:
		\begin{itemize}
			\item \( \Delta t \) is the pacing interval.
			\item \( \Omega \) is the evaluation window in which we compute how much spend deviation can be repaid given a fixed bid \( b \). 	Note that \( \Delta t \leq  \Omega  \leq T_{\tau} \).
			\item \( T_{\tau} \) is the remaining lifetime of the campaign.
		\end{itemize}
		Taking a closer look at \( \mathcal{P} (b, \Omega) \), we observe:
		\begin{itemize}
			\item \( \mathcal{R}(b) \) represents the spend deviation repaid in \textbf{one pacing interval} \( \Delta t \) for a given bid \( b \).
			\item \( \mathcal{R}(b) \cdot \frac{\Omega}{\Delta t} \) estimates the \textbf{total repayable spend deviation} over the next \( \Omega \) time window if we maintain the bid at \( b \).
			\item \( \mathcal{P} (b, \Omega) \) thus represents the \textbf{remaining spend deviation} after \( \Omega \) time, assuming a constant bid \( b \).
		\end{itemize}
		Our objective is now clear: We need to iterate over the following values:
		\[
		b \in \{ b_l , b_l + \Delta b, b_l + 2\cdot \Delta b, \dots, b_u \},  \quad 	\Omega \in \{ \Delta t,  2\cdot \Delta t, 3\cdot \Delta t, \dots, T_{\tau} \}
		\]
		to find the \textbf{maximum} \( b_{\tau}^* \) along with some optimal \( \Omega^* \) such that:
		\[
		\mathcal{P} (b_{\tau}^*, \Omega^*)  = 0.
		\]
		
		Theoretically, there are three possible scenarios:
		
		\begin{itemize}
			\item \textbf{Existence of \( b, \Omega \) such that \( \mathcal{P}(b, \Omega) = 0 \).} \\
			In this case, among all valid pairs \( \{b, \Omega\} \), we select the \textbf{highest} \( b \).
			
			\item \textbf{\( \mathcal{P}(b, \Omega) \) is always negative.} \\
			This implies that the actual CPA remains lower than \( C \) regardless of bid level or window selection. 
			To determine the optimal bid:
			\begin{enumerate}
				\item Find all \( \{b, \Omega\} \) where \( \mathcal{P} \) is maximized.
				\item Select the \textbf{largest} \( b \) among those candidates.
			\end{enumerate}
			
			\item \textbf{\( \mathcal{P}(b, \Omega) \) is always positive.} \\
			This implies that the actual CPA is always higher than \( C \). 
			To determine the optimal bid:
			\begin{enumerate}
				\item Find all \( \{b, \Omega\} \) where \( \mathcal{P} \) is minimized.
				\item Select the \textbf{smallest} \( b \) among those candidates, to avoid overspending when conversions are too expensive.
			\end{enumerate}
		\end{itemize}
		
		The algorithm we discussed above is summarized in \autoref{alg:mpc_target_cpa}. 
		
		\begin{algorithm}[H]
			\caption{MPC-Based Target CPA Bidding Algorithm}
			\label{alg:mpc_target_cpa}
			\begin{algorithmic}[1]
				\Require 
				$B$: Total budget; $C$: Target CPA; $\Delta t$: Pacing interval; $T$: Campaign lifetime; \\
				$[b_l, b_u]$: Bid range; $\Delta b$: Step size; $S_{\tau}$: Observed spend at $\tau$; \\
				$NC_{\tau}$: Observed conversions at $\tau$.
				\Ensure $b_{\tau}^*$: Optimal bid for the next pacing interval.
				
				\State \textbf{Step 1: Compute Remaining Budget and Spend Deviation}
				\State Compute remaining budget:
				\[
				B_{\tau} \gets B - S_{\tau}
				\]
				\State Compute accumulated spend deviation:
				\[
				D_{\tau} \gets S_{\tau} - C \cdot NC_{\tau}
				\]
				
				\State \textbf{Step 2: Construct Models $f(b)$ and $g(b)$}
				\State Collect the most recent $N$ bid-spend pairs $\{b_k, s_k\}$ and apply LIS or PAVA to construct $f(b)$ normalized to $\Delta t$.
				\State Collect the most recent $N$ bid-conversion pairs $\{b_k, n_k\}$ and apply LIS or PAVA to construct $g(b)$ normalized to $\Delta t$.
				\State Define the repay rate function:
				\[
				\mathcal{R}(b) \gets f(b) - C \cdot g(b)
				\]
				
				\State \textbf{Step 3: Search for Optimal Bid $b_{\tau}^*$}
				\State Initialize $b_{\tau}^* \gets b_l$.
				\For{$b$ from $b_l$ to $b_u$ with step size $\Delta b$}
				\For{$\Omega$ from $\Delta t$ to $T_{\tau}$ with step size $\Delta t$}
				\State Compute:
				\[
				\mathcal{P}(b, \Omega) \gets \mathcal{R}(b) \cdot \frac{\Omega}{\Delta t} + D_{\tau}
				\]
				\If{$\mathcal{P}(b, \Omega) = 0$}
				\State Update $b_{\tau}^* \gets \max(b_{\tau}^*, b)$.
				\ElsIf{$\mathcal{P}(b, \Omega) < 0$}
				\State Store $(b, \Omega)$ where $\mathcal{P}$ is maximized.
				\ElsIf{$\mathcal{P}(b, \Omega) > 0$}
				\State Store $(b, \Omega)$ where $\mathcal{P}$ is minimized.
				\EndIf
				\EndFor
				\EndFor
				
				\State \textbf{Step 4: Determine Final Bid}
				\If{at least one $(b, \Omega)$ satisfies $\mathcal{P}(b, \Omega) = 0$}
				\State Choose the largest $b$ among candidates.
				\ElsIf{$\mathcal{P}(b, \Omega)$ is always negative}
				\State Choose the largest $b$ where $\mathcal{P}$ is maximized.
				\ElsIf{$\mathcal{P}(b, \Omega)$ is always positive}
				\State Choose the smallest $b$ where $\mathcal{P}$ is minimized.
				\EndIf
				
				\State \textbf{Step 5: Return Optimal Bid}
				\State \Return $b_{\tau}^*$
			\end{algorithmic}
		\end{algorithm}
		
		\paragraph{Alternative way to construct \(	\mathcal{P} (b, \Omega)\)}   Based on supply instead of duration, TBA. 
		
		\subsection{Cost Cap vs Target CPA}
		As we  mentioned earlier, for target CPA, if the cost constraint is strictly satisfied—meaning that each conversion costs exactly \( C \) on average—then maximizing conversions is equivalent to maximizing overall spend delivery.
		
		This results in a fundamental difference between target CPA and cost cap. In the cost cap strategy:
		\begin{itemize}
			\item If the advertiser sets the cap \( C \) too high, the cost constraint becomes ineffective, reducing the problem to a max delivery problem. In this case, the optimal strategy is to allocate the budget across the pacing lifetime based on the supply pattern.
			\item If \( C \) is set too low, the cost constraint becomes the limiting factor, and the actual spend under the optimal strategy will be less than that of the max delivery problem with the same settings (except for the cost constraint). In this case, the delivery is still distributed across the campaign lifetime, though the budget may not be fully depleted by the end of the campaign.
		\end{itemize}
		
		However, target CPA behaves quite differently, especially when the target \( C \) is relatively high compared to the market level. Unlike cost cap, which slows pacing and adjusts bids to match market conditions, target CPA dynamically adjusts bids to maintain an average CPA near \( C \). 
		
		\begin{itemize}
			\item If the target \( C \) is set at a reasonable level relative to the market, the algorithm dynamically adjusts bids to maximize conversions while maintaining an average CPA close to \( C \). In this case, the pacing behavior is stable, and the budget is allocated efficiently throughout the campaign lifetime.
			\item If the target \( C \) is set too high, the algorithm may initially bid more aggressively to acquire conversions at a higher cost, potentially leading to faster budget depletion.
		\end{itemize}
			
		This is also the reason why we previously mentioned that target CPA is more suitable for advertisers who have a clearer understanding of their ideal CPA target. If the target CPA is set too high, advertisers may end up paying more than the competitive market level to acquire conversions, potentially leading to inefficient spending and faster budget depletion.
		
		We summarize the differences between cost cap and target CPA in \autoref{tab:costcap_vs_targetcpa}.

\begin{table}[h]
	\centering
	\renewcommand{\arraystretch}{1.3}
	\begin{tabular}{|p{5cm}|p{5cm}|p{5cm}|}
		\hline
		\textbf{} & \textbf{Cost Cap} & \textbf{Target CPA} \\
		\hline
		\textbf{Objective} & Maximizes conversions while ensuring the average cost per action (CPA) does not exceed a strict threshold \( C \). & Maximizes conversions while keeping the average CPA close to \( C \), allowing for some fluctuation. \\
		\hline
		\textbf{Bid Adjustment Behavior} & Adjusts bids dynamically to maintain cost efficiency, \textbf{slowing down delivery if necessary} to ensure CPA does not exceed \( C \). & Adjusts bids dynamically to acquire as many conversions as possible \textbf{without strictly enforcing an upper CPA limit}. May allow temporary deviations. \\
		\hline
		\textbf{ \( C \) Too High} & If \( C \) is set too high, the cost constraint becomes inactive, and the system behaves like a \textbf{max delivery} strategy, distributing spend based on supply. & If \( C \) is set too high, the system \textbf{may initially bid aggressively} to acquire expensive conversions, potentially leading to \textbf{faster budget depletion}. \\
		\hline
		\textbf{\( C \) Too Low} & If \( C \) is set too low, the system becomes \textbf{overly restrictive}, limiting auction participation and leading to \textbf{under-delivery} (budget may not be fully spent). & If \( C \) is set too low, the system \textbf{reduces bid competitiveness}, potentially limiting auction wins but still prioritizing conversion volume within constraints. \\
		\hline
		\textbf{Budget Consumption Behavior} & Budget pacing is \textbf{stable and controlled} to ensure delivery throughout the campaign lifetime. & Budget \textbf{may be depleted earlier than scheduled} if the system needs to bid higher to meet the target CPA. \\
		\hline
	\end{tabular}
	\caption{Comparison of Cost Cap vs. Target CPA}
	\label{tab:costcap_vs_targetcpa}
\end{table}

	\end{section}

	\section{Remarks}  \label{sec:mpc_remarks}
	
	\subsection{Applicability Across Funnel Stages}
	This online isotonic-regression approach is particularly natural and effective for upper-funnel campaigns, such as impression- and video-view objectives, because their result signals are dense and observed with little delay; recent bid--spend and bid--result pairs therefore provide fresh and sufficiently stable data for fitting $f(b)$ and $g(b)$. For lower-funnel objectives, realized clicks and especially conversions are typically sparser, noisier, and delayed, making a response curve fitted directly from recent outcomes unreliable. A practical alternative is to construct $g(b)$ from a calibrated proxy signal: for click-cost control, the sum of predicted click-through rates ($\mathrm{pCTR}$) over won opportunities can serve as the expected click count, while conversion objectives may use $\mathrm{pCTR}\times\mathrm{pCVR}$ or another calibrated expected-conversion signal. The controller can then enforce a proxy cost constraint through $h_{\mathrm{proxy}}(b)=f(b)/g_{\mathrm{proxy}}(b)$, with periodic calibration against realized outcomes and, where necessary, delayed-feedback correction, attribution, and variance-reduction mechanisms \cite{chen2026lightweight}.
	
	\subsection{Bidding Stability Considerations}
	
	For system stability, when designing a bid update algorithm, it is important to ensure that the overall bidding dynamics remain stable. Drastic fluctuations in bid levels can lead to performance issues. For instance, a sudden drop in bids may result in under-delivery, while a sharp increase in bids could cause rapid budget depletion within a short period. 
	
	To address this concern, the algorithms described above can be modified to maintain stability in a production environment. For example, when performing bid searches from \(b_l\) to \(b_u\) in cost cap and target CPA strategies, we can impose constraints on bid variation to prevent excessive deviation from the previous bid \(b_t\)	\footnote{The choice of this percentile depends on the pacing interval. For example, if the algorithm is expected to adjust bids 10 times per hour, the percentile can be determined based on this number (10) and the number of bid adjustment opportunities available (i.e., \(1 \text{ hour} / \text{pacing interval}\)).}. A possible approach is to set:
	
	\[
	b_l = (1 - 0.1) \cdot b_t, \quad b_u = (1 + 0.1) \cdot b_t. 
	\]
	
	This principle also applies to the max-delivery problem, where lower and upper bounds can be imposed based on the last bid \(b_t\) to prevent excessive fluctuations in the newly updated bid.

	\subsection{MPC Variants}  MPC is a versatile framework, and the methods presented in this chapter represent only one possible instantiation for solving max-delivery and cost-cap problems. In particular, the controllers described above use a one-step receding-horizon control (RHC) formulation: at each pacing cycle, the response over the next interval is predicted, a single bid is optimized and applied, and the optimization problem is solved again after new delivery feedback becomes available. This design is attractive in production because it requires only short-horizon response estimates and keeps both model serving and online optimization lightweight \cite{chen2026lightweight}.
	
	With greater modeling and computational capacity, the same framework can be extended to a multi-step RHC controller with horizon \(H>1\). Such a controller jointly optimizes a sequence of future bids using interval-specific forecasts of supply, spend, and results, while applying only the first bid before re-optimizing at the next pacing cycle. Multi-step planning can better anticipate intraday traffic variation, future budget scarcity, and interactions between budget and cost constraints, thereby improving planning when the forecasts are sufficiently accurate. This additional planning capability comes at the cost of greater modeling requirements, serving latency, optimization complexity, and sensitivity to forecast error.
	
	The predictive models used within MPC are also not limited to LIS or isotonic regression. Depending on data availability and production requirements, the bid-to-spend and bid-to-result functions may be estimated using parametric bid-landscape models, splines or other shape-constrained regression methods, tree-based machine-learning models, deep neural networks, sequence models, or auction simulators. Richer ML/DL models can condition on features such as time, campaign characteristics, placement, audience, and marketplace state, and can share information across campaigns when campaign-level observations are sparse. Hybrid designs are also possible, for example by combining an offline global model with lightweight online calibration or monotonic correction. These alternatives trade some of the simplicity, interpretability, and stability of the online PAVA approach for potentially greater predictive accuracy and more effective long-horizon planning.
	
		
	\subsection{Proof of Monotonicity}
	We prove that \( g(\cdot) \), as defined in \autoref{eq:ecpm_bid_cost}, is monotonically non-decreasing. It suffices to show that for any \( \tau \geq 0 \):
	
	\[
	g(x + \tau) - g(x) \geq 0.
	\]
	
	\paragraph{Step 1: Expressing \( g(x) \)}
	From the definition of \( g(x) \):
	
	\[
	g(x) = \frac{\int_{0}^{x} z p(z) \text{d}z}{\int_{0}^{x} p(z) \text{d}z}.
	\]
	
	Thus, for \( g(x+\tau) - g(x) \), we compute:
	
	\begin{equation*}
		\begin{aligned}
			g(x+\tau) - g(x) 
			&= \frac{\int_{0}^{x+\tau} z p(z) \text{d}z}{\int_{0}^{x + \tau} p(z) \text{d}z} - \frac{\int_{0}^{x} z p(z) \text{d}z}{\int_{0}^{x} p(z) \text{d}z} \\
			&= \frac{\left(\int_{0}^{x+\tau} z p(z) \text{d}z \right)\cdot \left(\int_{0}^{x} p(z) \text{d}z\right) - \left(\int_{0}^{x + \tau} p(z) \text{d}z\right) \cdot \left( \int_{0}^{x} z p(z) \text{d}z\right)}{\left(\int_{0}^{x + \tau} p(z) \text{d}z\right) \cdot \left(\int_{0}^{x} p(z) \text{d}z\right)}.
		\end{aligned}
	\end{equation*}
	
	\paragraph{Step 2: Defining the Numerator}
	Define the numerator as \( \mathcal{I}(\tau) \):
	
	\[
	\mathcal{I}(\tau) = \left(\int_{0}^{x+\tau} z p(z) \text{d}z \right)\cdot \left(\int_{0}^{x} p(z) \text{d}z\right) - \left(\int_{0}^{x + \tau} p(z) \text{d}z\right) \cdot \left( \int_{0}^{x} z p(z) \text{d}z\right).
	\]
	
	It suffices to show that for all \( \tau \geq 0 \):
	
	\[
	\mathcal{I}(\tau) \geq 0.
	\]
	
	\paragraph{Step 3: Computing the Derivative \( \mathcal{I}'(\tau) \)}
	Since \( \mathcal{I}(0) = 0 \), we show that \( \mathcal{I}'(\tau) \geq 0 \) whenever \( \tau \geq 0 \):
	
	\begin{equation*}
		\begin{aligned}
			\frac{\text{d} \mathcal{I}(\tau)}{\text{d} \tau} 
			&= (x+\tau) \cdot p(x+\tau) \cdot \int_{0}^{x} p(z) \text{d}z - p(x+\tau) \cdot \int_{0}^{x} z p(z) \text{d}z \\
			&= p(x + \tau) \cdot \left[ (x + \tau) \cdot \int_{0}^{x} p(z) \text{d}z - \int_{0}^{x} z p(z) \text{d}z \right] \\
			&= p(x + \tau) \cdot \left[ \int_{0}^{x} (x - z) \cdot p(z) \text{d}z + \tau \cdot \int_{0}^{x} p(z) \text{d}z \right] \\
			&= p(x + \tau) \cdot \left( \mathcal{I}_1 + \mathcal{I}_2 \right).
		\end{aligned}
	\end{equation*}
	
	\paragraph{Step 4: Showing \( \mathcal{I}'(\tau) \geq 0 \)}
	Since \( x - z \geq 0 \) for \( z \in [0, x] \), we obtain:
	
	\[
	\mathcal{I}_1 = \int_{0}^{x} (x - z) \cdot p(z) \text{d}z \geq 0.
	\]
	
	Moreover, since \( \tau \geq 0 \):
	
	\[
	\mathcal{I}_2 = \tau \cdot \int_{0}^{x} p(z) \text{d}z \geq 0.
	\]
	
	Since \( p(x+\tau) \geq 0 \), it follows that:
	
	\[
	\frac{\text{d} \mathcal{I}(\tau)}{\text{d} \tau} \geq 0 \quad \text{for } \tau \geq 0.
	\]
	
	\paragraph{Step 5: Conclusion}
	Since \( \mathcal{I}(0) = 0 \) and \( \mathcal{I}'(\tau) \geq 0 \) for all \( \tau \geq 0 \), we conclude that:
	
	\[
	\mathcal{I}(\tau) \geq 0.
	\]
	
	Thus, \( g(\cdot) \) is \textbf{monotonically non-decreasing}.

	\chapter{Dual Online Gradient Descent}
	
	\intro{
		In this chapter, we introduce an adaptive optimal control method called Dual Online Gradient Descent (DOGD). This method updates the bid price by adaptively updating the parameters in the dual space using a stochastic gradient descent algorithm. We demonstrate how this approach can be applied to solve both the max delivery and cost cap problems.
	}
	
	\begin{section}{Max Delivery}
	\subsection{Main Algorithm}
	The intuition behind controller-based approaches introduced in previous sections is to tweak the bid of a campaign by comparing the current delivery status to the target delivery schedule. This is based on the fact that the optimal budget consumption rate should be proportional to the distributional density of eligible auction opportunities for the campaign. We can take another perspective by directly solving this problem in the dual space. The key observation is as follows:
	
	Recall that the optimal bid per impression is given by:
	\[
	b_t^{*} = \frac{r_t}{\lambda^*}.
	\]
	Therefore, to find \(b_t^{*}\), it is sufficient to determine \(\lambda^*\). Note that the dual problem is given by \autoref{eq:max_delivery_dual}:
	\[
	\min_{\lambda \geq 0}  \mathcal{L}^*(\lambda) = \min_{\lambda \geq 0}  \sum_{t=1}^{T} \left[ (r_t - \lambda c_t)_{+} + \lambda \cdot \frac{B}{T} \right].
	\]
	Denoting \((r_t - \lambda c_t)_{+} + \lambda \cdot \frac{B}{T}\) by \(f_t(\lambda)\), we have:
	\[
	\min_{\lambda \geq 0}  \sum_{t=1}^{T} f_t(\lambda).
	\]
	Readers who are familiar with optimization should recognize that this is a standard one-dimensional convex optimization problem. As auction requests arrive online in a streaming manner, it can naturally be solved using the Stochastic Gradient Descent (SGD) method. The update rule for \(\lambda\) is given by:
	\begin{equation} \label{eq:dogd_md_update_rule}
	\lambda_t \gets \lambda_t - \epsilon_t \cdot \nabla_{\lambda} f_t(\lambda) = \lambda_t - \epsilon_t \cdot \left( \frac{B}{T} - \mathds{1}_{ \{r_t > \lambda_t c_t \}} \cdot c_t \right),
	\end{equation}
	where \(\epsilon_t\) is the step size(learning rate), \(\mathds{1}_{\{r_t > \lambda_t c_t\}}\) is the indicator function, and \(c_t\) is the highest competing bid per impression in the market. Note that \(\mathds{1}_{ \{r_t > \lambda_t c_t \}} \cdot c_t\) represents the observed spend in the \(t\)-th auction, while \(B/T\) is the expected spend per auction. The gradient thus quantifies the deviation from the expected spend. The bid at the \(t\)-th auction round is:
	\[
	b_t = \frac{r_t}{\lambda_t}.
	\]
	
	The discussion above highlights the core idea of the \textbf{D}ual \textbf{O}nline \textbf{G}radient \textbf{D}escent (\textbf{DOGD}) algorithm for the max delivery problem, which we summarize in \autoref{alg:dogd_md}:

	\begin{algorithm}[H]
		\caption{DOGD for Max Delivery}
		\label{alg:dogd_md}
		\begin{algorithmic}[1]
			\Require $B$: Total budget, $T$:  Predicted total number of auction opportunities, $\epsilon_t$: Step size schedule
			
			\State Initialize $\lambda_0 \gets \lambda_{\text{init}}$ \Comment{Initial dual variable}
			
			\For{all incoming auction requests indexed by $t$} \Comment{Iterate over auction rounds}
			\State Observe  $r_t$: pCTR, $c_t$: highest competing eCPM in auction $t$
			\State Compute the gradient of $f_t(\lambda)$:
			\[
			\nabla_{\lambda} f_t(\lambda) = \frac{B}{T} - \mathds{1}_{ \{r_t > \lambda_t c_t\}} \cdot c_t
			\]
			\State Update $\lambda_t$ using SGD:
			\[
			\lambda_t \gets \lambda_t - \epsilon_t \cdot \nabla_{\lambda} f_t(\lambda)
			\]
			\State Compute the bid per impression for the $t$-th auction:
			\[
			b_t \gets \frac{r_t}{\lambda_t}
			\]
			\State Submit $b_t$ for auction $t$
			\EndFor
		\end{algorithmic}
	\end{algorithm}
	
	As we discussed before, instead of updating bids for every auction, it is more common to update bids in a batch manner. Suppose the update interval is \(\Delta t\) and \(R(t)\) is the number of observed auction requests within this interval. The mini-batch gradient within \(\Delta t\) is given by:
	\[
	\sum_{s \in (t, t+\Delta t)} \nabla_{\lambda} f_s(\lambda) = \sum_{s \in (t, t+\Delta t)} \left( \frac{B}{T} - \mathds{1}_{ \{r_s > \lambda_s c_s\}} \cdot c_s \right) = \frac{R(t)}{T} \cdot B - \sum_{s \in (t, t+\Delta t)} \mathds{1}_{ \{r_s > \lambda_s c_s\}} \cdot c_s.
	\]
	Note that \(\sum_{s \in (t, t+\Delta t)} \mathds{1}_{ \{r_s > \lambda_s c_s\}} \cdot c_s\) represents the actual spend during \(\Delta t\), which we denote as \(S(t)\). The mini-batch gradient can then be written as:
	\[
	\sum_{s \in (t, t+\Delta t)} \nabla_{\lambda} f_s(\lambda) = \frac{R(t)}{T} \cdot B - S(t).
	\]
	The mini-batch update rule is given by:
	\[
	\lambda_t \gets \lambda_t - \epsilon_t \cdot \left( \frac{R(t)}{T} \cdot B - S(t) \right).
	\]
	The bid per click remains unchanged within \((t, t+\Delta t)\) and is given by:
	\[
	b_{\text{click}, t} = \frac{1}{\lambda_t}
	\]
	For each auction request \(s \in (t, t+\Delta t)\), the bid per impression is computed as:
	\[
	b_s = b_{\text{click}, t} \cdot r_s = \frac{r_s}{\lambda_t}
	\]

	We summarize this Mini-Batch DOGD algorithm in \autoref{alg:dogd_md_batch}:

	\begin{algorithm}[H]
		\caption{Mini-Batch DOGD for Max Delivery Problem}
		\label{alg:dogd_md_batch}
		\begin{algorithmic}[1]
			\Require $B$: Total budget, $T$: Predicted total number of auction opportunities, $\Delta t$: Mini-batch update interval, $\epsilon_t$: Step size schedule
			\Ensure Optimal dual variable $\lambda_t$ and corresponding bids per impression $b_s$
			
			\State Initialize $\lambda_0 \gets \lambda_{\text{init}}$ \Comment{Initial dual variable}
			
			\For{$t = 0$ to \texttt{EndOfDay} with step size $\Delta t$} \Comment{Iterate over mini-batches}
			\State Count the number of auction requests \(R(t)\) and observe the actual spend $S(t)$ during interval \((t, t+\Delta t)\)
			
			\State Compute the mini-batch gradient:
			\[
			\text{BatchGrad}_t = \sum_{ s \in (t, t+\Delta t)} \nabla_{\lambda} f_s(\lambda) = \frac{R(t)}{T} \cdot B - S(t)
			\]
			\State Update the dual variable using the mini-batch gradient:
			\[
			\lambda_t \gets \lambda_t - \epsilon_t \cdot \text{BatchGrad}_t 
			\]
			\State Compute the bid per click for all auctions in \((t, t+\Delta t)\):
			\[
			b_{click, t} = \frac{1}{\lambda_t}
			\]
			\State Compute bid per impression $b_s$ for all $s \in (t, t+\Delta t)$ with pCTR $r_s$:
			\[
			b_s = b_{click, t} \cdot r_s
			\]
			\EndFor
		
		\end{algorithmic}
	\end{algorithm}

	Some remarks on the DOGD algorithm: The optimization is performed at the campaign level. Interestingly, under certain regularity conditions, this campaign-level optimization leads to a marketplace Nash equilibrium. Furthermore, regret analysis can be conducted to demonstrate that this algorithm is theoretically optimal. For more technical details, one may refer to \cite{balseiro2020dual}, \cite{balseiro2019learning}, and \cite{gao2022bidding}.

	\subsection{Practical Considerations} 
		Someome practical considerations for implementing the DOGD algorithm for max delivery in real-world production sytem:
	\begin{itemize}
		\item \textbf{$\lambda$ Initialization.}
		When historical auction logs are available, they can be replayed or used to estimate a bid $b_{\mathrm{init}}$ whose expected spend per opportunity matches the target $B/T$, after which $\lambda_{\mathrm{init}}$ is set to $1/b_{\mathrm{init}}$ \cite{gao2022bidding}. If reliable historical data are unavailable, a bounded cold-start exploration procedure can test a small range of bids and initialize $\lambda$ from the observed bid--spend response. More details will be discussed in \autoref{init}.
		
		\item \textbf{Normalization}.
		\begin{itemize}
			\item \textbf{Normalization of the update rule.}
			Define the dimensionless delivery ratio
			\[
			\rho_t = \frac{S(t)/R(t)}{B/T}.
			\]
			The update can then use the normalized pacing error $1-\rho_t$ instead of the raw spend difference, making its scale more comparable across campaigns with different budgets and traffic volumes.
			
			\item \textbf{Normalization of $\lambda$.}
			Define $\widetilde{\lambda}_t=\lambda_t/\lambda_{\mathrm{norm}}$ and apply the normalized update
			\[
			\widetilde{\lambda}_{t+1}
			\gets \widetilde{\lambda}_t-\eta_t(1-\rho_t).
			\]
			A practical choice is $\lambda_{\mathrm{norm}}\approx\lambda_{\mathrm{init}}$: choosing it too small can slow convergence, whereas choosing it too large can amplify noise and produce oscillatory updates.
		\end{itemize}
	\end{itemize}
	
	\end{section}

	\begin{section}{Cost Cap}
		\subsection{Main Algorithm}
		The Cost Cap problem can also be solved using the DOGD algorithm. Recall that the optimal bid per click for the cost cap is given by \autoref{eq:cost_cap_bid_formula}:
		\begin{equation*}
			b^*_{\text{click}} = \frac{\lambda^*}{\lambda^* + \mu^*} \cdot \frac{1}{\lambda^*} + \frac{\mu^*}{\lambda^* + \mu^*} \cdot C,
		\end{equation*}
		where \(\lambda^*\) and \(\mu^*\) are the dual variables. 
		
		Similar to the method discussed in the Max Delivery problem in the previous section, solving this problem reduces to determining the dual variables \(\lambda^*\) and \(\mu^*\). The dual problem of the cost cap is given by \autoref{eq:cost_cap_dual}:
		\begin{equation*}
			\min_{\lambda \geq 0, \mu \geq 0} \mathcal{L}^*(\lambda, \mu) = \min_{\lambda \geq 0, \mu \geq 0} \sum_{t=1}^{T} \left[ \big(r_t - \lambda c_t - \mu c_t + \mu C r_t\big)_{+} + \lambda \cdot \frac{B}{T} \right].
		\end{equation*}
		Define the per-time step loss function as:
		\[
		f_t(\lambda, \mu) = \big(r_t - \lambda c_t - \mu c_t + \mu C r_t\big)_{+} + \lambda \cdot \frac{B}{T}.
		\]
		The dual problem can then be rewritten as:
		\[
		\min_{\lambda \geq 0, \mu \geq 0} \mathcal{L}^*(\lambda, \mu) = \min_{\lambda \geq 0, \mu \geq 0} \sum_{t=1}^{T} f_t(\lambda, \mu).
		\]
		Using stochastic gradient descent (SGD), the update rules for \(\lambda\) and \(\mu\) are:
		\begin{equation*}
			\begin{aligned}
				\lambda_{t+1} &  \gets  \lambda_t - \epsilon_t \cdot \nabla_\lambda f_t(\lambda, \mu)= \lambda_t - \epsilon_t \cdot \left( \frac{B}{T} - c_t \cdot \mathds{1}_{\{r_t - \lambda c_t - \mu c_t + \mu C r_t > 0\}} \right), \\
				\mu_{t+1} & \gets  \mu_t - \epsilon_t \cdot \nabla_\mu f_t(\lambda, \mu)=  \mu_t - \epsilon_t \cdot \big(C r_t - c_t\big) \cdot \mathds{1}_{\{r_t - \lambda c_t - \mu c_t + \mu C r_t > 0\}},
			\end{aligned}
		\end{equation*}
		where \(\epsilon_t\) is the learning rate, and \(\mathds{1}_{\{\cdot\}}\) is the indicator function. If we examine the update rules more closely, \(c_t / r_t\) represents the actual cost per click in the sense of expectation. From this, we can observe that the gradients of \(\lambda\) and \(\mu\) quantify the deviations from the target spend and the target cost per click (CPC), respectively.

		\begin{algorithm}[H]
			\caption{Dual Online Gradient Descent (DOGD) for Cost Cap}
			\label{alg:dogd_cost_cap}
			\begin{algorithmic}[1]
				\Require $B$: Total budget, $C$: Target cost per click (CPC), $T$: Predicted total number of auction opportunities, $\epsilon_t$: Step size schedule
				\Ensure Bid values for auctions
				\State Initialize $\lambda_0 \gets \lambda_{\text{init}}, \mu_0 \gets \mu_{\text{init}}$ \Comment{Initial dual variables}
				\For{all incoming auction requests indexed by $t$}
				\State Observe $r_t$: predicted click-through rate (pCTR), $c_t$: highest competing eCPM in auction $t$
				\State Compute the gradient of $f_t(\lambda, \mu)$:
				\[
				\nabla_\lambda f_t(\lambda, \mu) \gets \frac{B}{T} - c_t \cdot \mathds{1}_{\{r_t - \lambda c_t - \mu c_t + \mu C r_t > 0\}}
				\]
				\[
				\nabla_\mu f_t(\lambda, \mu) \gets \big(C r_t - c_t\big) \cdot \mathds{1}_{\{r_t - \lambda c_t - \mu c_t + \mu C r_t > 0\}}
				\]
				\State Update $\lambda_t$ and $\mu_t$ using SGD:
				\[
				\lambda_{t+1} \gets \lambda_t - \epsilon_t \cdot \nabla_\lambda f_t(\lambda, \mu)
				\]
				\[
				\mu_{t+1} \gets \mu_t - \epsilon_t \cdot \nabla_\mu f_t(\lambda, \mu)
				\]
				\State Compute the bid per impression for the $t$-th auction:
				\[
				b_t \gets \frac{1 + \mu_t \cdot C}{\lambda_t + \mu_t} \cdot r_t
				\]
				\State Submit $b_t$ for auction $t$
				\EndFor
			\end{algorithmic}
		\end{algorithm}
	
	As we discussed in the previous section, in practice, it is more common to implement the batch update algorithm. The batch update of $\lambda$ is quite similar to the formula used in max delivery. The mini-batch gradient is given by:
	\[
	\sum_{s \in (t, t+\Delta t)} \nabla_{\lambda} f_s(\lambda, \mu) = \frac{R(t)}{T} \cdot B - S(t),
	\]
	where $\Delta t$ is the update time interval, $R(t)$ is the number of observed auction requests, and $S(t)$ is the actual spend during $\Delta t$. As for $\mu$, note that:
	\[
	\sum_{s \in (t, t+\Delta t)} r_s \mathds{1}_{\{r_s - \lambda c_s - \mu c_s + \mu C r_s > 0\}}
	\]
	is the expected number of conversions (in this case, clicks) during $\Delta t$. If, within $\Delta t$, there are sufficient actual conversions (denoted as $N(t)$), then $N(t)$ is a good approximation of the sum above. Thus:
	\[
	\sum_{s \in (t, t+\Delta t)} \nabla_{\mu} f_s(\lambda, \mu) \approx C \cdot N(t) - S(t).
	\]
	
	The mini-batch update rule is then:
	\begin{equation*}
		\begin{aligned}
			\lambda_{t+1} & \gets \lambda_t - \epsilon_t \cdot \left( \frac{R(t)}{T} \cdot B - S(t) \right), \\
			\mu_{t+1} & \gets \mu_t - \epsilon_t \cdot \left( C \cdot N(t) - S(t) \right),
		\end{aligned}
	\end{equation*}
	where $\epsilon_t$ is the step size.
	
	Similar to max delivery, in the mini-batch update, the bid per click remains unchanged within $(t, t + \Delta t)$ and is given by:
	\[
	b_{\text{click}, t} = \frac{1 + \mu_t C}{\lambda_t + \mu_t}.
	\]
	
	The bid per impression for any $s \in (t, t+\Delta t)$ is:
	\[
	b_s = b_{\text{click}, t} \cdot r_s = \frac{1 + \mu_t C}{\lambda_t + \mu_t} \cdot r_s.
	\]
	
	\begin{algorithm}[H]
		\caption{Mini-Batch DOGD for Cost Cap Problem}
		\label{alg:minibatch_dogd_costcap}
		\begin{algorithmic}[1]
			\Require $B$: Total budget, $C$: Target cost per click (CPC), $T$: Predicted total number of auction opportunities, $\Delta t$: Mini-batch update interval, $\epsilon_t$: Step size schedule
			\Ensure Optimal dual variables $\lambda_t$, $\mu_t$, and corresponding bids per impression $b_s$
			\State Initialize $\lambda_0 \gets \lambda_{\text{init}}, \mu_0 \gets \mu_{\text{init}}$ \Comment{Initial dual variables}
			\For{$t = 0$ to EndOfDay with step size $\Delta t$} \Comment{Iterate over mini-batches}
			\State Count the number of auction requests $R(t)$ and observe the actual spend $S(t)$ during interval $(t, t + \Delta t)$
			\State Count the number of conversions (clicks) $N(t)$ during interval $(t, t + \Delta t)$
			\State Compute the mini-batch gradients:
			\[
			\text{BatchGrad}_{\lambda, t} = \frac{R(t)}{T} \cdot B - S(t)
			\]
			\[
			\text{BatchGrad}_{\mu, t} = C \cdot N(t) - S(t)
			\]
			\State Update the dual variables using the mini-batch gradients:
			\[
			\lambda_{t+1} \gets \lambda_t - \epsilon_t \cdot \text{BatchGrad}_{\lambda, t}
			\]
			\[
			\mu_{t+1} \gets \mu_t - \epsilon_t \cdot \text{BatchGrad}_{\mu, t}
			\]
			\State Compute the bid per click for all auctions in $(t, t + \Delta t)$:
			\[
			b_{\text{click}, t} = \frac{1 + \mu_t C}{\lambda_t + \mu_t}
			\]
			\State Compute bid per impression $b_s$ for all $s \in (t, t + \Delta t)$ with pCTR $r_s$:
			\[
			b_s = b_{\text{click}, t} \cdot r_s = \frac{1 + \mu_t C}{\lambda_t + \mu_t} \cdot r_s
			\]
			\EndFor
		\end{algorithmic}
	\end{algorithm}

	Related algorithms can be found in \cite{gao2022bidding}.

	
	\end{section}

 \section{Remarks}

	\subsection{Other Applications of DOGD}
	DOGD is a versatile framework and is not restricted to a single-campaign max-delivery or cost-cap controller. Its essential structure extends to separable online allocation problems with shared resource constraints: a dual variable is introduced for each shared constraint and updated from the corresponding observed constraint residual \cite{balseiro2020dual}. As demonstrated in the following application chapters, a common dual variable can coordinate delivery across onsite and offsite channels, even when the channels use different auction mechanisms, while channel-specific bidding formulas translate the shared dual price into bids. The same idea applies to campaign-group optimization, where multiple campaigns share a common budget and the group-level dual variable is updated from their aggregate spend. Additional dual variables may similarly be introduced for campaign-level cost caps, minimum-delivery requirements, or other business constraints.
	
	\subsection{DOGD and PID}
	DOGD can be viewed as dual mirror descent: a quadratic mirror map produces the additive update used above, while an entropic mirror map produces a multiplicative update that naturally preserves the positivity of $\lambda$ \cite{gao2022bidding}. In a dual-based pacing controller, the pacing error is also a subgradient of the dual objective. Applying suitable weights to the current and past subgradients yields convolutional mirror descent, which is equivalent to a PID-style update; the proportional, integral, and derivative terms correspond to different filters of the error history \cite{balseiro2022analysis}.

	\subsection{Nash Equilibrium}
	The DOGD controller described above is implemented independently at the campaign level: each campaign learns its own dual price from its budget and expenditure path, without explicit coordination with other campaigns. Nevertheless, the interaction of campaign-level adaptive
	pacing policies can produce a stable marketplace outcome in repeated auctions. Balseiro and Gur show that, when all bidders adopt their adaptive pacing strategies and the market satisfies the paper's regularity and large-market conditions, the induced dynamics converge
	and the resulting strategy profile forms an approximate Nash equilibrium in dynamic strategies. As the numbers of auctions and competitors grow, the benefit from a unilateral deviation---including a strategy with access to complete information---becomes negligible
	\cite{balseiro2019learning}. 
	
	\subsection{Duality Gap}
	Let $P^\star$ denote the optimal value of the original max-delivery problem with binary decisions $x_t\in\{0,1\}$, and let $P_{\mathrm{LP}}^\star$ denote the value of its linear relaxation with $0\leq x_t\leq 1$. The dual objective optimized by DOGD is exactly the
	dual of this linear relaxation:
	\[
	D^\star
	=
	\min_{\lambda\geq 0}
	\left\{
	\lambda B + \sum_{t=1}^{T}\bigl(r_t-\lambda c_t\bigr)_{+}
	\right\}.
	\]
	Strong duality for linear programming gives $D^\star=P_{\mathrm{LP}}^\star$, while weak duality gives $P^\star\leq D^\star$. Therefore, the gap relative to the original binary problem is an \emph{integrality gap}, rather than a failure of strong duality:
	\[
	0\leq D^\star-P^\star
	=P_{\mathrm{LP}}^\star-P^\star
	\leq r_{\max},
	\qquad
	r_{\max}=\max_{1\leq t\leq T} r_t.
	\]
	For a one-budget knapsack relaxation, an optimal extreme point has at most one fractional auction; discarding that auction yields a feasible binary solution and loses at most $r_{\max}$. Amar and Renegar use this LP--dual threshold structure to show that, under a low-individual-impact setting, a linear bid of the form $r_t/\lambda^\star$ obtains a feasible solution whose value is within one maximum-item value of the integer optimum \cite{amar2018second}. Consequently, the relative relaxation gap is small when no individual auction contributes a material fraction of the campaign's total value.
	


	\part{Applications}
	\label{part:misc}
	
	\chapter{Initialization of Campaign Bid}
	 \label{init}
	
	\intro{
		The initial bid of a campaign is crucial in the pacing problem, as it influences how quickly the bidding algorithm converges to the optimal level. In this chapter, we present two approaches—parametric and non-parametric methods—to demonstrate how algorithms can be designed to derive reasonable initial bids, thereby accelerating the convergence of the bidding algorithm.
	}
	
	Most of the algorithms discussed in \autoref{part:pacing_algorithms} focus on how to update bids for pacing during a campaign's delivery. However, setting the initial bid properly is equally important for both individual campaign performance and the overall stability of the marketplace. 
	
	A poorly chosen initial bid can significantly impact the performance of the campaign:
	\begin{itemize}
		\item If the initial bid is \textbf{too low}, the campaign might take longer to adjust its bid to a reasonable market level, delaying its ability to win auction opportunities and meet delivery objectives.
		\item If the initial bid is \textbf{too high}, the campaign might win auction opportunities at unreasonably high prices, leading to rapid budget depletion without maximizing conversions or other objectives.
	\end{itemize}
	
	From a macroscopic perspective, poorly set initial bids across campaigns can influence the stability of the entire marketplace. If many campaigns start with bids significantly deviating from the equilibrium market level, it can cause fluctuations in auction dynamics, leading to inefficient resource allocation.
	
	In this chapter, we discuss how to compute a \textbf{reasonable initial bid} for a given ad campaign, balancing the need to quickly achieve competitive performance while avoiding excessive costs or instability.

	
	\section{Parametric Approach}
	In this section, we discuss the computation of the optimal initial bid using parametric methods. The algorithms described below rely on certain model predictions as input parameters. We do not delve too deeply into the underlying modeling techniques; instead, we primarily focus on the methodologies for determining the initial bids, assuming that all relevant signals are ready for use.
	
	\subsection{Max Delivery}
	We first discuss how to compute the initial bid for a cold start max delivery campaign. As shown in \autoref{part:pacing_algorithms}, the optimal bid for a max delivery campaign is a constant that exactly depletes the campaign's budget. For a cold start campaign, where no prior data is available, we need to leverage models that estimate market conditions and campaign-specific parameters to make the best guess for the initial bid.
	
	More specifically, suppose the highest eCPMs for this campaign among other bidders follow a log-normal distribution with parameters \(\mu\) and \(\sigma\). Additionally, assume the conversion rate\footnote{Conversion rate refers to the probability of achieving the campaign's objective conditioned on an impression. For example, for CPC campaigns, it corresponds to CTR, while for CPA campaigns, it corresponds to PVR, etc.} of the ad (independently) follows a log-normal distribution with parameters \(\mu'\) and \(\sigma'\).\footnote{Both eCPM and conversion rate are non-negative. It is common in practice to assume these data are subject to log-normal distributions. Empirical evidence suggests that the log-normal distribution is a good candidate for fitting these data.} Given a budget \(B\) and a total opportunity forecast \(T\), under a second price auction, we claim that the optimal bid per conversion \(b^*\) can be determined by solving the following equation:
	\begin{equation} \label{eq:init_bid_formula}
		e^{\mu + \frac{\sigma^2}{2}} \Phi \left( \frac{\mu' - \mu + \ln b^* - \sigma^2}{\sqrt{(\sigma')^2 + \sigma^2}} \right) = \frac{B}{T},
	\end{equation}
	where \(\Phi\) denotes the cumulative distribution function (CDF) of a standard Gaussian distribution \(\mathcal{N}(0, 1)\).
	
	We now break down \autoref{eq:init_bid_formula} to interpret the meaning of each term:
	
	\begin{itemize} 
		\item \textbf{\( \exp (\mu + \sigma^2/2)\)}:  
		Suppose the clearing price(the highest eCPM among other bidders) \(Z\) follows a log-normal distribution:  
		\[
		Z \sim \mathcal{LN}(\mu, \sigma^2).
		\]
		Then, the expected clearing price per impression is given by:  
		\[
		\mathbb{E} [Z] = \exp \left( \mu + \frac{\sigma^2}{2} \right).
		\]  
		Thus, the term \( \exp (\mu + \sigma^2/2) \) represents the expected clearing price per impression.
		
		\item \textbf{\(\Phi \left( \frac{\mu' - \mu + \ln b^* - \sigma^2}{\sqrt{(\sigma')^2 + \sigma^2}} \right)\)}:  
		Suppose the conversion rate is represented by a log-normal random variable:  
		\[
		R \sim \mathcal{LN}(\mu', (\sigma')^2).
		\]
		Given a bid \( b^* \), the auction is won if and only if:  
		\[
		b^* \cdot R > Z.
		\]
		The probability of winning an auction, conditioned on \( R, Z \), and \( b^* \), is given by:  
		\[
		P(b^* \cdot R > Z).
		\]
		Since \( R \) and \( Z \) are independent, we can prove:  
		\[
		P(b^* \cdot R > Z) = \Phi \left( \frac{\mu' - \mu + \ln b^* - \sigma^2}{\sqrt{(\sigma')^2 + \sigma^2}} \right).
		\]
		Therefore, the \(\Phi\) term represents the probability that the advertiser's eCPM bid, given \( b^* \), exceeds the highest competing eCPM bid.
		
		\item \textbf{\(B/T\)}:  
		This term represents the target cost per impression for the campaign.
	\end{itemize}
	
	In summary, \autoref{eq:init_bid_formula} implies that \( b^* \) is the bid level at which the expected cost per impression matches the target cost per impression, which is consistent with our previous derivations. We will prove this result in \autoref{ch:init_bid_remarks} of Remarks. For now, we assume this is correct. The problem is then equivalent to finding \(\mu, \sigma, \mu', \sigma'\), and \(T\). Below, we describe methods for estimating these parameters:
	
	\begin{itemize}
		\item \(\mu, \sigma\): These represent the distribution of eCPMs for this campaign. There are two main approaches to estimate these parameters:
		\begin{enumerate}
			\item \textbf{Sampling-Based Approach:} 
			Sample auctions from the target audience and record the highest eCPMs observed for each auction. Fit these observed eCPMs to a log-normal distribution using methods such as maximum likelihood estimation (MLE) to obtain \(\mu\) and \(\sigma\).
			\item \textbf{Historical Data Aggregation:}
			If historical data is available for similar campaigns with comparable targeting criteria, aggregate the eCPMs and fit a log-normal distribution to derive \(\mu\) and \(\sigma\).
		\end{enumerate}
		
		\item \(\mu', \sigma'\): These parameters describe the distribution of conversion rates, reflecting the quality of the ad campaign. Possible approaches include:
		\begin{enumerate}
			\item \textbf{Regression-Based Prediction:}
			Leverage features such as the campaign's ad creatives, targeting criteria, and historical performance data to train a regression model (e.g., linear regression, gradient boosting, or neural networks). Use this model to predict \(\mu'\) and \(\sigma'\).
			\item \textbf{Lookalike Campaigns:}
			Analyze the conversion rates of similar past campaigns to estimate \(\mu'\) and \(\sigma'\). Use transfer learning techniques if datasets from lookalike campaigns are small.
		\end{enumerate}
		
		\item \(T\): This represents the predicted number of auction requests and we may agregate historical supply data for the given targeting criteria of the campaign to forecast the total opportunity.
		
	\end{itemize}
	Once these parameters are determined, we can compute \(b^*\) directly by applying the inverse cumulative distribution function \(\Phi^{-1}\) of the standard Gaussian distribution to solve \autoref{eq:init_bid_formula}. The algorithm is summarized in \autoref{alg:init_bid_max_delivery}.
	
	\begin{algorithm}
		\caption{Optimal Initial Bid For Max Delivery Problem}
		\label{alg:init_bid_max_delivery}
		\begin{algorithmic}[1]
			\Require 
			$B$: Campaign budget; $T$: Predicted total opportunity forecast; \\
			$\mu, \sigma$: Parameters of eCPM distribution (log-normal); \\
			$\mu', \sigma'$: Parameters of campaign-specific conversion rate distribution (log-normal).
			\Ensure $b^*$: Optimal initial bid per conversion.
			
			\State \textbf{Step 1: Compute Auxiliary Values}
			\State Compute the expected spend per impression:
			\[
			\text{Spend} \gets e^{\mu + \frac{\sigma^2}{2}}
			\]
			\State Compute the standard deviation for combined log-normal distributions:
			\[
			\text{SD} \gets \sqrt{(\sigma')^2 + \sigma^2}
			\]
			\State Compute the RHS of the CDF equation:
			\[
			\text{RHS} \gets \frac{B}{T \cdot \text{Spend}}
			\]
			
			\State \textbf{Step 2: Solve for Optimal Bid}
			\State Compute the inverse CDF value:
			\[
			z \gets \Phi^{-1}(\text{RHS})
			\]
			\State Compute the logarithm of the optimal bid:
			\[
			\ln b^* \gets \mu - \mu' + \sigma^2 + \text{SD} \cdot z
			\]
			\State Compute the optimal bid:
			\[
			b^* \gets \exp(\ln b^*)
			\]
			
			\State \textbf{Step 3: Return Optimal Bid}
			\Return $b^*$
		\end{algorithmic}
	\end{algorithm}
	
	\subsection{Cost Cap}
	Deciding the optimal initial bid for a cost cap campaign is more complex than for a max delivery campaign. Recall that the objective of cost cap bidding is to maximize conversions while ensuring that the cost per result does not exceed a specified cap \(C\) set by the advertiser. The initial bid computed in \autoref{alg:init_bid_max_delivery} only considers the budget constraint. However, the posterior cost per result when bidding with this initial value may exceed the cap threshold. To address this issue, a natural approach is to cap the initial bid in \autoref{alg:init_bid_max_delivery} by the specified cost cap. Formally, we define the initial bid for a cost cap campaign as:
	\[
	\min \left(b^*, C\right).
	\]
	
	If we assume that the ratio \( \sigma \) between the highest eCPM and the second-highest eCPM remains stable over time, we can further refine the formula as:
	\[
	\min \left(b^*, \frac{C}{\sigma}\right).
	\]
	
	For additional details, readers may refer to the methodologies described in the "Cost-Min" algorithm.
	
	\section{Non-Parametric Approach}
	For campaigns with historical auction data (e.g., daily pacing campaigns that reset the initial bid at the beginning of each new pacing cycle), non-parametric methods can be employed to derive the initial bid directly from past auction data. 
	
	Unlike parametric methods, which rely heavily on the accuracy of prediction models, non-parametric approaches are less dependent on model quality. When sufficient auction data are available, non-parametric methods provide a robust alternative for determining the initial bid.

	\subsection{Auction Replay}
	Suppose we want to compute the initial bid for a daily pacing campaign and have access to its historical auction data from the past \(N\) days. If we are confident that the auction data from a particular day reflects a similar auction environment—such as having the same budget and a comparable supply or traffic pattern—with the upcoming pacing day,\footnote{In practice, the traffic pattern might be similar to the previous day or the same day from the prior week.} then this historical auction data can be leveraged to run simulations(auction replay) and estimate the optimal initial bid.

	More specifically, assume that the historical data from the selected day includes all the auctions in which the campaign has participated. For each auction opportunity \(i\), we have access to the highest eCPM \( c_i \) and the predicted click-through rate (pCTR) \( r_i \) for the campaign in the given auction slot. Suppose there are \( T \) auction opportunities. Given the set of pairs \( \{c_i, r_i\}_{i=1}^{T} \), our objective is to determine a constant bid that ensures the daily budget \( B \) is fully depleted by the end of the day.
	
	The simulation can be executed as follows: for a given bid price \( b \), we iterate over all auction opportunities. For each auction opportunity \( i \), we compute the eCPM as:
	
	\[
	eCPM_i = b \cdot r_i.
	\]
	
	This computed \( eCPM_i \) is then compared to the highest competing eCPM \( c_i \). If \( eCPM_i > c_i \), we win the auction and pay the second-highest price \( c_i \); otherwise, we lose the auction and incur no cost.
	
	The total cost is accumulated across all \( T \) auctions and compared to the daily budget \( B \). If the total spend is less than \( B \), the campaign is under-delivered, indicating that the bid needs to be increased. Conversely, if the spend exceeds \( B \), the bid should be decreased. 
	
	This process is repeated iteratively until we determine the optimal bid \( b^* \), where the budget is fully spent by the end of the day. Since a higher bid generally leads to higher spending, a \textbf{binary search} approach can be employed to accelerate the search for \( b^* \). We predefine a search interval \([b_l, b_u]\) and apply binary search to efficiently converge on the optimal bid. We summarize this idea in \autoref{alg:auction_replay_optimal_bid}. 

	\begin{algorithm}
		\caption{Optimal Initial Bid via Auction Replay}
		\label{alg:auction_replay_optimal_bid}
		\begin{algorithmic}[1]
			\Require Historical auction data $\{(c_i, r_i)\}_{i=1}^{T}$, budget $B$, search interval $[b_l, b_u]$, convergence threshold $\epsilon$
			\Ensure Optimal initial bid $b^*$
			
			\While{$b_u - b_l > \epsilon$}
			\State $b \gets \frac{b_l + b_u}{2}$ \Comment{Midpoint of search interval}
			\State $\text{total\_cost} \gets 0$
			
			\For{$i = 1$ to $T$}
			\State $eCPM_i \gets b \cdot r_i$
			\If{$eCPM_i > c_i$}
			\State $\text{total\_cost} \gets \text{total\_cost} + c_i$
			\EndIf
			\EndFor
			
			\If{$\text{total\_cost} < B$}
			\State $b_l \gets b$ \Comment{Increase bid to spend more}
			\Else
			\State $b_u \gets b$ \Comment{Decrease bid to spend less}
			\EndIf
			\EndWhile
			
			\State \Return $b^* \gets \frac{b_l + b_u}{2}$ \Comment{Final bid approximation}
		\end{algorithmic}
	\end{algorithm}

	\subsection{Converged Bids Average}
	The simulation method discussed above relies on auction data, including second-highest prices and predicted click-through rates (pCTRs). An alternative approach to estimating the initial bid requires only the historical bids from the campaign itself. 
	
	The key assumption behind this method is that the bidding algorithm dynamically adjusts bids online to reach an optimal level. If we observe that the bidding dynamics have converged over time, the average of these converged bids provides a good approximation of the optimal bid.
	
	Suppose we have a total of \( T \) bid data points, denoted as \( \{b_i\}_{i=1}^{T} \), collected throughout the day. To determine whether a consecutive subsequence of bids has converged, a simple approach is to compute the variance of the subsequence. If the variance is below a predefined threshold \( \delta \), we consider the sequence to be convergent. 
	
	To further reduce variance fluctuations, we impose a constraint that the subsequence length must be at least a predefined value, say \( K \). The objective, therefore, is to find the longest consecutive subsequence with a minimum length of \( K \) and variance less than \( \delta \). 
	
	A brute-force approach would require \( O(T^3) \) time complexity in the worst-case scenario—since there are \( O(T^2) \) possible consecutive subsequences, and computing the variance for each subsequence takes \( O(m) \) time (where \( m \) is the length of the subsequence). This becomes computationally expensive when the bid sequence is long; for instance, if the pacing interval is 30 seconds, there could be up to 2,880 bid data points per day.
	
	A more efficient approach utilizes a sliding window combined with two pointers and prefix sums:
	\begin{itemize}
		\item Prefix sums allow us to compute the mean and variance of a subsequence efficiently.
		\item Sliding window and two-pointer techniques help search for the longest valid subsequence while maintaining the variance constraint.
	\end{itemize}
	
	This optimized method significantly improves computational efficiency to \(O(T)\) compared to the brute-force approach, making it more efficient for large-scale bidding data computation. We summarize this approach in the following \autoref{alg:init_bid_bid_average}:

	\begin{algorithm}
		\caption{Optimal Initial Bid via Converged Bids Average}
		\label{alg:init_bid_bid_average}
		\begin{algorithmic}[1]
			\Require Sequence $\{b_i\}_{i=1}^{T}$, variance threshold $\delta$, minimum length $K$
			\Ensure Longest consecutive subsequence with variance $\leq \delta$ and length $\geq K$, and its average bid
			
			\State Compute prefix sums: 
			\[
			S(i) = \sum_{j=1}^{i} b_j, \quad Q(i) = \sum_{j=1}^{i} b_j^2
			\]
			
			\State Initialize $L \gets 1$, $max\_length \gets 0$, $best\_interval \gets (0,0)$
			
			\For{$R = 1$ to $T$}
			\While{$R - L + 1 \geq K$}
			\State Compute mean: 
			\[
			\mu = \frac{S(R) - S(L-1)}{R - L + 1}
			\]
			\State Compute variance:
			\[
			\sigma^2 = \frac{Q(R) - Q(L-1)}{R - L + 1} - \mu^2
			\]
			
			\If{$\sigma^2 \leq \delta$}
			\If{$R - L + 1 > max\_length$}
			\State $max\_length \gets R - L + 1$
			\State $best\_interval \gets (L, R)$
			\State $best\_average \gets \mu$ \Comment{Store the best average bid}
			\EndIf
			\State \textbf{break} \Comment{Expand the window further}
			\Else
			\State $L \gets L + 1$ \Comment{Shrink the window}
			\EndIf
			\EndWhile
			\EndFor
			
			\State \Return $best\_interval, best\_average$
		\end{algorithmic}
	\end{algorithm}
		
	In practice, bid scales can vary significantly across different campaigns. To enhance the robustness of the algorithm, we can rescale the bid data points \( \{b_i\}_{i=1}^{T} \) using a reference value, such as the initial bid \( b_1 \) or the average bid over the dataset. By normalizing bids across different campaigns to a comparable scale, we reduce variability and improve the effectiveness of the algorithm discussed above. This normalization ensures that the variance-based convergence detection remains consistent across campaigns with different bidding magnitudes, making the approach more robust and generalizable.

	\section{Remarks} \label{ch:init_bid_remarks}
	
	\subsection{Proof of \autoref{eq:init_bid_formula}}
	
		We assume:
		\begin{itemize}
			\item The clearing price $Z$ follows a lognormal distribution $\mathrm{Lognormal}(\mu,\sigma^2)$ 
			with PDF
			\[
			p_{\mu,\sigma}(z) 
			\;=\; \frac{1}{z\,\sigma\sqrt{2\pi}}
			\exp\!\Bigl(-\tfrac{(\ln z - \mu)^{2}}{2\,\sigma^{2}}\Bigr),
			\quad z>0.
			\]
			\item A random factor $R$ follows a lognormal distribution $\mathrm{Lognormal}(\mu',\sigma'^2)$ 
			and scales our bid so that the \emph{actual} bid is $b\,R$.  
			\item We pay the second price, i.e.\ we pay $Z$ if $Z < b\,R$, and we pay $0$ otherwise.
		\end{itemize}
		Define $c(b)$ to be the expected cost per impression when bidding $b$ (per unit of $R$).  
		Formally,
		\[
		c(b)
		\;=\; \iint_{\{\,z < b\,r\,\}} z\,p_{\mu,\sigma}(z)\,p_{\mu',\sigma'}(r)\,dz\,dr
		\;=\;
		\int_{0}^{\infty}\!\!\biggl(\int_{0}^{\,b\,r} z\,p_{\mu,\sigma}(z)\,dz\biggr)
		p_{\mu',\sigma'}(r)\,dr.
		\]
		
		\noindent \textbf{Step 1. Evaluate the inner integral.}
		Since $Z \sim \mathrm{Lognormal}(\mu,\sigma^2)$, one has the standard identity:
		\[
		\int_{0}^{x} z\,p_{\mu,\sigma}(z)\,dz
		\;=\;
		\exp\!\Bigl(\mu + \tfrac12\,\sigma^2\Bigr)\,
		\Phi\!\Bigl(\tfrac{\ln x - \mu - \sigma^2}{\sigma}\Bigr),
		\]
		where $\Phi(\cdot)$ is the CDF of the standard normal distribution.  
		By setting $x = b\,r$, it follows that
		\[
		\int_{0}^{b\,r} z\,p_{\mu,\sigma}(z)\,dz
		\;=\;
		\exp\!\bigl(\mu + \tfrac12\,\sigma^2\bigr)\,
		\Phi\!\Bigl(\tfrac{\ln(b\,r) - \mu - \sigma^2}{\sigma}\Bigr).
		\]
		Thus
		\[
		c(b)
		\;=\;
		\exp\!\bigl(\mu + \tfrac12\,\sigma^2\bigr)\;
		\int_{0}^{\infty}
		\Phi\!\Bigl(\tfrac{\ln(b\,r) - \mu - \sigma^2}{\sigma}\Bigr)\,
		p_{\mu',\sigma'}(r)
		\,dr.
		\]
		
		\noindent \textbf{Step 2. Substitute for the outer integral (the lognormal $R$).}
		Since $R \sim \mathrm{Lognormal}(\mu',\sigma'^2)$,
		\[
		p_{\mu',\sigma'}(r)\,dr
		\;=\;
		\frac{1}{r\,\sigma'\sqrt{2\pi}}
		\exp\!\Bigl(-\tfrac{(\ln r - \mu')^2}{2\,\sigma'^2}\Bigr)\,dr.
		\]
		Make the change of variable $x := \ln r$.  Then $r = e^x$, $dr = e^x\,dx$, and
		\[
		p_{\mu',\sigma'}(r)\,dr
		\;=\;
		\frac{1}{\sigma'\sqrt{2\pi}}
		\exp\!\Bigl(-\tfrac{(x - \mu')^2}{2\,\sigma'^2}\Bigr)\,dx.
		\]
		Hence
		\[
		c(b)
		\;=\;
		\exp\!\bigl(\mu + \tfrac12\,\sigma^2\bigr)\!
		\int_{-\infty}^{\infty}
		\Phi\!\Bigl(\tfrac{\ln b + x - \mu - \sigma^2}{\sigma}\Bigr)\,
		\frac{1}{\sigma'\sqrt{2\pi}}
		\exp\!\Bigl(-\tfrac{(x-\mu')^2}{2\,\sigma'^2}\Bigr)\,dx.
		\]
		
		\noindent \textbf{Step 3. Apply a bivariate‐normal identity.}
		Set $x = \sigma'\,y + \mu'$, i.e.\ $y=\tfrac{x-\mu'}{\sigma'}$.  Then $dx=\sigma'\,dy$, and 
		\[
		\ln b + x - \mu - \sigma^2
		\;=\;
		\bigl(\ln b \;+\;\mu' \;-\;\mu \;-\;\sigma^2\bigr)
		\;+\;\sigma'\,y.
		\]
		Inside the integral, 
		\[
		\Phi\!\Bigl(\tfrac{\ln b + x - \mu - \sigma^2}{\sigma}\Bigr)
		\quad\text{and}\quad
		\exp\!\Bigl(-\tfrac{(x-\mu')^2}{2\,\sigma'^2}\Bigr)\,\frac{dx}{\sigma'\sqrt{2\pi}}
		\]
		become a classic form:
		\[
		\int_{-\infty}^{\infty}
		\Phi(\alpha + \beta\,y)\,
		\frac{1}{\sqrt{2\pi}}\,e^{-\tfrac{y^2}{2}}\;dy
		\;=\;
		\Phi\!\Bigl(\tfrac{\alpha}{\sqrt{1+\beta^2}}\Bigr),
		\]
		with $\alpha = \tfrac{\ln b + \mu' - \mu - \sigma^2}{\sigma}$ and $\beta = \tfrac{\sigma'}{\sigma}$.
		This well‐known identity yields
		\[
		c(b)
		\;=\;
		\exp\!\bigl(\mu + \tfrac12\,\sigma^2\bigr)\;
		\;\Phi\!\Bigl(
		\tfrac{\ln b + \mu' - \mu - \sigma^2}{\sqrt{\sigma^2 + \sigma'^2}}
		\Bigr).
		\]
		By slightly rearranging parameters, one typically writes
		\[
		c(b)
		\;=\;
		\exp\!\Bigl(\mu + \tfrac12\,\sigma^2\Bigr)
		\,\Phi\!\Bigl(\tfrac{\mu' - \mu + \ln b - \sigma^2}{\sqrt{\sigma'^2 + \sigma^2}}\Bigr),
		\]
		which is the standard \emph{double‐lognormal} closed‐form expression.
		
		\noindent \textbf{Budget constraint and conclusion.}
		If the total budget is $B$ for $T$ impressions, \emph{exhausting} that budget exactly means 
		\[
		c(b^*)\,T \;=\; B
		\quad\Longrightarrow\quad
		c(b^*) \;=\;\frac{B}{T}.
		\]
		Thus we set
		\[
		\exp\!\Bigl(\mu + \tfrac12\,\sigma^2\Bigr)
		\,\Phi\!\Bigl(\tfrac{\mu' - \mu + \ln(b^*) - \sigma^2}{\sqrt{\sigma'^2 + \sigma^2}}\Bigr)
		\;=\;
		\frac{B}{T},
		\]
		which is exactly Equation (1.1).  Hence there is a unique $b^*$ solving 
		$c(b^*) = B/T$, and the proof is complete.

	\subsection{Bid Cap Suggestion Problem}
	When an advertiser creates a cost cap campaign in an Ad Manager\footnote{An Ad Manager is a tool or platform that enables businesses, advertisers, and marketers to create, manage, and optimize their advertising campaigns across various digital channels. Examples include Facebook Ads Manager, Google Ads, LinkedIn Campaign Manager, and Amazon DSP.}, the platform typically provides a suggested bid for the cap threshold to prevent misconfiguration by the advertiser. 
	
	This suggested bid plays a critical role in the ad platform's business. If the suggested bid is too low, the cost cap campaign may suffer from low delivery, resulting in revenue loss for the platform. Conversely, if the suggested bid is too high, it may deter advertisers, leading to a lower adoption rate. 
	
	From the platform's perspective, the optimal strategy is to suggest a bid that allows the campaign budget to be fully spent by the end of its lifetime. Consequently, the bid cap suggestion problem can be formulated as an instance of the initial bid optimization problem for max delivery with the same budget. All approaches discussed for computing the initial bid in max delivery campaigns can therefore be applied to derive an appropriate bid suggestion.

	\chapter{Bid Response Prediction}
	
	\intro{
		In this chapter, we demonstrate how to model response prediction—how different metrics (e.g., spend, conversions) respond to bid changes—by leveraging real-time auction data alongside prediction models. This approach helps address the data sparsity and signal delay issues encountered in deep-funnel conversion products when applying the MPC algorithm in previous chapters.
	}
	
	In this chapter, we discuss Bid Response Prediction. The bid response prediction model aims to understand how changes in bid prices can potentially impact the market and key campaign metrics. The bid landscape forecasting model is one example of such models, as it predicts how bid adjustments affect spending by analyzing the winning price distribution. We have also briefly described other models in \autoref{part:pacing_algorithms}. For instance, in the MPC controller, we construct bid-to-spend, bid-to-conversions, and bid-to-cost-per-result functions by leveraging real-time auction data. In this chapter, we explore additional approaches that utilize predictive models to achieve this goal.
	
	\section{Bid Cost Prediction}
	In the previous chapter, when modeling bid response prediction, we assumed that the campaign follows an oCPM(optimized Cost Per Mille) charging model, meaning the campaign is charged per impression. Under this assumption, there is no need to account for signal delay issues, and we can directly use real-time bid-cost pairs to build the model.
	
	However, as previously discussed, this assumption does not hold when the charging rule is based on actual results, particularly for ads with deep-funnel objectives such as post-click conversions, app installs, and lead generation. In such cases, conversions are sparse, and it may take hours or even days before a new conversion is observed.
	
	In this section, we introduce a new method for predicting the bid-cost relationship in real-time bidding for in-house campaigns with deep conversion objectives. To ensure broader applicability, we no longer assume a second-price auction. Instead, we consider the generalized second-price (GSP) auction, in which multiple ad slots are auctioned simultaneously.

	Let's first review how the GSP auction works in practice. For simplicity, we use pay-per-click(PPC) model as an example. Assume that for a given auction request, there are \(N\) campaigns optimizing for clicks and \(k\) (\(k < N\)) available ad slots. Let \(s_j\) denote the \(j\)-th slot from the top, \(p_i\) be the predicted click-through rate (CTR) for campaign \(i\) in the first ad slot. The position bias discount factor for slot \(j\) IS \(\alpha_j\), where:
		\[
		\alpha_1 = 1 \geq \alpha_2 \geq \dots \geq \alpha_k.
		\]
	The predicted CTR for campaign \(i\) in slot \(j\) is then  \(p_i \cdot \alpha_j\). Suppose that  bidder \(i\) submits a bid \(b_i\) per click. The GSP auction operates as follows:
	\begin{itemize}
		\item Compute the eCPM for the first ad slot for all campaigns based on their bids and predicted CTRs:
		\[
		eCPM_i = b_i \cdot p_i.
		\]
		
		\item Rank all campaigns in descending order based on eCPM values. If necessary, reindex them so that:
		\[
		eCPM_1 \geq eCPM_2 \geq \dots \geq eCPM_N.
		\]
		
		\item Assign ad slots: The top-ranked campaign receives the first slot, the second-ranked campaign receives the second slot, and so on.
		
		\item Compute the pay-per-click (PPC) for each advertiser \( i \) and charge the bidder only if a click event occurs:
		\[
		c_i = \frac{eCPM_{i+1}}{p_i}.
		\]
		Here, \( eCPM_{i+1} \) represents the next highest eCPM in the ranking, ensuring that each advertiser pays a price determined by the bidder ranked below them. This ensures that the cost-per-click (CPC) aligns with the generalized second-price rule, where advertisers are charged based on the minimum bid required to maintain their slot.
	\end{itemize}

	In the previous discussion, we assumed that the campaign follows an oCPM charging model. Under this model, advertisers are charged per impression, regardless of whether a click event occurs. As a result, it is feasible to collect bid-cost pairs without concern that most of the cost values will be zero. Under pay-per-click model, this does not work as most of cost values are 0. One workaround to get rid of this issue is to use accumulated pay per impression cost as the approximation to the pay-per-click cost by running the auction replay with real-time auction data, if the pCTR is unbiased, such approximation will be accurate when number of impressions is sufficient. More specifically, for all auction requests of the given campaign in the last pacing interval(suppose there are \(T\) requests), we collect the auction data \(\{A_t\}_{t=1}^T\) with each \(A_t\) as: 
	\[
		A_t = \{ p_t, eCPM_{t, 1}, \cdots, eCPM_{t, k}\}
	\]
	where \(k\) is the number of ad slots, \(p_t\) is the pCTR of this campaign for \(t\)-th auction request at the first ad slot, \(\{eCPM_{t, j}\}_{j=1}^{k}\) are the first \(k\) eCPMs in the ranking stage(i.e., the eCPMs computed for the first ad slot) in \(t\)-th auction. We also assume the position bias decaying factors \(\{\alpha_i\}_{i=1}^k\) are available and fixed across all requests. 
		
	We now compute the bid-cost prediction function \( c = f(b) \), where \( f \) is the pay-per-impression spend rate function of bid price \( b \). This function represents the expected spend of the campaign within one pacing interval\footnote{For simplicity, we assume that two adjacent pacing intervals have similar traffic patterns, predicted click-through rates (pCTR), and winning price distributions. Under this assumption, it is reasonable to use data from one interval to infer metrics in the subsequent interval.}. 
	
	For a fixed bid price \( b \), the corresponding spend is computed by iterating through all auction requests \(\{A_t\}_{t=1}^T\). For each auction \( A_t \), we compute the eCPM of this campaign for the first ad slot as:
	
	\[
	eCPM = b \cdot p_t.
	\]
	
	If the campaign participates in auction \( A_t \), it wins a slot if and only if:
	
	\[
	eCPM \geq eCPM_{t,k}.
	\]
	
	Otherwise, it loses the auction. If it wins, we determine the lowest slot \( j(t) \) (\(1 \leq j(t) \leq k\)) that the campaign qualifies for, such that:
	
	\[
	eCPM_{t, j(t)-1} > eCPM > eCPM_{t, j(t)}.
	\]
	
	This means that the campaign secures the \( j(t) \)-th ad slot in the \( t \)-th auction. The corresponding pay-per-impression cost for winning this slot is:
	
	\[
	eCPM_{t, j(t)} \cdot \alpha_{j(t)},
	\]
	where \( \alpha_{j(t)} \) accounts for position bias adjustment.
		
	The accumulated pay-per-impression cost over the pacing interval is then:
	
	\[
	\sum_{t=1}^{T} eCPM_{t, j(t)} \cdot \alpha_{j(t)}.
	\]
	
	This sum represents the expected cost for the pay-per-click (PPC) campaign over the pacing interval and defines the function \( f(b) \), assuming that the predicted click-through rate \( p_t \) is unbiased for this campaign. 
	
\begin{figure}[H]
		\centering
		\includegraphics[width=0.99\textwidth]{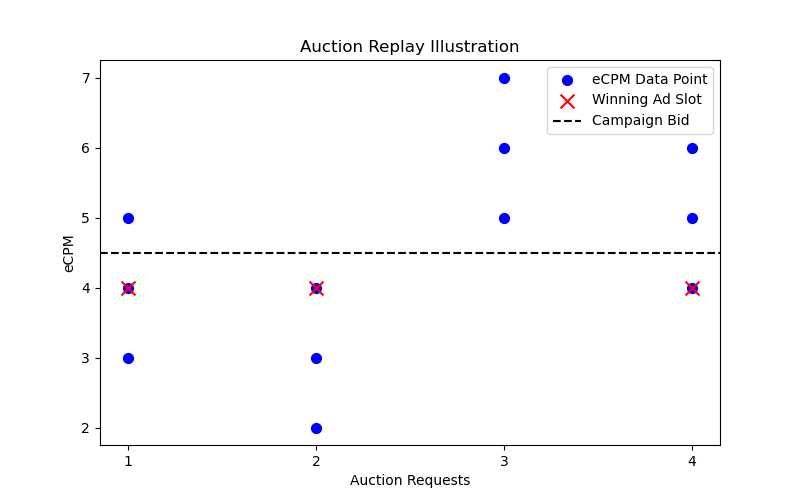}
		\caption{Illustration of Deriving Bid-Cost Relation via Auction Replay}
		\label{fig:bid_cost_relation}
	\end{figure}

	The procedure is illustrated in \autoref{fig:bid_cost_relation}. Each group of blue dots along the vertical lines represents the eCPMs of different ad slots in an auction. The dotted line indicates the eCPM of the campaign for a given bid. 
	
	The dots marked with a red "X" indicate the ad slots the campaign wins in each auction. The expected cost for the given bid is then computed as the accumulated sum of the pay-per-impression costs for each winning ad slot.
	
	We summarize the idea in the following algorithm

	\begin{algorithm}[H]
		\caption{Bid-Cost Prediction Algorithm}
		\label{alg:bid_cost_prediction}
		\begin{algorithmic}[1]
			\Require  
			\begin{itemize}
				\item \( \{A_t\}_{t=1}^{T} \) — Set of auction requests over the pacing interval, where each auction request \( A_t \) contains:
				\begin{itemize}
					\item \( p_t \) — Predicted click-through rate (pCTR) for the campaign in auction \( A_t \).
					\item \( k \) — Number of available ad slots in the auction.
					\item \( \{eCPM_{t,j}\}_{j=1}^{k} \) — Sorted eCPMs of the top \( k \) advertisers who win ad slots.
				\end{itemize}
				\item \( \{\alpha_j\}_{j=1}^{k} \) — Position bias discount factors for each ad slot, where:
				\[
				\alpha_1 = 1 \geq \alpha_2 \geq \dots \geq \alpha_k.
				\]
				\item \( b_l, b_u \) — Lower and upper bounds of the bid range.
				\item \( \Delta b \) — Bid increment step size.
			\end{itemize}
			
			\Ensure Predicted bid-cost function \( f(b) \) for \( b \in [b_l, b_u] \).
			
			\For{\( b = b_l \) to \( b_u \) with step size \( \Delta b \)} \Comment{Iterate over bid values}
			\State Initialize \( f(b) \gets 0 \) \Comment{Initialize total cost for bid \( b \)}
			
			\For{\( t = 1 \) to \( T \)} \Comment{Iterate over all auction requests}
			\State Extract \( A_t = (p_t, \{eCPM_{t,j}\}_{j=1}^{k}) \)
			\State Compute \( eCPM \gets b \cdot p_t \) \Comment{Calculate campaign eCPM for the first slot}
			
			\If{\( eCPM \geq eCPM_{t,k} \)} \Comment{Check if the campaign wins a slot}
			\State Find \( j(t) \) such that \( eCPM_{t, j(t)-1} > eCPM > eCPM_{t, j(t)} \)
			\State Compute impression cost: \( c_t \gets eCPM_{t, j(t)} \cdot \alpha_{j(t)} \)
			\State Update total cost: \( f(b) \gets f(b) + c_t \)
			\EndIf
			\EndFor
			\EndFor
			
			\State \Return \( f(b) \) \Comment{Return the bid-cost function \( f(b) \)}
		\end{algorithmic}
	\end{algorithm}

	\section{Bid Conversion Prediction}
		
	For completeness, we briefly discuss how to use the same method to model the bid-to-conversion relationship. The principle described in the previous section applies with only a slight modification. 
	
	We adopt the exact same problem configuration as in the previous section, except that in this case, we aim to compute the expected number of conversions (clicks) for a given bid level \( b \). The procedure remains identical, with the only difference being that instead of computing the impression cost, we use the predicted click-through rate (pCTR) per impression to approximate the total number of clicks when the campaign wins the auction.
	
	The algorithm is presented as follows:

		\begin{algorithm}[H]
		\caption{Bid-Conversion Prediction Algorithm}
		\label{alg:bid_conversion_prediction}
		\begin{algorithmic}[1]
			\Require  
			\begin{itemize}
				\item \( \{A_t\}_{t=1}^{T} \) — Set of auction requests over the pacing interval, where each auction request \( A_t \) contains:
				\begin{itemize}
					\item \( p_t \) — Predicted click-through rate (pCTR) for the campaign in auction \( A_t \).
					\item \( k \) — Number of available ad slots in the auction.
					\item \( \{eCPM_{t,j}\}_{j=1}^{k} \) — Sorted eCPMs of the top \( k \) advertisers who win ad slots.
				\end{itemize}
				\item \( \{\alpha_j\}_{j=1}^{k} \) — Position bias discount factors for each ad slot, where:
				\[
				\alpha_1 = 1 \geq \alpha_2 \geq \dots \geq \alpha_k.
				\]
				\item \( b_l, b_u \) — Lower and upper bounds of the bid range.
				\item \( \Delta b \) — Bid increment step size.
			\end{itemize}
			
			\Ensure Predicted bid-conversion function \( g(b) \) for \( b \in [b_l, b_u] \).
			
			\For{\( b = b_l \) to \( b_u \) with step size \( \Delta b \)} \Comment{Iterate over bid values}
			\State Initialize \( g(b) \gets 0 \) \Comment{Initialize total cost for bid \( b \)}
			
			\For{\( t = 1 \) to \( T \)} \Comment{Iterate over all auction requests}
			\State Extract \( A_t = (p_t, \{eCPM_{t,j}\}_{j=1}^{k}) \)
			
			\If{\( eCPM \geq eCPM_{t,k} \)} \Comment{Check if the campaign wins a slot}
			\State Find \( j(t) \) such that \( eCPM_{t, j(t)-1} > eCPM > eCPM_{t, j(t)} \)
			\State Compute position bias adjusted pCTR: \( \tilde{p}_t \gets p_t \cdot \alpha_{j(t)} \)
			\State Update total cost: \( g(b) \gets g(b) + \tilde{p}_t \)
			\EndIf
			\EndFor
			\EndFor
			
			\State \Return \( g(b) \) \Comment{Return the bid-conversion function \( g(b) \)}
		\end{algorithmic}
	\end{algorithm}

	\section{Remarks}
	
	Bid landscape forecasting provides a closely related, but distinct, approach to constructing bid-response functions. In the classical formulation, the model estimates the conditional distribution of the market (or winning) price from auction and campaign features. The distribution can then be converted into a winning-probability curve and, together with traffic and user-response predictions, into expected spend and conversion curves as functions of the bid. Early work constructed campaign-level landscapes by decomposing eligible inventory into traffic segments \cite{cui2011bid}; later work learned the full feature-conditioned market-price distribution without prescribing a fixed parametric form \cite{wang2016functional}, including deep sequential models over the price space \cite{ren2019deep}. From this perspective, bid landscape forecasting is a model-based alternative or complement to the auction-replay estimator in this chapter: replay uses the realized top-\(k\) competitive eCPMs from a recent interval, whereas landscape forecasting smooths and generalizes competition patterns across auctions.
	
	A central difficulty in learning a bid landscape from bidder-side logs is censoring. For a lost auction, the exact market price is generally not observed; the submitted bid provides only a lower bound. Wu et al. address this issue using censored regression and a win-rate-weighted mixture model \cite{wu2015predicting}, while functional and deep landscape models incorporate survival-analysis ideas \cite{wang2016functional,ren2019deep}. Censoring is less problematic for an in-house platform that records the top-\(k\) competing eCPMs required by the replay procedure. Nevertheless, both replay- and model-based response curves remain sensitive to changes in traffic and competition, errors in position-bias factors, and miscalibration of pCTR or pCVR. A bid landscape models market competition rather than user response, so conversion prediction still requires a calibrated response model. Periodically rebuilding the curves and re-optimizing with fresh feedback is therefore important, consistent with the adaptive treatment of forecast errors in \cite{lang2012handling}. Finally, most of the cited work considers single-slot second-price RTB; extending it to the multi-slot GSP setting here requires slot-specific competitive thresholds and their dependence across positions, rather than a single scalar market-price distribution.
	

	\chapter{Bid Shading}
	
	\intro{
		In this chapter, we discuss bid shading techniques, which are crucial for bidding algorithm design under the first-price auction model. We begin by examining the industry's shift from waterfall bidding to header bidding, which has made bid shading a necessary component of bidding algorithm design.  Next, we demonstrate how bid shading algorithms can be developed under various campaign configurations. Additionally, we include a section outlining the structure of bidding algorithms under arbitrary auction mechanisms. Finally, we conclude the chapter by listing several winning probability estimation methods, an indispensable component of bid shading.
	}
	
	\section{From Waterfall Bidding to Header Bidding}
	First price auction became more and more popular for  real-time bidding due to the shift from the traditional waterfall bidding to header bidding. 
	
	\subsection{Waterfall Bidding}
	Waterfall bidding is an early method used in programmatic advertising where ad impressions are sequentially offered to demand partners (e.g., DSPs or ad exchanges) based on a predefined priority order set by the publisher. If the first demand partner does not fill the impression, the request moves down the chain to the next partner, continuing until the impression is filled or no suitable ad is found.
	
	For a given publisher (e.g., a website), when a user visits the platform, an ad request is triggered from the publisher's ad server. The ad server then calls the first demand partner (e.g., an SSP or an ad exchange) based on predefined priorities. If the demand partner provides an ad that meets or exceeds the publisher’s floor price, the impression is filled, and the process stops. Otherwise, the ad server sends the request to the next available demand partner, continuing sequentially until the ad slot is filled. This method was dominant in programmatic advertising for a long time.
	
	Although waterfall bidding is simple to implement, it has several significant limitations:
	
	\begin{itemize}
		\item \textbf{Sequential Process Delays Auction Speed:} Since requests are processed one by one, the decision-making process is inherently slow.
		\item \textbf{Lack of Fair Competition:} From a bidder's perspective, not all bidders get to participate simultaneously, as the auction operates based on a predefined priority list.
		\item \textbf{Potential Revenue Loss:} From a publisher's perspective, a lower-priority bidder may be willing to pay more than the first bidder but never gets the opportunity to compete, leading to potential revenue loss.
	\end{itemize}

	\subsection{Header Bidding}
	Header bidding is a bidding strategy introduced around 2014–2015 to address the limitations of waterfall bidding. Unlike waterfall bidding, which runs auctions sequentially based on predefined priorities of demand partners, header bidding allows all demand partners to participate in the auction simultaneously. This ensures that the demand partner with the highest bid wins the ad slot, thereby increasing competition and boosting revenue for publishers.
	
	\paragraph{Challenges for DSPs Under Header Bidding:}
	With the advent of header bidding, the traditional second-price auction strategy used by DSPs faces new challenges. To illustrate the issue, consider the following example:
	
	Suppose two DSPs, \( D_1 \) and \( D_2 \), compete for an ad request from a publisher. Within each DSP \( i \), there are two campaigns, \( C_{i,1} \) and \( C_{i,2} \), interested in this impression. For simplicity, assume all campaigns submit bids for the impression as follows:
	
	\[
	C_{1,1} = \$1, \quad C_{1,2} = \$4, \quad C_{2,1} = \$2, \quad C_{2,2} = \$3.
	\]
	
	When the ad request arrives, each DSP conducts an internal auction among its campaigns using a second-price auction:
	
	\begin{itemize}
		\item \textbf{DSP 1:} Campaign \( C_{1,2} \) wins the internal auction but, under the second-price rule, submits the second-highest price within DSP 1, which is \textbf{\$1}, to the ad server.
		\item \textbf{DSP 2:} Campaign \( C_{2,2} \) wins the internal auction and submits \textbf{\$2}, the second-highest price within DSP 2, to the ad server.
	\end{itemize}
	
	Under header bidding, DSP 1 and DSP 2 submit bids of \textbf{\$1} and \textbf{\$2}, respectively, to the ad server. Regardless of whether the ad server runs a \textbf{first-price} or \textbf{second-price} auction, DSP 2 will ultimately win the impression.
	
	\paragraph{The Problem:}
	This scenario highlights a fundamental inefficiency when we run second price auction with DSPs in header bidding:
	
	\begin{itemize}
		\item \textbf{From the bidder’s perspective:} The campaign with the highest original bid, \( C_{1,2} \) (\$4), does not win the auction. This contradicts the expected outcome of an efficient auction system, where the highest bidder should secure the impression.
		\item \textbf{From the publisher’s perspective:} There is a clear revenue loss. If all campaigns participated in a unified auction directly, the second-highest bid among them would be \textbf{\$3}, meaning the publisher could have earned \textbf{\$3} instead of \textbf{\$2} (under a first-price auction) or \textbf{\$1} (under a second-price auction at the ad server level).
	\end{itemize}
	
	This example illustrates how the traditional second-price auction model within DSPs leads to inefficiencies in the presence of header bidding, ultimately resulting in suboptimal outcomes for both advertisers and publishers. 
	
	Another scenario arises when a publisher sends requests to multiple SSPs, and each SSP, in turn, forwards these requests to multiple DSPs. If the SSP runs a second-price auction, the same issues persist, reinforcing the challenges inherent in this model. 
	
	These inefficiencies have been a driving force behind the industry's shift in recent years from second-price auctions to first-price auctions, aiming to enhance transparency and improve fairness in programmatic advertising. An interesting discussion on this trend can be found in \cite{despotakis2021first}. In this chapter, we will explore strategies for designing optimal bidding approaches under first-price auctions.

	\section{Bid Shading under First Price Auction}
	In this section, we explore the design of bidding strategies under first-price auctions. The general formulation follows a similar structure to that used in second-price auctions, as discussed in \autoref{part:pacing_algorithms}. The bidding problem can be framed as an optimization problem that seeks to maximize a specific objective while satisfying given constraints, such as a budget constraint in the case of the max delivery problem.
	
	The objective can take different forms: 
	\begin{itemize}
		\item \textbf{Welfare Maximization}: The goal is to maximize the total number of conversions for a given campaign.
		\item \textbf{Utility Maximization}: The objective is to maximize the campaign’s surplus.
	\end{itemize}
	
	Additionally, other factors must be considered. For instance, if a DSP bids on behalf of a campaign for external impressions, the cost incurred by the campaign represents only part of the equation. The DSP often applies a markup to generate additional profit for the platform itself. In such cases, margin profit optimization becomes an important factor in the bidding strategy.
	
	This chapter will discuss these variations in detail, including strategies to handle different optimization objectives and practical challenges in first-price auction environments.
	
	\subsection{Welfare Maximization}
	We discuss how to solve the max delivery problem in a first-price auction (FPA) setting. One possible formulation of the bidding problem is to optimize the total welfare (conversions) of a campaign for a given budget, similar to our previous discussion in \autoref{part:pacing_algorithms}.  
	\paragraph{Problem Formulation}
	We assume that a campaign from a DSP is bidding for external ad impressions through an ad exchange running a first-price auction. For simplicity, we consider a CPM campaign. The problem can be formulated as:
	
	\begin{equation}  \label{eq:fpa_welfare}
		\begin{aligned}
			\max_{b_t \geq 0} \quad & \sum_{t=1}^T P_t(b_t) \\
			\text{s.t.} \quad &  \sum_{t=1}^{T} P_t(b_t) \cdot b_t \leq B. 
		\end{aligned}
	\end{equation}
	where:
	\begin{itemize}
		\item \( P_t(b_t) \) \footnote{Recall that under a second-price auction, we write \(P_t(b_t)\) as \(\mathds{1}_{\{ b_t > c_t \}}\), where \(c_t\) represents the supporting price. For a Demand-Side Platform (DSP) or an ad network (adNet), the value of \(c_t\) is not always available, as the data are often censored. Ad exchanges often do not disclose losing bids or even the winning bid if the DSP is not the winner. In practice, DSPs often only see partial information about the auction results. The DSP typically knows if they won or lost. If they win, they know the price they paid (their own bid). If they lose, they usually don't know the winning bid.} represents the probability of winning the impression at the \(t\)-th auction given bid per impression \(b_t\).
		\item \( B \) is the total budget and \( T \) is the total number of auction opportunities.
	\end{itemize}
	
	If at time \( t \), the highest competing bid from other bidders(winning price) is \( c_t \), then the probability of winning is given by:
	\[
	P_t(b_t) = \mathbb{P} (b_t \geq c_t).
	\]
	\( P_t(b_t) \) is typically increasing and nonlinear with respect to \( b_t \), making direct optimization challenging.
	
	\paragraph{Derivation of the Optimal Bid}
	To solve for the optimal bid \( b_t^* \), we introduce the Lagrangian function with multiplier \( \lambda \geq 0 \):
	
	\[
	\mathcal{L}(b_t, \lambda) = \sum_{t=1}^{T} P_t(b_t) - \lambda \left( \sum_{t=1}^T P_t(b_t) b_t - B \right).
	\]
	Differentiating \( \mathcal{L} \) with respect to \( b_t \):
	\begin{equation*}
		\begin{aligned}
			\frac{\partial } {\partial b_t} \mathcal{L} (b_t, \lambda) &= \frac{\partial}{\partial b_t} \left[ P_t(b_t) - \lambda P_t(b_t) b_t\right] \\
			&= P_t^{'}(b_t) - \lambda b_t P_t^{'}(b_t) - \lambda P_t(b_t).
		\end{aligned}
	\end{equation*}
	To find the optimal bid \( b_t^* \), we set \( \frac{\partial \mathcal{L}(b_t^*, \lambda^*)}{\partial b_t}= 0 \):
	\begin{equation*} 
		P_t^{'}(b_t^*) - \lambda^* b_t^* P_t^{'}(b_t^*) - \lambda^* P_t(b_t^*) = 0.
	\end{equation*}
	Rearranging for \( b_t^* \):
	
	\begin{equation} \label{eq:fpa_welfare_b_start}
		b_t^*   + \frac{P_t(b_t^*)}{P_t^{'}(b_t^*)} = \frac{1}{\lambda^*}
	\end{equation}

	Since \( P_t(b_t) \) is generally nonlinear, \autoref{eq:fpa_welfare_b_start} has no closed-form solution in most cases. Typically, numerical methods such as Newton's method or other root-finding algorithms are used to compute \( b_t^* \) iteratively. However, in some special cases where \( P_t(b_t) \) has a tractable form, an analytical solution may exist. 
		
	To solve for \(b_t^*\), we need to determine \(\lambda^*\) and the winning probability function \(P_t(b_t)\) (hence its derivative \(P_t'(b_t)\)):
	
	\begin{itemize}
		\item \textbf{Determining \(\lambda^*\) \footnote{ We may also derive the update rule for \(\lambda\) using the DOGD method, which we discuss in \autoref{ch:bid_shading_dogd_lambda}.}. } To determine \(\lambda^*\), note that if the budget constraint is active, the KKT condition implies that the optimal value of \(\lambda\) must satisfy the budget exactly:
		\[
		\sum_{t=1}^{T} P_t(b_t(\lambda^*)) \cdot b_t(\lambda^*) = B.
		\]
		where \(b_t(\lambda^*)\) is obtained from \eqref{eq:fpa_welfare_b_start}. If we assume that \(P_t(\cdot)\) is log-concave (which holds for most distributions discussed later), we can show that the function:
		\[
		b + \frac{P_t(b)}{P_t^{'}(b)}
		\]
		is strictly increasing with respect to \(b\). Therefore, to find \(\lambda^*\), we can leverage the PID controller discussed in \autoref{part:pacing_algorithms} to iteratively update \(\lambda\) based on the pacing status (i.e., whether delivery is ahead or behind schedule).
		
		\item \textbf{Determining \(P_t\):} \(P_t(b_t)\) is the winning probability function, which depends on both \(b_t\) and the auction environment (e.g., the bids submitted by other bidders). It can be derived using either parametric methods (where we assume some parametrized distributions a priori and fit the distribution using real data) or non-parametric methods (e.g., quantile regression). We will discuss these methods in detail in \autoref{ch:bid_shading_win_prob}.
		
	\end{itemize}
	
	Assuming that both \(\lambda^*\) and \(P_t\) are determined and that \(P_t\) is log-concave, we solve for \(b_t^*\) using a numerical method. In practice, we search for \(b_t^*\) within a predefined range \([b_l, b_u]\). Since the left-hand side of \autoref{eq:fpa_welfare_b_start} is monotonically increasing with respect to \(b_t^*\), a binary search method can be applied to efficiently determine the solution to \autoref{eq:fpa_welfare_b_start}. The algorithm we discuss here is summarized in \autoref{alg:bid_shading_fpa}.
	
	\begin{algorithm}
		\caption{Bid Strategy for Max Delivery Welfare Maximization Problem under FPA}
		\label{alg:bid_shading_fpa}
		\begin{algorithmic}[1]
			\Require 
			\( B \): Total budget; \( T \): Total auction opportunities; \\
			\(\lambda_0\): Initial value for Lagrange multiplier; \\
			\( [b_l, b_u] \): Search range for bid price; \(\epsilon\): threshold for binary search; \\
			\( P_t(b), P_t^{'}(b) \): Winning probability function and its derivative;  \\
			\(PID\):  PID controller in \autoref{alg:pid_md};
			\Ensure \( b_t^* \): Optimal bid at auction opportunity \(t\).
			
			\State \textbf{Step 1: Initialize Parameters}
			\State Set initial Lagrange multiplier \(\lambda \gets \lambda_0\)
			\State Define budget pacing target: \( \frac{B}{T} \)
			
			\State \textbf{Step 2: Solve for Optimal \( b_t^* \)}
			\For{each pacing interval}
			\State Observe actual spend \(S\)
			\State Update \(\lambda\) using PID controller:  \(\lambda \gets PID(S, B, T, \lambda)\)
			\For{each auction opportunity \( t \)}
			\State \textbf{Perform Binary Search to Solve for \( b_t^* \)}
			\State Initialize search range: \( b_{\text{low}} \gets b_l \), \( b_{\text{high}} \gets b_u \)
			\While{\( b_{\text{high}} - b_{\text{low}} > \epsilon  \)}
			\State Set \( b_{\text{mid}} \gets \frac{b_{\text{low}} + b_{\text{high}}}{2} \)
			\State Compute \( F(b_{\text{mid}}) = P_t^{'}(b_{\text{mid}}) - \lambda b_{\text{mid}} P_t^{'}(b_{\text{mid}}) - \lambda P_t(b_{\text{mid}}) \)
			\If{ \( F(b_{\text{mid}}) = 0 \) }
			\State \( b_t^* \gets b_{\text{mid}} \)
			\State \textbf{break}
			\ElsIf{ \( F(b_{\text{mid}}) > 0 \) }
			\State Set \( b_{\text{low}} \gets b_{\text{mid}} \)
			\Else
			\State Set \( b_{\text{high}} \gets b_{\text{mid}} \) 
			\EndIf
			\EndWhile
			\State Set \( b_t^* \gets b_{\text{mid}} \)
			\EndFor
			\EndFor
			
			\State \textbf{Step 3: Return Optimal Bids}
			\State \Return \( b_t^* \) for all \( t \)
		\end{algorithmic}
	\end{algorithm}
	
	\paragraph{Interpretation of the Bidding Formula} 
	Rearranging \autoref{eq:fpa_welfare_b_start}, we obtain:
	\[
	b_t^*   = \frac{1}{\lambda^*} - \frac{P_t(b_t^*)}{P_t^{'}(b_t^*)}.
	\]
	Since \( P_t(b) = \mathbb{P}(b \geq c_t) \) represents the cumulative distribution function (CDF) of the winning price \( c_t \), its derivative \( P_t^{'}(b) \) corresponds to the probability density function (PDF). Typically, \( P_t^{'}(b) \) is positive, implying that the ratio \( P_t(b)/P_t^{'}(b) \) is also positive.
	
	Comparing this to the optimal bidding formula for the max delivery problem under second-price auctions (see \eqref{eq:md_optimal_formula}), we observe that:
	\begin{itemize}
		\item The term \( \frac{1}{\lambda^*} \) acts as a \textbf{base bid} determined by the pacing multiplier.
		\item The positive correction term \( \frac{P_t(b_t^*)}{P_t^{'}(b_t^*)} \) serves as a \textbf{bid shading factor}, which depends on the winning probability distribution \( P_t(\cdot) \) and adjusts the base bid accordingly.
	\end{itemize}
	This formulation captures the strategic nature of bid adjustments under first-price auctions, where bidders optimize their bids based on market competition.

	\subsection{Utility Maximization}
	We analyze the utility maximization problem for a maximum delivery campaign under a first-price auction in this section.
	\paragraph{Problem Formulation}
	The optimization framework presented above seeks to maximize the total number of conversions. In this context, the advertiser aims to secure as many conversions as possible while ensuring that the total expenditure remains within the allocated budget, without explicitly considering the individual valuation of each conversion.
	
	However, in many scenarios, the advertiser assigns a specific valuation \( v_t \) to each conversion. As established in the Vickrey-Clarke-Groves (VCG) auction framework, an advertiser may wish to optimize for \textit{utility} (or \textit{surplus}), which is defined as the difference between the valuation and the corresponding payment. Consequently, the objective shifts to maximizing the expected surplus, leading to the following optimization problem:
	\begin{equation}  \label{eq:fpa_utility}
		\begin{aligned}
			\max_{b_t \geq 0} \quad & \sum_{t=1}^T P_t(b_t) \left[ v_t - b_t \right] \\
			\text{s.t.} \quad & \sum_{t=1}^{T} P_t(b_t) \cdot b_t \leq B.
		\end{aligned}
	\end{equation}
	\paragraph{Derivation of the Optimal Bid}
	We employ a Lagrangian approach similar to the one used previously, introducing \(\lambda \geq 0\) as the Lagrange multiplier for the budget constraint:
	
	\[
	\mathcal{L}(b_t, \lambda) = \sum_{t=1}^{T} P_t(b_t) \left[v_t - b_t\right] - \lambda \left( \sum_{t=1}^T P_t(b_t) b_t - B \right).
	\]
	Differentiating with respect to \(b_t\):
	\begin{equation*}
		\begin{aligned}
			\frac{\partial } {\partial b_t} \mathcal{L} (b_t, \lambda) &= \frac{\partial}{\partial b_t} \left[ P_t(b_t)\left[v_t - b_t\right] - \lambda P_t(b_t) b_t\right] \\
			&=  P_t^{'}(b_t) \left[ v_t - b_t(1+\lambda)\right] - \left[ 1+\lambda\right] P_t(b_t).
		\end{aligned}
	\end{equation*}
	Setting \( \frac{\partial \mathcal{L}(b_t^*, \lambda^*)}{\partial b_t}= 0 \):
	\[
	P_t^{'}(b_t) \left[ v_t - b_t(1+\lambda)\right] - \left[ 1+\lambda\right] P_t(b_t) = 0.
	\]
	Rearranging for \( b_t^* \), we obtain:
	\begin{equation} \label{eq:fpa_utility_b_star}
		b_t^*   + \frac{P_t(b_t^*)}{P_t^{'}(b_t^*)} = \frac{v_t}{1+\lambda^*}.
	\end{equation}
	
	To determine \(b_t^*\), we first solve for \(\lambda^*\) and define \(P_t(\cdot)\) analogously to the procedure outlined for \eqref{eq:fpa_welfare_b_start}. We then apply a binary search to identify the optimal bid \(b_t^*\) within a predefined range \([b_l, b_u]\). The complete procedure is summarized in \autoref{alg:bid_shading_fpa_utility}.
	
	\begin{algorithm}
		\caption{Bid Strategy for Max Delivery Utility Maximization Problem under FPA}
		\label{alg:bid_shading_fpa_utility}
		\begin{algorithmic}[1]
			\Require 
			\( B \): Total budget; \( T \): Total auction opportunities; \\
			\(\lambda_0\): Initial value for Lagrange multiplier; \\
			\( [b_l, b_u] \): Search range for bid price; \(\epsilon\): threshold for binary search; \\
			\( P_t(b), P_t^{'}(b) \): Winning probability function and its derivative;  \\
			\(PID\):  PID controller in \autoref{alg:pid_md};
			\Ensure \( b_t^* \): Optimal bid at auction opportunity \(t\).
			
			\State \textbf{Step 1: Initialize Parameters}
			\State Set initial Lagrange multiplier \(\lambda \gets \lambda_0\)
			\State Define budget pacing target: \( \frac{B}{T} \)
			
			\State \textbf{Step 2: Solve for Optimal \( b_t^* \)}
			\For{each pacing interval}
			\State Observe actual spend \(S\)
			\State Update \(\lambda\) using PID controller:  \(\lambda \gets PID(S, B, T, \lambda)\)
			\For{each auction opportunity \( t \)}
			\State \textbf{Perform Binary Search to Solve for \( b_t^* \)}
			\State Compute value of auction opportunity: \(v_t\)
			\State Initialize search range: \( b_{\text{low}} \gets b_l \), \( b_{\text{high}} \gets b_u \)
			\While{\( b_{\text{high}} - b_{\text{low}} > \epsilon  \)}
			\State Set \( b_{\text{mid}} \gets \frac{b_{\text{low}} + b_{\text{high}}}{2} \)
			\State Compute \( F(b_{\text{mid}}) = P_t^{'}(b_{\text{mid}}) v_t - (1+ \lambda ) b_{\text{mid}} P_t^{'}(b_{\text{mid}}) - (1+\lambda) P_t(b_{\text{mid}}) \)
			\If{ \( F(b_{\text{mid}}) = 0 \) }
			\State \( b_t^* \gets b_{\text{mid}} \)
			\State \textbf{break}
			\ElsIf{ \( F(b_{\text{mid}}) > 0 \) }
			\State Set \( b_{\text{low}} \gets b_{\text{mid}} \)
			\Else
			\State Set \( b_{\text{high}} \gets b_{\text{mid}} \) 
			\EndIf
			\EndWhile
			\State Set \( b_t^* \gets b_{\text{mid}} \)
			\EndFor
			\EndFor
			
			\State \textbf{Step 3: Return Optimal Bids}
			\State \Return \( b_t^* \) for all \( t \)
		\end{algorithmic}
	\end{algorithm}

	\paragraph{Interpretation of the Bidding Formula}
	Rearranging \eqref{eq:fpa_utility_b_star}, we obtain:
	\[
	b_t^* = \frac{v_t}{1+\lambda^*} - \frac{P_t(b_t^*)}{P_t^{'}(b_t^*)}.
	\]
	This formulation closely parallels the conversion maximization formula \eqref{eq:fpa_welfare_b_start}. In fact, under the same settings, except replacing the first-price auction with a second-price auction, the optimal bid simplifies to:
	\[
	b_{t, \text{second}}^* = \frac{v_t}{1+ \lambda^*}.
	\]
	Thus, the first term, \( \frac{v_t}{1 + \lambda^*} \), can be interpreted as a base bid governed by the pacing multiplier. The second term, \( \frac{P_t(b_t^*)}{P_t^{'}(b_t^*)} \), serves as a shading factor that adjusts the bid downward, reflecting the bidder’s strategic behavior in a first-price auction environment.

	\subsection{Marginal Profit Optimization}
	The bidding problems discussed so far assume that all payments (revenues) remain within the platform itself. However, consider a scenario where a Demand-Side Platform (DSP) or an Ad Network (AdNet) bids for external ad exchange or Supply-Side Platform (SSP) ad slots on behalf of an advertiser. In such cases, additional pricing adjustments are necessary to ensure profitability for the bidding platform (DSP or AdNet). This marginal profit requirement introduces additional complexities beyond the standard formulations discussed earlier, necessitating a revised optimization framework that accounts for both the advertiser's objectives and the platform's profitability.
	
	\paragraph{Problem Formulation}
	To incorporate the platform's profit margin, we introduce a markup factor \( m > 0 \) and modify the optimization problem in \eqref{eq:fpa_welfare} as follows:
	
	\begin{equation} \label{eq:fpa_welfare_maginal}
		\begin{aligned} 
			\max_{b_t \geq 0} \quad & \sum_{t=1}^T P_t(b_t) \\
			\text{s.t.} \quad & \sum_{t=1}^{T} P_t(b_t) \cdot b_t \cdot (1+m) \leq B. 
		\end{aligned}
	\end{equation}
	
	Compared to \eqref{eq:fpa_welfare}, an additional markup \( b_t \cdot m \) is applied to each impression, ensuring that the platform (DSP or AdNet) earns a margin on top of the campaign's bid price.
	
	\paragraph{Derivation of the Optimal Bid} 
	The approach to solving \eqref{eq:fpa_welfare_maginal} follows the same methodology as \eqref{eq:fpa_welfare}. We introduce the Lagrangian:
	
	\[
	\mathcal{L}(b_t, \lambda) = \sum_{t=1}^{T} P_t(b_t) - \lambda \left( \sum_{t=1}^T P_t(b_t) \cdot b_t \cdot (1+m) - B \right).
	\]
	
	By differentiating with respect to \( b_t \) and following the same steps as before, we obtain the counterpart of \eqref{eq:fpa_welfare_b_start}:
	
	\begin{equation} \label{eq:fpa_welfare_marginal_formula}
	b_t^* + \frac{P_t(b_t^*)}{P_t^{'}(b_t^*)} = \frac{1}{\lambda^* (1+m)}.
	\end{equation}
	
	The procedure for determining \( \lambda^* \) and estimating \( P_t \) remains unchanged from previous formulations. For completeness, we summarize the bidding strategy with marginal profit in \autoref{alg:bid_shading_fpa_margin}.

	\begin{algorithm}
		\caption{Bid Strategy for Max Delivery Welfare Maximization Problem with Marginal Profit under FPA}
		\label{alg:bid_shading_fpa_margin}
		\begin{algorithmic}[1]
			\Require 
			\( B \): Total budget; \( T \): Total auction opportunities; \(m\): Markup factor;\\
			\(\lambda_0\): Initial value for Lagrange multiplier; \\
			\( [b_l, b_u] \): Search range for bid price; \(\epsilon\): threshold for binary search; \\
			\( P_t(b), P_t^{'}(b) \): Winning probability function and its derivative;  \\
			\(PID\):  PID controller in \autoref{alg:pid_md};
			\Ensure \( b_t^* \): Optimal bid at auction opportunity \(t\).
			
			\State \textbf{Step 1: Initialize Parameters}
			\State Set initial Lagrange multiplier \(\lambda \gets \lambda_0\)
			\State Define budget pacing target: \( \frac{B}{T} \)
			
			\State \textbf{Step 2: Solve for Optimal \( b_t^* \)}
			\For{each pacing interval}
			\State Observe actual spend \(S\)
			\State Update \(\lambda\) using PID controller:  \(\lambda \gets PID(S, B, T, \lambda)\)
			\For{each auction opportunity \( t \)}
			\State \textbf{Perform Binary Search to Solve for \( b_t^* \)}
			\State Initialize search range: \( b_{\text{low}} \gets b_l \), \( b_{\text{high}} \gets b_u \)
			\While{\( b_{\text{high}} - b_{\text{low}} > \epsilon  \)}
			\State Set \( b_{\text{mid}} \gets \frac{b_{\text{low}} + b_{\text{high}}}{2} \)
			\State Compute \( F(b_{\text{mid}}) = P_t^{'}(b_{\text{mid}}) - \lambda \cdot (1+m) \cdot b_{\text{mid}} P_t^{'}(b_{\text{mid}}) - \lambda \cdot (1+m) \cdot  P_t(b_{\text{mid}}) \)
			\If{ \( F(b_{\text{mid}}) = 0 \) }
			\State \( b_t^* \gets b_{\text{mid}} \)
			\State \textbf{break}
			\ElsIf{ \( F(b_{\text{mid}}) > 0 \) }
			\State Set \( b_{\text{low}} \gets b_{\text{mid}} \)
			\Else
			\State Set \( b_{\text{high}} \gets b_{\text{mid}} \) 
			\EndIf
			\EndWhile
			\State Set \( b_t^* \gets b_{\text{mid}} \)
			\EndFor
			\EndFor
			
			\State \textbf{Step 3: Return Optimal Bids}
			\State \Return \( b_t^* \) for all \( t \)
		\end{algorithmic}
	\end{algorithm}

	\section{Bidding under Arbitrary Auction}
	In previous sections, we explored how different auction mechanisms influence bidding dynamics. Beyond first-price and second-price auctions, other widely used auction mechanisms in practice include Generalized Second Price (GSP) and Vickrey-Clarke-Groves (VCG). In this section, we present a framework for designing an optimal bidding strategy under an arbitrary auction mechanism. A similar treatment can be found in \cite{gao2022bidding}. 
	
	\paragraph{Problem Formulation}
	We consider the max delivery problem for an oCPM ad campaign that aims to maximize total conversions. Suppose there are \(T\) auction opportunities and a total budget \(B\). The optimization problem under an arbitrary auction mechanism can be formulated as follows:
	\begin{equation} \label{eq:bidding_general_auction}
		\begin{aligned}
			\max_{b_t \geq 0} \quad & \sum_{t=1}^{T} r_t \cdot G_t(b_t) \\
			\text{s.t.} \quad & \sum_{t=1}^{T} H_t(b_t) \leq B.
		\end{aligned}
	\end{equation}
	where:
	\begin{itemize}
		\item \( G_t(b_t) \) denotes the probability of winning the \(t\)-th auction given a per-impression bid \( b_t \).
		\item \( H_t(b_t) \) represents the expected cost incurred for the \(t\)-th auction opportunity.
		\item \( r_t \) represents the predicted conversion rate for the \(t\)-th auction.
	\end{itemize}
	
	For analytical convenience, we impose the following regularity conditions:
	\begin{itemize}
		\item Both \( G_t \) and \( H_t \) are differentiable functions with first derivatives defined as:
		\[
		g_t = G_t'(b_t), \quad h_t = H_t'(b_t).
		\]
		\item Boundary conditions: \( G_t(0) = 0 \), \( G_t(+\infty) = 1 \), and \( g_t \geq 0 \);  \( H_t(0) = 0 \) and \( h_t \geq 0 \).
	\end{itemize}
	
	These conditions hold for most auction mechanisms in practice, including first-price and second-price auctions, GSP, Myerson's optimal auction, and VCG. 
	
	\paragraph{Derivation of the Optimal Bid}
	To solve \eqref{eq:bidding_general_auction}, we introduce the Lagrangian function with multiplier \( \lambda \geq 0 \):
	
	\begin{equation*}
		\mathcal{L}(b_t, \lambda) = \sum_{t=1}^{T} r_t \cdot G_t(b_t) - \lambda \left( \sum_{t=1}^{T} H_t(b_t) - B \right).
	\end{equation*}
	
	The optimal bid \( b_t^* \) satisfies the first-order condition:
	
	\[
	\frac{\partial}{\partial b_t} \mathcal{L}(b_t^*, \lambda^*) = 0.
	\]
	
	Expanding the derivative, we obtain:
	
	\[
	r_t \cdot g_t(b_t^*) - \lambda^* h_t(b_t^*) = 0.
	\]
	
	Solving for \( b_t^* \), we derive:
	
	\begin{equation} \label{eq:optimal_bid_general_auction}
		b_t^{*} = \left( \frac{h_t}{g_t} \right)^{-1} \left( \frac{r_t}{\lambda^*} \right),
	\end{equation}
	where \( (h_t/g_t)^{-1} \) denotes the inverse function of \( h_t/g_t \).
	
	\paragraph{Interpretation of the Bidding Formula}
	This formulation provides a general method for determining the optimal bid across various auction mechanisms. The exact computation of \( b_t^* \) depends on the specific forms of \( G_t(b_t) \) and \( H_t(b_t) \), which we will analyze below:

	\begin{itemize}
		\item \textbf{First Price Auction:} In the first price auction under the welfare maximization problem, we have:
		\[
		G_t(b_t) = P_t(b_t), \quad H_t(b_t) = P_t(b_t) \cdot b_t.
		\]
		Computing the derivatives, we obtain:
		\[
		\frac{H_t^{'}(b_t)}{G_t^{'}(b_t)} = \frac{p_t(b_t) \cdot b_t + P_t(b_t)}{p_t(b_t)}.
		\]
		Substituting into \eqref{eq:optimal_bid_general_auction}, we get:
		\[
		\frac{p_t(b_t) \cdot b_t + P_t(b_t)}{p_t(b_t)} = \frac{r_t}{\lambda^*}.
		\]
		For CPM campaigns where \( r_t = 1.0 \), this equation exactly matches \eqref{eq:fpa_welfare_b_start}.
		
		Similarly, for the utility maximization problem, \eqref{eq:optimal_bid_general_auction} recovers \eqref{eq:fpa_utility_b_star} through analogous derivations, which we leave as an exercise for the reader.
		
		\item \textbf{Second Price Auction:} By Myerson’s Lemma, for any dominant strategy incentive-compatible (DSIC) auction, we have:
		\[
		H_t(b_t) = b_t \cdot G_t(b_t) - \int_{0}^{b_t} G_t(z) \text{d}z.
		\]
		Taking the derivative on both sides:
		\[
		h_t(b_t) = g_t(b_t) \cdot b_t + G_t(b_t) - G_t(b_t),
		\]
		which simplifies to:
		\[
		\frac{h_t(b_t)}{g_t(b_t)} = b_t.
		\]
		Since \( h_t/g_t \) is an identity function, its inverse is also the identity function. Therefore, \eqref{eq:optimal_bid_general_auction} reduces to:
		\[
		b_t^* = \frac{r_t}{\lambda^*}.
		\]
		We recover the well-known optimal bidding strategy for second-price auctions. More generally, since Myerson’s Lemma applies to any DSIC auction mechanism, we can assert that the same bidding formula is optimal, e.g., for the VCG auction.
		
		\item \textbf{General Auction:} The general form of the auction mechanism determines the exact expressions for \( G_t(b_t) \) and \( H_t(b_t) \). By applying the optimal bidding framework, we can derive bid adjustments tailored to specific auction environments. The key challenge lies in correctly estimating \( P_t(b_t) \) and \( Q_t(b_t) \), which we discuss in \autoref{ch:bid_shading_win_prob}.
	\end{itemize}

	\section{Winning Probability Estimation}  \label{ch:bid_shading_win_prob}
	From previous discussions we can see that one indispensible piece of information we need in bid shading strategy is the probability winning function \(P_t(b)\) for a given bid level \(b\). In this section, we briefly introduce two main approaches to estimate \(P_t(b)\): parametric methods and non-parametric methods.
	
	\subsection{Parametric Methods}
	Parametric methods assume that the clearing price \( c_t \) follows a known distribution or that the winning probability function \( P_t(\cdot) \) can be represented by a specific parametric form. Historical auction data are then used to estimate the parameters of this distribution or function.
	
	Here, we discuss an approach proposed by \cite{pan2020bid}, which assumes that \( P_t(\cdot) \) can be approximated by a sigmoid function:
	\[
	\hat{P}(b \mid X_t; \omega, \beta)
	= \frac{1}{1 + \exp\!\left[-\left(w_0 + \sum_{i=1}^k w_i x_{i,t} + \beta \log b \right)\right]},
	\]
	where:
	\begin{itemize}
		\item \( X_t = \{x_{i,t}\} \) is the feature vector describing the auction context of the \(t\)-th opportunity (e.g., campaign type, device type, or temporal features such as day of week and hour of day),
		\item \( \omega = \{w_i\} \) and \( \beta > 0 \) are model parameters to be estimated using historical data.
	\end{itemize}
	The use of \( \log(b) \) instead of \( b \) ensures that \( \hat{P}(b \mid X_t; \omega, \beta) \to 0 \) as \( b \to 0 \) and \( \hat{P}(b \mid X_t; \omega, \beta) \to 1 \) as \( b \to \infty \), consistent with the expected behavior of a winning probability function.
	
	To learn the model parameters \( (\omega, \beta) \), suppose the auction data take the form
	\[
	\{ \mathcal{A}_t \}_{t=1}^T = \{b_t, X_t, y_t\}_{t=1}^T,
	\]
	where:
	\begin{itemize}
		\item \( b_t \) is the bid price submitted in the \(t\)-th auction,
		\item \( X_t \) is the feature vector describing the auction context,
		\item \( y_t \in \{0,1\} \) indicates whether the campaign won the \(t\)-th auction (\(y_t=1\) for a win, \(y_t=0\) otherwise).
	\end{itemize}
	For each sample \( \mathcal{A}_t \), the negative log-likelihood loss is defined as
	\[
	\ell\big(\hat{P}(b_t \mid X_t; \omega, \beta); y_t\big)
	= -y_t \log \hat{P}(b_t \mid X_t; \omega, \beta)
	- (1 - y_t) \log \big(1 - \hat{P}(b_t \mid X_t; \omega, \beta)\big).
	\]
	
	The model parameters are obtained by minimizing the total loss:
	\[
	\omega^*, \beta^* = \arg\min_{\omega, \beta} \sum_{t=1}^{T}
	\ell\big(\hat{P}(b_t \mid X_t; \omega, \beta); y_t\big).
	\]

	Once \( \omega^* \) and \( \beta^* \) are estimated(e.g., using gradient descent method), the winning probability and its derivative for any auction context \( X_t \) are given by:
	\[
	P_t(b_t)
	= \frac{1}{1 + \exp\!\left[-\left(w_0 + \sum_{i=1}^k w_i^{*} x_{i,t} + \beta^{*} \log b_t \right)\right]},
	\]
	and
	\[
	P_t'(b_t)
	= P_t(b_t)\big(1 - P_t(b_t)\big) \frac{\beta^{*}}{b_t}.
	\]
	
	The idea we discuss above can be summarized as in \autoref{alg:parametric_winning_prob}. 
	
	\begin{algorithm}[H]
		\caption{Parametric Estimation of Winning Probability \(P(b \mid X)\) and Its Derivative \(P'(b \mid X)\)}
		\label{alg:parametric_winning_prob}
		\begin{algorithmic}[1]
			\Require Training data and model parameters 
			\begin{itemize}
				\item Training data $\{ \mathcal{A}_t \}_{t=1}^T = \{b_t, X_t, y_t\}_{t=1}^T$  
				\Statex \quad where $b_t$: bid price, $X_t$: feature vector, $y_t \in \{0,1\}$: win indicator
				\item Model parameters \(\omega = \{w_0, w_1, \dots, w_k\}\) and \(\beta > 0\).
			\end{itemize}
			
			\Ensure Estimated winning probability function \(P(b \mid X)\) and its derivative \(P'(b \mid X)\).
			
			\Statex
			
			\State \textbf{Step 1: Define the parametric model.}
			\[
			\hat{P}(b \mid X_t; \omega, \beta)
			= \frac{1}{1 + \exp\!\left[-\left(w_0 + \sum_{i=1}^{k} w_i x_{i,t} + \beta \log b \right)\right]}.
			\]
			
			\State \textbf{Step 2: Compute the negative log-likelihood loss for each training sample.}
			\[
			\ell_t = -y_t \log \hat{P}(b_t \mid X_t; \omega, \beta)
			- (1 - y_t) \log \!\big(1 - \hat{P}(b_t \mid X_t; \omega, \beta)\big).
			\]
			
			\State \textbf{Step 3: Estimate parameters by minimizing the total loss.}
			\[
			(\omega^*, \beta^*) = \arg\min_{\omega, \beta} \sum_{t=1}^{T} \ell_t.
			\]
			
			\State \textbf{Step 4: Compute the winning probability and its derivative for new auction data.}
			\[
			P(b \mid X) = \frac{1}{1 + \exp\!\left[-\left(w_0^* + \sum_{i=1}^{k} w_i^* x_i + \beta^* \log b \right)\right]},
			\]
			\[
			P'(b \mid X) = P(b \mid X)\big(1 - P(b \mid X)\big)\cdot \frac{\beta^*}{b}.
			\]
			
			\State \Return \(P(b \mid X)\), \(P'(b \mid X)\).
		\end{algorithmic}
	\end{algorithm}

	We will prove in the Remarks section of this chapter that \( P_t(b) \) defined in this manner is log-concave with respect to \( b \). This property then allows us to directly apply the bid shading strategies from \autoref{alg:bid_shading_fpa}, \autoref{alg:bid_shading_fpa_utility}, and \autoref{alg:bid_shading_fpa_margin}.
	
	Other parametric methods have been proposed based on various probability distributions. We list a few notable examples below; readers interested in further details can refer to the corresponding papers.
	
	\begin{itemize}
		\item \textbf{Gamma Distribution:} \cite{zhu2017gamma}
		\item \textbf{Gaussian Distribution:} \cite{wu2015predicting}
		\item \textbf{Gumbel Distribution:} \cite{wu2018deep}
		\item \textbf{Lognormal Distribution:} \cite{zhou2021efficient}
		\item \textbf{Mixture of Gaussians:} \cite{ghosh2019scalable}
	\end{itemize}

	\subsection{Non-Parametric Methods}
	An alternative approach is to use non-parametric methods to estimate the winning probability.  
	Quantile regression is a natural fit for estimating the cumulative distribution function (CDF) in scenarios where data are censored, such as in first-price auction environments in RTB. We outline the core idea here; for a more comprehensive introduction, see \cite{gimenes2022quantile} and \cite{koenker2001quantile}.  
	
	In our setting, quantile regression aims to estimate the inverse of the winning probability function \(P(b \mid X)\).  
	We partition the interval \([0, 1]\) into buckets \(0 \leq \tau_0 < \tau_1 < \cdots < \tau_i < \cdots < \tau_n = 1\). For each quantile \(\tau\), define
	\[
	q_{\tau}(X) = \inf \{ b : P(b \mid X) \geq \tau \}.
	\]
	The goal is to learn an estimator \(\hat{q}_{\tau}(X \mid \omega)\) of \(q_{\tau}(X)\), parameterized by \(\omega\), based on the observed auction data.  
	
	Suppose the auction data take the form as in the previous section:
	\[
	\{ \mathcal{A}_t \}_{t=1}^T = \{b_t, X_t, y_t\}_{t=1}^T,
	\]
	Let the supporting price \(c_t\) follow some unknown i.i.d. distribution. For each training sample \(\mathcal{A}_t\), the \textit{pinball loss} for quantile regression is computed as
	\[
	\ell_\tau\big(\hat{q}_\tau(X_t \mid \omega); b_t, y_t\big) 
	= (1-y_t)\cdot \tau \cdot [\,b_t - \hat{q}_\tau(X_t \mid \omega)\,]_+ 
	+ y_t \cdot (1-\tau) \cdot [\,\hat{q}_\tau(X_t \mid \omega) - b_t\,]_+,
	\]
	where \([x]_+ = \max(0, x)\). The model parameters \(\omega\) are then learned by solving
	\[
	\omega^* = \arg\min_{\omega} \sum_{t=1}^{T} \ell_\tau\big(\hat{q}_\tau(X_t \mid \omega); b_t, y_t\big).
	\]
	
	Given the estimated quantile functions \(\{ \hat{q}_{\tau}(X \mid \omega^*)\}_{\tau = 0}^n\), we can recover the winning probability \(P(b \mid X)\) as follows: find the bucket \(k\) such that
	\[
	\hat{q}_{\tau_k}(X \mid \omega^*) \leq b < \hat{q}_{\tau_{k+1}}(X \mid \omega^*),
	\]
	and linearly interpolate to approximate
	\[
	P(b \mid X) \approx \hat{P}(b \mid X) = \tau_k + (\tau_{k+1} - \tau_k) \cdot 
	\frac{b - \hat{q}_{\tau_k}(X \mid \omega^*)}{\hat{q}_{\tau_{k+1}}(X \mid \omega^*) - \hat{q}_{\tau_k}(X \mid \omega^*)}.
	\]
	The derivative can be similarly approximated as
	\[
	P'(b \mid X) \approx \frac{\tau_{k+1} - \tau_k}{\hat{q}_{\tau_{k+1}}(X \mid \omega^*) - \hat{q}_{\tau_k}(X \mid \omega^*)}.
	\]
	
	We summarize the quantile regression algorithm in the following \autoref{alg:quantile_regression}:
	
	\begin{algorithm}[H]
		\caption{Estimation of Winning Probability \(P(b \mid X)\) and Its Derivative \(P'(b \mid X)\) via Quantile Regression}
		\label{alg:quantile_regression}
		\begin{algorithmic}[1]
			\Require Training data $\{ \mathcal{A}_t \}_{t=1}^T = \{b_t, X_t, y_t\}_{t=1}^T$  
			\Statex \quad where $b_t$: bid price, $X_t$: feature vector, $y_t \in \{0,1\}$: win indicator
			\Require Quantile levels $0 \leq \tau_0 < \tau_1 < \dots < \tau_n = 1$
			
			\State Train quantile regression models:
			\For{each quantile $\tau_i$}
			\State Compute pinball loss for auction data:
			\[
			\ell_{\tau_i}\!\left(\hat{q}_{\tau_i}(X_t \mid \omega); b_t, y_t\right) 
			= (1-y_t)\cdot \tau_i \cdot [\, b_t - \hat{q}_{\tau_i}(X_t \mid \omega)\, ]_+ 
			+ y_t \cdot (1-\tau_i) \cdot [\, \hat{q}_{\tau_i}(X_t \mid \omega) - b_t \, ]_+ ,
			\]
			where $[x]_+ = \max(0, x)$.
			\State Learn parameters by minimizing empirical loss:
			\[
			\omega^* = \arg\min_{\omega} \sum_{t=1}^{T} 
			\ell_{\tau_i}\big(\hat{q}_{\tau_i}(X_t \mid \omega); b_t, y_t\big).
			\]
			\EndFor
			
			\State Given a bid $b$ and feature vector $X$:  
			\State Find bucket $k$ such that
			\[
			\hat{q}_{\tau_k}(X \mid \omega^*) \leq b < \hat{q}_{\tau_{k+1}}(X \mid \omega^*).
			\]
			
			\State Estimate winning probability:
			\[
			P(b \mid X) \approx \tau_k + (\tau_{k+1} - \tau_k) \cdot
			\frac{b - \hat{q}_{\tau_k}(X \mid \omega^*)}{\hat{q}_{\tau_{k+1}}(X \mid \omega^*) - \hat{q}_{\tau_k}(X \mid \omega^*)}.
			\]
			
			\State Estimate derivative:
			\[
			P'(b \mid X) \approx \frac{\tau_{k+1} - \tau_k}{\hat{q}_{\tau_{k+1}}(X \mid \omega^*) - \hat{q}_{\tau_k}(X \mid \omega^*)}.
			\]
			
			\Ensure Approximations of $P(b \mid X)$ and $P'(b \mid X)$
		\end{algorithmic}
	\end{algorithm}

	For other non-parametric approaches, readers may refer to \cite{delnoij2020competing}, \cite{gligorijevic2020bid},  \cite{kaplan1958nonparametric},  \cite{ren2019deep}, \cite{xu2024simultaneous}, \cite{zhang2021meow} and \cite{zhou2021efficient}.
	


	\section{Remarks}
	\subsection{\(\lambda\) Update with DOGD} \label{ch:bid_shading_dogd_lambda}
	In solving \eqref{eq:fpa_welfare}, we previously mentioned that the PID controller method can be used to update \(\lambda\) online, as the optimal \(\lambda^*\) is the value that exactly depletes the total budget. Here, we introduce an alternative approach by applying the DOGD (Dual Online Gradient Descent) method. We apply this method to the more general problem \eqref{eq:bidding_general_auction}: 
	
	\begin{equation*}
		\begin{aligned}
			\max_{b_t \geq 0} \quad & \sum_{t=1}^{T} r_t \cdot G_t(b_t) \\
			\text{s.t.} \quad & \sum_{t=1}^{T} H_t(b_t) \leq B.
		\end{aligned}
	\end{equation*}
	
	The Lagrangian for this problem is:
	\[
	\mathcal{L}(b_t, \lambda) = \sum_{t=1}^{T} \left[ r_t \cdot G_t(b_t) - \lambda H_t(b_t) + \lambda \cdot \frac{B}{T} \right]
	\]
	The corresponding dual problem is:
	\[
	\mathcal{L}(\lambda) = \max_{b_t \geq 0} \mathcal{L}(b_t, \lambda)
	\]
	As derived in \eqref{eq:optimal_bid_general_auction}, the maximum is attained when:
	\[
	b_t = b_t(\lambda) = \left(\frac{h_t}{g_t} \right)^{-1} \left(\frac{r_t}{\lambda}\right)
	\]
	where \(h_t\) and \(g_t\) represent the derivative functions of \(H_t\) and \(G_t\), respectively. Substituting this into the Lagrangian gives:
	\[
	\mathcal{L}(\lambda) = \sum_{t=1}^T \left[ r_t \cdot G_t(b_t(\lambda)) - \lambda H_t(b_t(\lambda)) + \lambda \cdot \frac{B}{T} \right] = \sum_{t=1}^T L_t(\lambda)
	\]
	with:
	\[
	L_t(\lambda) = r_t \cdot G_t(b_t(\lambda)) - \lambda H_t(b_t(\lambda)) + \lambda \cdot \frac{B}{T}
	\]
	To update \(\lambda\), we apply gradient descent in the dual space:
	\[
	\lambda \gets \lambda - \epsilon_t \cdot \frac{\partial}{\partial \lambda} \mathcal{L}(\lambda)
	\]
	Since auction opportunities arrive in a streaming manner, we use a stochastic gradient descent approach based on each incoming auction:
	\[
	\lambda \gets \lambda - \epsilon_t \cdot \frac{\partial}{\partial \lambda} L_t(\lambda)
	\]
	The derivative is computed as:
	\[
	\frac{\partial}{\partial \lambda} L_t(\lambda) = b_t'(\lambda) \left[ r_t \cdot g_t(b_t(\lambda)) - \lambda \cdot h_t(b_t(\lambda)) \right] + \frac{B}{T} - H_t(b_t(\lambda))
	\]
	Recall that:
	\[
	\frac{h_t(b_t)}{g_t(b_t)} = \frac{r_t}{\lambda}
	\]
	implying:
	\[
	r_t \cdot g_t(b_t(\lambda)) - \lambda \cdot h_t(b_t(\lambda)) = 0
	\]
	Therefore:
	\[
	\frac{\partial}{\partial \lambda} L_t(\lambda) = \frac{B}{T} - H_t(b_t)
	\]
	The update rule becomes:
	\begin{equation} \label{eq:fpa_general_update_rule}
		\lambda \gets \lambda - \epsilon_t \cdot \left(\frac{B}{T} - H_t(b_t)\right)
	\end{equation}
	
	In this update rule, \(\frac{B}{T}\) represents the target spend per auction opportunity, while \(H_t(b_t)\) denotes the actual spend per auction opportunity. The update step for \(\lambda\) adjusts according to the deviation between the actual spend and the target spend, mirroring the update rule derived in \eqref{eq:dogd_md_update_rule} for the maximum delivery problem under the standard second-price auction.

	\subsection{Monotonicity of the Ratio of PDF to CDF under Log-Concavity}
	We prove the following statement:
	
		Let \( F(x) \) be the CDF and \( f(x) \) the PDF of a log-concave distribution. Then the ratio
		\[
		R(x) = \frac{F(x)}{f(x)}
		\]
		is non-decreasing.

	\begin{proof}
		We proceed in the following steps:
		
		\noindent
		\textbf{Step 1: Compute the derivative of \( R(x) \).}  
		Define \( R(x) = \frac{F(x)}{f(x)} \). By the quotient rule:
		\begin{align*}
			R'(x) &= \frac{d}{dx} \left(\frac{F(x)}{f(x)}\right) = \frac{F'(x) f(x) - F(x) f'(x)}{f(x)^2} \\
			&= \frac{f(x)^2 - F(x) f'(x)}{f(x)^2} \quad \text{(since \( F'(x) = f(x) \))} \\
			&= 1 - \frac{F(x) f'(x)}{f(x)^2}.
		\end{align*}
		For \( R(x) \) to be non-decreasing, we need to show:
		\begin{equation}
			f(x)^2 \geq F(x) f'(x) \quad \forall x. \label{eq:key_inequality}
		\end{equation}
		
		\noindent
		\textbf{Step 2: Log-Concavity and Its Consequence.}  
		Since \( f(x) \) is log-concave, its logarithm \( \log f(x) \) is concave, meaning:
		\[
		\frac{d^2}{dx^2} \log f(x) \leq 0.
		\]
		Define:
		\[
		g(x) = \frac{f'(x)}{f(x)} = \frac{d}{dx} \log f(x).
		\]
		Since \( \log f(x) \) is concave, its derivative \( g(x) \) is non-increasing, meaning:
		\[
		g'(x) = \frac{d}{dx} \left(\frac{f'(x)}{f(x)}\right) \leq 0.
		\]
		This key fact ensures that the relative change in \( f(x) \) is decreasing.
		
		\noindent
		\textbf{Step 3: Define \( h(x) = f(x) - F(x) g(x) \).}  
		To analyze the key inequality \eqref{eq:key_inequality}, define:
		
		\[
		h(x) = f(x) - F(x) g(x).
		\]
		
		Taking its derivative:
		\begin{equation*}
			\begin{aligned}
				h'(x) &= f'(x) - \left[ F'(x) g(x) + F(x) g'(x) \right] \\
				&= f'(x) - \left[ f(x) g(x) + F(x) g'(x) \right] \\
				&= f'(x) - f'(x) - F(x) g'(x) \quad \text{(since \( f'(x) = f(x) g(x) \))} \\
				&= -F(x) g'(x).
			\end{aligned}
		\end{equation*}
		Since \( g'(x) \leq 0 \) (log-concavity implies \( g(x) \) is non-increasing), we get:
		
		\[
		h'(x) = -F(x) g'(x) \geq 0.
		\]
		
		Thus, \( h(x) \) is non-decreasing.
		
		\noindent
		\textbf{Step 4: Boundary Condition Analysis.}  
		To ensure \( h(x) \geq 0 \) for all \( x \), consider the boundary conditions:
		
		- If \( x \to -\infty \), then \( F(x) \to 0 \) and \( f(x) \to 0 \), and it is known from Mill's ratio for log-concave distributions that:
		\[
		\lim_{x \to -\infty} \frac{F(x)}{f(x)} = 0.
		\]
		This implies that at the lower bound, \( h(x) \geq 0 \).
		
		- If the distribution has bounded support, i.e., starting at \( x = a \), then \( F(a) = 0 \) and thus:
		\[
		h(a) = f(a) > 0.
		\]
		
		Since \( h(x) \) is non-decreasing, it follows that:
		
		\[
		h(x) \geq 0 \quad \forall x.
		\]
		
		\noindent
		\textbf{Step 5: Final Conclusion.}  
		Since \( h(x) = f(x) - F(x) g(x) \) is non-negative, this implies:
		
		\[
		f(x)^2 \geq F(x) f'(x).
		\]
		
		Thus:
		
		\[
		R'(x) = \frac{f(x)^2 - F(x) f'(x)}{f(x)^2} \geq 0.
		\]
		
		which proves that \( R(x) \) is non-decreasing.
		
		\[
		\frac{F(x)}{f(x)} \text{ is increasing under log-concavity.}
		\]
		
	\end{proof}
	
	\subsection{Log-concavity of Winning Probability Function under Sigmoid Assumption}
	We consider the winning probability function given by:
	
	\[
	P_t(b) = \frac{1}{1 + e^{-(w_0 + \sum_{i=1}^{k} w_i x_i + \beta \log b)}}
	\]
	which can be rewritten as:
	\[
	P_t(b) = \frac{1}{1 + e^{-z}}, \quad \text{where} \quad z = w_0 + \sum_{i=1}^{k} w_i x_i + \beta \log b.
	\]
	Our goal is to establish the log-concavity of \( P_t(b) \), i.e., to show that:
	\[
	\frac{d^2}{db^2} \log P_t(b) \leq 0.
	\]
	
	\paragraph{Step 1: First Derivative of \( P_t(b) \)}
	Differentiating \( P_t(b) \) with respect to \( b \):
	\[
	\frac{d}{db} P_t(b) = P_t(b) (1 - P_t(b)) \frac{d}{db} z.
	\]
	Since:
	\[
	\frac{d}{db} z = \frac{\beta}{b},
	\]
	it follows that:
	\[
	\frac{d}{db} P_t(b) = P_t(b) (1 - P_t(b)) \frac{\beta}{b}.
	\]
	
	\paragraph{Step 2: Second Derivative of \( P_t(b) \)}
	Differentiating again:
	\[
	\frac{d^2}{db^2} P_t(b) = \frac{d}{db} \left[ P_t(b) (1 - P_t(b)) \frac{\beta}{b} \right].
	\]
	Applying the product rule:
	\[
	\frac{d^2}{db^2} P_t(b) = (1 - 2P_t(b)) P_t(b) (1 - P_t(b)) \frac{\beta^2}{b^2} + P_t(b) (1 - P_t(b)) \left(-\frac{\beta}{b^2}\right).
	\]
	Factoring \( P_t(b) (1 - P_t(b)) \frac{1}{b^2} \):
	\[
	\frac{d^2}{db^2} P_t(b) = P_t(b) (1 - P_t(b)) \frac{1}{b^2} \left[ (1 - 2P_t(b)) \beta^2 - \beta \right].
	\]
	\paragraph{Step 3: Second Derivative of \( \log P_t(b) \)}
	By the identity:
	\[
	\frac{d^2}{db^2} \log P_t(b) = \frac{P_t''(b)}{P_t(b)} - \left( \frac{P_t'(b)}{P_t(b)} \right)^2,
	\]
	we substitute \( P_t''(b) \) and \( P_t'(b) \) to obtain:
	\[
	\frac{d^2}{db^2} \log P_t(b) = \frac{1 - P_t(b)}{b^2} \left[ (1 - 2P_t(b)) \beta^2 - \beta - (1 - P_t(b)) \beta^2 \right].
	\]
	Simplifying:
	\[
	\frac{d^2}{db^2} \log P_t(b) = \frac{1 - P_t(b)}{b^2} \left[ \beta^2 - 2P_t(b) \beta^2 - \beta - \beta^2 + P_t(b) \beta^2 \right].
	\]
	\[
	= \frac{1 - P_t(b)}{b^2} \left[ - P_t(b) \beta^2 - \beta \right].
	\]
	
	\paragraph{Step 4: Sign Analysis and Conclusion}
	Since \( 1 - P_t(b) \) is positive and \( b^2 \) is positive, the sign of \( \frac{d^2}{db^2} \log P_t(b) \) depends on:
	\[
	-P_t(b) \beta^2 - \beta.
	\]
	\begin{itemize}
	\item If \( \beta > 0 \), then \( -P_t(b) \beta^2 - \beta \) is negative for all \( P_t(b) \), ensuring \( \frac{d^2}{db^2} \log P_t(b) \leq 0 \).
	\item  If \( \beta < 0 \), the expression could be positive, violating log-concavity.
	\end{itemize}
	Thus, we conclude that
	\[
	P_t(b) \text{ is log-concave if and only if } \beta > 0.
	\]
		
	\subsection{Feedback loops and Distribution Shift when Estimating Winning Probability}
	TBA, \cite{kim2023addressing}

	\chapter{Multi-Channel Delivery}
	
	\intro{
	We discuss how bidding algorithms should be designed when a campaign can be delivered across multiple channels with different auction mechanisms. This is particularly important for platforms that extend their ad delivery to audience networks, external SSPs, and ad exchanges.
	}
	
	A number of digital advertising platforms (e.g., Meta and LinkedIn) offer advertisers the option to extend their sponsored content (campaigns) to third-party, offsite platforms, thereby enabling campaigns to reach a broader audience. One key motivation for this extension is that these platforms possess vast amounts of first-party, onsite data, which can be leveraged to more effectively target the desired audience and predict conversion rates compared to traditional Demand-Side Platforms (DSPs). By extending delivery to offsite traffic, the platform can increase its available supply while continuing to capitalize on the value of its onsite data.
	
	In such cases, campaigns can be delivered across multiple placements. For onsite placements, the auction mechanism might use a second-price or generalized second-price auction, whereas for offsite placements (e.g., ad exchanges or Supply-Side Platforms (SSPs)), the auction is more likely to follow a first-price auction model (as discussed in the previous chapter). When designing a bid optimization strategy, we need to take these hybrid auction mechanisms into consideration.
	
	\section{Multi-Channel Max Delivery Problem}
	 We use a max-delivery example to illustrate how to manage multi-channel delivery under hybrid auction mechanisms. As before, we aim to maximize total conversions under a given budget constraint. Suppose there are \(M\) channels, with the first one representing the onsite platform and the remaining \(M-1\) channels representing offsite platforms. On the onsite platform, ad slots are auctioned using a standard second-price mechanism, and all auction data are fully available. In contrast, on offsite platforms, ad slots are auctioned using a first-price mechanism with only partial information visibility (i.e., the data are censored; the campaign only learns whether it won the auction or not). With these assumptions, the problem can be formulated as follows:
	
	\begin{equation} \label{eq:multi_channel_delivery_formulation}
		\begin{aligned}
			\max_{b_{i,t} \geq 0} \quad & \sum_{i=1}^{M} \sum_{t=1}^{T_i} r_{i,t} \cdot G_{i,t}(b_{i,t}) \\
			\text{s.t.} \quad & \sum_{i=1}^{M} \sum_{t=1}^{T_i} H_{i,t}(b_{i,t}) \leq B.
		\end{aligned}
	\end{equation}
	where:
	
	\begin{itemize}
		\item \(B\) is the total campaign budget.
		\item \(T_i\) is the number of auction opportunities for channel \(i\).
		\item \(i = 1, \cdots, M\) indicates different channels (e.g., \(i=1\) represents traffic from onsite sources, \(i=2\) represents traffic from an ad exchange, etc.).
		\item \(t = 1, \cdots, T_i\) indicates the \(t\)-th auction opportunity from channel \(i\).
		\item \(r_{i,t}\) is the estimated conversion rate for the \(t\)-th auction opportunity from channel \(i\).
		\item \(b_{i,t}\) is the bid amount per impression.
		\item \(G_{i,t}(b_{i,t})\) is the winning probability (or a binary indicator) given the bid price \(b_{i,t}\).
		\item \(H_{i,t}(b_{i,t})\) is the expected payment (or cost) given the bid price \(b_{i,t}\).
	\end{itemize}
	We apply the standard primal-dual method to derive the bidding formula. First, we introduce the Lagrangian with dual parameter \(\lambda\):
	
	\begin{equation*}
		\begin{aligned}
			\mathcal{L}(b_{i,t}, \lambda) = & \sum_{i=1}^{M} \sum_{t=1}^{T_i} r_{i,t} \cdot G_{i,t}(b_{i,t}) - \lambda \left( \sum_{i=1}^{M} \sum_{t=1}^{T_i} H_{i,t}(b_{i,t}) - B \right) \\
			= & \sum_{i=1}^{M} \underbrace{\sum_{t=1}^{T_i} \left[r_{i,t} \cdot G_{i,t}(b_{i,t}) - \lambda H_{i,t}(b_{i,t})\right]}_{S_i(b_{i,t})} + \lambda B.
		\end{aligned}
	\end{equation*}
	The dual problem is then obtained by maximizing over \(b_{i,t}\):
	
	\[
	\mathcal{L}(\lambda) = \max_{b_{i,t} \geq 0} \mathcal{L}(b_{i,t}, \lambda).
	\]
	
	We handle this maximization problem separately for different channels:
	
	\begin{itemize}
		\item \textbf{Onsite traffic \(i=1\)}: In this case, the auction follows a standard second-price mechanism. We have  
		\[
		G_{1,t}(b_{1,t}) = \mathds{1}_{\{b_{1,t} > c_t \}} \quad \text{and} \quad  H_{1, t}(b_{1,t}) = \mathds{1}_{\{b_{1,t} > c_t \}} \cdot c_t,
		\]
		where \(c_t\) is the supporting price (the highest bid among other bidders) for the \(t\)-th auction opportunity. Therefore, we get:
		\[
		\begin{aligned}
			S_1(b_{1,t}) = & \sum_{t=1}^{T_1} \left[ r_{1,t} \cdot G_{1,t}(b_{1,t})  - \lambda \cdot H_{1,t}(b_{1,t}) \right] \\
			= & \sum_{t=1}^{T_1} \left[ r_{1,t} \cdot \mathds{1}_{\{b_{1,t} > c_t\}} - \lambda \cdot \mathds{1}_{\{b_{1,t} > c_t\}} \cdot c_t \right] \\ 
			= & \sum_{t=1}^{T_1} \mathds{1}_{\{b_{1,t} > c_t\}} \left(r_{1,t} - \lambda \cdot c_t \right).
		\end{aligned}
		\]
		To maximize the sum \(S_1\) over \(b_{1,t}\), as we do for the max-delivery case in Part \(\mathrm{II}\)  \autoref{sec:md_optimal}, we choose \(b_{1,t} > c_t\) if \( r_{1,t} > \lambda \cdot c_t\) and \(b_{1,t} \leq c_t\) otherwise. The maximal value is then:
		\begin{equation}\label{eq:multi_channel_md_onsite_dual}
		\max_{b_{1,t} \geq 0} S_1(b_{1,t}) = \sum_{t=1}^{T_1} \left(r_{1,t} - \lambda \cdot c_t \right)_{+}.
		\end{equation}
		For a fixed dual parameter \(\lambda\), the optimal bid for onsite traffic is given by:
		\begin{equation} \label{eq:multi_channel_md_onsite_bid}
			b_{1,t}^{*} = \frac{r_{1,t}}{\lambda}.
		\end{equation}
		This bidding formula is identical to the one derived in Part \(\mathrm{II}\) \autoref{sec:md_optimal}.

		\item \textbf{Offsite traffic \( 2 \leq i \leq M\)}: In this case, the auction follows a first-price mechanism. Assume that the marginal profit percentile \(m\) is collected for these offsite auctions. We then have: 
		\[
		G_{i,t}(b_{i,t}) = P_{i,t}(b_{i,t}) \quad \text{and} \quad H_{i,t}(b_{i,t}) = P_{i,t}(b_{i,t}) \cdot b_{i,t} \cdot (1+m),
		\]
		where \(P_{i,t}(b_{i,t})\) denotes the winning probability for the \(t\)-th auction opportunity from the \(i\)-th offsite channel given bid \(b_{i,t}\). For the \(i\)-th channel, we have: 
		\[
		\begin{aligned}
			S_i(b_{i,t}) = & \sum_{t=1}^{T_i} \left[ r_{i,t} \cdot G_{i,t}(b_{i,t}) - \lambda \cdot H_{i,t}(b_{i,t}) \right] \\
			= & \sum_{t=1}^{T_i} \left[r_{i,t} \cdot P_{i,t}(b_{i,t}) - \lambda \cdot P_{i,t}(b_{i,t}) \cdot b_{i,t} \cdot (1+m) \right].
		\end{aligned}
		\]
		
		Similar to the derivation in \eqref{eq:fpa_welfare_marginal_formula}, for a fixed dual parameter \(\lambda\), \(S_i\) attains its maximum when \(b_{i,t} = b_{i,t}^*\) such that:
		
		\begin{equation} \label{eq:multi_channel_md_offsite_bid}
			b_{i,t}^* + \frac{P_{i,t}(b_{i,t}^*)}{P_{i,t}^{'}(b_{i,t}^*)} = \frac{r_{i,t}}{\lambda (1+m)}.
		\end{equation}
		
		The maximal value of \(S_i\) over \(b_{i,t}\) is then:
		
		\begin{equation} \label{eq:multi_channel_md_offsite_dual}
			\max_{b_{i,t} \geq 0} S_i(b_{i,t}) = \sum_{t=1}^{T_i} \left[ r_{i,t} \cdot G_{i,t}(b_{i,t}^{*}(\lambda)) - \lambda \cdot H_{i,t}(b_{i,t}^{*}(\lambda)) \right],
		\end{equation}
	where \(b_{i,t}^{*}(\lambda)\) is the solution of \(b_{i,t}^{*}\) derived from equation \eqref{eq:multi_channel_md_offsite_bid}.
	\end{itemize}
	
	Combining \eqref{eq:multi_channel_md_onsite_bid} and \eqref{eq:multi_channel_md_offsite_bid}, we have:
	
	\begin{equation} \label{eq:multi_channel_md_dual}
		\begin{aligned}
			\mathcal{L}(\lambda) & = \max_{b_{i,t} \geq 0} \mathcal{L}(b_{i,t}, \lambda) \\
			& = \max_{b_{1,t} \geq 0} S_1(b_{1,t})  + \sum_{i =2}^{M} \max_{b_{i,t} \geq 0} S_{i}(b_{i,t})  + \lambda B \\
			& =  \sum_{t=1}^{T_1} \left(r_{1,t} - \lambda \cdot c_t \right)_{+} +  \sum_{i =2}^{M}  \sum_{t=1}^{T_i} \left[ r_{i,t} \cdot G_{i,t}(b_{i,t}^{*}(\lambda)) - \lambda \cdot H_{i,t}(b_{i,t}^{*}(\lambda)) \right] + \lambda B \\
			& = \sum_{t=1}^{T_1} \left[ \underbrace{\left(r_{1,t} - \lambda \cdot c_t \right)_{+}  + \lambda \cdot \frac{B}{T} }_{L_{1,t}(\lambda)} \right] \\
			& \quad +  \sum_{i =2}^{M}  \sum_{t=1}^{T_i} \left[ \underbrace{ r_{i,t} \cdot G_{i,t}(b_{i,t}^{*}(\lambda)) - \lambda \cdot H_{i,t}(b_{i,t}^{*}(\lambda)) + \lambda \cdot \frac{B}{T}}_{L_{i,t}(\lambda)} \right],
		\end{aligned}
	\end{equation}
	
	where \(T = \sum_{i=1}^M T_i\) is the total traffic.  To find the optimal \(\lambda\), we apply the dual gradient descent method:
	\[
	\lambda \gets \lambda - \epsilon \cdot \frac{\partial }{\partial \lambda} \mathcal{L}(\lambda).
	\]
	For streaming data, we use stochastic gradient descent, and the update rule for \(\lambda\) is:
	\[
	\lambda \gets \lambda - \epsilon \cdot \frac{\partial}{\partial \lambda} L_{i,t}(\lambda).
	\]
	There are two cases:
	\begin{itemize}
		\item \textbf{For onsite traffic (\(i=1\))}: 
		\[
		L_{1,t}(\lambda) = (r_{1,t} - \lambda \cdot c_t)_{+} + \lambda \cdot \frac{B}{T},
		\]
		so we have:
		
		\[
		\frac{\partial}{\partial \lambda} L_{1,t}(\lambda) = \frac{B}{T} - \mathds{1}_{\{r_{1,t} > \lambda \cdot c_t\}} \cdot c_t.
		\]
		
		\item \textbf{For offsite traffic (\(i = 2, \cdots, M\))}: 
		\[
		L_{i,t}(\lambda) = r_{i,t} \cdot G_{i,t}(b_{i,t}^{*}(\lambda)) - \lambda \cdot H_{i,t}(b_{i,t}^{*}(\lambda)) + \lambda \cdot \frac{B}{T}.
		\]
		From the discussion in Part \(\mathrm{III}\) in \autoref{ch:bid_shading_dogd_lambda}, we know that:
		\[
		\frac{\partial}{\partial \lambda}L_{i,t}(\lambda) = \frac{B}{T} - H_{i,t}(b_{i,t}(\lambda)).
		\]
	\end{itemize}
	Note that both \(\mathds{1}_{\{r_{1,t} > \lambda \cdot c_t\}} \cdot c_t\) and \(H_{i,t}(b_{i,t}(\lambda))\) represent the expected cost for the corresponding auction opportunity. Additionally, \(\frac{B}{T}\) denotes the target spend per auction opportunity for this campaign. From this, we can conclude that regardless of the platform type (onsite or offsite), the gradient \(\frac{\partial}{\partial \lambda} L_{i,t}(\lambda)\) indicates the deviation of the actual spend from the target spend.
	
	The algorithm to solve the multi-channel max-delivery problem \eqref{eq:multi_channel_delivery_formulation} is now clear: 
	
	\begin{itemize}
		\item Initialize the dual parameter \(\lambda\).
		\item Submit bids and observe the actual cost for each auction opportunity.
		\item Update \(\lambda\) by applying gradient descent, where the gradient is the difference between the target spend \(\frac{B}{T}\) and the actual spend.
		\item Update the bid using either \eqref{eq:multi_channel_md_onsite_bid} or \eqref{eq:multi_channel_md_offsite_bid}, depending on whether the traffic is onsite or offsite.
	\end{itemize}
	
	Essentially, for the hybrid auction mechanisms in the multi-channel problem, \(\lambda\) is the parameter responsible for the overall delivery control (including both onsite and offsite traffic). Once \(\lambda\) is updated, the method for determining the bid level follows the same principles as discussed previously, with the specific bid calculation depending on whether the traffic originates from an onsite (second-price) or offsite (first-price) channel.

	As we mentioned before, in practice, we may implement the algorithm in a batch manner, the main idea of this approach can be summarized in \autoref{alg:multi_channel_md_algorithm}:
	
	\begin{algorithm}[h]
		\caption{DOGD-Based Bid Strategy for Multi-Channel Max Delivery Optimization}
		\label{alg:multi_channel_md_algorithm}
		\begin{algorithmic}[1]
			\Require 
			\(B\): Total campaign budget; 
			\(T\): Total auction opportunities; 
			\(\lambda_0\): Initial value for the dual parameter; 
			\(\epsilon\): Learning rate; 
			\(m\): Marginal profit percentile; 
			\(P_{i,t}(b)\), \(P_{i,t}'(b)\): Winning probability function and its derivative.
			\Ensure 
			\(b_{i,t}^*\): Optimal bid at auction opportunity \(t\).
			
			\State \textbf{Step 1: Initialize Parameters}
			\begin{itemize}
				\item Initialize dual parameter: \(\lambda \leftarrow \lambda_0\)
				\item Define target spend: \(\frac{B}{T}\)
			\end{itemize}
			
			\State \textbf{Step 2: Solve for Optimal \(b_{i,t}^*\)}
				\For{each pacing interval}
				\State Observe actual spend \(S\) and update \(\lambda\) using gradient descent:
				\[
				\lambda \gets \lambda - \epsilon \left(\frac{B}{T} - S\right)
				\]
				\For{each auction opportunity \(t\)}
				\If{traffic is onsite (\(i=1\))}
				\State Compute bid:
				\[
				b_{1,t}^* = \frac{r_{1,t}}{\lambda}
				\]
				\Else
				\State Perform Binary Search to Solve for \(b_{i,t}^*\)
				\State Initialize search range: \(b_{\text{low}} \leftarrow b_l, b_{\text{high}} \leftarrow b_u\)
				\While{\(b_{\text{high}} - b_{\text{low}} > \epsilon\)}
				\State Set \(b_{\text{mid}} \leftarrow \frac{b_{\text{low}} + b_{\text{high}}}{2}\)
				\State Compute:
				\[
				F(b_{\text{mid}}) = P_{i,t}'(b_{\text{mid}}) - \lambda \cdot (1+m) \cdot b_{\text{mid}}P_{i,t}'(b_{\text{mid}}) - \lambda \cdot (1+m) \cdot P_{i,t}(b_{\text{mid}})
				\]
				\If{\(F(b_{\text{mid}}) = 0\)}
				\State \(b_{i,t}^* \leftarrow b_{\text{mid}}\)
				\State break
				\ElsIf{\(F(b_{\text{mid}}) > 0\)}
				\State \(b_{\text{low}} \leftarrow b_{\text{mid}}\)
				\Else
				\State \(b_{\text{high}} \leftarrow b_{\text{mid}}\)
				\EndIf
				\EndWhile
				\EndIf
				\EndFor
				\EndFor
			
			\State \textbf{Step 3: Return Optimal Bids}
			\begin{itemize}
				\item Return \(b_{i,t}^*\).
			\end{itemize}
	\end{algorithmic}
	\end{algorithm}



	\chapter{Campaign Group Optimization}
	
	\intro{
		     In this chapter, we introduce the problem of campaign group optimization, where multiple campaigns share a common budget. We present a mathematical formulation of this problem and explore principled approaches for bid optimization, including control-based methods and learning-based strategies, to adaptively adjust bids in response to changing market conditions. Additionally, we discuss key practical considerations to enhance the robustness of the bidding algorithm in real-world applications.
	}
	
	A number of platforms provide advertisers with the option to create a campaign group consisting of multiple campaigns with the same objective, sharing a total budget. Some platforms even use this as the default configuration. 
	
	From the advertiser's perspective, if the delivery algorithm is well-designed, the budget will be automatically allocated across different campaigns to achieve optimal performance. This simplifies the campaign creation process, as advertisers no longer need to manually split the budget across campaigns based on estimations.
	
	From the advertising platform's perspective, allowing budget sharing across different campaigns provides greater flexibility in designing delivery algorithms. This enables the platform to allocate budget more intelligently, optimizing overall performance while mitigating the risk of under-delivery when certain campaigns have lower quality (i.e., lower conversion rates, leading to less competitive bids).

	\section{Max Delivery Problem for Campaign Group Optimization}
	
	We first discuss the max-delivery problem. The basic setting is very similar to the max-delivery problem for a single campaign. When advertisers create a campaign group, they specify the lifetime and input the budget. The only difference is that multiple campaigns with different creatives are created under this group, and these campaigns share the group budget.
	
	\paragraph{Problem Formulation}
	Suppose there are \(N\) campaigns within the campaign group sharing a total budget \(B\). We aim to maximize the total conversions of all campaigns within the given budget. Under a standard second-price auction, the problem can be formulated as:
	
	\begin{equation} \label{eq:campaign_group_md}
		\begin{aligned}
			\max_{x_{i,t} \in \{0,1\}} \quad & \sum_{i=1}^{N} \sum_{t=1}^{T_i} x_{i,t} \cdot r_{i,t}  \\
			\text{s.t.} \quad & \sum_{i=1}^{N} \sum_{t=1}^{T_i} x_{i,t} \cdot c_{i,t} \leq B.
		\end{aligned}
	\end{equation}
	
	where \(x_{i,t} = \mathds{1}_{\{b_{i,t} > c_{i,t}\}}\), and:
	
	\begin{itemize}
		\item \(B\) is the total campaign group budget.
		\item \(T_i\) is the number of eligible auction opportunities for campaign \(i\).
		\item \(x_{i,t}\) is the binary decision variable indicating whether campaign \(i\) wins its \(t\)-th auction opportunity.
		\item \(r_{i,t}\) is the estimated conversion rate for the \(t\)-th auction opportunity for campaign \(i\).
		\item \(c_{i,t}\) is the highest competing bid per impression for the \(t\)-th auction opportunity for campaign \(i\).
		\item \(b_{i,t}\) is the bid amount per impression for the \(t\)-th auction opportunity for campaign \(i\).
	\end{itemize}
	
	\paragraph{Derivation of the Optimal Bid}
	As always, we introduce the Lagrangian with the dual multiplier \(\lambda \geq 0\):
	
	\[
	\begin{aligned}
		\mathcal{L}(x_{i,t}, \lambda) & =  \sum_{i=1}^{N} \sum_{t=1}^{T_i} x_{i,t} \cdot r_{i,t}  - \lambda \cdot \left( \sum_{i=1}^{N} \sum_{t=1}^{T_i} x_{i,t} \cdot c_{i,t} - B \right) \\
		& =  \sum_{i=1}^{N} \sum_{t=1}^{T_i} \left( x_{i,t} \cdot r_{i,t} -\lambda \cdot x_{i,t} \cdot c_{i,t} \right) + \lambda \cdot B \\
		& = \sum_{i=1}^{N} \sum_{t=1}^{T_i} \left[x_{i,t} \cdot \left( r_{i,t} - \lambda \cdot c_{i,t}\right) + \lambda \cdot \frac{B}{T} \right]
	\end{aligned}
	\]
	
	where \(T = \sum T_i\) is the total traffic of the campaign group. As reasoned in Part \(\mathbb{II}\) \autoref{sec:md_optimal}, to maximize \(\mathcal{L}(x_{i,t}, \lambda)\), we set \(x_{i,t} = 1\) whenever \(r_{i,t} - \lambda c_{i,t} > 0\) and \(x_{i,t} = 0\) otherwise. Then:
	
	\[
	\mathcal{L}^{*}(\lambda) = \max_{x_{i,t} \in \{0,1\}} \mathcal{L}(x_{i,t}, \lambda) = \sum_{i=1}^{N} \sum_{t=1}^{T_i} \left[ \underbrace{ \left( r_{i,t} - \lambda \cdot c_{i,t}\right)_{+} + \lambda \cdot \frac{B}{T}}_{f_{i,t}(\lambda)} \right].
	\]
	
	where \((\cdot)_{+}\) is the ReLU function. The dual problem is:
	
	\[
	\min_{\lambda \geq 0} \mathcal{L}^{*}(\lambda) = \min_{\lambda \geq 0} \sum_{i=1}^{N} \sum_{t=1}^{T_i} \left[\left( r_{i,t} - \lambda \cdot c_{i,t}\right)_{+} + \lambda \cdot \frac{B}{T} \right].
	\]
	
	Assuming the problem is feasible, we find the optimal dual parameter:
	
	\[
	\lambda^{*} = \argmin_{\lambda \geq 0} \mathcal{L}^{*}(\lambda).
	\]
	
	By the same argument as for the single campaign max-delivery problem, we see that \(\lambda^{*} > 0\) is the value that just depletes the budget such that:
	
	\[
	\sum_{i=1}^{N} \sum_{t=1}^{T_i} x_{i,t} \cdot c_{i,t} = B.
	\]
	
	The corresponding optimal bid per impression is:
	
	\[
	b_{i,t}^{*} = \frac{r_{i,t}}{\lambda^{*}}.
	\]
	
	The optimal bid per click (objective) is given by the following constant bid:
	
	\[
	b_{\text{click}}^{*} = \frac{1}{\lambda^{*}}. 
	\]
	
	\paragraph{Determining \(b_{\text{click}}^{*}\)}
	Most of the algorithms discussed in \autoref{part:pacing_algorithms} can be applied to determine the optimal bid level. We briefly discuss PID control and DOGD methods:
	
	\begin{itemize}
		\item \textbf{PID Controller:}  \(\lambda^*\) is the value of \(\lambda\) that depletes the total budget \(B\). Based on overall traffic patterns across all campaigns, we construct a target spend per pacing interval for the campaign group. The actual spend within each pacing interval is collected and compared to the target as the error signal \(e(t)\). A PID controller, as in \autoref{alg:pid_md}, is applied to update \(b_{\text{click}}\) adaptively. The bid per impression for campaign \(i\) is then computed as:
		\[
		b_{i,t} = b_{\text{click}} \cdot r_{i,t}.
		\]
		\item \textbf{DOGD:} We update \(\lambda\) in dual space using DOGD. The update rule is:
		\[
		\lambda \gets \lambda - \epsilon \cdot \frac{\partial}{\partial \lambda} \mathcal{L}^{*}(\lambda).
		\]
		For streaming data, we apply stochastic gradient descent to \(f_{i,t}\):
		\[
		\lambda \gets \lambda - \epsilon \cdot \left( \frac{B}{T} - \mathds{1}_{\{ r_{i,t} - \lambda \cdot c_{i,t} > 0 \}}  \cdot c_{i,t} \right).
		\]
		Note that \(\frac{B}{T} - \mathds{1}_{\{ r_{i,t} - \lambda \cdot c_{i,t} \}}  \cdot c_{i,t}\) represents the gap between the target spend and the actual spend for each auction request.  The update rule we derived here, to some extent, is identical to the update rule derived in \autoref{alg:dogd_md}. This similarity suggests that we can implement either the standard stochastic gradient descent (SGD) or a mini-batch version of the DOGD algorithm to update the bids.
	\end{itemize}
	
	One important consideration is that by applying these methods, we implicitly assume that all campaigns within the group share the same distributions for predicted click-through rate (pCTR) and the second-highest price. However, this assumption does not always hold in practice, as campaign quality can vary significantly. Some high-quality campaigns within the group may have substantially higher pCTR compared to others. 
	
	We will discuss how to handle this heterogeneous distribution later in this chapter. For now, we summarize the implementation of the PID controller and Dual Online Gradient Descent (DOGD) in \autoref{alg:campaign_group_md_pid} and \autoref{alg:campaign_group_md_dogd}, respectively.

	\begin{algorithm}
		\caption{PID Controller for Max Delivery in Campaign Group Optimization}
		\label{alg:campaign_group_md_pid}
		\begin{algorithmic}[1]
			\Require \\
			\(\Delta t\): Time interval for bid updates (e.g., 1 minute), 	\(\Delta \tau\): Time interval of target budget per bucket \\
			\(B\): Total campaign group budget \\
			\(\{s_k\}\): Target spend for each pacing interval \(k\) for the campaign group\\
			\(K_p, K_i, K_d\): PID controller gains \\
			\(N\): Number of campaigns within the campaign group \\
			\(M\): Number of bid updates in a day (\(M = \text{MinutesInOneDay} / \Delta t\))
			
			\Ensure \(b_{i,t}^*\): Optimal bid for each campaign \(i\) and auction opportunity \(t\)
			
			\State Initialize \(u(t_0) \gets 0\), cumulative\_error \(\gets 0\), previous\_error \(\gets 0\),  \( b_{click} \gets \)  initBidPerClick 
			\For{\(k = 1\) to \(M\)} \Comment{Loop over pacing intervals}
			\State Measure observed total spend across all campaigns during \(k\)-th pacing interval:
			\[
			r_k \gets \sum_{i=1}^{N} \text{observed\_spend}_i(k)
			\]
			
			\State Compute the error factor:
			\[
			e_k \gets r_k - s_k
			\]
			
			\State Update the control signal using the PID formula:
			\[
			u_k \gets K_p \cdot e_k + K_i \cdot \text{cumulative\_error} + K_d \cdot \frac{e_k - \text{previous\_error}}{\Delta t}
			\]
			
			\State Update the cumulative error:
			\[
			\text{cumulative\_error} \gets \text{cumulative\_error} + e_k \cdot \Delta t
			\]
			\State Update the bid per click price:
			\[
			b_{click} \gets b_{clikc} \cdot \exp(u(k))
			\]
			
			\For{each campaign \(i = 1, \dots, N\)}
			\For{each auction opportunity \(t\) in campaign \(i\)}
			\State Get conversion rate \(r_{i,t}\) and submit the new bid:
			\[
			b_{i,t} = b_{click} \cdot r_{i,t}
			\]
			\EndFor
			\EndFor
			
			\State Update the previous error:
			\[
			\text{previous\_error} \gets e_k
			\]
			\EndFor
		\end{algorithmic}
	\end{algorithm}

	\begin{algorithm}
		\caption{DOGD for Max Delivery in Campaign Group Optimization}
		\label{alg:campaign_group_md_dogd}
		\begin{algorithmic}[1]
			\Require \\
			$B$: Total budget, $T$: Predicted total number of auction opportunities \\
			$\Delta t$: Mini-batch update interval, $\epsilon_t$: Step size \\
			\(N\): Number of campaigns within the campaign group \\
			\(M\): Number of bid updates in a day (\(M = \text{MinutesInOneDay} / \Delta t\))
			\Ensure Optimal bid for each campaign \(i\) and auction opportunity \(t\)
			
			\State Initialize $\lambda \gets \lambda_{\text{init}}$ \Comment{Initial dual variable}
			
			\For{\(k = 1\) to \(M\)} \Comment{Loop over pacing intervals}
				
				\State Count the number of total auction requests \(R_k\) across all campaigns
				\State Observe the actual spend $S_k$ during the pacing interval \(\mathcal{I}_k\)
				
				\State Compute the mini-batch gradient:
				\[
				\text{BatchGrad}_k = \sum_{ s \in \mathcal{I}_k} \nabla_{\lambda} f_{i,t}(\lambda) = \frac{R_k}{T} \cdot B - S_k
				\]
				\State Update the dual variable using the mini-batch gradient:
				\[
				\lambda \gets \lambda - \epsilon \cdot \text{BatchGrad}_k
				\]
				\State Compute the bid per click:
				\[
				b_{click} = \frac{1}{\lambda}
				\]
			
				\For{each campaign \(i = 1, \dots, N\)}
					\For{each auction opportunity \(t \in \mathcal{I}_k\) in campaign \(i\)}
						\State Get conversion rate \(r_{i,t}\) and submit the new bid:
						\[
						b_{i,t} = b_{click} \cdot r_{i,t}
						\]
					\EndFor
				\EndFor
			\EndFor
		
		\end{algorithmic}
	\end{algorithm}
	
	\paragraph{Interpretation of the Bidding Formula}
	If we examine the optimal bid formula:
	
	\[
	b_{i,t}^* = \frac{r_{i,t}}{\lambda^*}
	\]
	and note that \(\lambda^*\) is updated based on the overall delivery of the campaign group, we can see that this formula is identical to the one used when treating the campaign group holistically. The only difference is that, for each request, the conversion rate is determined by the quality of the creative associated with the specific campaign within the group. 
	
	Since the denominator (\(\lambda^*\)) is the same across all campaigns, we can expect that the final budget allocation for each campaign will be roughly proportional to its average conversion rate. This is because the average bid per impression for each campaign is determined solely by its conversion rate. As a result, campaigns with higher-quality creatives (i.e., those with higher conversion rates) will consume a larger share of the budget and achieve more conversions. 
	
	Thus, the bidding algorithm inherently allocates more budget to higher-quality campaigns. From another perspective, this validates the motivation behind the campaign group product, as it automatically selects and prioritizes better-performing ads for delivery to the target audience.

	\section{Cost Cap Problem for Campaign Group Optimization}
	
	We now discuss the cost cap problem for campaign group optimization. The only difference between cost cap and max delivery is that, for each campaign, there is an additional cost control constraint. Specifically, at the end of the campaign’s lifetime, the average cost per result should not exceed a pre-specified cap.
	
	\paragraph{Problem Formulation}
	Suppose there are \(N\) campaigns within the campaign group sharing a total budget \(B\). For each campaign \(i = 1, \dots, N\), the advertiser imposes a cap \(C_i\) on the average cost per conversion. Under a standard second-price auction, the problem can be formulated as:
	
	\begin{equation} \label{eq:campaign_group_cost_cap}
		\begin{aligned}
			\max_{x_{i,t} \in \{0,1\}} \quad & \sum_{i=1}^{N} \sum_{t=1}^{T_i} x_{i,t} \cdot r_{i,t}  \\
			\text{s.t.} \quad & \sum_{i=1}^{N} \sum_{t=1}^{T_i} x_{i,t} \cdot c_{i,t} \leq B, \\
			&  \frac{ \sum_{t=1}^{T_i} x_{i,t} \cdot c_{i,t}}{  \sum_{t=1}^{T_i} x_{i,t} \cdot r_{i,t}} \leq C_i, \quad i = 1, \dots, N.
		\end{aligned}
	\end{equation}
	
	where:
	
	\begin{itemize}
		\item \(B\) is the total campaign group budget.
		\item \(T_i\) is the number of eligible auction opportunities for campaign \(i\).
		\item \(x_{i,t}\) is the binary decision variable indicating whether campaign \(i\) wins its \(t\)-th auction opportunity.
		\item \(r_{i,t}\) is the estimated conversion rate for the \(t\)-th auction opportunity for campaign \(i\).
		\item \(c_{i,t}\) is the highest competing bid per impression for the \(t\)-th auction opportunity for campaign \(i\).
		\item \(b_{i,t}\) is the bid amount per impression for the \(t\)-th auction opportunity for campaign \(i\).
		\item \(C_i\) is the cost cap for campaign \(i\).
	\end{itemize}
	
	\paragraph{Derivation of the Optimal Bid}
	We use the primal-dual method. First, we write the Lagrangian:
	
	\[
	\begin{aligned}
		\mathcal{L}(x_{i,t}, \lambda, \mu_i)  = &  \sum_{i=1}^{N} \sum_{t=1}^{T_i} x_{i,t} \cdot r_{i,t}  - \lambda \left(\sum_{i=1}^{N} \sum_{t=1}^{T_i} x_{i,t} \cdot c_{i,t} - B \right)  \\ 
		& \qquad - \sum_{i=1}^{N} \mu_i \cdot \left[ \sum_{t=1}^{T_i} x_{i,t} \cdot c_{i,t} - C_i \cdot \left(  \sum_{t=1}^{T_i} x_{i,t} \cdot r_{i,t} \right)\right] \\
		= &    \sum_{i=1}^{N} \sum_{t=1}^{T_i} \left( x_{i,t} \cdot r_{i,t} - \lambda \cdot x_{i,t} \cdot c_{i,t} + \lambda \cdot \frac{B}{T} - \mu_i \cdot x_{i,t} \cdot c_{i,t} + \mu_i \cdot C_i \cdot x_{i,t} \cdot r_{i,t} \right) \\
		= &   \sum_{i=1}^{N} \sum_{t=1}^{T_i} \left[ x_{i,t} \cdot \left(r_{i,t} - \lambda \cdot c_{i,t} - \mu_i \cdot c_{i,t} + \mu_i \cdot C_i \cdot r_{i,t} \right) + \lambda \cdot \frac{B}{T} \right]
	\end{aligned}
	\]
	
	Similar to the max delivery problem, to maximize \(\mathcal{L}(x_{i,t}, \lambda, \mu_i)\), we set \(x_{i,t} = 1\) whenever:
	
	\[
	r_{i,t} - \lambda \cdot c_{i,t} - \mu_i \cdot c_{i,t} + \mu_i \cdot C_i \cdot r_{i,t} > 0,
	\]
	
	and \(x_{i,t} = 0\) otherwise. The dual objective \(\mathcal{L}^{*}(\lambda, \mu_i)\) then becomes:
	
	\[
	\mathcal{L}^{*}(\lambda, \mu_{i}) = \sum_{i=1}^{N} \sum_{t=1}^{T_i} \left[  \left(r_{i,t} - \lambda \cdot c_{i,t} - \mu_i \cdot c_{i,t} + \mu_i \cdot C_i \cdot r_{i,t} \right)_{+} + \lambda \cdot \frac{B}{T} \right].
	\]
	
	Suppose we find a feasible solution to the dual:
	
	\[
	\lambda^*, \mu_{i}^{*} = \argmin_{\lambda \geq 0, \mu_i \geq 0} \mathcal{L}^{*}(\lambda, \mu_i).
	\]
	
	The optimal bid per impression is then given by:
	
	\[
	b_{i,t}^{*} = \frac{1+\mu_{i}^{*} C_i}{\lambda^{*} + \mu_{i}^{*}} \cdot r_{i,t}.
	\]

	\paragraph{Determining \(b_{i,t}^{*}\)}  
	There are various approaches to updating bids. In this section, we use Model Predictive Control (MPC) as an example to solve the problem. Other methods, such as PID control and Dual Online Gradient Descent (DOGD), are also applicable. Readers are encouraged to explore and implement these alternative algorithms themselves.
	
	For simplicity, we assume that all campaigns within the group follow the same distribution for predicted click-through rate (pCTR) and the second-highest price. As discussed earlier, at a specific time \(\tau\), the MPC method formulates a new optimization problem to solve over a receding time horizon \(H\). 
	
	For each campaign \(i\), we leverage prediction models to estimate the remaining budget and the number of remaining auction opportunities. These estimates are then used to compute the target spend and the target cost per result for each campaign. Once these targets are established, bid landscape models—such as bid-to-spend, bid-to-number-of-conversions, and bid-to-cost-per-conversion (CPX)—can be utilized to evaluate each bid candidate and determine whether it satisfies the constraints within the receding horizon. The optimal bid is then selected as the new update.

	\paragraph{Interpretation of the Bidding Formula}
	The optimal bid formula above is essentially identical to the one derived for the single-campaign cost cap problem. Here, \(\lambda\) is the dual parameter responsible for overall delivery control, while \(\mu_i\) is the dual parameter responsible for cost control specific to campaign \(i\).
	
	\begin{itemize}
	
	\item  If the cost constraint is not active, then \(\mu_i^* = 0\), and the bidding formula simplifies to:
	
	\[
	b_{i,t}^{*} = \frac{r_{i,t}}{\lambda^*},
	\]
	
	which corresponds to the max delivery problem.
	
	\item  If the budget constraint is not active, then \(\lambda^* = 0\), and the bidding formula reduces to:
	
	\[
	b_{i,t}^{*} = \left( \frac{1}{\mu^*} + C_i \right) \cdot r_{i,t},
	\]
	
	which only incorporates the cost cap constraint.
	\end{itemize}
	
	This formulation allows for a flexible allocation of budget while ensuring that each campaign meets its cost cap constraints.

	We summarize the above discussion in the following algorithm:
	
		\begin{algorithm}
		\caption{MPC for Cost Cap in Campaign Group Optimization}
		\label{alg:campaign_group_cost_cap_mpc}
		\begin{algorithmic}[1]
			\Require  \\
			$B_{\tau, r}$: Remaining budget; $B$: Total budget; $C_i$: Cost cap for campaign \(i\);  $N$: number of campaigns\\
			$T_i$: Total predicted requests for campaign \(i\) \\
			$H$: Receding time horizon; $NC_{\tau, i}$: Observed conversions for campaign \(i\); $TR_{\tau, i}$: Predicted total remaining requests for campaign \(i\); \\
			$NR_{\tau, H, i}$: Predicted requests in $(\tau, \tau + H)$ for campaign \(i\); $[b_l, b_u]$: Bid bounds; $\Delta b$: Search step size.
			\Ensure $b_i^*$: Optimal bid for $(\tau, \tau + H)$ for campaign \(i\).
			
			\State \textbf{Step 1: Compute Budget and Cost Cap Constraints}
			
			\State Compute the budget allocation for each campaign \(i\):
			\[
			B_i \gets \frac{T_i}{ \sum_{j=1}^{N} T_j}\cdot B
			\]
			
			\State Compute the budget allocation for the receding horizon for each campaign \(i\):
			\[
			B_{\tau, H, i} \gets \frac{NR_{\tau, H, i}}{\sum_{i=1}^N TR_{ \tau, i}} \cdot B_{\tau, r}
			\]
			
			\State Compute the cost per result upper bound for the remaining time:
			\[
			C_{\tau, r, i} \gets \frac{B_{\tau, r, i}}{\frac{B_i}{C_i} - NC_{\tau}}
			\]

			\State \textbf{Step 2: Construct Models $f_i(b)$ and $g_i(b)$}
			\State Collect the most recent $N$ bid-spend pairs $\{b_k, s_k\}$ and apply LIS or PAVA to construct $f_i(b)$ normalized to $H$.
			\State Collect the most recent $N$ bid-conversion pairs $\{b_k, n_k\}$ and apply LIS or PAVA to construct $g_i(b)$ normalized to $H$.
			\State Compute $h_i(b)$ as:
			\[
			h_i(b) \gets \frac{f_i(b)}{g_i(b)}
			\]
			
			\State \textbf{Step 3: Search for Optimal Bid $b^*$}
			\State Initialize $b^* \gets b_l$.
			\For{$b$ from $b_l$ to $b_u$ with step size $\Delta b$}
			\If{$\sum_{i=1}^N f_i(b) \leq B_{\tau, H}$ \textbf{and} $h_i(b) \leq C_{\tau, r, i}$}
			\State Update $b^* \gets b$.
			\EndIf
			\EndFor
			
			\State \textbf{Step 4: Return Optimal Bid}
			\State \Return $b^*$
		\end{algorithmic}
	\end{algorithm}
	
	\section{Remarks} 
	\subsection{Minimum Delivery Constraints in Campaign Group Optimization}
	
	Some advertisers wish to ensure that each campaign in the group receives a certain amount of impressions to prevent the entire budget from being spent on a single campaign. To achieve this, they may specify a minimum portion of the total budget that must be allocated to each campaign. 
	
	In this case, the optimization problem can be formulated as follows:
	
	\begin{equation} \label{eq:campaign_group_md_min_delivery}
		\begin{aligned}
			\max_{x_{i,t} \in \{0,1\}} \quad & \sum_{i=1}^{N} \sum_{t=1}^{T_i} x_{i,t} \cdot r_{i,t}  \\
			\text{s.t.} \quad & \sum_{i=1}^{N} \sum_{t=1}^{T_i} x_{i,t} \cdot c_{i,t} \leq B, \\
			& \sum_{t=1}^{T_i} x_{i,t} \cdot c_{i,t} \geq s_i \cdot B, \quad i = 1, \dots, N, \\
			& \sum_{i=1}^{N} s_i \leq 1.
		\end{aligned}
	\end{equation}
	
	where:
	\begin{itemize}
		\item \(B\) is the total campaign group budget.
		\item \(T_i\) is the number of eligible auction opportunities for campaign \(i\).
		\item \(x_{i,t}\) is the binary decision variable indicating whether campaign \(i\) wins its \(t\)-th auction opportunity.
		\item \(r_{i,t}\) is the estimated conversion rate for the \(t\)-th auction opportunity for campaign \(i\).
		\item \(c_{i,t}\) is the highest competing bid per impression for the \(t\)-th auction opportunity for campaign \(i\).
		\item \(b_{i,t}\) is the bid amount per impression for the \(t\)-th auction opportunity for campaign \(i\).
		\item \(s_i\) is the minimum fraction of the total budget that must be allocated to campaign \(i\).
	\end{itemize}
	
	The additional constraint \( \sum_{i=1}^{N} s_i \leq 1 \) ensures that the budget allocation remains feasible, preventing an over-constrained system where the sum of required allocations exceeds the available budget. 
	
	For completeness, we present a bid update algorithm using the Dual Online Gradient Descent (DOGD) method. Readers interested in alternative approaches may also explore control-based methods, such as Model Predictive Control (MPC).

	\paragraph{Lagrangian Formulation}
	To solve this problem, we introduce dual multipliers:
	\begin{itemize}
		\item \(\lambda \geq 0\) for the overall budget constraint.
		\item \(\gamma_i \geq 0\) for the minimum budget constraint of each campaign.
	\end{itemize}
	
	The Lagrangian function is:
	
	\[
	\begin{aligned}
		\mathcal{L}(x_{i,t}, \lambda, \gamma_i) &= \sum_{i=1}^{N} \sum_{t=1}^{T_i} x_{i,t} \cdot r_{i,t}  
		- \lambda \left( \sum_{i=1}^{N} \sum_{t=1}^{T_i} x_{i,t} \cdot c_{i,t} - B \right) \\
		&\quad - \sum_{i=1}^{N} \gamma_i \left( s_i B - \sum_{t=1}^{T_i} x_{i,t} \cdot c_{i,t} \right).
	\end{aligned}
	\]
	
	Expanding and rearranging terms:
	
	\[
	\begin{aligned}
		\mathcal{L}(x_{i,t}, \lambda, \gamma_i) &= \sum_{i=1}^{N} \sum_{t=1}^{T_i} \left( x_{i,t} \cdot r_{i,t} - \lambda \cdot x_{i,t} \cdot c_{i,t} + \lambda \cdot \frac{B}{T} + \gamma_i \cdot x_{i,t} \cdot c_{i,t} - \gamma_i \cdot s_i B \right).
	\end{aligned}
	\]
	
	\paragraph{Optimal Selection Condition}
	To maximize \(\mathcal{L}(x_{i,t}, \lambda, \gamma_i)\), we select \(x_{i,t} = 1\) whenever:
	
	\[
	r_{i,t} - \lambda \cdot c_{i,t} + \gamma_i \cdot c_{i,t} > 0.
	\]
	
	Rearranging:
	
	\[
	r_{i,t} > (\lambda - \gamma_i) \cdot c_{i,t}.
	\]
	
	Thus, campaign \(i\) wins the auction if:
	
	\[
	\frac{r_{i,t}}{c_{i,t}} > (\lambda - \gamma_i).
	\]
	
	\paragraph{Optimal Bid Formula}
	From the standard bidding mechanism, the optimal bid is given by:
	
	\begin{equation} \label{eq:optimal_bid_min_delivery}
		b_{i,t}^* = \frac{r_{i,t}}{\lambda - \gamma_i}.
	\end{equation}
	
	where:
	\begin{itemize}
		\item \(\lambda\) controls the overall budget pacing.
		\item \(\gamma_i\) ensures that each campaign receives at least \(s_i B\) of the budget.
	\end{itemize}
	
	\paragraph{Dual Gradient Descent Updates}
	
	We update \(\lambda\) and \(\gamma_i\) using stochastic gradient descent (SGD):
	
	\begin{equation*} 
		\lambda \gets \lambda - \epsilon_{\lambda} \left( \frac{B}{T} -  x_{i,t} \cdot c_{i,t} \right).
	\end{equation*}
	
	\begin{equation*}
		\gamma_i \gets \gamma_i - \epsilon_{\gamma} \left(  x_{i,t} \cdot c_{i,t} - s_i \cdot \frac{B}{T_i} \right).
	\end{equation*}
	
	These updates ensure:
	\begin{itemize}
		\item \(\lambda\) adjusts the overall budget pacing.
		\item \(\gamma_i\) enforces the minimum budget constraint for each campaign.
	\end{itemize}

	\paragraph{Interpretation of the Formula}
	\begin{itemize}
		\item If \(\gamma_i = 0\) (i.e., the minimum budget constraint is inactive for campaign \(i\)), then:
		\[
		b_{i,t}^* = \frac{r_{i,t}}{\lambda}.
		\]
		which is the standard max delivery bid formula.
		
		\item If \(\gamma_i > 0\) (i.e., campaign \(i\) is underfunded and needs more budget), then:
		\begin{itemize}
			\item The bid increases (since \(\lambda - \gamma_i\) is reduced).
			\item The campaign becomes more competitive and wins more auctions.
		\end{itemize}
		
		\item If \(\gamma_i\) is too high, the campaign overcompensates, leading to potential inefficiencies. 
	\end{itemize}
	
	\subsection{Some Practical Considerations}
	1. heterogeneous distribution of campaigins in the group, e.g., pctr distributions are different  
	
	\subsection{Another Formulation of Cost Cap}
	An alternative way to formulate the cost-cap problem is to maximize total conversions with campaign-specific
	weights derived from the caps specified by advertisers. The motivation is simple: a higher cap $C_i$ is often a
	signal that conversions in campaign $i$ are more valuable. Therefore, instead of treating conversions from all
	campaigns equally, we up-weight conversions from campaigns with larger caps.
	
	Under this consideration, we modify~\eqref{eq:campaign_group_cost_cap} as follows:
	\begin{equation}\label{eq:campaign_group_cost_cap_weighted}
		\begin{aligned}
			\max_{x_{i,t} \in \{0,1\}} \quad
			& \sum_{i=1}^{N} \sum_{t=1}^{T_i} x_{i,t}\, r_{i,t}\, C_i \\
			\text{s.t.} \quad
			& \sum_{i=1}^{N} \sum_{t=1}^{T_i} x_{i,t}\, c_{i,t} \le B, \\
			& \frac{\sum_{t=1}^{T_i} x_{i,t}\, c_{i,t}}{\sum_{t=1}^{T_i} x_{i,t}\, r_{i,t}} \le C_i,
			\quad i = 1,\ldots,N.
		\end{aligned}
	\end{equation}

	\chapter{Deep Retention Problem}
	
	\intro{
		In this chapter, we discuss the deep retention optimization problem, which arises in a special business scenario where post-conversion retention is critical for advertisers. 
	}

	\section{Deep Retention Problem}
	For some particular objective such as APP ads, user acquisition alone is not enough—retaining high-value users who engage deeply with the app is critical for long-term success. For example, traditional Cost-Per-Install (CPI) bidding focuses only on getting users to install an app, but many of these users never return or contribute to revenue. 
	
	To address this issue, Deep Retention bidding is introduced to optimize not just for installs, but for long-term user engagement and monetization.Deep retention focuses on in-app behaviors that signal lasting value, such as:
	
	\begin{itemize}
	\item Frequent app reopens (D3, D7, D14, D30 retention rates)
	\item In-app purchases (IAPs) and subscription activations
	\item Product views, add-to-cart, and purchases in e-commerce apps
	\item Level completion, engagement with premium content, or social interactions in gaming and entertainment apps
	\end{itemize}
	
	\paragraph{Problem Formulation}
	
	Under standard second price auction, we can formulate the deep conversion problem as follows:
	
	\begin{equation} \label{eq:deep_conversion}
		\begin{aligned}
			\max_{x_{t} \in \{0,1\}} \quad & \sum_{t=1}^{T} x_{t} \cdot c_t  \\
			\text{s.t.} \quad & \sum_{t=1}^{T} x_{t} \cdot c_{t} \leq B \\
			\quad & \frac{ \sum_{t=1}^{T} x_{t} \cdot c_{t}  }{ \sum_{t=1}^{T} x_{t} \cdot r_t} \leq C \\
			\quad & \frac{ \sum_{t=1}^{T} x_{t} \cdot c_{t}  }{ \sum_{t=1}^{T} x_{t} \cdot r_t \cdot d_t} \leq D. 
		\end{aligned}
	\end{equation}
	where 
	\begin{itemize}
		\item \(B\) is the total budget.
		\item \(T\) is the (predicted) number of auction opportunities.
		\item \(x_{t}\) is the binary decision varaible indicating whether campaign wins \(t\)-th auction opportunity.
		\item \(c_t\) is the highest competing bid per impression for the \(t\)-th auction opportunity.
		\item \(r_t\) is the impression to action(e.g., App install) conversion rate for \(t\)-th auction opportunity for the campaign.
		\item  \(d_t\) is the deep conversion rate conditioned on action for \(t\)-th auciton opportunity for the campaign.
		\item \(C\) is the cap for average cost per action.  
		\item \(D\) is the cap for average cost per deep conversion.
	\end{itemize}
	
	\paragraph{Derivation of the Optimal Bid}
	
	We first introduce the Lagrangian with dual parameters:
	
	\[
	\begin{aligned}
		\mathcal{L}(x_t, \lambda, \mu, \gamma) &= \sum_{t=1}^T  x_t \cdot c_t  
		- \lambda \left(\sum_{t=1}^{T} x_t \cdot c_t - B \right) \\
		&\quad - \mu \left(\sum_{t=1}^{T} x_t \cdot c_t - C \cdot \sum_{t=1}^{T} x_t \cdot r_t \right) 
		- \gamma \left( \sum_{t=1}^{T} x_t \cdot c_t - D \cdot \sum_{t=1}^{T} x_t \cdot r_t \cdot d_t \right)  \\
		&= \sum_{t=1}^{T}  \left\{ x_t \cdot \left[ c_t \cdot \left( 1 - \lambda - \mu - \gamma \right) 
		+ \mu \cdot C \cdot r_t + \gamma \cdot D \cdot r_t \cdot d_t \right] + \lambda \cdot \frac{B}{T} \right\}.
	\end{aligned}
	\]
	
	To maximize \(\mathcal{L}(x_t, \lambda, \mu, \gamma)\), we set \(x_t = 1\) whenever:
	
	\[
	c_t \cdot \left( 1 - \lambda - \mu - \gamma \right) + \mu \cdot C \cdot r_t + \gamma \cdot D \cdot r_t \cdot d_t > 0, 
	\]
	
	and \(x_t = 0\) otherwise. The dual objective \(\mathcal{L}^{*}(\lambda, \mu, \gamma)\) then becomes:
	
	\[
	\mathcal{L}^{*}(\lambda, \mu, \gamma) = \sum_{t=1}^{T}  \left\{  \left( c_t \cdot \left( 1 - \lambda - \mu - \gamma \right) 
	+ \mu \cdot C \cdot r_t + \gamma \cdot D \cdot r_t \cdot d_t \right)_{+} + \lambda \cdot \frac{B}{T} \right\}.
	\]
	
	Suppose we find a feasible solution to the dual problem:
	
	\[
	\lambda^*, \mu^*, \gamma^* = \argmin_{\lambda \geq 0, \mu \geq 0, \gamma \geq 0} \mathcal{L}^{*}(\lambda, \mu, \gamma).
	\]
	
	The optimal bid per impression is then given by:
	
	\begin{equation}\label{eq:deep_conversion_bid_impression}
	b_t^{*} = \frac{\mu^* \cdot C  + \gamma^* \cdot D  \cdot d_t}{\lambda^* + \mu^* + \gamma^* -1} \cdot r_t.
	\end{equation}
	
	The bid per conversion can be expressed as:
	
	\begin{equation}\label{eq:deep_conversion_bid_conversion}
		b_{conversion, t}^* =  \frac{\mu^* \cdot C  + \gamma^* \cdot D  \cdot d_t}{\lambda^* + \mu^* + \gamma^* -1}.
	\end{equation}
	
	\paragraph{Interpretation of the Bidding Formula}
	
	Examining \eqref{eq:deep_conversion_bid_conversion}, we observe the roles of the dual parameters:
	\begin{itemize}
		\item \(\lambda\) governs controls budget pacing.
		\item \(\mu\) controls the cost per conversion.
		\item \(\gamma\) controls the cost per deep conversion.
	\end{itemize}
	
	If the deep conversion rate is unbiased, we define \(C^{'} = D\cdot d_t\) as the normalized cost per conversion target based on the deep conversion target. The term \(\mu \cdot C + \gamma \cdot C^{'}\) represents a linear combination of the original conversion cost target and the normalized deep conversion cost target.
	
	In theory, only the most restrictive constraint should be active, meaning either \(\mu\) or \(\gamma\) should be positive while the other remains zero. A well-designed pacing algorithm should iteratively update \(\mu\) and \(\gamma\) based on real-time observed data, allowing the active constraint to emerge dynamically.

	\paragraph{Algorithm Design} 
	
	There are multiple approaches to designing algorithms for bid updates. In this section, we provide a detailed discussion on using a PID controller for this scenario. We will briefly mention alternative approaches based on Model Predictive Control (MPC) and Dual Online Gradient Descent (DOGD) in the section of Remarks.
	
	For simplicity, we rewrite \eqref{eq:deep_conversion_bid_impression} as:
	
	\[
	b_t^{*} = \alpha \cdot  \left(\beta_1 \cdot C \cdot r_t + \beta_2 \cdot D \cdot d_t \cdot r_t  \right),
	\]
	
	where:
	
	\[
	\alpha = \frac{1}{\lambda^*}, \quad 
	\beta_1 = \frac{\lambda^*}{\lambda^* + \mu^* + \gamma^* -1}  \cdot \mu^*, \quad 
	\beta_2 = \frac{\lambda^*}{\lambda^* + \mu^* + \gamma^* -1}  \cdot \gamma^*.
	\]
	
	Here, \(\alpha\) is responsible for budget delivery control, while \(\beta_1\) and \(\beta_2\) regulate conversion cost control and deep conversion cost control, respectively.\footnote{By adopting this approach, we simplify the nonlinear interactions between delivery and cost control. Note that both \(\beta_1\) and \(\beta_2\) depend on \(\lambda^*, \mu^*,\) and \(\gamma^*\). The rewritten formula can be regarded as a linear approximation of the actual optimal bid, which may be suboptimal in some cases.} 
	
	With this formulation in mind, we can design a PID control mechanism as follows: At each pacing cycle, we observe the actual budget spend, cost per conversion, and cost per deep conversion. These values are then compared to their respective targets—namely, the target spend, target cost per conversion \(C\), and target cost per deep conversion \(D\). The differences between observed and target values are used as error signals for the PID controller, which subsequently updates the dual parameters accordingly. 
	
	We summarize the discussion above in the following \autoref{alg:deep_retention_pid}:
	
	\begin{algorithm}
		\caption{PID-Based Bid Algorithm for Deep Retention Problem}
		\label{alg:deep_retention_pid}
		\begin{algorithmic}[1]
		\Require \\
		\(B\): Total budget, \(T\): Total (predicted) number of auction opportunities, \(C\): Target cost per conversion, \(D\): Target cost per deep conversion, 
		PID gains \(K_p, K_i, K_d\), positive initial values \( \alpha_0, \beta_{1,0}, \beta_{2,0}\).
		
		\Ensure 
		\(b_t^*\): Optimal bid per impression.
		
		\textbf{Step 1: Initialize Parameters}
			\State Initialize \(\alpha, \beta_1, \beta_2 \gets  \alpha_0, \beta_{1,0}, \beta_{2,0}\)
			\State Initialize \(I_{\alpha}, I_{\beta_1}, I_{\beta_2} \gets 0, 0, 0\)
			\State Initialize previous errors \(e_{\alpha}^{prev}, e_{\beta_1}^{prev}, e_{\beta_2}^{prev} \gets 0, 0, 0\)

		\textbf{Step 2: Iterative Bid Updates}

			\For{each pacing cycle}
			\State Observe spend \(S_{obs}\), cost per conversion \(C_{obs}\), and cost per deep conversion \(D_{obs}\)
			\State Count the number of auction opportunities \(N\)
			\State Compute errors: \(e_{\alpha} = \frac{B}{T} \cdot N - S_{obs}\), \(e_{\beta_1} = C - C_{obs}\), \(e_{\beta_2} = D - D_{obs}\)
			\State Update integral terms:
			\[
			I_{\alpha} \gets I_{\alpha} + e_{\alpha}, \quad
			I_{\beta_1} \gets I_{\beta_1} + e_{\beta_1}, \quad
			I_{\beta_2} \gets I_{\beta_2} + e_{\beta_2}
			\]
			\State Compute derivative terms: 
			\[
			D_{\alpha} = e_{\alpha} - e_{\alpha}^{prev}, \quad
			D_{\beta_1} = e_{\beta_1} - e_{\beta_1}^{prev}, \quad
			D_{\beta_2} = e_{\beta_2} - e_{\beta_2}^{prev}
			\]
			\State Update parameters using PID:
			\[
			\alpha \gets \alpha \cdot \exp \left( K_p e_{\alpha} + K_i I_{\alpha} + K_d D_{\alpha} \right)
			\]
			\[
			\beta_1 \gets \beta_1 \cdot \exp \left( K_p e_{\beta_1} + K_i I_{\beta_1} + K_d D_{\beta_1} \right)
			\]
			\[
			\beta_2 \gets \beta_2 \cdot \exp \left( K_p e_{\beta_2} + K_i I_{\beta_2} + K_d D_{\beta_2} \right)
			\]
			\For{each auction opportunity \(t\)}
			\State Get coversion rate \(r_t\) and deep conversion rate \(d_t\) from prediction models
			\State Compute optimal bid:
			\[
			b_t^{*} = \alpha \cdot (\beta_1 \cdot C \cdot r_t + \beta_2 \cdot D \cdot d_t \cdot r_t)
			\]
			\State Submit \(b_t^*\)
			\EndFor
			\State Update previous errors:
			 \[
			e_{\alpha}^{prev} \gets e_{\alpha}, \quad e_{\beta_1}^{prev} \gets e_{\beta_1}, \quad e_{\beta_2}^{prev} \gets e_{\beta_2}
			\]
			\EndFor
		\end{algorithmic}
	\end{algorithm}

	\paragraph{Practical Considerations}
	
	We previously mentioned that the error signals for (deep) conversion cost control are derived from the differences between the target (deep) conversion cost and the actual (deep) conversion cost. However, in practice, each pacing cycle operates on a minute-scale interval, meaning that conversion events (e.g., app installs, in-app purchases) are unlikely to occur in most pacing intervals. If we rely solely on observed data as error signal inputs, the bid dynamics may become highly unstable. 
	
	A common workaround is to use predicted (deep) conversion rates as approximations for actual observed events. To obtain these approximations, for each winning auction opportunity \(t\) within a pacing cycle, we collect the tuple \((c_t, r_t, d_t)\), where:
	\begin{itemize}
		\item \(c_t\) is the winning price,
		\item \(r_t\) is the predicted conversion rate,
		\item \(d_t\) is the predicted deep conversion rate.
	\end{itemize}
	Using these values, we approximate the actual cost per conversion and cost per deep conversion as:
	
	\[
	\tilde{C}_{obs} = \frac{\sum_{t} c_t}{\sum_t r_t}, \quad 
	\tilde{D}_{obs} = \frac{\sum_{t} c_t}{\sum_t r_t \cdot d_t}.
	\]
	
	These approximations help smooth bid updates by reducing the impact of sparse conversion events, ensuring a more stable bidding dynamics.

	\section{Remarks}
	
	\subsection{MPC and DOGD Approaches for Deep Retention Problem}
	TBA
	
	\subsection{A Variant Formulation of Deep Retention Problem}
	
	We present an alternative formulation of the deep retention problem.
	
	\paragraph{Problem Formulation}
	We formulate the problem as:
	
	\begin{equation} \label{eq:deep_conversion_variant}
		\begin{aligned}
			\max_{x_{t} \in \{0,1\}} \quad & \sum_{t=1}^{T} x_{t} \cdot c_t  \\
			\text{s.t.} \quad & \sum_{t=1}^{T} x_{t} \cdot c_{t} \leq B, \\
			& \frac{ \sum_{t=1}^{T} x_{t} \cdot c_{t}  }{ \sum_{t=1}^{T} x_{t} \cdot r_t} \leq C, \\
			& \frac{\sum_{t=1}^{T} x_{t} \cdot r_t \cdot C  }{ \sum_{t=1}^{T} x_{t} \cdot r_t \cdot d_t} \leq D.
		\end{aligned}
	\end{equation}
	
	It is easy to see that if the constraints in \eqref{eq:deep_conversion_variant} hold, then the constraints in the original formulation \eqref{eq:deep_conversion} also hold automatically. Comparing \eqref{eq:deep_conversion_variant} to \eqref{eq:deep_conversion}, the only difference lies in the deep conversion cost control constraint, where we replace:
	
	\begin{equation} \label{eq:deep_conversion_deep_constraint}
		\frac{ \sum_{t=1}^{T} x_{t} \cdot c_{t}  }{ \sum_{t=1}^{T} x_{t} \cdot r_t \cdot d_t} \leq D
	\end{equation}
	
	with:
	
	\begin{equation}  \label{eq:deep_conversion_deep_constraint_variant}
		\frac{\sum_{t=1}^{T} x_{t} \cdot r_t \cdot C  }{ \sum_{t=1}^{T} x_{t} \cdot r_t \cdot d_t} \leq D.
	\end{equation}
	
	Rewriting \eqref{eq:deep_conversion_variant}, we obtain:
	
	\[
	\frac{ \sum_{t=1}^{T} x_{t} \cdot r_t \cdot d_t}{\sum_{t=1}^{T} x_{t} \cdot r_t } \geq \frac{C}{D}.
	\]
	
	Here, \(\frac{C}{D}\) represents the desired deep conversion rate given a shallow conversion, while the left-hand side corresponds to the (predicted) deep conversion rate. Therefore, constraint \eqref{eq:deep_conversion_deep_constraint_variant} enforces that the actual deep conversion rate is no less than the target deep conversion rate. This formulation focuses on optimizing deep conversions conditioned on conversion events occurring, whereas \eqref{eq:deep_conversion_deep_constraint} focuses on the end-to-end optimization of deep conversion events.
	
	The optimal bid formula for \eqref{eq:deep_conversion_variant} is given by:
	
	\[
	b_{t}^{*} = \frac{(\mu^* - \gamma^*)\cdot C \cdot r_t - \gamma^* \cdot D \cdot r_t \cdot d_t}{\lambda^* + \mu^* - 1},
	\]
	
	where \(\lambda^*, \mu^*\), and \(\gamma^*\) are the dual parameters corresponding to the budget delivery constraint, conversion constraint, and deep conversion constraint, respectively. The derivation of this formula is left as an exercise for the reader.  
	
	\paragraph{PID Controller Design}
	
	We now design a PID controller to update the bids. First, we rewrite the above formula as:
	
	\begin{equation} \label{eq:deep_conversion_variant_bid_formula}
		b_{t}^{*} = \underbrace{\alpha}_{\text{Delivery Control}} \cdot  
		\underbrace{\beta_1 \cdot C \cdot r_t}_{\text{Conversion Cost Control}} \cdot  
		\underbrace{ \left[1 + \beta_2 \left( \frac{d_t}{ \frac{C}{D} } -1 \right) \right]}_{\text{Deep Conversion Adjustment}}
	\end{equation}
	where:
	
	\[
	\alpha = \frac{1}{\lambda^*}, \quad 
	\beta_1 = \frac{\lambda^* \cdot \mu^*}{\lambda^* + \mu^* -1}, \quad  
	\beta_2 = \frac{\gamma^*}{\mu^*}.
	\]
	
	From \eqref{eq:deep_conversion_variant_bid_formula}, we can see that the bid consists of three components: Delivery control (\(\alpha\)),  Conversion cost control (\(\beta_1\)), and  Deep conversion adjustment (\(\beta_2\)).  
	
	The PID controller is designed for each part as follows:
	
	\begin{itemize}
		\item \textbf{Delivery Control \(\alpha\):}  
		The goal of this component is to ensure that the overall budget delivery remains on target.  
		 At each pacing interval, we collect the actual spend and compare it to the target spend.  
		The difference serves as the error signal for \(\alpha\) modulation in the PID controller.
		
		\item \textbf{Conversion Cost Control \(\beta_1\):}  
		The objective is to maintain the cost per conversion at the target level \(C\).  
		The error signal for \(\beta_1\) is defined as the difference between the target cost \(C\) and the observed cost per conversion within the pacing interval.
		
		\item \textbf{Deep Conversion Adjustment \(\beta_2\):}  
		The goal is to ensure that the observed deep conversion rate meets or exceeds the target deep conversion rate \(\frac{C}{D}\).  
		Intuitively, if the deep conversion rate is already higher than the target, the deep conversion constraint is inactive, and no adjustment is needed. In this case, \(\beta_2\) should approach zero, meaning no additional bid adjustment.  If the deep conversion rate is below the target, we need to increase \(\beta_2\).  
		The term:
		
		\[
		1 + \beta_2 \left( \frac{d_t}{ \frac{C}{D} } -1 \right)
		\]
		
		ensures that when \(d_t > \frac{C}{D}\), the bid price increases, whereas when \(d_t < \frac{C}{D}\), the bid price is suppressed.  This aligns with the intuition that when the deep conversion target is missed, we boost bids for auction requests with a higher deep conversion rate while lowering bids for those with a lower deep conversion rate.  This analysis suggests that the error signal for \(\beta_2\) should be the difference between the target deep conversion rate \(\frac{C}{D}\) and the observed deep conversion rate in the past pacing interval.
	\end{itemize}
	
	\paragraph{PID Control Implementation}
	
	Based on the above discussion, we design the PID control algorithm for solving \eqref{eq:deep_conversion_variant}.  
	At each pacing interval, we retrieve: the actual spend \(S_{obs}\),the observed cost per conversion \(C_{obs}\) and the observed deep conversion rate \(d_{obs}\). The error signals for \(\alpha, \beta_1\), and \(\beta_2\) are computed as:
	
	\[
	e_{\alpha} = S_{target} - S_{obs}, \quad 
	e_{\beta_1} = C - C_{obs}, \quad 
	e_{\beta_2} = \frac{C}{D} - d_{obs}.
	\]
	The PID control equations are then applied to modulate these parameters accordingly. We summarize the discussion here in the following \autoref{alg:deep_conversion_variant_pid}. In practice, the observed conversions and cost per result can be replaced with predicted values from prediction models to mitigate data sparsity issues, as discussed in the previous section.

	\begin{algorithm}
		\caption{PID-Based Bid Algorithm for Varint Deep Conversion Problem}
		\label{alg:deep_conversion_variant_pid}
		\begin{algorithmic}[1]
		\Require 
		\(B\): Total budget,  \(T\): Total (predicted) number of auction opportunities, \(C\): Target cost per conversion, \(D\): Target cost per deep conversion,  
		 PID gains \(K_p, K_i, K_d\), initial values \(\alpha_0, \beta_{1,0}, \beta_{2,0}\).
		
		\Ensure 
		\(b_t^*\): Optimal bid per impression.
		
		\textbf{Step 1: Initialize Parameters}
			\State Initialize \(\alpha, \beta_1, \beta_2\) and integral terms \(I_{\alpha}, I_{\beta_1}, I_{\beta_2}\)
			\State Initialize previous errors \(e_{\alpha}^{prev}, e_{\beta_1}^{prev}, e_{\beta_2}^{prev}\)
		
		\textbf{Step 2: Iterative Bid Updates}
			\For{each pacing interval}
			\State Count the number of auction opportunities \(N\)
			\State Observe actual spend \(S_{obs}\), cost per conversion \(C_{obs}\), and deep conversion rate \(d_{obs}\)
			\State Compute error signals:
			\[
			e_{\alpha} = \frac{B}{T} \cdot N - S_{obs}, \quad e_{\beta_1} = C - C_{obs}, \quad e_{\beta_2} = \frac{C}{D} - d_{obs}
			\]
			\State Update integral terms:
			\[
			I_{\alpha} \gets I_{\alpha} + e_{\alpha}, \quad
			I_{\beta_1} \gets I_{\beta_1} + e_{\beta_1}, \quad
			I_{\beta_2} \gets I_{\beta_2} + e_{\beta_2}
			\]
			\State Compute derivative terms:
			\[
			D_{\alpha} = e_{\alpha} - e_{\alpha}^{prev}, \quad
			D_{\beta_1} = e_{\beta_1} - e_{\beta_1}^{prev}, \quad
			D_{\beta_2} = e_{\beta_2} - e_{\beta_2}^{prev}
			\]
			\State Update PID-controlled parameters:
			\[
			\alpha \gets \alpha \cdot \exp \left( K_p e_{\alpha} + K_i I_{\alpha} + K_d D_{\alpha} \right)
			\]
			\[
			\beta_1 \gets \beta_1 \cdot \exp \left( K_p e_{\beta_1} + K_i I_{\beta_1} + K_d D_{\beta_1} \right)
			\]
			\[
			\beta_2 \gets \beta_2 \cdot \exp \left( K_p e_{\beta_2} + K_i I_{\beta_2} + K_d D_{\beta_2} \right)
			\]
			\For{each auction opportunity \(t\)}
			\State Get conversion rate \(r_t\) and deep conversion rate \(d_t\) from prediction models
			\State Compute optimal bid:
			\[
			b_t^{*} = \alpha \cdot (\beta_1 \cdot C \cdot r_t) \cdot \left[1 + \beta_2 \left( \frac{d_t}{ \frac{C}{D} } -1 \right)\right]
			\]
			\State Submit \(b_t^*\)
			\EndFor
			\State Update previous errors:
			\[
			e_{\alpha}^{prev} \gets e_{\alpha}, \quad e_{\beta_1}^{prev} \gets e_{\beta_1}, \quad e_{\beta_2}^{prev} \gets e_{\beta_2}
			\]
			\EndFor
		\end{algorithmic}
	\end{algorithm}

	\chapter{Reach \& Frequency and Guaranteed Delivery}
	
	\intro{
	In this chapter, we discuss two brand bidding products: Reach \& Frequency and Guaranteed Delivery. For these products, advertisers specify branding requirements related to reach and frequency metrics or demand guarantees on impression delivery. We demonstrate how these requirements can be formulated as optimization problems and how control methods can be leveraged to design corresponding pacing algorithms.
	}

In this chapter, we discuss two popular brand awareness bidding products: Reach \& Frequency and Guaranteed Delivery. These products provide advertisers with greater control over ad delivery, audience reach, and frequency management, helping them achieve their marketing objectives more effectively.

\section{Reach and Frequency Problem}
\paragraph{Problem Formulation}

Reach and Frequency(R\&F) is a product designed to help advertisers optimize brand ad campaigns by focusing on two key metrics:
\begin{itemize}
	\item Reach: The number of unique users exposed to the ad.
	\item Frequency: The number of times each user sees the ad within a specific timeframe.
\end{itemize}
This bidding product ensures that advertisers can plan and predict their campaign's outcomes more accurately, making it ideal for brands that need measurable and scalable visibility while managing overexposure and staying within a fixed budget.

R\&F can be formulated in terms of the target frequency. Suppose there are $M$ targeting users for this ad campaign. For each user $m$, the number of eligible auction opportunities is $T_m$. The desired frequency for each user within the campaign lifetime is between $F_l$ and $F_u$. The reach and frequency can be formulated as the following optimization problem:
\begin{equation} \label{eq:r_n_f}
	\begin{aligned}
		\max_{x_{m,t} \in \{0,1\}} \quad & \sum_{m=1}^M \sum_{t=1}^{T_m} x_{m,t} \\
		\text{s.t.} \quad 
		& \sum_{m=1}^{M} \sum_{t=1}^{T_m} x_{m,t} \cdot c_{m,t} \leq B, \\
		& \sum_{t=1}^{T_m} x_{m,t} \geq F_{l}, \quad \forall m = 1, 2, \dots, M, \\
		& \sum_{t=1}^{T_m} x_{m,t} \leq F_{u}, \quad \forall m = 1, 2, \dots, M,
	\end{aligned}
\end{equation}
where $x_{m,t} \in \{0,1\}$ is an indicator of whether the campaign wins the $t$-th auction for user $m$, $c_{m,t}$ is the cost for the $t$-th auction for user $m$, and $B$ is the total budget of the campaign.

\paragraph{Derivation of the Optimal Bid}
We use the DOGD method to solve this problem. The Lagrangian is defined as:
\begin{equation*}
	\begin{aligned}
		& \mathcal{L}(\lambda, \mu_m, \gamma_m, x_{m,t}) \\
		= & \sum_{m=1}^{M} \sum_{t=1}^{T_m} x_{m,t} + \lambda \cdot \left(B - \sum_{m=1}^{M} \sum_{t=1}^{T_m} x_{m,t} \cdot c_{m,t} \right) + \sum_{m=1}^{M} \mu_m \cdot \left( F_u - \sum_{t=1}^{T_m} x_{m,t} \right) \\
		& \quad + \sum_{m=1}^{M} \gamma_m \cdot \left(\sum_{t=1}^{T_m} x_{m,t} - F_l \right) \\
		= & \sum_{m=1}^{M} \sum_{t=1}^{T_m} \left[1 - \lambda c_{m,t} - \mu_m + \gamma_m \right] \cdot x_{m,t} + \lambda B + \sum_{m=1}^{M} \mu_m F_u - \sum_{m=1}^{M} \gamma_m F_l.
	\end{aligned}
\end{equation*}

To maximize $\mathcal{L}(\lambda, \mu_m, \gamma_m, x_{m,t})$, we set $x_{m,t} = 1$ whenever $1 - \lambda c_{m,t} - \mu_m + \gamma_m > 0$, and $x_{m,t} = 0$ otherwise. This gives the dual problem's objective function:
\begin{equation*}
	\begin{aligned}
		\mathcal{L}^*(\lambda, \mu_m, \gamma_m) = & \max_{x_{m,t} \in \{0,1\}} \mathcal{L}(\lambda, \mu_m, \gamma_m, x_{m,t}) \\
		= & \sum_{m=1}^{M} \sum_{t=1}^{T_m} \left(1 - \lambda c_{m,t} - \mu_m + \gamma_m \right)_{+} + \lambda B + \sum_{m=1}^{M} \mu_m F_u - \sum_{m=1}^{M} \gamma_m F_l,
	\end{aligned}
\end{equation*}
where $(z)_{+} = \max(0, z)$.

Under the Second Price Auction (SPA), the optimal bid for user $m$ is given by:
\[
b_m^* = \frac{1 - \mu_m^* + \gamma_m^*}{\lambda^*},
\]
where $\lambda^*$, $\mu_m^*$, and $\gamma_m^*$ are the solutions to the dual problem.

We can apply stochastic gradient descent (SGD) to iteratively update these parameters:
\begin{equation*}
	\begin{aligned}
		\lambda_t & \gets \lambda_{t-1} - \epsilon_t \cdot \frac{\partial}{\partial \lambda} \mathcal{L}^*, \\
		\mu_{m,t} & \gets \mu_{m,t-1} - \epsilon_t \cdot \frac{\partial}{\partial \mu_m} \mathcal{L}^*, \\
		\gamma_{m,t} & \gets \gamma_{m,t-1} - \epsilon_t \cdot \frac{\partial}{\partial \gamma_m} \mathcal{L}^*.
	\end{aligned}
\end{equation*}

The gradients are computed as:
\begin{equation*}
	\begin{aligned}
		\frac{\partial}{\partial \lambda} \mathcal{L}^* & = \frac{B}{T} - c_{m,t} \cdot x_{m,t}, \\
		\frac{\partial}{\partial \mu_m} \mathcal{L}^* & = \frac{F_u}{T_m} - x_{m,t}, \\
		\frac{\partial}{\partial \gamma_m} \mathcal{L}^* & = x_{m,t} - \frac{F_l}{T_m}.
	\end{aligned}
\end{equation*}
where $T= \sum T_m$ is the total number of auction opportunities for this campaign cross all targeting users.  The dual parameter $\lambda$ is responsible for the overall delivery control. The gradient of $\lambda$ is simply the gap between the expected target cost per auction and the actual spend per auction. The other two sets of dual parameters, $\mu_m$ and $\gamma_m$, are responsible for controlling the impression frequency and cadence for each user $m$. The gradients of these parameters compare the actual impressions with the expected lower and upper bounds per auction. This comparison determines how to tweak the bid to achieve the desired frequency for the ad campaign.

\paragraph{Algorithm Design} To implement this reach and frequency algorithm, we initialize $\lambda_0$ for overall delivery control, and a set of $\mu_{m,0}$ and $\gamma_{m,0}$ for frequency control for each user $m$. At each step, we observe the actual spend and impressions. The parameter $\lambda$ is always updated based on the update rule described above, while $\mu_m$ and $\gamma_m$ are updated only for the user $m$ who triggers the auction request. The idea discussed above can be summarized as the following \autoref{alg:reach_frequency}:

\begin{algorithm}
	\caption{Reach and Frequency Algorithm with Dual Parameters}
	\label{alg:reach_frequency}
	\begin{algorithmic}[1]
		\Require $B$: Total budget, $F_l$: Minimum frequency, $F_u$: Maximum frequency
		\Require $T_m$: Expected number of auction opportunities from user $m$
		\Require $\lambda_0$: Initial dual parameter for delivery control
		\Require $\mu_{m,0}, \gamma_{m,0}$: Initial dual parameters for frequency control for each user $m$
		\Require $\epsilon_t$: Learning rates for $\lambda$, $\mu$, and $\gamma$
		\Ensure  Bids $b_{m,t}$ for each user $m$ 
		\State Compute the total auction opportunites
		\[
			T = \sum T_m
		\]
		\State Initialize $\lambda_0 \gets \lambda_0$, $\mu_{m, 0} \gets \mu_{m,0}$, $\gamma_{m,0} \gets \gamma_{m,0}$ for all $m = 1, \dots, M$
		\For{each auction request at time $t$}
		\State Observe the user $m$ triggering the auction and the auction cost $c_{m,t}$
		\State Observe the impression $x_{m,t}$
		\State Update dual parameters for delivery control
		\[
		\lambda_t \gets \lambda_{t-1} - \epsilon_t \cdot \left(\frac{B}{T} - c_{m,t}\right)
		\]
		\State Update dual parameters of user $m$ for frequency control
		\[
		\mu_{m,t} \gets \mu_{m, t-1} - \epsilon_t \cdot \left(\frac{F_u}{T_m} - x_{m,t}\right)
		\]
		\[
		\gamma_{m, t} \gets \gamma_{m, t-1} - \epsilon_t \cdot \left(x_{m,t} - \frac{F_l}{T_m}\right)
		\]
		
		\State Compute the bid price:
		\[
		b_{m,t} = \frac{1 - \mu_{m, t} + \gamma_{m,t}}{\lambda_t}
		\]
		\EndFor

	\end{algorithmic}
\end{algorithm}

As we mentioned before, in practice, it's more common to implement the algorithm in a mini-batch manner, the bids stay unchanged within a time As we mentioned before, in practice, it is more common to implement the algorithm in a mini-batch manner, where the bids remain unchanged within a time duration $\Delta t$. We first compute the mini-batch gradients for all auction opportunities within $(t, t+\Delta t)$. 

For $\lambda$, the gradient is:
\begin{equation*}
	\begin{aligned}
		\sum_{s \in (t, t+\Delta t)} \frac{\partial }{\partial \lambda} \mathcal{L}^* 
		= & \sum_{s \in (t, t+\Delta t)} \left( \frac{B}{T} - c_{m,s} \cdot x_{m,s} \right) \\
		= & \frac{R(t)}{T} B - S(t),
	\end{aligned}
\end{equation*}
where $R(t)$ is the number of observed auction requests, and $S(t)$ is the actual spend during $\Delta t$.

For $\mu_m$, the gradient is:
\begin{equation*}
	\begin{aligned}
		\sum_{s \in (t, t+\Delta t)} \frac{\partial }{\partial \mu_m} \mathcal{L}^* 
		= & \sum_{s \in (t, t+\Delta t)} \left( \frac{F_u}{T_m} - x_{m,s} \right) \\
		= & \frac{R_m(t)}{T_m} F_u - I_m(t),
	\end{aligned}
\end{equation*}
where $R_m(t)$ is the number of observed auction requests from user $m$, and $I_m(t)$ is the number of impressions shown to this user.

For $\gamma_m$, the gradient is:
\begin{equation*}
	\begin{aligned}
		\sum_{s \in (t, t+\Delta t)} \frac{\partial }{\partial \gamma_m} \mathcal{L}^* 
		= & \sum_{s \in (t, t+\Delta t)} \left( x_{m,s} - \frac{F_l}{T_m} \right) \\
		= & I_m(t) - \frac{R_m(t)}{T_m} F_l.
	\end{aligned}
\end{equation*}

We summarize the mini-batch algorithm as follows:

\begin{algorithm}[H]
	\caption{Mini-Batch Reach and Frequency Algorithm}
	\label{alg:mini_batch_r_n_f}
	\begin{algorithmic}[1]
		\Require $B$: Total budget, $F_l$: Minimum frequency, $F_u$: Maximum frequency
		\Require $\lambda_0$, $\mu_{m,0}$, $\gamma_{m,0}$: Initial dual parameters
		\Require $\Delta t$: Mini-batch interval, $\epsilon_\lambda$, $\epsilon_\mu$, $\epsilon_\gamma$: Learning rates
		\Ensure Optimal bids $b_m^*$ for each user $m$
		\State Initialize $\lambda \gets \lambda_0$, $\mu_m \gets \mu_{m,0}$, $\gamma_m \gets \gamma_{m,0}$ for all $m = 1, \dots, M$
		\For{each mini-batch interval $(t, t+\Delta t)$}
		\State Observe $R(t)$, $S(t)$, $R_m(t)$, and $I_m(t)$ for all users
		\State Compute mini-batch gradients:
		\[
		\lambda \gets \lambda - \epsilon_\lambda \cdot \left(\frac{R(t)}{T} B - S(t)\right)
		\]
		\[
		\mu_m \gets \mu_m - \epsilon_\mu \cdot \left(\frac{R_m(t)}{T_m} F_u - I_m(t)\right), \quad \forall m
		\]
		\[
		\gamma_m \gets \gamma_m - \epsilon_\gamma \cdot \left(I_m(t) - \frac{R_m(t)}{T_m} F_l\right), \quad \forall m
		\]
		\State Compute the bid for each user $m$:
		\[
		b_{m,t } = \frac{1 - \mu_m + \gamma_m}{\lambda}
		\]
		\EndFor
	\end{algorithmic}
\end{algorithm}

\section{Guaranteed Delivery Problem}
	Guaranteed Delivery (GD) ads, also referred to as programmatic guaranteed ads or reserved media buys, are advertising deals in which advertisers purchase a predetermined volume of impressions (or another agreed-upon metric, such as video views) directly from a publisher or via a platform. The price is determined upfront, and a certain portion of the inventory is reserved for these ads. 
	
	While traditional GD ads are sold at a fixed price for reserved inventory without participating in real-time auctions, some ad platforms incorporate an internal auction mechanism to enhance efficiency and reduce costs. This approach allows the system to optimize inventory allocation by identifying lower-cost opportunities rather than always serving GD ads at a fixed high CPM. 
	
	\paragraph{Problem Formulation} 
	We present a simple formulation of the Guaranteed Delivery (GD) problem:
	
	\begin{equation} \label{eq:gd}
		\begin{aligned}
			\min_{x_{t} \in \{0,1\}} \quad & \sum_{t=1}^T x_t \cdot c_t \\
			\text{s.t.} \quad & \sum_{t=1}^{T}  x_t  \geq G.
		\end{aligned}
	\end{equation}
	
	where:
	\begin{itemize}
		\item \(T\) represents the total available inventory for the campaign, i.e., the number of eligible auction opportunities.
		\item \(G\) is the guaranteed delivery goal of impressions specified in the deal. We assume \(G \leq T\) to ensure sufficient inventory for meeting the delivery goal.
		\item \(c_t\) is the highest competing bid at the \(t\)-th auction opportunity. Under a second-price auction, this is also the winning price.
		\item \(x_t\) is a binary decision variable indicating whether the campaign wins the \(t\)-th auction opportunity. Under a second-price auction, \(x_t = \mathds{1}_{ \{ b_t > c_t\}}\), where \(b_t\) is the bid for the impression.
	\end{itemize}
	
	\paragraph{Derivation of the Optimal Bid}  
	We apply the primal-dual method to solve \eqref{eq:gd}. The first step is to formulate the Lagrangian:
	
	\[
	\begin{aligned}
		\mathcal{L}(x_t, \lambda) = &  \sum_{t=1}^T x_t \cdot c_t  + \lambda \cdot \left(G -  \sum_{t=1}^{T}  x_t   \right) \\
		= & \sum_{t=1}^{T} \left[ x_t \cdot \left( c_t - \lambda \right) + \lambda \cdot \frac{G}{T} \right].
	\end{aligned}
	\]
	
	To minimize \(\mathcal{L}(x_t, \lambda)\), we set \(x_t = 0\) whenever \( c_t - \lambda > 0\) and \(x_t = 1\) otherwise. Then, we obtain:
	
	\[
	\mathcal{L}^{*}(\lambda) = \min_{x_{t} \in \{0,1\}} \mathcal{L}(x_t, \lambda) = \sum_{t=1}^{T} \left[ \underbrace{(c_t - \lambda) \cdot \mathds{1}_{\{ c_t < \lambda\}} + \lambda \cdot \frac{G}{T}}_{f_t( \lambda)} \right].
	\]
	
	The corresponding dual problem is:
	
	\[
	\max_{\lambda \geq 0} \mathcal{L}^{*}(\lambda) = \max_{\lambda \geq 0}   \sum_{t=1}^{T} \left[ (c_t - \lambda) \cdot \mathds{1}_{\{ c_t < \lambda\}} + \lambda \cdot \frac{G}{T} \right].
	\]
	
	Assuming the problem is feasible, we determine the optimal dual parameter:
	
	\[
	\lambda^* = \arg\max_{\lambda \geq 0}  \mathcal{L}^{*}(\lambda).
	\]
	
	By the KKT condition, we have:
	
	\[
	\sum_{t=1}^{T}  x_t  = G.
	\]
	
	Thus, the optimal bid per impression is:
	
	\[
	b^* = \lambda^*.
	\]
	
	\paragraph{Algorithm Design}  
	We design a bid update rule using the Dual Online Gradient Descent (DOGD) algorithm. Other approaches, such as PID control and Model Predictive Control (MPC), are left for interested readers.
	
	Applying stochastic gradient ascent (since the dual problem is a maximization problem), the update rule for \(\lambda\) is:
	
	\[
	\lambda \gets \lambda + \epsilon \cdot \frac{\partial }{\partial \lambda} f_t(\lambda) = \lambda + \epsilon \cdot \left( \frac{G}{T} - \mathds{1}_{\{ c_t < \lambda\}} \right).
	\]
	
	Since we bid using \(\lambda\), the term \(\mathds{1}_{\{ c_t < \lambda\}}\) corresponds to \(x_t\), which indicates whether the \(t\)-th auction opportunity is won. The term \(\frac{G}{T}\) represents the expected number of impressions the campaign should win per auction opportunity.
	
	For a mini-batch update within each pacing interval, the rule is:
	
	\[
	\lambda \gets \lambda + \epsilon \cdot \left( \frac{G}{T} \cdot N - W \right),
	\]
	
	where:
	\begin{itemize}
		\item \(W\) is the number of impressions won by the campaign within the pacing interval.
		\item \(N\) is the number of auction opportunities.
		\item \(\frac{G}{T} \cdot N\) represents the target number of impressions.
	\end{itemize}
	
	This update rule ensures that if the actual impressions are below the target, we increase the bid, and if the impressions exceed the target, we decrease the bid. This guarantees smooth budget pacing and delivery alignment.

	We summarize the idea discussed above in the following \autoref{alg:dogd_gd}:
	
	\begin{algorithm}[H]
		\caption{DOGD-Based Bid Algorithm for Guaranteed Delivery}
		\label{alg:dogd_gd}
		\begin{algorithmic}[1]
		\Require 
		\(G\): Guaranteed impressions goal, \(T\): Total available inventory, 
		\(\epsilon\): Learning rate, 
		Initial bid parameter \(\lambda_0\).
		
		\Ensure 
		\(b^*\): Optimal bid per impression.
		
		\textbf{Step 1: Initialize Parameters}
			\State Initialize \(\lambda \gets \lambda_0\)
		
		\textbf{Step 2: Iterative Bid Updates}
			\For{each pacing interval}
			\State Observe actual impressions won \(W\) and number of auction opportunities \(N\)
			\State Compute target impressions per interval:
			\[
			G_{target} = \frac{G}{T} \cdot N
			\]
			\State Compute update step:
			\[
			\lambda \gets \lambda + \epsilon \cdot (G_{target} - W)
			\]
			\State Update bid:
			\[
			b^* = \lambda
			\]
			\State Submit \(b^*\) for the auction opportunities in the next pacing interval 
			\EndFor
		\end{algorithmic}
	\end{algorithm}
	
	For more details of guaranteed delivery problems, one may refer to the related papers, such as \cite{fang2019large}, \cite{bharadwaj2012shale}, \cite{feldman2009online}, \cite{chen2014dynamic}. 
	
\section{Remarks}
\begin{itemize}
	\item Marketplace level formulation of R\&F problem
	\item Sliding window optimization for R\&F.
	\item GD cost cap problem: \cite{zhang2022control}
	\item GD for unique reach
	\item Implementation tips(e.g., rescaling)
	\item prediction of expected number of auction opportunity per use
	\item Compare different frequency control setups:  lower bound + upper bound vs single median target 
	
\end{itemize}
	
	\subsection{Fixed Frequency Problem}
	Fixed frequency target problem: Sometimes advertisers want to target a specific frequency (e.g., $F$). In this case, we simply set $F_l = F_u = F$, and \autoref{eq:r_n_f} simplifies to:
	\begin{equation*}
		\begin{aligned}
			\max_{x_{m,t} \in \{0,1\}} \quad & \sum_{m=1}^M \sum_{t=1}^{T_m} x_{m,t} \\
			\text{s.t.} \quad 
			& \sum_{m=1}^{M} \sum_{t=1}^{T_m} x_{m,t} \cdot c_{m,t} \leq B, \\
			& \sum_{t=1}^{T_m} x_{m,t} = F, \quad \forall m = 1, 2, \dots, M.
		\end{aligned}
	\end{equation*}
	
	The corresponding bid formula is:
	\[
	b_{m, t} = \frac{1 - \mu_{m, t}}{\lambda_t}.
	\]
	
	The update rules for the dual parameters are:
	\[
	\lambda_t \gets \lambda_{t-1} - \epsilon_t \cdot \left(\frac{B}{T} - c_{m,t}\right),
	\]
	\[
	\mu_{m,t} \gets \mu_{m, t-1} - \epsilon_t \cdot \left(\frac{F}{T_m} - x_{m,t}\right).
	\]
	
	The mini-batch algorithm for this specific target frequency can be derived in a similar way, where gradients are computed over all auction opportunities within each interval $\Delta t$, and the dual parameters are updated iteratively.

	\chapter{Enhanced Objective Bidding}
	
	\intro{
		In this chapter, we discuss how to design strategy for enhanced objective bidding. Major platforms offer "Enhanced" bidding (e.g., Enhanced CPC) where the system modifies a manual bid based on the predicted conversion probability. This chapter explains the hybrid control logic required to respect a manual bid cap while still optimizing for conversion value, bridging the gap between manual advertiser control and algorithmic optimization. We also show how the same framework can power bidding strategies for a specific type of consideration campaigns, helping brand advertisers drive stronger downstream actions and deeper-funnel results.
	}
	
	\section{Enhanced CPC Bidding}
	Enhanced CPC (eCPC) is a hybrid bidding approach that sits between manual CPC and fully automated “Smart Bidding.” Advertisers provide a manual Max CPC bid as a baseline, and the system adjusts that bid up or down at auction time based on the predicted likelihood that a click will lead to a valuable downstream action (e.g., a post-click conversion). These per-auction adjustments use contextual signals such as device, location, time, and audience. The goal is to retain the simplicity and guardrails of manual bidding while reallocating spend toward higher-quality clicks, improving conversion volume and/or efficiency without requiring a fixed CPA/ROAS target. For advertisers who do not have a clear target CPA but still want to optimize post-click conversions, eCPC is a practical alternative.
	
	Let $b_{\text{click}}$ be the manual CPC bid, i.e., the value per \emph{click} that the advertiser is willing to pay.
	Define the (average) post-click conversion rate as
	\[
	pCVR \;=\; \Pr(\mathrm{conv}\mid \mathrm{click}),
	\]
	and for a specific request (auction context) $i$, define
	\[
	pCTR_i \;=\; \Pr(\mathrm{click}\mid \mathrm{imp}, i),
	\qquad
	pCVR_i \;=\; \Pr(\mathrm{conv}\mid \mathrm{click}, i).
	\]
	Let $V$ denote the value (in dollars) of a conversion.
	
	Under standard CPC bidding, the implied bid per impression is
	\[
	b_{\text{click}} \cdot pCTR_i.
	\]
	Under enhanced CPC (eCPC), the bid per impression is adjusted to
	\[
	b_{\text{click}} \cdot pCTR_i \cdot \frac{pCVR_i}{pCVR}.
	\]
	Relative to regular CPC bidding, eCPC applies a multiplicative adjustment factor $\tfrac{pCVR_i}{pCVR}$ that
	rescales the bid on a per-request basis. Intuitively, for two requests with the same predicted click probability,
	we would like to bid more aggressively on the one that is more likely to convert after a click (larger $pCVR_i$),
	and bid less on the one with lower post-click conversion likelihood. 
	
	Let us use a simple toy example to illustrate how eCPC works. Consider a campaign with a manual Max CPC bid of
	\$2. Suppose there are three auction requests (contexts) that would each result in a click if the campaign wins
	the auction and receives an impression. Assume the predicted post-click conversion rates are $0.015$, $0.020$,
	and $0.025$ for requests 1, 2, and 3, respectively.
	
	Under manual CPC bidding, the campaign bids \$2 per click for all three requests. Under eCPC, we first compute
	the average predicted post-click conversion rate:
	\[
	\frac{0.015 + 0.020 + 0.025}{3} \;=\; 0.020.
	\]
	The eCPC-adjusted bids per click are then
	\[
	\text{request 1: } \$2 \times \frac{0.015}{0.020} = \$1.5, \quad
	\text{request 2: } \$2 \times \frac{0.020}{0.020} = \$2.0, \quad
	\text{request 3: } \$2 \times \frac{0.025}{0.020} = \$2.5.
	\]

	\paragraph{How to derive the eCPC bid formula}
	We now show how to derive the eCPC bid adjustment formula. A conversion on impression $i$ happens via a click and then a post-click conversion, so
	\[
	\Pr(\mathrm{conv}\mid \mathrm{imp}, i)
	\;=\;
	\Pr(\mathrm{click}\mid \mathrm{imp}, i)\,\Pr(\mathrm{conv}\mid \mathrm{click}, i)
	\;=\;
	pCTR_i \cdot pCVR_i.
	\]
	Hence the expected conversion value per impression is
	\[
	b^{(\mathrm{imp})}_i
	\;=\;
	V \cdot \Pr(\mathrm{conv}\mid \mathrm{imp}, i)
	\;=\;
	V \cdot pCTR_i \cdot pCVR_i.
	\]
	
	If the manual CPC bid $b_{\text{click}}$ is calibrated to the \emph{average} post-click conversion rate,
	then the expected conversion value \emph{per click} is
	\[
	\mathbb{E}[\text{value}\mid \mathrm{click}]
	\;=\;
	V \cdot \Pr(\mathrm{conv}\mid \mathrm{click})
	\;=\;
	V \cdot pCVR,
	\]
	so we can write the manual CPC anchor as
	\[
	b_{\text{click}} \;=\; V \cdot pCVR
	\qquad\Longrightarrow\qquad
	V \;=\; \frac{b_{\text{click}}}{pCVR}.
	\]
	
	Plugging $V = b_{\text{click}}/pCVR$ into $b^{(\mathrm{imp})}_i$ gives
	\[
	b^{(\mathrm{imp})}_i
	\;=\;
	\frac{b_{\text{click}}}{pCVR}\cdot pCTR_i \cdot pCVR_i
	\;=\;
	b_{\text{click}}\cdot pCTR_i \cdot \frac{pCVR_i}{pCVR}.
	\]
	
	\paragraph{Remark (CPC vs. CPM view).}
	If one instead writes the \emph{CPC} bid adjustment (what eCPC conceptually scales) as
	\[
	b^{(\mathrm{CPC})}_i \;=\; b_{\text{click}}\cdot \frac{pCVR_i}{pCVR},
	\]
	then the corresponding impression-level value is
	\[
	pCTR_i \cdot b^{(\mathrm{CPC})}_i
	\;=\;
	b_{\text{click}}\cdot pCTR_i \cdot \frac{pCVR_i}{pCVR}.
	\]

	\paragraph{Intuition (one line).}
	The adjustment $b_i \propto r_i$ equalizes \emph{marginal conversions per marginal dollar} across contexts:
	increasing $b_i$ slightly increases wins at rate $f_i(b_i)$; this yields marginal conversions $\propto r_i f_i(b_i)$
	and marginal spend $\propto b_i f_i(b_i)$, so the optimality condition is $r_i / b_i = \lambda$ for all $i$,
	i.e., $b_i \propto r_i$.

	\paragraph{Practical Considerations}
	We list several practical considerations for implementing Enhanced CPC in a real-world production system:
	
	\begin{itemize}
		\item \textbf{Guardrails for the objective-adjustment factor.}
		Enhanced CPC introduces a multiplicative adjustment factor on top of the base CPC bid. In production, it is
		standard to clip this factor within a predefined range to avoid extreme bids caused by prediction noise,
		calibration drift, or distribution shift. Concretely, the adjustment factor used online is
		\[
		\max\!\left( L,\; \min\!\left( U,\; \frac{pCVR_i}{pCVR} \right)\right),
		\]
		where $L$ and $U$ are lower and upper bounds satisfying $0 < L < 1 < U$. In practice, the best choice of
		$(L,U)$ depends on model quality and traffic dynamics, and is typically tuned during online A/B testing to achieve the best performance..

		\item \textbf{Service placement and feature availability.}
		In many ad stacks, the pacing component operates at the campaign level and is intentionally agnostic to request-level features. As a result, the Enhanced CPC adjustment is usually applied in a downstream  service, where per-request signals (e.g., $pCVR_i$) are available.
		
		\item \textbf{Model training and the selection bias.}
		Care is needed to avoid feedback loops that cause selection bias when estimating the reference $pCVR$. For example, if we only log $pCVR$ for impressions that a campaign actually wins, then enabling ECPC changes what gets logged: the campaign is more likely to win when $pCVR$ (and hence the bid) is high. Averaging these selectively high $pCVR$ values will inflate the reference $pCVR$. Since the ECPC adjustment is inversely proportional to the reference $pCVR$, this inflated reference pushes bids downward, causing the campaign to lose even more low-$pCVR$ impressions. This is a typical selection bias problem, the reference $pCVR$ then drifts upward further, creating a self-reinforcing loop.
		
		To prevent this, the reference $pCVR$ should ideally be computed from \emph{all matched impressions}, not only the won ones. In practice, this requires logging $pCVR$ for every match (rather than only a small subset such as the top-$K$ candidates per request). While one could consider applying a correction factor to partially debias win-only logging, designing such a correction that reliably eliminates feedback effects is non-trivial in production.

		\item \textbf{CPC control and conversion optimization tradeoff.}
		With eCPC, one can show that if the conversion model is well-calibirate, the realized CPA and the total number of conversions are not simultaneously worse than under the original CPC bidding rule. However, improving post-click conversions does not come for free. Because eCPC dynamically adjusts bids at the request level based on predicted conversion likelihood, the system may pay more for some higher-quality clicks. As a result, it is possible to observe a higher realized CPA under eCPC than under a fixed manual CPC bid, even when the total number of post-click conversions increases. This tradeoff should be kept in mind when designing and deploying enhanced-objective bidding methods such as eCPC in production.
		
	\end{itemize}

	\paragraph{Simulation}
	We run a simple simulation to compare the performance between manual CPC and eCPC.

	\section{Optimizing Consideration in Brand Advertising}

	The framework described above generalizes to other business settings where we need tight control over a specific upper-funnel objective while simultaneously optimizing a deeper-funnel outcome.
	
	As a concrete example, consider brand advertising. Traditionally, brand campaigns optimize for impressions with CPM-based pacing and delivery control. In recent years, however, advertisers have increasingly demanded optimization toward consideration signals. For example, on e-commerce platforms, even for brand ads, advertisers might care about user behaviors such as saves, comments, image/video interactions, and add-to-cart events. While these actions may not immediately translate into purchases, they are strong leading indicators of future purchase intent and therefore provide a useful intermediate objective for optimization.
	
	\paragraph{Algorithm Design}
	We use a CPM brand campaign as an example to show how to design the algorithm. The principle is similar to enhanced CPC bidding. Suppose the (predicted) average post-impression consideration conversion rate for this campaign is $pConAvg$, for an incoming request $i$, the per-impression bid is $b_i$ and (predicted) post-impression consideration conversion rate is $pCon_i$, the adjusted per-impression bid is 
	\[
		b_i \cdot \frac{pCon_i}{pConAvg}
	\]
	Of course, if we clip the adjustment factor with lower bound $L$ and upper bound $U$ , the clipped adjusted bid is then
	\[
		b_i \cdot \max \left( L, \min \left(U, \frac{pCon_i}{pConAvg} \right) \right)
	\]

	\paragraph{Practical Considerations}
	Apart from the considerations we listed for enhanced CPC in the previous section, some specifc points for consideration should be taken care of  as well
	 \begin{itemize}
	 	\item \textbf{Multi Consideration Signal Modeling.}
	 	As we have seen, there are multiple consideration signals that advertisers may care about (e.g., in the e-commerce example above: saves, comments, image/video interactions, and add-to-cart events). There are two common approaches to training a consideration (engagement) prediction model in this case.
	 	
	 	If we treat all consideration signals as equally important, the solution is straightforward: label an example as positive whenever any consideration signal occurs, regardless of type. This reduces to standard binary classification.
	 	
	 	In practice, however, different signals often carry different value. Continuing with the e-commerce example,
	 	advertisers typically place more weight on add-to-cart events than on image/video interactions, since add-to-cart
	 	is a more direct behavioral indicator of purchase intent. In this setting, treating all signals as equally
	 	important is inappropriate.
	 	
	 	One approach is to keep binary labels for each event type and train a multi-task model with a weighted
	 	Multi-Task Binary Cross-Entropy (BCE) loss. Each signal type (save, comment, image/video interaction,
	 	add-to-cart, etc.) has its own binary label and prediction head. The overall loss is a weighted sum of the
	 	per-task cross-entropy losses:
	 	\[
	 	\sum_{k} \alpha_k \cdot \Big( -y_k \log p_k - (1-y_k)\log(1-p_k) \Big),
	 	\]
	 	where $y_k \in \{0,1\}$ and $p_k \in (0,1)$ are the label and predicted probability for signal $k$, and
	 	$\alpha_k \geq 0$ is its relative importance weight such that $\sum_k \alpha_k = 1$. This formulation preserves per-action calibration. At inference,
	 	we can aggregate the predicted probabilities into a single consideration score via
	 	\[
	 	\sum_{k} \alpha_k \cdot p_k.
	 	\]
	 	
	 	A second approach is to collapse multiple event types into a single normalized composite label and train a
	 	single-head model using a soft-label BCE loss. Specifically, define the soft label
	 	\[
	 	\tilde{y} \;=\; \sum_k \alpha_k \cdot y_k,
	 	\]
	 	and optimize
	 	\[
	 	-\tilde{y}\log p \;-\; (1-\tilde{y})\log(1-p).
	 	\]
	 	At inference time, the model outputs an expected consideration engagement probability that is implicitly
	 	weighted by the relative importance of the underlying signals.
	 	
	 	For both approaches, the hyper-parameter weights $\{\alpha_k\}$ can be set based on business priorities, or
	 	tuned online to optimize the relevant production metrics.

	 	The detailed modeling and system choices are beyond the scope of this book; we provide references for readers
	 	who are interested. Multi-task BCE training and weighted labeling techniques are discussed in recommendation
	 	systems literature such as \cite{hu2008collaborative,ma2018modeling}. The use of soft labels and related
	 	training objectives is closely related to ideas in knowledge distillation \cite{hinton2015distilling}.

	 \end{itemize}
	
	\section{Summary}
	
	This chapter introduced \emph{enhanced objective bidding}, a class of hybrid strategies that bridge manual advertiser control and algorithmic optimization. The core principle is to keep an advertiser-provided \emph{anchor bid} (e.g., a manual Max CPC or a base CPM bid) while applying a \emph{per-request multiplicative adjustment} based on predicted downstream quality. This preserves upper-funnel guardrails (such as CPC/CPM control and pacing behavior) while reallocating delivery toward traffic that is more likely to produce the desired lower-funnel outcome.
	
	We used Enhanced CPC (eCPC) as the canonical example. Starting from a manual CPC bid $b_{\text{click}}$, eCPC rescales the effective bid using predicted post-click conversion likelihood. With campaign-level average post-click conversion rate $pCVR$ and request-level predictions $(pCTR_i, pCVR_i)$, manual CPC implies an impression-level value of $b_{\text{click}}\cdot pCTR_i$, while eCPC adjusts it to
	\[
	b_{\text{click}} \cdot pCTR_i \cdot \frac{pCVR_i}{pCVR}.
	\]
	We provided a toy example and a derivation that interprets the manual CPC bid as an average conversion value per click and converts it to an expected value per impression. We also highlighted the key intuition: the adjustment favors contexts with higher post-click conversion propensity and de-emphasizes those with lower propensity.
	
	We then discussed practical production considerations: clipping the adjustment factor within $[L,U]$ to control risk, applying the adjustment in a service layer where request-level features are available, and avoiding selection bias when estimating the reference $pCVR$ (e.g., win-only logging can create feedback loops). Finally, we emphasized a fundamental tradeoff: eCPC can increase conversion volume by paying more for higher-quality clicks, which may lead to higher realized CPA even when conversions improve.
	
	The same framework generalizes beyond performance campaigns. Using brand advertising as an example, we showed how a CPM-style campaign can optimize toward consideration signals by scaling the per-impression bid with a predicted consideration rate ratio, optionally with clipping. When multiple consideration signals are present, we outlined two modeling approaches: multi-task learning with a weighted BCE objective, and soft-label training using a single composite target. Overall, enhanced objective bidding provides a simple and extensible design pattern for achieving stronger downstream outcomes without fully relinquishing control of the campaign's primary delivery constraints.

	\chapter{Even Budget Pacing}
	
	\intro{
		In this chapter, we demonstrate how to approach the even budget pacing problem. This problem arises when advertisers require their campaign budget to be distributed evenly across its lifetime. We show how to use the MPC controller to address this challenge. 		
	}
	
	\section{Budget Pacing with Intra-Period Limits}
	Some advertisers prefer to pace their budget evenly over the campaign's duration rather than spending the majority within a short period. For example, in a daily pacing campaign, advertisers may set constraints to ensure that no more than 50\% of the total daily budget is spent within a single hour. Similarly, in a lifetime pacing campaign, they may aim to prevent a significant portion of the budget from being spent within just one or two days.  
	
	\paragraph{Problem Formulation}  
	One approach to addressing this requirement is to impose a spending limit for each intra-period. Suppose we divide the campaign's lifetime \(I\) into \(N\) consecutive and mutually exclusive sub-intervals, denoted as \(\{I_i\}_{i=1}^{N}\). The revised pacing problem can then be formulated as follows:
	
	\begin{equation}\label{even_pacing}
		\begin{aligned}
			\max_{x_t \in \{0,1\}} \quad & \sum_{t=1}^T x_t \cdot r_t \\
			\text{s.t.} \quad &  \sum_{t=1}^{T} x_t \cdot c_t \leq B, \\
			& \sum_{t \in I_i} x_t \cdot c_t \leq \sigma B, \quad i = 1, \dots, N.
		\end{aligned}
	\end{equation}
	where:
	\begin{itemize}
		\item \(B\) is the total budget of the campaign.
		\item \(T\) is the number of auction opportunities for this campaign.
		\item \(r_t\) is the estimated conversion rate for the \(t\)-th auction opportunity.
		\item \(c_t\) is the expected payment (cost) per impression for the \(t\)-th auction.
		\item \(x_t\) is a binary decision variable indicating whether the campaign wins the \(t\)-th auction opportunity.
		\item \(\sigma\) is the intra-period cap, ensuring that the spending within each period \(I_i\) does not exceed \(\sigma B\).
	\end{itemize}
	
	Under a standard second-price auction, \(x_t\) can be expressed as \(\mathds{1}_{\{b_t > c_t\}}\), where \(b_t\) is the bid amount per impression. The time intervals \(\{I_i\}\) are often chosen to be of equal duration, satisfying \(I_i \cap I_j = \emptyset\) for \(i \neq j\) and \(\bigcup_{i=1}^N I_i = I\).

	\paragraph{Derivation of the Optimal Bid} The Lagrangian is:
	\[
		\begin{aligned}
			\mathcal{L}(x_t, \lambda, \lambda_i) & = \sum_{t=1}^{T} x_t \cdot r_t - \lambda \cdot \left(\sum_{t=1}^{T}x_t \cdot c_t -B \right) - \sum_{i=1}^{N}\left[ \lambda_i \cdot \left(\sum_{t \in I_i}  x_t \cdot c_t - \sigma \cdot B  \right)\right] \\
			& = \sum_{t=1}^{T}
			 \left[ 
			 	x_t \cdot \left( r_t - \lambda \cdot c_t -  \left(\sum_{i=1}^N \lambda_i \cdot \mathds{1}_{\{t \in I_i \}} \right) \cdot c_t  \right) + \lambda \cdot \frac{B}{T} + \left(\sum_{i=1}^N \lambda_i\right) \cdot \frac{\sigma B}{T}
			\right].
		\end{aligned}
	\]
	To maximize \(\mathcal{L}(x_t, \lambda, \lambda_i)\), we set \(x_t = 1\) whenever:
	\[
		r_t - \lambda \cdot c_t -  \left(\sum_{i=1}^N \lambda_i \cdot \mathds{1}_{\{t \in I_i \}} \right) \cdot c_t > 0,
	\]
	and \(x_t=0\) otherwise. The dual \(\mathcal{L}^{*}(\lambda, \lambda_i)\) then becomes:
	\[
		\mathcal{L}^{*}(\lambda, \lambda_i) =  \sum_{t=1}^{T}
		\left[ \underbrace{
		 \left( r_t - \lambda \cdot c_t -  \left(\sum_{i=1}^N \lambda_i \cdot \mathds{1}_{\{t \in I_i \}} \right) \cdot c_t  \right)_{+} + \lambda \cdot \frac{B}{T} + \left(\sum_{i=1}^N \lambda_i\right) \cdot \frac{\sigma B}{T}}_{f_t(\lambda, \lambda_i)}
		\right].
	\]
	Suppose we find a feasible solution to the dual:
	\[
		\lambda, \lambda_i = \argmin_{\lambda \geq 0, \lambda_i \geq 0} \mathcal{L}^{*}(\lambda, \lambda_i). 
	\]
	The optimal bid per impression is then given by:
	\begin{equation} \label{eq:even_pacing_bid_formula}
		b_t^{*} = \frac{r_t}{\lambda^* - \sum_{i=1}^{N} \lambda_i^* \cdot \mathds{1}_{\{ t \in I_i\}}}.
	\end{equation}

	\paragraph{Interpretation of the Bidding Formula}  
	Since \(I_i \cap I_j = \emptyset\) whenever \(i \neq j\), each auction opportunity \(t\) belongs to exactly one interval, say \(I_i\). The bid formula \eqref{eq:even_pacing_bid_formula} is thus equivalent to:
	
	\begin{equation} \label{eq:even_pacing_bid_formula_i}
		b_t^{*} = \frac{r_t}{\lambda^* - \lambda_i^*}.
	\end{equation}
	
	If the intra-period constraint for \(I_i\) is not active, meaning that the budget allocated based on the overall traffic pattern is less than \(\sigma \cdot \frac{B}{T}\), then the corresponding dual parameter satisfies \(\lambda_i^{*} = 0\). In this case, equation \eqref{eq:even_pacing_bid_formula_i} simplifies to:
	
	\[
	b_t^{*} = \frac{r_t}{\lambda^*},
	\]
	which corresponds to the standard bid formula without an intra-period constraint.  
	
	On the other hand, if the intra-period constraint is active—meaning that \(\sigma \cdot \frac{B}{T}\) imposes a stricter limit on \(I_i\)—then \(\lambda_i^* > 0\), and the optimal bid \(b_t^*\) is lower than the bid level in the absence of the intra-period constraint, i.e., \(\frac{r_t}{\lambda^*}\).  
	
	This pacing principle aligns with our intuition: when the new constraint for \(I_i\) is not active, the campaign follows the standard pacing strategy. However, if the constraint is active, the bid must be reduced to ensure compliance with the stricter intra-period spending limit.

	\paragraph{Algorithm Design}  
	We can design the pacing algorithm using any of the approaches discussed in \autoref{part:pacing_algorithms}. For example, if we adopt the DOGD framework, we can compute the gradient to derive the stochastic gradient descent update rule:
	
	\[
	\lambda \gets \lambda - \epsilon \cdot \frac{\partial}{\partial \lambda} f_t(\lambda, \lambda_i) = \lambda - \epsilon \cdot \left(\frac{B}{T} - c_t \cdot x_t \right),
	\]
	
	\[
	\lambda_i \gets \lambda_i - \epsilon \cdot \frac{\partial}{\partial \lambda_i} f_t(\lambda, \lambda_i) = \lambda_i - \epsilon \cdot \left( \sigma \cdot \frac{B}{T} - \mathds{1}_{t\in I_i} \cdot c_t \cdot x_t\right).
	\]
	
	Readers should already be familiar with the update rule for \(\lambda\). For \(\lambda_i\), the update occurs only within \(I_i\), where the gradient is the difference between the target spend rate \(\sigma \cdot \frac{B}{T}\) and the actual spend per impression \( \mathds{1}_{t\in I_i} \cdot c_t \cdot x_t\) in the \(i\)-th intra-period. This update rule provides the foundation for designing the corresponding DOGD-based pacing algorithm.
	
	A more suitable framework for this problem is Model Predictive Control (MPC). Since each intra-period has its own constraint, MPC naturally fits as an adaptive receding horizon control approach. At the beginning of each pacing interval within \(I_i\), say at time \(\tau\), the receding horizon control problem can be formulated as:
	
	\begin{equation}\label{eq:even_pacing_mpc}
	\begin{aligned}
		\max_{x_t \in \{0,1\}} \quad & \sum_{t \in I_{\tau}} x_t \cdot r_t \\
		\text{s.t.} \quad &  \sum_{t \in I_{\tau} } x_t \cdot c_t \leq \min \{B_{\tau}, B_{\tau, i}\}.
	\end{aligned}
	\end{equation}
	where:
	\begin{itemize}
		\item \(I_{\tau}\) represents the remaining time in \(I_i\) starting from \(\tau\).
		\item \(B_{\tau}\) is the budget allocated for \(I_{\tau}\), derived from the remaining overall budget (normalized\footnote{Here, we normalize the budget based on the duration of pacing intervals under the assumption that traffic is uniformly distributed. However, a more accurate approach would be to normalize the budget based on the actual traffic pattern.} for \(I_{\tau}\)).
		\item \(B_{\tau, i}\) is the remaining budget from the intra-period constraint, computed as \(\sigma \cdot \frac{B}{T}\) minus the amount spent up to \(\tau\) since the beginning of \(I_i\).
	\end{itemize}
	
	The effective budget for \(I_{\tau}\) is then adaptively set as the minimum of \(B_{\tau}\) and \(B_{\tau, i}\), ensuring compliance with both the overall and intra-period constraints for the campaign. This receding horizon control problem effectively reduces to a standard max delivery problem, allowing us to apply previously discussed techniques to determine the optimal bid for the next update. 
	
	We summarize the MPC approach in \autoref{alg:even_pacing_mpc}. 
	
	\begin{algorithm}
		\caption{MPC-Based Bidding Algorithm for Budget Pacing with Intra-Period Limits}
		\label{alg:even_pacing_mpc}
		\begin{algorithmic}[1]
			\Require \\
			 Total budget \(B\), total time horizon \(T\) \\
			 Intra-periods \(\{I_i\}_{i=1}^{N}\),  intra-period cap \(\sigma\), 
			\Ensure Computes optimal bid per impression \(b_t^*\)
			
			\For{each pacing interval starting at \(\tau\)}
			\State Identify the current intra-period \(I_i\) such that \(\tau \in I_i\)
			\State Compute remaining time in \(I_i\) from \(\tau\) to the end of \(I_i\), denoted as \(I_{\tau}\)
			\State Observe budget spent from the beginning of \(I_i\) up to \(\tau\), denoted as \(B_{\text{spent}, i}\)
			\State Compute the remaining budget for the intra-period:
			\[
			B_{\tau, i} = \sigma B - B_{\text{spent}, i}
			\]
			\State Observe total remaining budget \(B_{\text{remain}}\) and normalize it for \(I_{\tau}\):
			\[
			B_{\tau} = \frac{|I_{\tau}|}{|I_{\text{remain}}|} B_{\text{remain}}
			\]
			where \(|\cdot|\) denotes the duration of the interval, \(I_{\text{remain}}\) is the remaining lifetime of this campaign
			\State Compute the effective budget for \(I_{\tau}\) as 
			\[
				B_{\text{effective}} = \min \{B_{\tau, i}, B_{\tau}  \}
			\]
			
			\State Solve the max-delivery problem \eqref{eq:even_pacing_mpc} with effective budget \(B_{\text{effective}}  \) to get the optimal bid per conversion for \(I_{\tau}\): 	\(b_{\tau}^{*}\)
		
			\For{each auction request \(t\) in pacing interval \(I_{\tau}\)}
			\State Get predicted conversion rate \(r_t\) from the prediction model
			\State Compute bid per impression:
			\[
			b_t^{*} = b_{\tau}^{*} \cdot r_t
			\]
			\State Submit \(b_t^{*}\) for auction request \(t\)
			\EndFor
			\EndFor
		\end{algorithmic}
	\end{algorithm}

	\section{Remarks}

	\subsection{Throttling-based Approach}
	In practice, some traffic spikes may not be captured by prediction models (e.g., unusual traffic surges due to sudden events), and such fluctuations can occur within just a few seconds. Within such a short time frame, there is often no opportunity to adjust bids in response to these traffic spikes—especially considering that the pacing interval is typically around 30 seconds to a few minutes. Consequently, the budget may be consumed much faster than expected or even be depleted within seconds.  
	
	In such cases, a bid-based pacing algorithm alone may not be sufficient to address the issue. One potential workaround is to employ the throttling techniques discussed earlier to temporarily prevent the campaign from participating in auctions.  
	
	The throttling-based method can serve as a safeguard for even pacing. For instance, consider the pacing formulation in \eqref{even_pacing}. Every \(X\) seconds, we compute the total budget spent, \(B_s\), since the beginning of \(I_i\) up to the current time. If \(B_s\) exceeds a predefined threshold (e.g., \(0.8 \cdot \sigma \cdot B\)), the throttling mechanism is triggered. In this case, the throttling probability increases as \(B_s\) approaches the budget limit \(\sigma \cdot B\), thereby reducing participation in auctions to prevent overspending.
	
	Such a throttling mechanism requires only the accumulated budget data of the campaign. For oCPM campaigns, these signals can be collected with negligible delays, making it highly suitable for second-level granularity control.

	\subsection{Comparison of Different Budget Allocation Patterns}
\begin{figure}[H]
		\centering
		\includegraphics[width=0.99\textwidth]{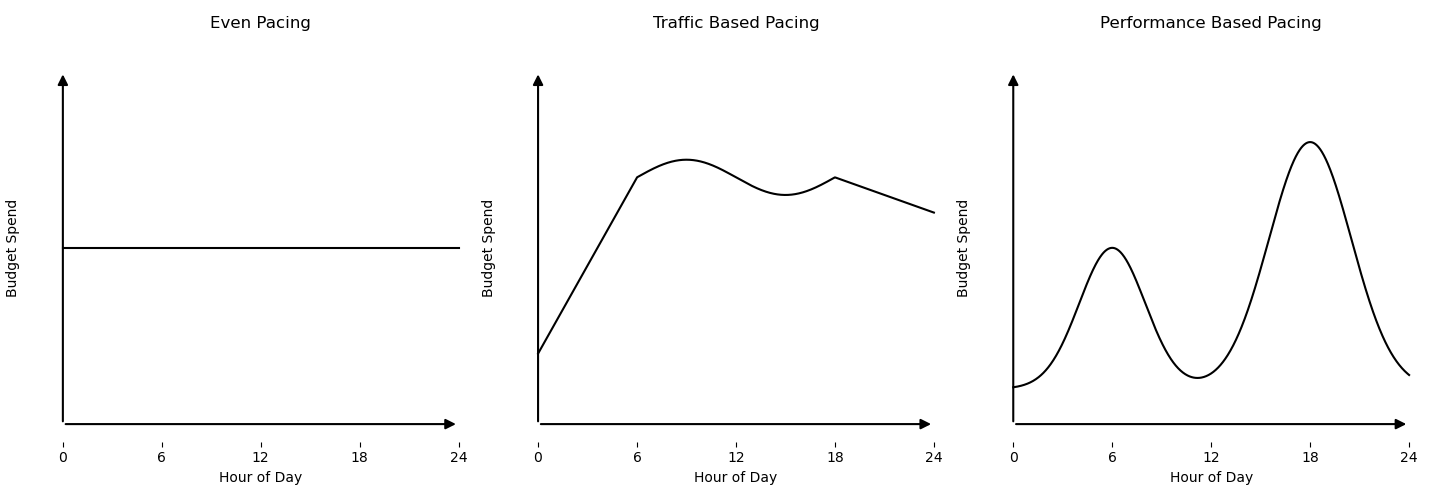}
		\caption{Different Budget Pacing Patterns}
		\label{fig:budget_pacing_pattern}
	\end{figure}

	We compare three different budget allocation strategies:  
	\begin{itemize}
		\item \textbf{Even Pacing}: This strategy is similar to the one discussed in this chapter, where the budget is evenly distributed throughout the day. As shown in the leftmost plot of \autoref{fig:budget_pacing_pattern}, the budget spending targets remain constant within each hour.
		
		\item \textbf{Traffic-Based Pacing}: In this strategy, the budget is allocated based on traffic volume. This approach has been discussed throughout the book and is optimal under the assumption that conversion rates and supporting prices follow some i.i.d. distributions. The middle plot in \autoref{fig:budget_pacing_pattern} illustrates this pattern. Since traffic volume is relatively low at night and higher during the day, the budget allocation forms a bell-shaped curve, peaking around noon.
		
		\item \textbf{Performance-Based Pacing}: Unlike the previous strategy, this approach does not assume an i.i.d. distribution. Instead, it considers variations in online traffic quality throughout the day when allocating the budget. As shown in the rightmost plot of \autoref{fig:budget_pacing_pattern}, the allocation curve now has two peaks—one around 6 AM and another around 6 PM—corresponding to periods when traffic exhibits higher CTRs compared to other times of the day.
	\end{itemize}
	
	A detailed discussion of these budget allocation patterns, along with additional strategies, can be found in \cite{lee2013real}. Readers interested in this topic may refer to this source for further technical details.


	\printbibliography

\end{document}